\documentclass[10pt]{iopart}
\usepackage{graphicx}
\usepackage{cite}
\usepackage[T1]{fontenc}
\usepackage{lmodern}

\usepackage{xspace}
\usepackage[bookmarksnumbered,bookmarksopen,bookmarksopenlevel=1,
colorlinks=true,pdfborder={0 0 0},plainpages=false,pdfpagelabels,
citecolor=blue]{hyperref}
\usepackage{subfigure}
\usepackage{multirow}
\usepackage{tikz}
\usepackage{siunitx}
\usepackage{wasysym}

\newcommand*{\antibar}[1]{\ensuremath{#1\bar{#1}}\xspace}
\newcommand*{\bfig}{\begin{figure*}[htbp]\centering}
\newcommand*{\efig}[2]{\caption{#2}\label{fig:#1}\end{figure*}}

\newcommand*{\pcite}[1]{\protect{\cite{#1}}}

\newcommand*{\ttbar}{\antibar{t}}
\newcommand*{\bbbar}{\antibar{b}}
\newcommand*{\ppbar}{\antibar{p}}
\newcommand*{\qqbar}{\antibar{q}}
\newcommand*{\Wboson}{\ensuremath{W} boson}
\newcommand*{\Wjets}{\ensuremath{W}+jets}
\newcommand*{\Zboson}{\ensuremath{Z} boson}
\newcommand*{\Zjets}{\ensuremath{Z}+jets}
\newcommand*{\topquark}{top quark}
\newcommand*{\ljets}{single-lepton channel}
\newcommand*{\emu}{\ensuremath{e\mu}\xspace}
\newcommand*{\ttH}{\ensuremath{\ttbar H}\xspace}
\newcommand*{\ttV}{\ensuremath{\ttbar V}\xspace}
\newcommand*{\ttW}{\ensuremath{\ttbar W}\xspace}
\newcommand*{\ttZ}{\ensuremath{\ttbar Z}\xspace}
\newcommand*{\ttga}{\ensuremath{\ttbar \gamma}\xspace}
\newcommand*{\ttj}{\ensuremath{\ttbar}+jets}
\newcommand*{\tauhad}{\ensuremath{\tau_{\mathrm{had}}}\xspace}
\newcommand*{\taulep}{\ensuremath{\tau_{\ell}}\xspace}
\newcommand*{\MSbar}{\ensuremath{\overline{\textrm{MS}}}\xspace}
\newcommand*{\TeV}{\tera\electronvolt}
\newcommand*{\GeV}{\giga\electronvolt}
\newcommand*{\MeV}{\mega\electronvolt}
\newcommand{\ifb}[1]{\SI{#1}{\per\femto\barn}}
\newcommand{\ipb}[1]{\SI{#1}{\per\pico\barn}}
\newcommand*{\Vtb}{\ensuremath{\vert V_{tb} \vert}\xspace}
\newcommand*{\Vtd}{\ensuremath{\vert V_{td} \vert}\xspace}
\newcommand*{\Vts}{\ensuremath{\vert V_{ts} \vert}\xspace}

\newcommand*{\pT}{\ensuremath{p_{\mathrm{T}}}\xspace}
\newcommand*{\ET}{\ensuremath{E_{\mathrm{T}}}\xspace}
\newcommand*{\HT}{\ensuremath{H_{\mathrm{T}}}\xspace}

\newcommand*{\Rt}{\ensuremath{R_t}\xspace}

\renewcommand*{\MET}{\ensuremath{E_{\mathrm{T}}^{\mathrm{miss}}}\xspace}
\newcommand*{\met}{\MET}
\newcommand*{\ETmiss}{\MET}
\newcommand*{\Atop}{\ensuremath{A_{\mathrm{C}}^{\ttbar}}\xspace}
\newcommand*{\Alep}{\ensuremath{A_{\mathrm{C}}^{\ell\ell}}\xspace}
\newcommand*{\AFB}{\ensuremath{A_{\mathrm{FB}}}\xspace}
\newcommand*{\RHF}{\ensuremath{R_{\mathrm{HF}}}\xspace}
\renewcommand*{\to}{\ensuremath{\rightarrow}\xspace}

\newcommand{\subfig}[3]{\raisebox{#3pt}{\subfigure{\includegraphics[width=#2\textwidth]{#1}}}}
\newcommand{\aerr}[3]{\ensuremath{#1\:^{+\,#2}_{-\,#3}}}
\newcommand*{\signif}[1]{\ensuremath{#1\,\sigma}}
\newcommand*{\xs}{cross-section}
\newcommand*{\topmass}{top-quark mass}
\newcommand*{\RunOne}{\mbox{Run-1}}
\newcommand{\sglept}{single lept.}
\newcommand*{\AcerMC}{\textsc{AcerMC}}
\newcommand*{\Herwig}{\textsc{Herwig}}
\newcommand*{\Pythia}{\textsc{Pythia}}
\newcommand*{\MadGraph}{\textsc{MadGraph}}
\newcommand*{\Alpgen}{\textsc{Alpgen}}
\newcommand*{\Powheg}{\textsc{Powheg}}
\newcommand*{\MCatNLO}{\textsc{MC@NLO}}
\newcommand*{\TopReX}{\textsc{TopReX}}
\newcommand*{\Protos}{\textsc{Protos}}
\newcommand*{\MCFM}{\textsc{MCFM}}
\newcommand*{\thetastar}{\ensuremath{\theta^\star}\xspace}

\begin{document}
\topical[Top-quark physics at the LHC]{Top-quark physics at the Large 
Hadron Collider}
\author{Markus Cristinziani} 
\address{Physikalisches Institut, Universit{\"a}t Bonn, Germany}
\ead{Markus.Cristinziani@cern.ch, cristinz@uni-bonn.de}
\author{Martijn Mulders} 
\address{CERN, Geneva, Switzerland}
\ead{Martijn.Mulders@cern.ch}

\begin{abstract}
This experimental review gives an overview of top-quark measurements performed 
by the two general purpose-detectors ATLAS and CMS during the first few years 
of running of the Large Hadron Collider. In the years $2010-2012$ each experiment
collected \ifb{5} of $pp$ collision data at $\sqrt{s} = \SI{7}{\TeV}$ and \ifb{20} 
at $\sqrt{s} = \SI{8}{\TeV}$, allowing detailed studies of top-quark production 
and decays, and measurements of the properties of the top quark with unprecedented 
precision. 
\end{abstract}
{\vspace{28pt plus 10pt minus 18pt}\noindent{\small\rm Accepted for publication by: {\it \jpg}\par}

\section{Introduction}

The top quark is the heaviest known elementary particle, and with its unique
properties has long been suspected of potentially carrying key information that
may lead to the solution of some of the paramount open questions in particle
physics. 

After the first observation of top quarks~\cite{Abe:1995hr,Abachi:1995iq} by the
CDF and D0 collaborations at the Tevatron proton--antiproton collider at
Fermilab in 1995, its properties were studied with up to \ifb{10} of data at
centre-of-mass energies of $1.8$ and \SI{1.96}{\TeV}. A comprehensive set of
measurements confirmed that the new particle behaved according to the Standard
Model (SM) predictions for the top quark, albeit with a precision that in most
cases was limited by the available sample sizes and important backgrounds from
$W$+jets and multijet events.

With the start of the Large Hadron Collider (LHC)~\cite{Evans:2008zzb} at CERN
in 2009, much larger samples of top-quark events have become available, produced
in proton--proton collisions at centre-of-mass energies of $\sqrt{s} = 7$ and
\SI{8}{\TeV} during the years 2009 to 2012, referred to as the LHC \RunOne. This
has allowed measurements of top pair and single-top production in much more
detail than before, and studies of top-quark properties with unprecedented
precision. The increased luminosity and higher $\sqrt{s}$ have also opened up
the possibility to observe much rarer processes such as associated production of
top quarks and vector bosons.

However, the enigmatic top quark remains as puzzling as ever. The discovery of a
Higgs boson during the LHC \RunOne~\cite{Aad:2012tfa,Chatrchyan:2012xdj}
completed the Standard Model, but did not answer pressing questions related to
the top quark.  Why is the top quark so much heavier than the other fermions? It
is $40$ times heavier than the bottom quark, at a mass scale similar to the $W$,
$Z$ and Higgs bosons.  Does the top quark play a special role in the mechanism
of electroweak symmetry breaking? Why is the mass of the Higgs boson so much
lighter than the GUT scale, without a symmetry that can protect the Higgs
potential against destabilizing radiative corrections, mostly due to top quarks?
A new question emerged with the discovery of the Higgs boson: why do the
Higgs-boson mass and the \topmass\ have exactly the values they have, in a small
window allowing the electroweak potential of the Standard Model to be stable up
to high energy scales and predicting a meta-stable state for the Universe? For
more details we refer to several excellent recent reviews about the
phenomenology of top quarks~\cite{Boos:2012hi,delDuca:2015gca}.

The Standard Model does not give answers to any of these questions. But since
the top quark as main destabilizer of the Higgs potential and main offender in
the unexplained mass hierarchy plays a central role in this puzzle, it may well
be that indications of possible answers in the form of new physics phenomena can
be found in the top-quark sector. Thus, it is essential to study top-quark
production, decays, couplings and other properties in detail with the best
possible precision.

Other recent experimental reviews exist that include or focus specifically on
legacy results of the Tevatron
experiments~\cite{Deliot:2014uua,Gerber:2015upa,Boos:2015bta}, or cover a
specific area of top-quark physics at the
LHC~\cite{Cortiana:2015rca,Giammanco:2015bxk,Aguilar-Saavedra:2014kpa,
Cristinziani:2015ksy,Meyer:2015euo}.
Earlier general reviews of experimental LHC top-quark results include
Refs.~\cite{Schilling:2012dx,Chierici:2014eqa,Kroninger:2015oma,Hawkings2015424}.

In this experimental review we summarize top physics measurements performed by
the LHC experiments (ATLAS~\cite{Aad:2008zzm}, CMS~\cite{Chatrchyan:2008aa} and
LHCb~\cite{Alves:2008zz}) during \RunOne, including all results submitted to
journals before June 2016.  At this point more than one hundred LHC \RunOne\ top
physics measurements have been published. A few analyses are still in progress,
including for example some of the measurements of the \topmass\ at \SI{8}{\TeV},
expected to be completed soon.

\section{Top-quark pair production}
\label{sec:ttbar}

\subsection{Inclusive \ttbar \xs s}

The measurement of the inclusive production \xs\ $\sigma$ of a top
quark--antiquark pair (\ttbar) is an important test of the theory of quantum
chromodynamics (QCD).  Calculations of the cross-section are performed at
next-to-next-to-leading order (NNLO) including the resummation of soft gluon
terms (NNLL)~\cite{Beneke:2011mq, Cacciari:2011hy, Baernreuther:2012ws,
Czakon:2012zr, Czakon:2012pz, Czakon:2013goa}, the main uncertainties stemming
from scales, parton distribution functions (PDFs) and the strong coupling
constant $\alpha_S$.

Top quarks decay almost exclusively via the flavour-changing charged current
decay $t \to Wb$. Several \ttbar final states are available for the measurement
of $\sigma_{\ttbar}$, depending on the decay of each of the \Wboson s from the
\topquark s.  It is customary to distinguish three \Wboson\ decay categories:
$W\to \qqbar$ (jets), $W\to \ell \nu_{\ell}$ (leptonic), where $\ell$ stands for
electrons and muons, including those from the leptonic decays of tau leptons,
and $W \to \tau_{\rm had} \nu_{\tau}$ (tau), where $\tau_{\rm had}$ indicates a
hadronically decaying tau lepton. Thus the \ttbar events are classified as
dilepton, single-lepton, all-hadronic and categories with tau leptons.  In
general these are complementary as they are affected by different systematic
uncertainties and possible inconsistencies between the \xs s would point to new
physics contributions.  The most precise measurements of the inclusive \ttbar
\xs\ in each channel by each of the general-purpose LHC experiments are shown in
\Fref{tt_xsec_summary}.

\begin{figure}[htbp]
\includegraphics[width=0.5\textwidth]{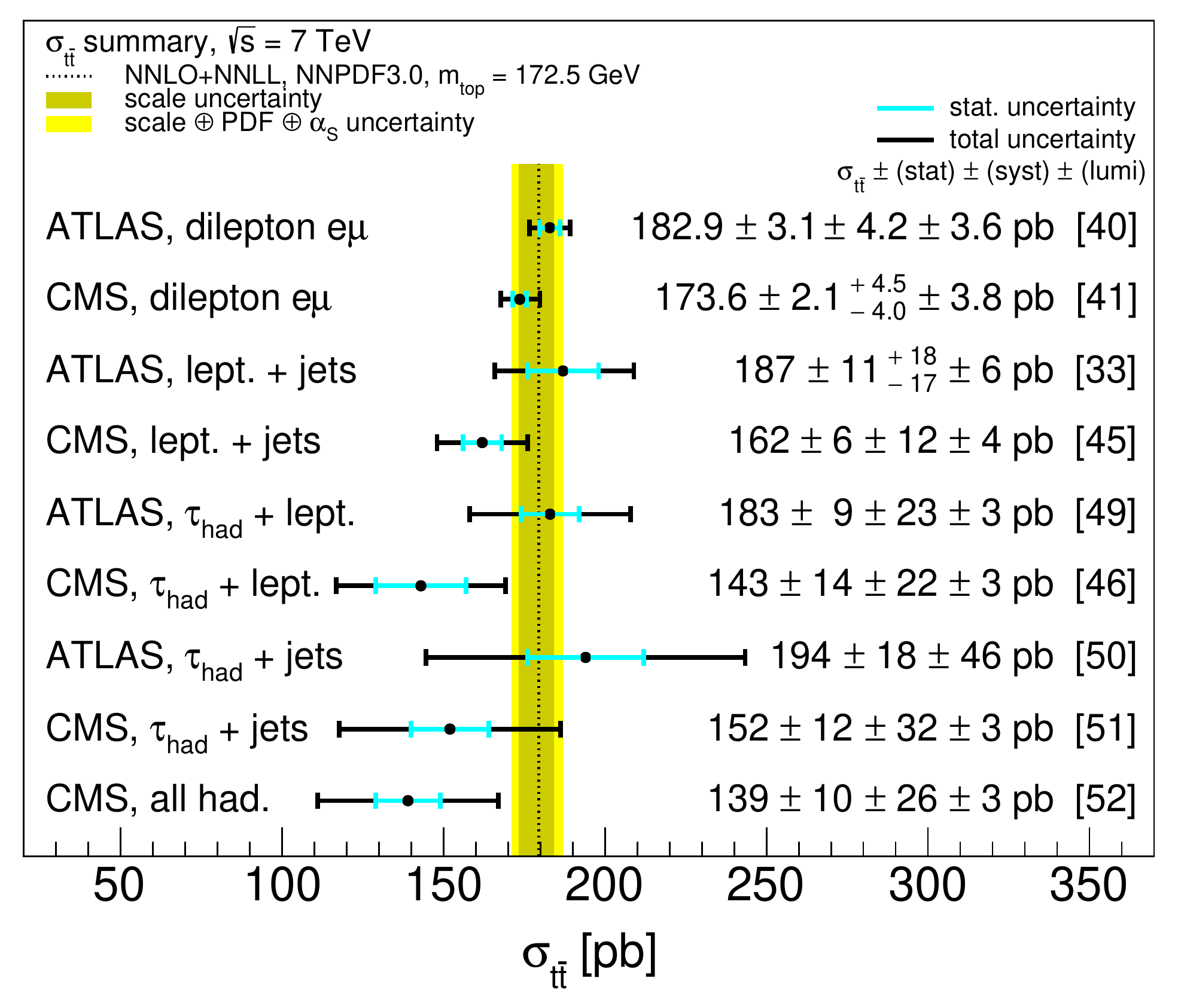}
\includegraphics[width=0.5\textwidth]{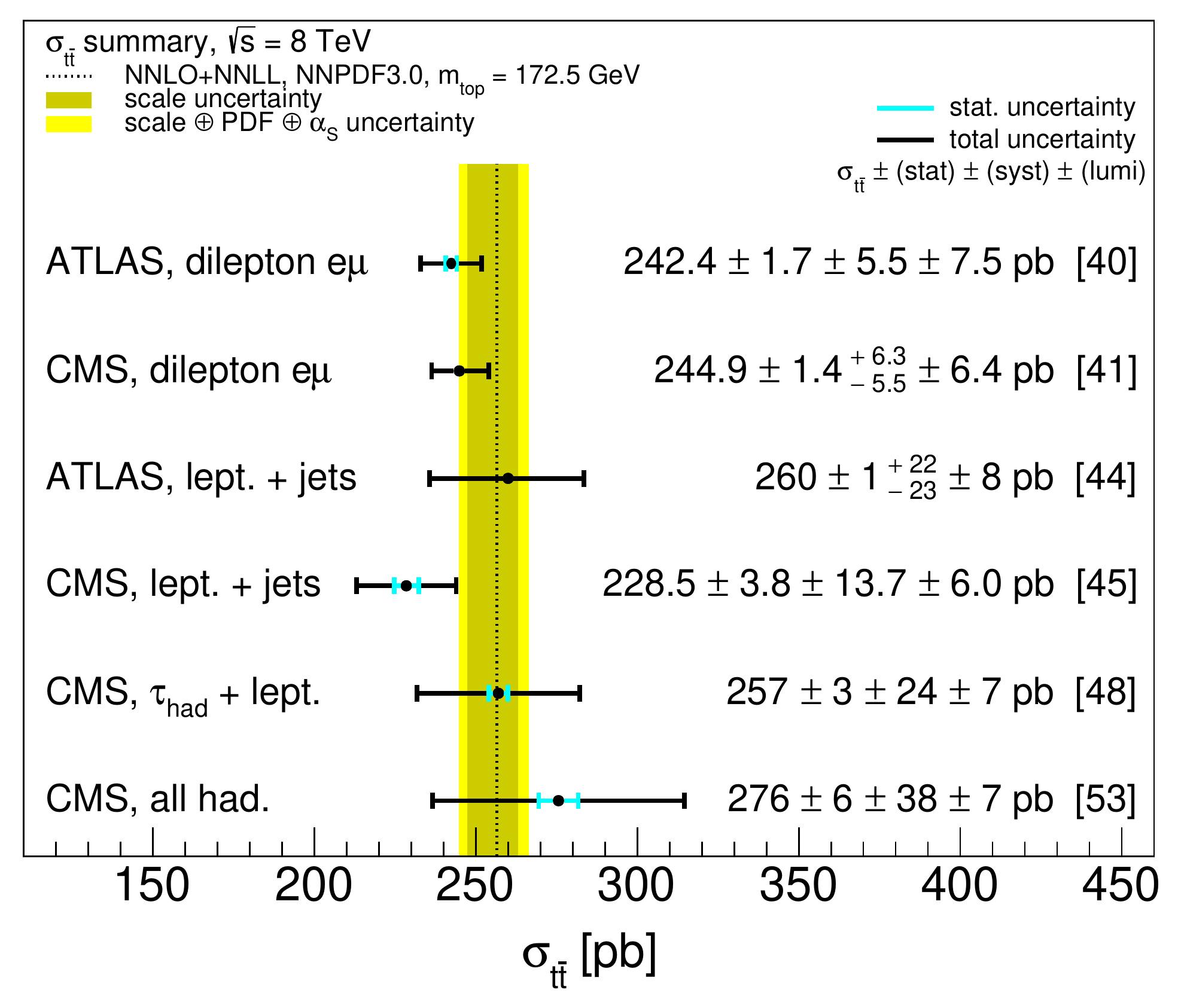}
\caption{Summary of the most precise inclusive \ttbar \xs\ measurements at 
\SI{7}{\TeV} (left) and \SI{8}{\TeV} (right) compared to calculations at 
NNLO+NNLL~\cite{Czakon:2013goa}.
An additional uncertainty of 3 -- \SI{4}{pb} due to the LHC beam energy
uncertainty is not included in the figure.}
\label{tt_xsec_summary}
\end{figure}

First measurements of \ttbar production were reported with $\sim\!\!\ipb{3}$ of
data at \SI{7}{\TeV} by CMS~\cite{Khachatryan:2010ez} and
ATLAS~\cite{Aad:2010ey} and established basic event reconstruction and selection
criteria, as well as data-driven background estimation techniques for the
$W$+jets and $Z$+jets processes and the contribution of events with at least one
{\em fake} (i.e.~non-prompt or misidentified) lepton.  The latter originate from
events where at least one lepton is not produced from a $W$ or $Z$ boson decay
from the hard process, but in secondary processes such as semileptonic $B$
hadron decays, photon conversions, or misidentified jets.  Both the
single-lepton and dilepton channels were used for these first measurements, the
former being the more precise one.  With the complete 2010 dataset (\ipb{35})
the inclusive \ttbar \xs\ was then determined with better precision in the
dilepton~\cite{Aad:2011yb, Chatrchyan:2011nb} and \ljets s~\cite{Aad:2012qf,
Chatrchyan:2011ew, Chatrchyan:2011yy}.  Taking advantage of the larger sample
size the \xs\ extraction techniques were refined, including the use of kinematic
likelihood discriminants and the information of jets identified as coming from
$b$-flavoured hadrons ($b$-tagging). These measurements were further improved
using \ifb{0.7} (ATLAS~\cite{ATLAS:2012aa}) and \ifb{2.3} of \SI{7}{\TeV} $pp$
collisions (CMS~\cite{Chatrchyan:2012bra, Chatrchyan:2012ria}), respectively.
With a partial \SI{8}{\TeV} dataset of \ifb{5.3} CMS~\cite{Chatrchyan:2013faa}
extracted the inclusive \xs\ at \SI{8}{\TeV} employing all three dilepton
channels. 

With more data the dilepton channel, and in particular the \emu final state,
being essentially background free, allowed for the most precise measurements.
In order to reduce the systematic uncertainty, ATLAS extracted simultaneously
the \ttbar \xs\ and the efficiency to reconstruct and tag
$b$-jets~\cite{Aad:2014kva} from the number of opposite charge \emu events with
exactly one and exactly two $b$-tagged jets (\Fref{AEmuJets}, left).
Backgrounds with fake leptons were estimated from the measured number of events
with two same-sign leptons and from simulation, while the $Z \to \tau \tau \to
\emu$ process was estimated from comparing $Z \to \ell \ell$ in data and
simulation.  In a reanalysis of the \RunOne\ dataset,
CMS~\cite{Khachatryan:2016mqs} also analysed the \emu channel but with a
slightly different approach. In this case events with any number of $b$-jets
were retained in the selection (\Fref{AEmuJets}, right), and a template fit to
multi-differential binned distributions related to the number of $b$-jets and
the transverse momentum (\pT) and multiplicity of other jets was performed to
extract the number of signal and background events.  Both ATLAS and CMS obtain
the most precise inclusive \xs\ measurements in the \emu channel by using this
method on the complete $7$ and \SI{8}{\TeV} datasets, reaching a relative
precision of $3-4\%$, similar to the uncertainty of the most accurate
theoretical predictions.  Measurements of the \ttbar \xs\ in the dilepton
channel have also been used to extract the top-quark pole mass (see
\Sref{sec:mass}), to determine $\alpha_S$~\cite{Chatrchyan:2013haa}, and to
search for pairs of supersymmetric top squarks with masses close to the
top-quark mass.  Furthermore, the ATLAS \emu dilepton sample was used for a more
comprehensive test of the Standard Model by measuring simultaneously the
production of \ttbar, $WW$ and $Z \to \tau \tau$~\cite{Aad:2014jra}.
Measurements at \SI{8}{\TeV} are available in the \ljets\ as well, albeit with
larger uncertainties~\cite{Aad:2015pga, Khachatryan:2016yzq}.

\begin{figure}[htbp]
\subfig{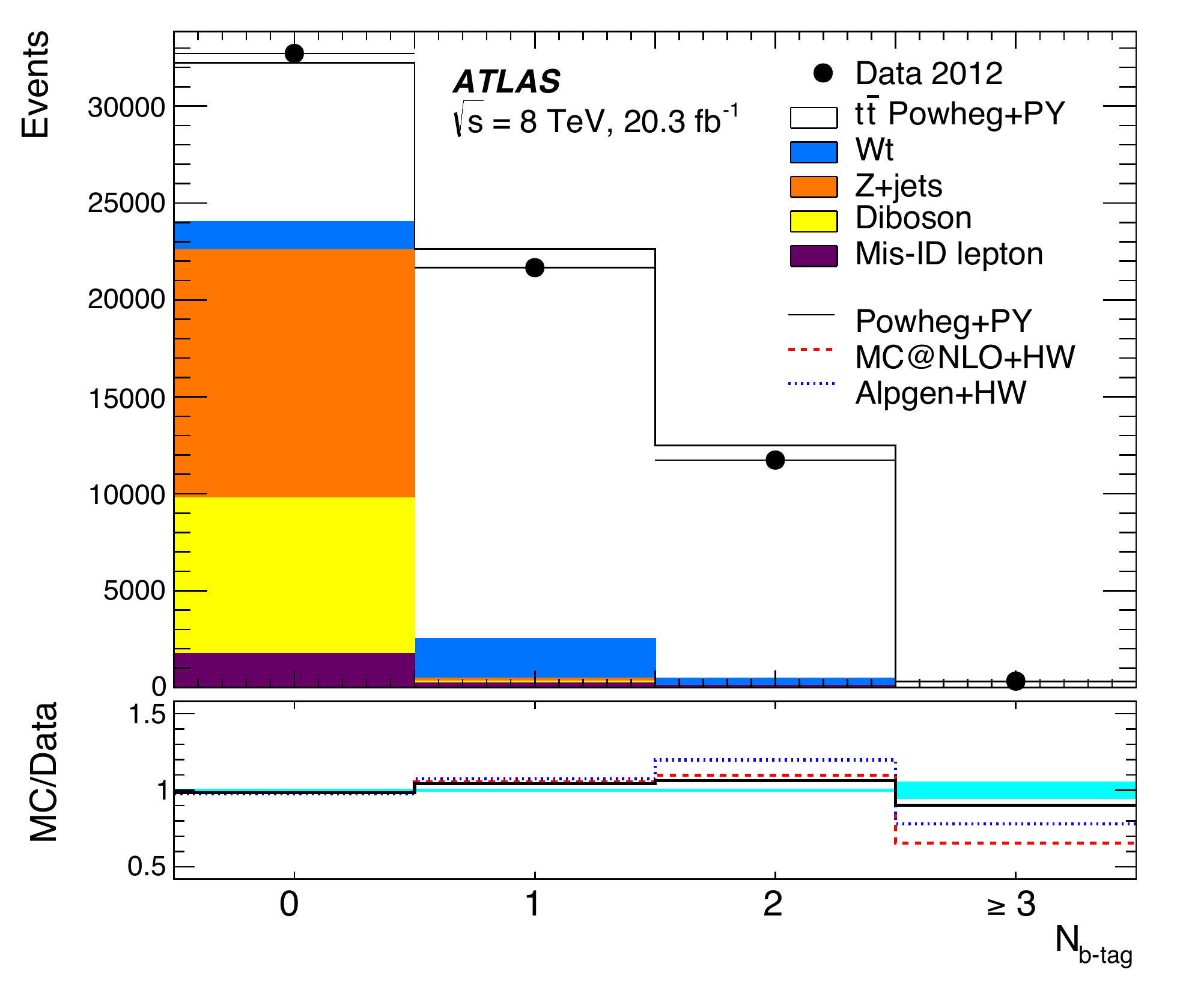}{0.48}{0}
\subfig{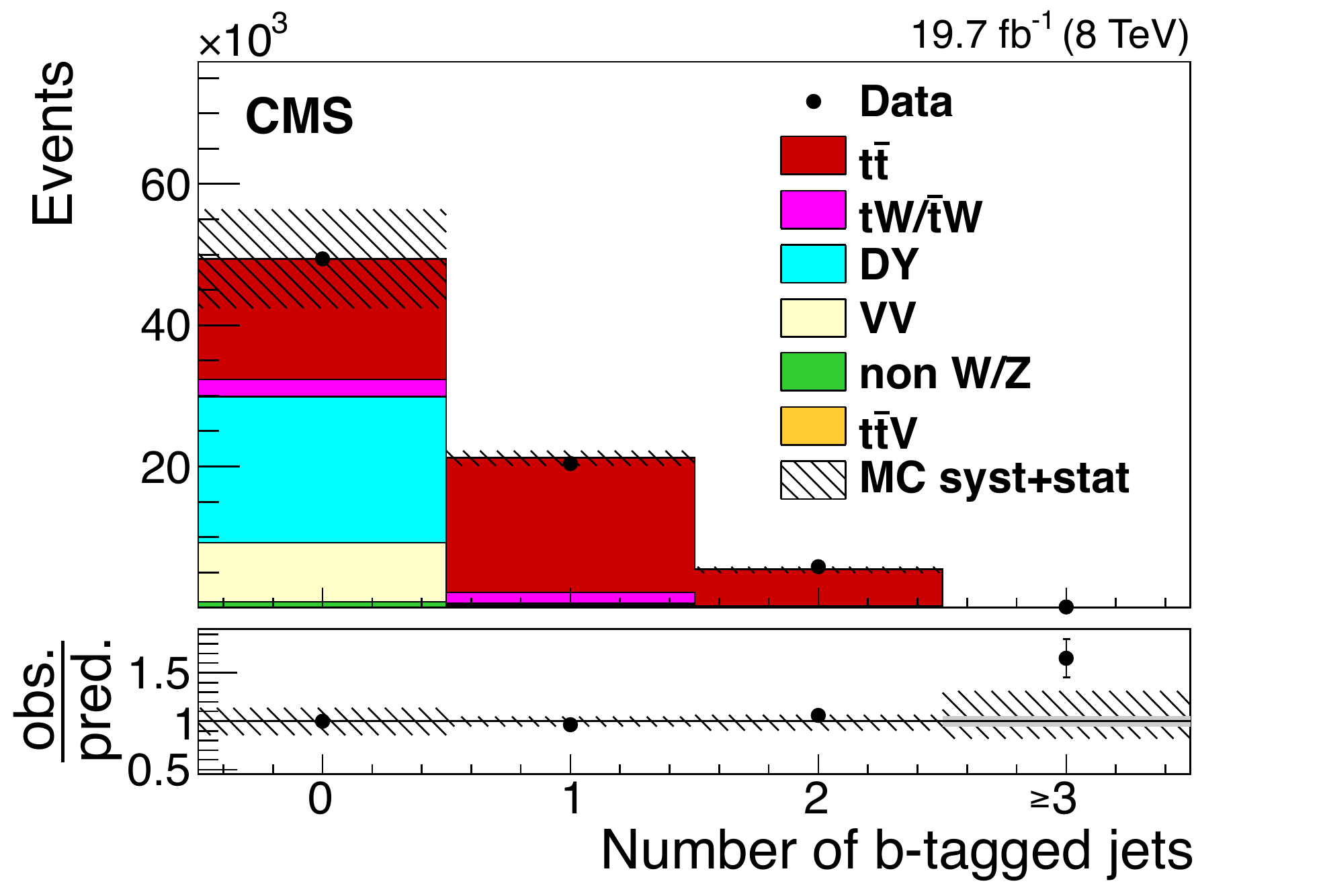}{0.52}{15}
\caption{Number of $b$-tagged jets in a \ttbar \emu selection for the inclusive \xs\
measurement by ATLAS~\pcite{Aad:2014kva} (left) and CMS~\pcite{Khachatryan:2016mqs} 
(right) for the measurement of the inclusive cross-section at \SI{8}{\TeV}.}
\label{AEmuJets}
\end{figure}

The inclusive \xs\ has also been measured by the two experiments in events with
one identified hadronically decaying tau lepton, both in the channels $\tau_{\rm
had}$+$\ell$~\cite{Chatrchyan:2012vs, Aad:2012mza, Khachatryan:2014loa,
Aad:2015dya} and $\tau_{\rm had}$+jets~\cite{Aad:2012vip, Chatrchyan:2013kff}.
Beyond verifying the consistency of the \ttbar \xs\ measurements in different
channels, the particular interest here also comes from the fact that
hypothetical charged Higgs bosons could be produced in top-quark decays, further
decay through $H^+ \to \tau^+ \nu_{\tau}$, and thus modify the apparent
branching fractions. Top-quark branching ratios were determined explicitly in
Ref.~\cite{Aad:2015dya} and found to be in good agreement with SM predictions,
with a precision of $2-8$\% depending on the decay channel.

In the all-hadronic channel there are no high-\pT neutrinos that escape
detection and there is a larger yield of \ttbar events compared to the other
channels. However, since there are at least six jets expected, an overwhelming
background of multijet production and the many possible jet combinations make
the reconstruction challenging. By employing kinematic fits or
neural-network-based selections CMS measured the inclusive \ttbar \xs\ in the
all-hadronic channel at \SI{7}{\TeV}~\cite{Chatrchyan:2013ual} and
\SI{8}{\TeV}~\cite{Khachatryan:2015fwh}.

With the precision measurements of the \ttbar \xs\ at both centre-of-mass
energies of \RunOne, the ratio of \xs\ at 8 and \SI{7}{\TeV} is a powerful test
of perturbative QCD, as the systematic uncertainties partially cancel, while
this ratio can be predicted with uncertainties below 1\%, dominated by PDF
uncertainties.  The experimental determinations have uncertainties of about
5\%, with better precision in the \emu channel~\cite{Aad:2014kva,
Khachatryan:2016mqs} than in the \ljets~\cite{Khachatryan:2016yzq}. 

\subsection{Production of \ttbar with additional jets}

At LHC \RunOne\ energies about half of the \ttbar events are expected to be
produced in association with one or more extra jets with $\pT > \SI{30}{\GeV}$.
As these jets are due to initial and final state QCD radiation, the measurement
of \ttj\ events allows higher-order QCD calculations to be tested. Such
calculations are available for up to two extra jets at
NLO~\cite{Dittmaier:2007wz, Bevilacqua:2011aa}.

The jet activity in \ttbar events can be probed by studying the {\em gap
fraction} variable~\cite{ATLAS:2012al,Chatrchyan:2014gma, Khachatryan:2015mva}
in dilepton events.  Events with a dilepton topology with exactly two $b$-tagged
jets are selected, and the fractions of such events that do not contain an
additional jet within a given central rapidity region, above a certain \pT
threshold ($Q_0$), or for which the scalar sum of the \pT of all additional jets
is not larger than a certain threshold \HT, are extracted.  The data were fully
corrected for detector effects and presented in a fiducial region at {\em
particle} level, defined using the same event selection criteria applied to the
reconstructed data, to avoid extrapolation to regions that are not
experimentally accessible.  The gap fraction as a function of the threshold is
compared with different simulation settings and can help to constrain their
allowed variations, as for instance the amount of initial state radiation, QCD
scales or parton-shower matching parameters (\Fref{jet_veto}). 

\begin{figure}[h]
\subfig{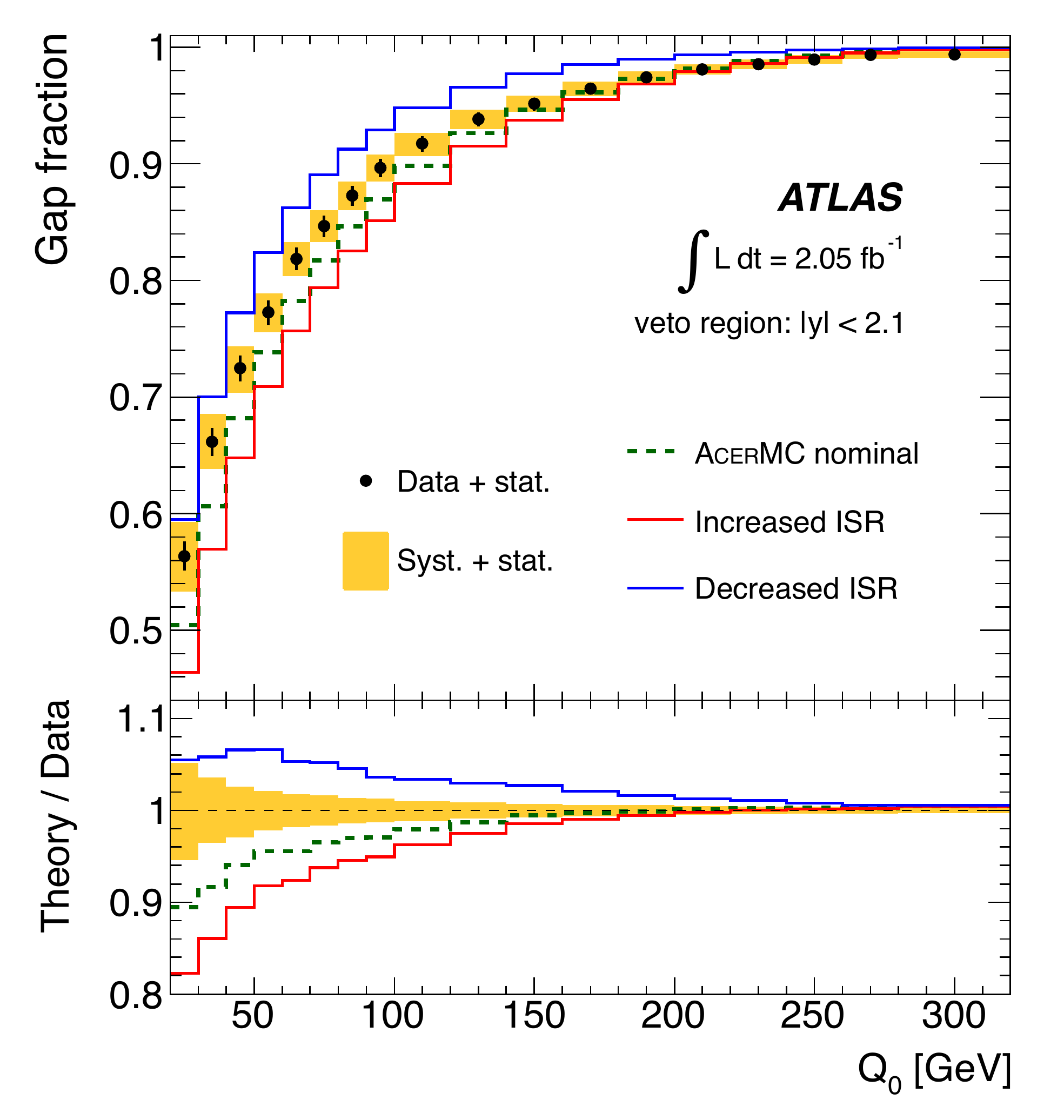}{0.48}{0}
\subfig{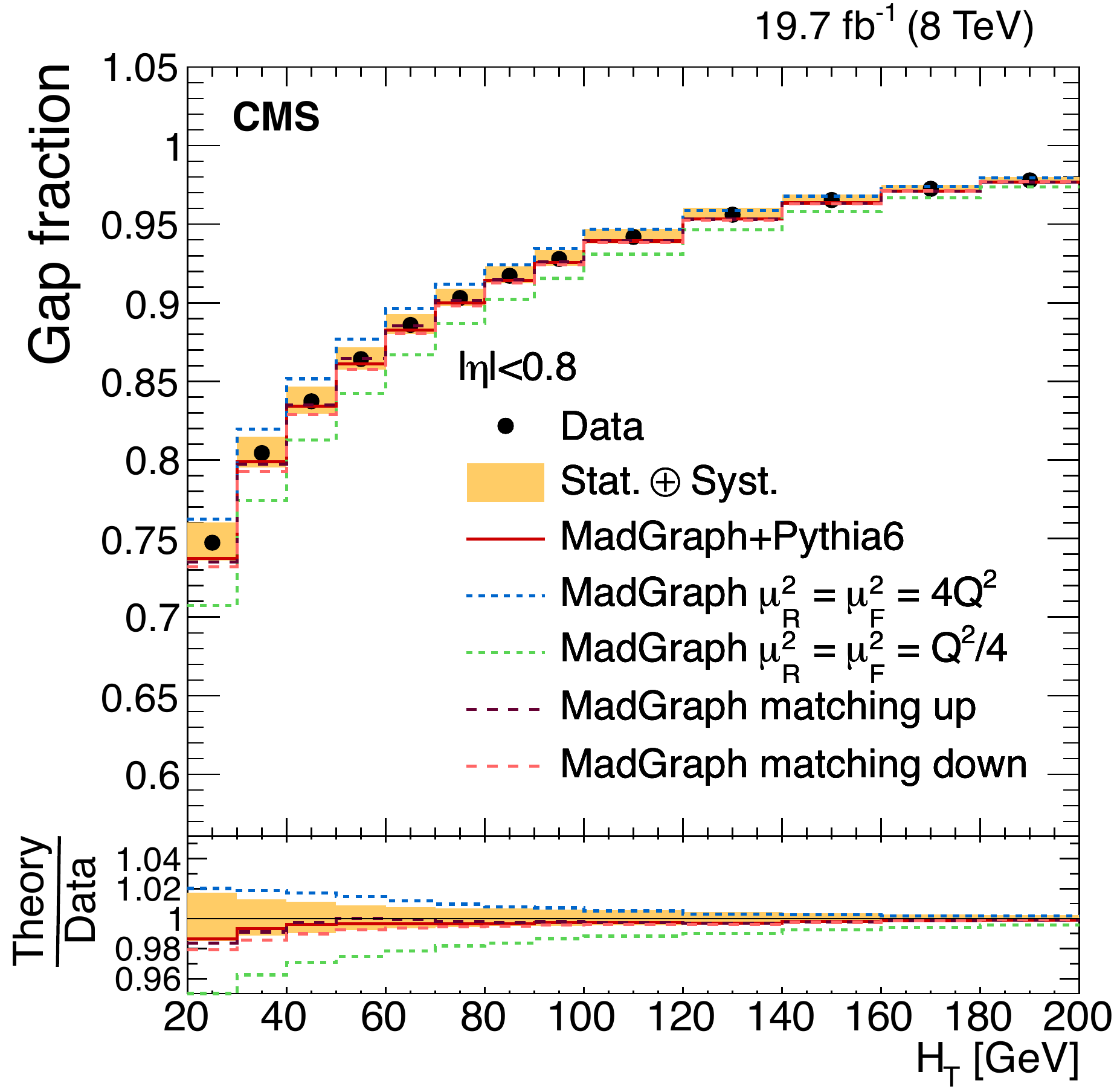}{0.48}{10}
\caption{The measured gap fraction as a function of the threshold $Q_0$ for $|y|
< 2.1$ compared to predictions by the \AcerMC~\cite{acermc} generator, where
different settings of the \Pythia~\cite{pythia6} parton-shower parameters were
used to produce samples with nominal, increased and decreased initial state
radiation (ISR)~\pcite{ATLAS:2012al} (left). Data are compared to predictions by
\MadGraph~\cite{madgraph} in the central region $|\eta|<0.8$ with modified
renormalization and factorization scales, and jet-parton matching thresholds as
a function of the threshold $H_T$~\pcite{Khachatryan:2015mva} (right).}
\label{jet_veto}
\end{figure}

Measurements of the number of reconstructed extra jets in \ttbar events have
been performed both in the dilepton and \ljets\ and compared to
expectation~\cite{Chatrchyan:2014gma, Aad:2014iaa}.  They provide important
tests of higher-order QCD effects, and serve as inputs for the validation and
development of generators and NLO QCD calculations of \ttj\ matched to
parton-shower algorithms.  In order to compare with predictions from different
generators, parton-shower models and parameter settings, the spectra were
extracted at particle level, by correcting the data for detector efficiency and
resolution effects. By defining the fiducial volume as close as possible to the
reconstructed objects these corrections are kept small, reducing the associated
uncertainties.  As a result of these studies ATLAS and CMS disfavour the
\MCatNLO~\cite{mcatnlo}+\Herwig~\cite{herwig} model with the tuning that was
used, as it predicts too few events with many high-\pT jets, while the allowed
parameter space for \Alpgen~\cite{alpgen}, \Powheg~\cite{powheg} and \MadGraph\
in combination with \Pythia\ or \Herwig\ can be constrained (\Fref{fig_njets},
left).

\begin{figure}[htbp]
\centering
\subfig{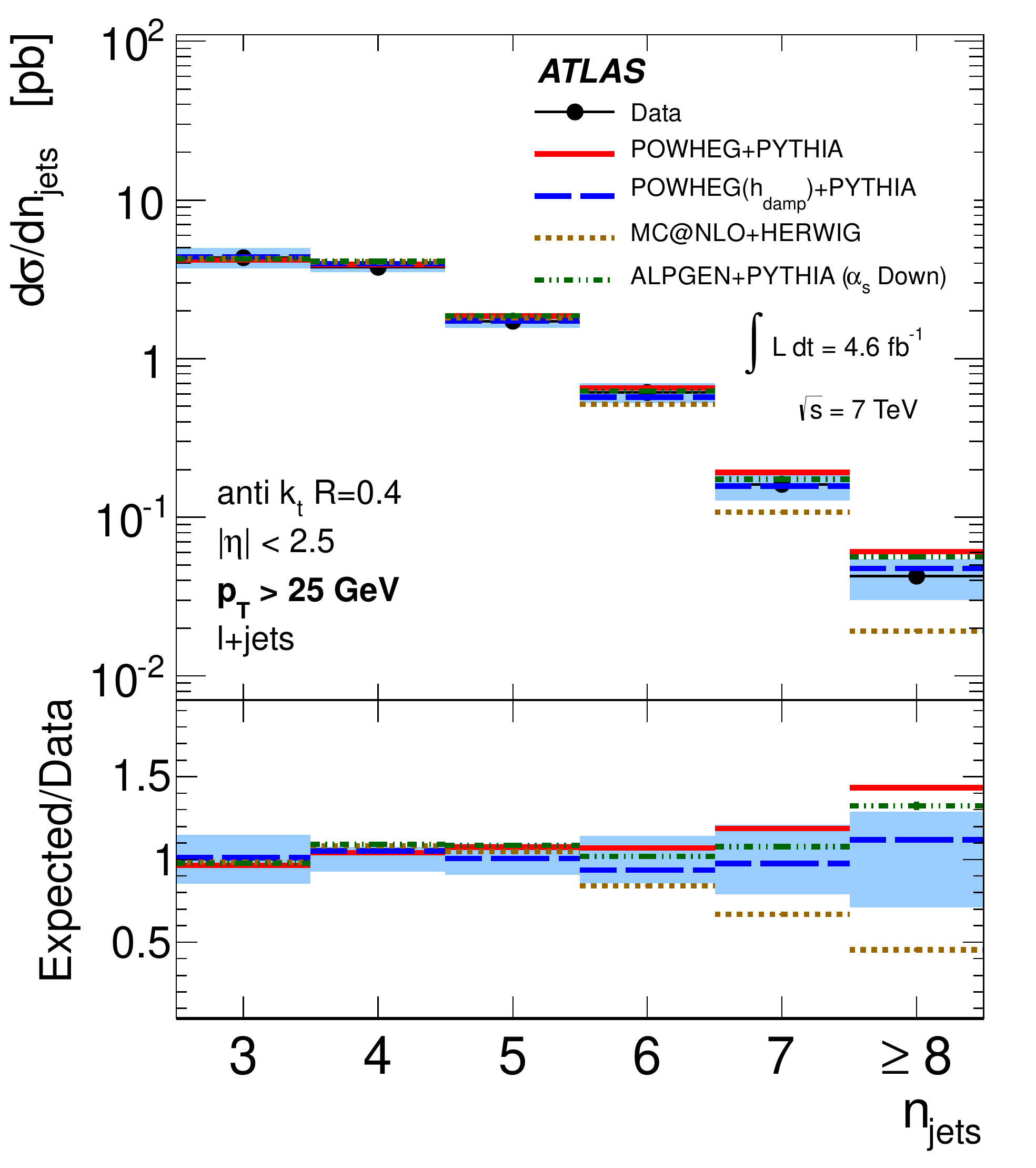}{0.47}{0}
\subfig{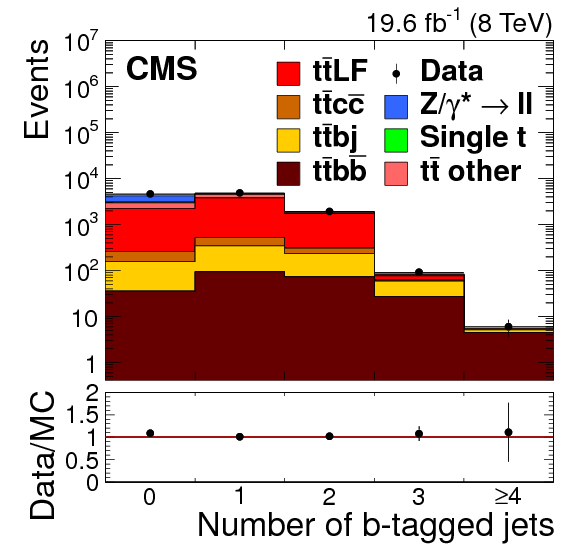}{0.485}{22}
\caption{Number of jets after unfolding compared to several generators and
parton-shower models~\pcite{Aad:2014iaa} (left). Measured number of $b$-tagged
jets compared to expectation, separating the different contributions depending
on the flavour of the additional jets~\pcite{CMS:2014yxa} (right).}
\label{fig_njets}
\end{figure}

The production of additional $b$ quarks in \ttbar events has also been studied.
The $\ttbar \bbbar$ final state is an irreducible non-resonant background to the
\ttH process and is difficult to model in simulation because of ambiguities in
matching matrix-element calculations to the parton shower.  From a fit to the
vertex mass distribution of the extra $b$-tagged jets beyond the two associated
to the \ttbar process in dilepton events, evidence of production of \ttbar in
association with $c$- or $b$-jets has been reported by ATLAS at
\SI{7}{\TeV}~\cite{Aad:2013tua}.  With the \SI{8}{\TeV} dataset CMS measured the
total \xs\ $\sigma_{\ttbar \bbbar}$ and the quantity $\RHF \equiv \sigma_{\ttbar
\bbbar}/\sigma_{\ttbar\!j\!j}$ in the dilepton channel~\cite{CMS:2014yxa} in the
full phase space at {\em parton} level, in order to compare with theoretical
calculations.  Events with two well-identified $b$-jets and at least two
additional jets were required. The $b$-tagging algorithm discriminator of the
third and fourth jet were used to separate $\ttbar \bbbar$ events from the
background, including $\ttbar\!j\!j$, with a template fit (\Fref{fig_njets},
right).

ATLAS performed four measurements of heavy-flavour production in top-quark pair
events~\cite{Aad:2015yja} in a fiducial volume at \SI{8}{\TeV}: a fit-based and
a cut-based measurement of $\sigma_{\ttbar \bbbar}$ in the dilepton channel, and
cut-based measurements of $\sigma_{\ttbar b}$ in the dilepton and \ljets s.  The
fit-based analysis used very tight selection criteria, including the requirement
of four $b$-tagged jets, relied on simulation for the background determination
and featured a high signal-to-background ratio.  A looser selection was applied
in the second analysis where the signal was extracted from a fit to the
multivariate $b$-jet identification discriminant.  The ratio \RHF was determined
at particle level to be $0.013 \pm 0.004$ with comparable systematic and
statistical uncertainties.  More events are produced in the \ljets; however this
channel is affected by additional backgrounds, where a $W$ boson can produce a
$c$ quark, resulting in a similar sensitivity to the dilepton channel. The
measurements were also presented after subtracting the expected contributions
from electroweak processes (\ttW, \ttZ and \ttH) in order to allow for
comparison with NLO QCD predictions.  Measurements were compared to predictions
using different $g\to\bbbar$ splitting rates in the parton shower: one of the
two most extreme \Pythia\ 8~\cite{pythia8} models is disfavoured by the
measurements (\Fref{ttbb}, left). Measurements were also compared to different
LO multileg and NLO generators, showing that the production of extra
heavy-flavoured jets is underestimated by up to a factor of two.

Differential \xs s in the fiducial phase space of the $\ttbar$ system and the
additional $b$-jets as a function of the \pT and $|\eta|$ of the leading and
subleading additional $b$-jets, the angular distance $\Delta R_{bb}$ between
them and the invariant mass $m_{bb}$ of these two $b$-jets have been measured at
\SI{8}{\TeV} in the dilepton channel by CMS~\cite{Khachatryan:2015mva}.  The
overall relative normalization of processes with a \ttbar and at least one
$b$-jet in the final state with respect to expectations from simulation
(\MadGraph+\Pythia) was determined to be $1.66 \pm 0.43$ using a template fit,
identifying the $b$-jets not associated to the \ttbar decay with a boosted
decision tree.  The distributions considered were well modelled by simulation
(\Fref{ttbb}, right).

\begin{figure}[htbp]
\subfig{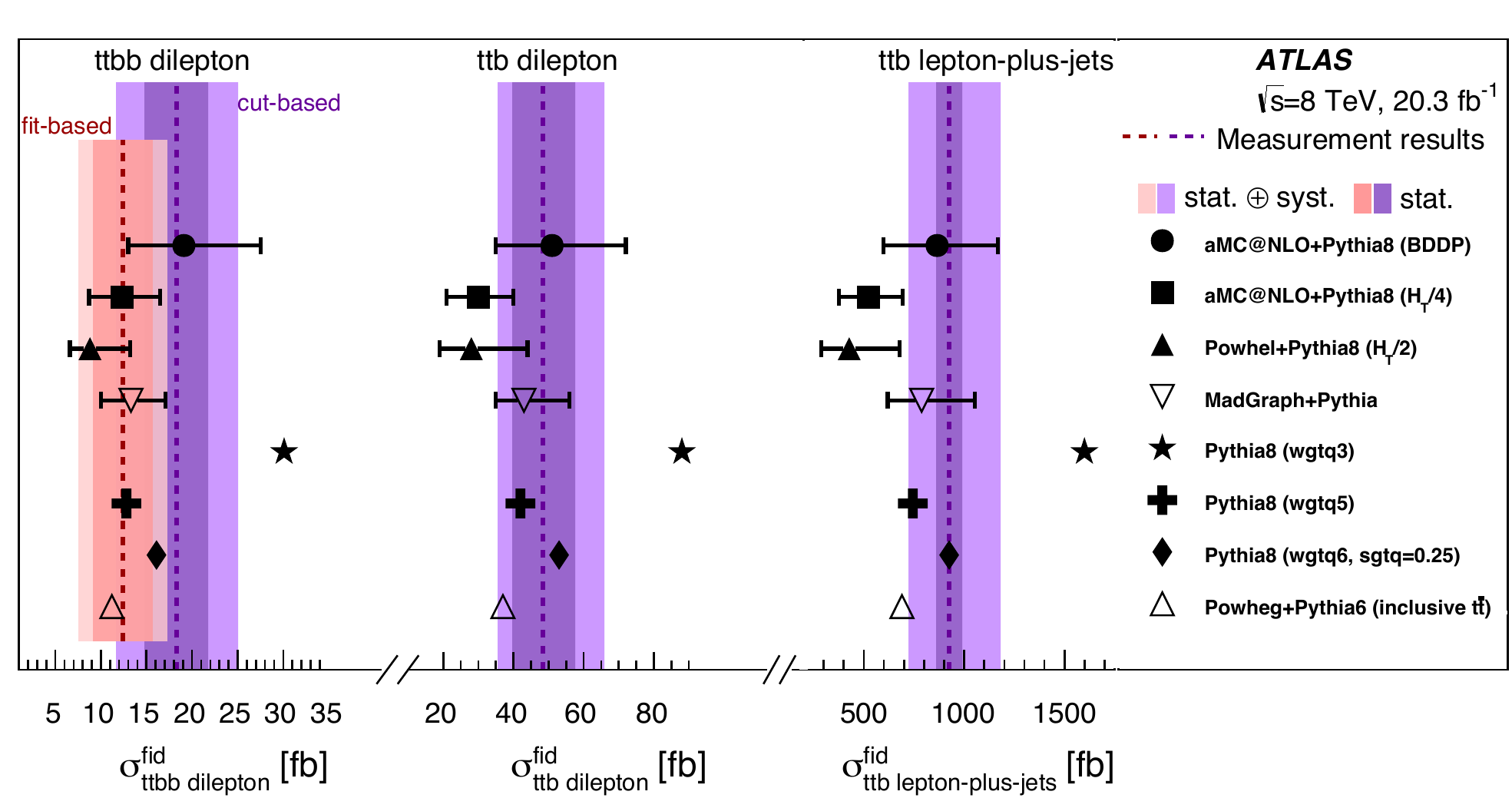}{0.67}{0}
\subfig{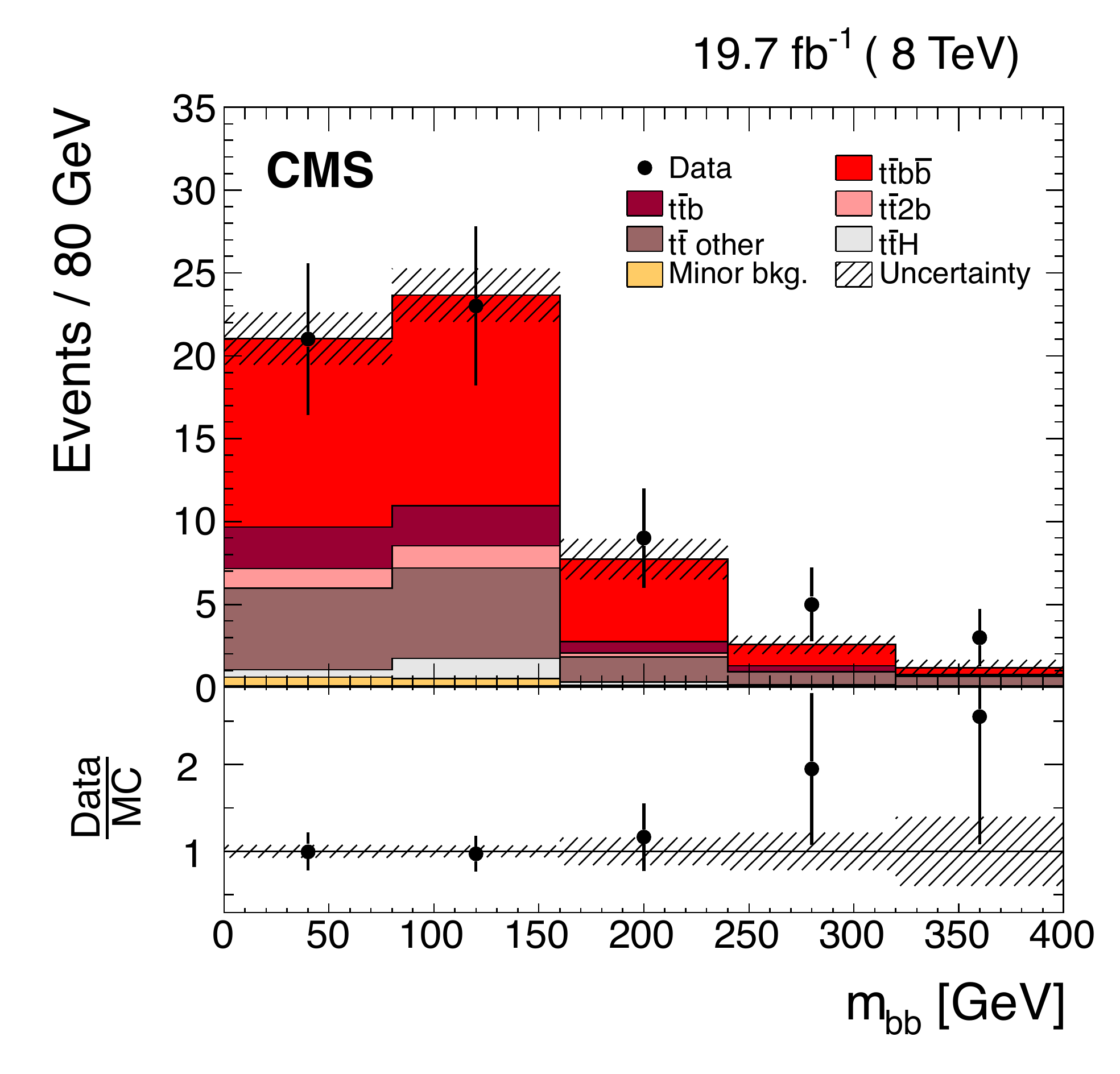}{0.33}{15}
\caption{Comparison of the measured cross-sections in three fiducial phase-space
regions with theoretical predictions obtained from a variety of different
generators. The measurements are shown with the contributions from \ttV and \ttH
removed to allow direct comparison to the predictions containing only the pure
QCD matrix elements~\pcite{Aad:2015yja} (left).  Distribution of the invariant
mass of the two leading additional $b$-jets in $\ttbar\bbbar$ events from data
(points) and from signal and background simulation
(histograms)~\pcite{Khachatryan:2015mva} (right).} \label{ttbb}
\end{figure}

\newpage

Within the Standard Model, $\ttbar\ttbar$ production is very small, with a \xs\
of $\sim\!\!\SI{1}{fb}$~\cite{Barger:1991vn, Barger:2010uw, Bevilacqua:2012em}
at \RunOne\ energies, but could be significantly enhanced in extended models.
By selecting events with pairs of leptons with the same electric charge using
\SI{8}{\TeV} data CMS and ATLAS placed upper limits on the Standard Model
$\ttbar\ttbar$ production \xs\ at the 95\% confidence level of \SI{49}{fb}
\cite{Chatrchyan:2013fea} and \SI{70}{fb} \cite{Aad:2015gdg}, respectively. In
the \ljets\ the very rare signal is difficult to discern from the overwhelming
\ttbar production and therefore multivariate techniques have been employed.
Using the full \SI{8}{\TeV} dataset CMS and ATLAS obtained stronger limits of
\SI{32}{fb} \cite{Khachatryan:2014sca} and \SI{23}{fb} \cite{Aad:2015kqa},
respectively.

Associated production of \ttbar with heavy bosons (\ttW, \ttZ and \ttH) is 
starting to be within reach of the LHC experiments and this is of
particular interest as it allows to probe directly the couplings 
of the top quark to the $Z$ and Higgs bosons. These studies are discussed 
in more detail in \Sref{sec:couplings}.

\subsection{Differential \ttbar \xs s}

With the large top-quark samples available at the LHC after \RunOne, it is
possible to study differential distributions in detail. On one hand this allows
for more thorough tests of perturbative QCD, to constrain the parameters of the
Monte-Carlo simulation programs and the proton PDFs, on the other hand it allows
for a better understanding of a major background in Higgs boson physics, rare
processes and searches for beyond-the-SM (BSM) effects.  The strategy for
differential measurements is to start with a tight event selection to obtain a
high purity \ttbar sample, often enhanced with dedicated event reconstruction
techniques.  After the estimated background is subtracted, the effects of
detector acceptance and resolution are accounted for by means of unfolding.
Distributions of various kinematic quantities of the top quark are then
presented at parton or particle level, and compared to predictions from various
simulation programs or perturbative QCD calculations.

ATLAS measured differential \ttbar distributions with \ifb{2.05} of \SI{7}{\TeV}
data~\cite{Aad:2012hg} using a kinematic likelihood fit to reconstruct the top
quarks in the \ljets.  In this first differential analysis, the bin size for the
unfolding procedure was optimized to reduce the total uncertainties and a simple
matrix inversion was preferred to a regularized unfolding in order to preserve
the sensitivity to potential isolated deviations in any single bin from the
Standard Model expectation. The typical measurement uncertainties ranged between
$10$ and $20\%$, already dominated by systematic effects.  Results were
presented as normalized \ttbar \xs s as a function of kinematic quantities of
the \ttbar pair, namely the invariant mass, $m_{\ttbar}$ (\Fref{diff-first},
left), the transverse momentum, $p_{\text{T},\ttbar}$ and the rapidity
$y_{\ttbar}$, and compared to calculations at NLO+NNLL~\cite{Ahrens:2010zv},
predictions by \MCFM~\cite{mcfm} at NLO, and the generators \MCatNLO\ and
\Alpgen\ with particular choices of parameter settings.

\begin{figure}[htbp]
\subfig{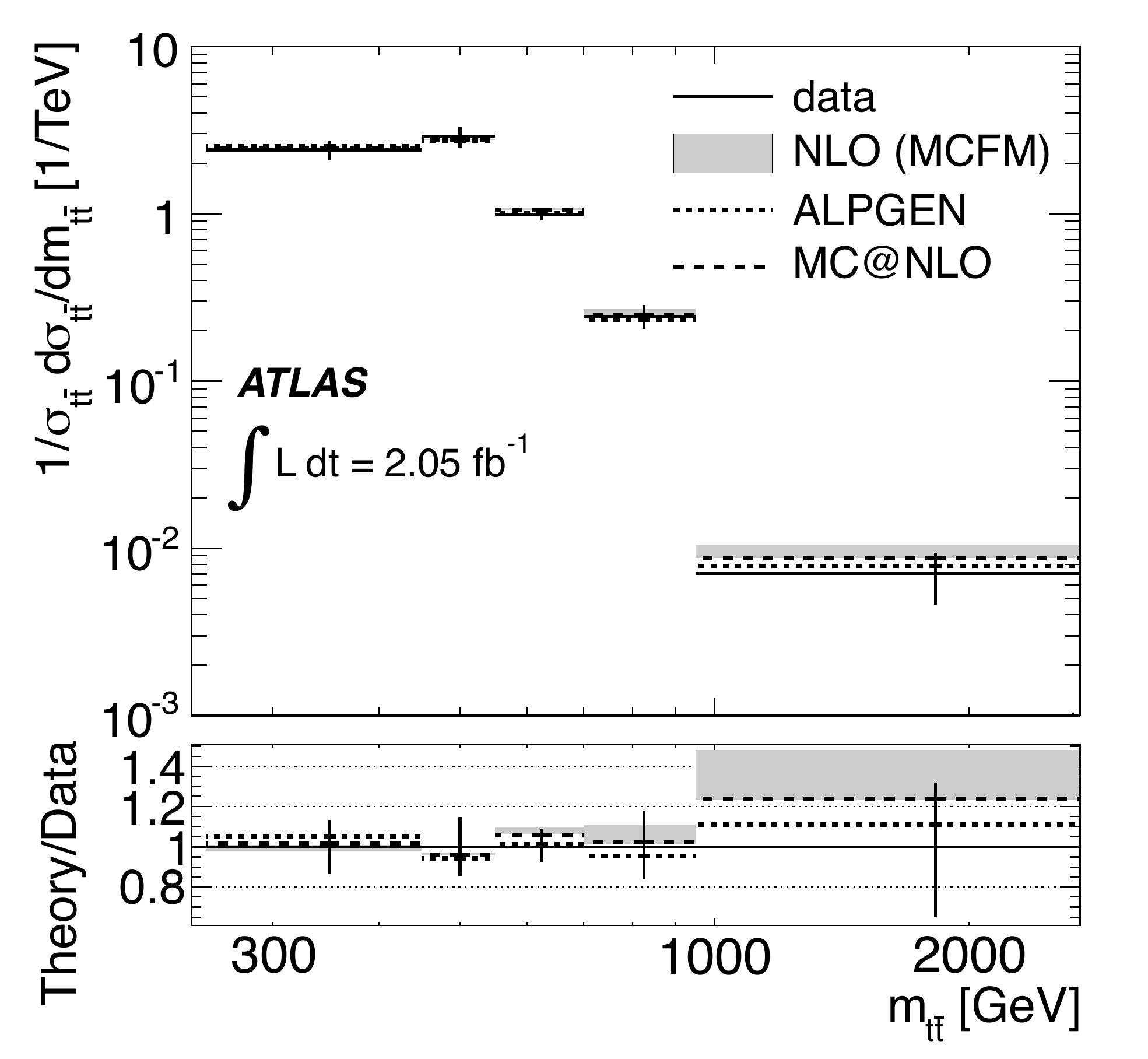}{0.50}{0}
\subfig{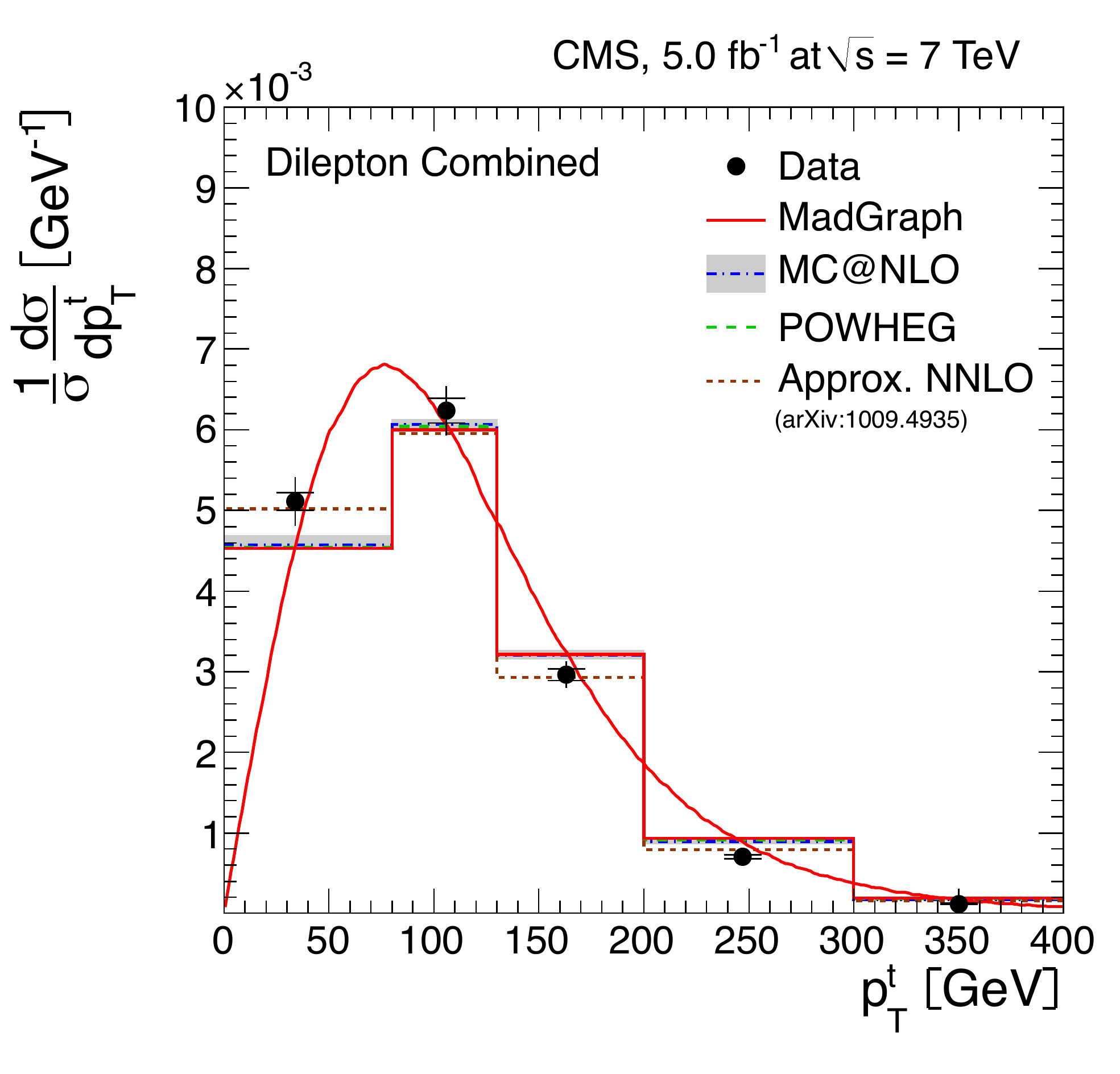}{0.50}{0}
\caption{First normalized differential \ttbar production \xs\ measurements: the
invariant mass of the \ttbar system in the \ljets, measured by
ATLAS~\pcite{Aad:2012hg} (left) and the transverse momentum of the top quark in
dilepton \ttbar events as measured by CMS~\pcite{Chatrchyan:2012saa} (right).}
\label{diff-first}
\end{figure}

Differential measurements using the full \SI{7}{\TeV} dataset were performed by
CMS in the single-lepton and dilepton channels~\cite{Chatrchyan:2012saa} and by
ATLAS in the single-lepton channel~\cite{Aad:2014zka}.  The kinematic properties
of the top quarks were obtained using kinematic likelihood fits (single-lepton
channel) and kinematic reconstruction (dilepton channel) algorithms. The data
was corrected for detector effects and acceptance by using a regularized
unfolding procedure with a bin width that was optimized to keep bin-to-bin
migrations small.  The results for the top-quark and \ttbar distributions were
presented at parton level and extrapolated to the full phase space.  CMS also
presented directly measured observables at particle level (leptons, $b$-jets) in
a fiducial volume as function of the transverse momentum, (pseudo)rapidity, and
invariant mass.

The normalized \ttbar \xs\ distributions were compared to various generators and
parton-shower programs, while the top-quark and \ttbar distributions were also
compared with calculations beyond NLO accuracy.  The measurements among the
different decay channels are in agreement with each other and with Standard
Model predictions in all distributions that were studied, with one exception:
the top-quark transverse momentum (\Fref{diff-first}, right) is poorly described
by any of the MC simulations that were considered.  ATLAS compared various NLO
PDF sets to the extracted \ttbar distributions, indicating that the data can be
used to improve the precision of future PDF fits, with
\textsc{HERAPDF1.5}~\cite{Aaron:2009aa} being able to describe the $m_{\ttbar}$
and $|y_{t}|$ distributions best, as shown in~\Fref{diff-7TeV} (left).

\begin{figure}[htbp]
\subfig{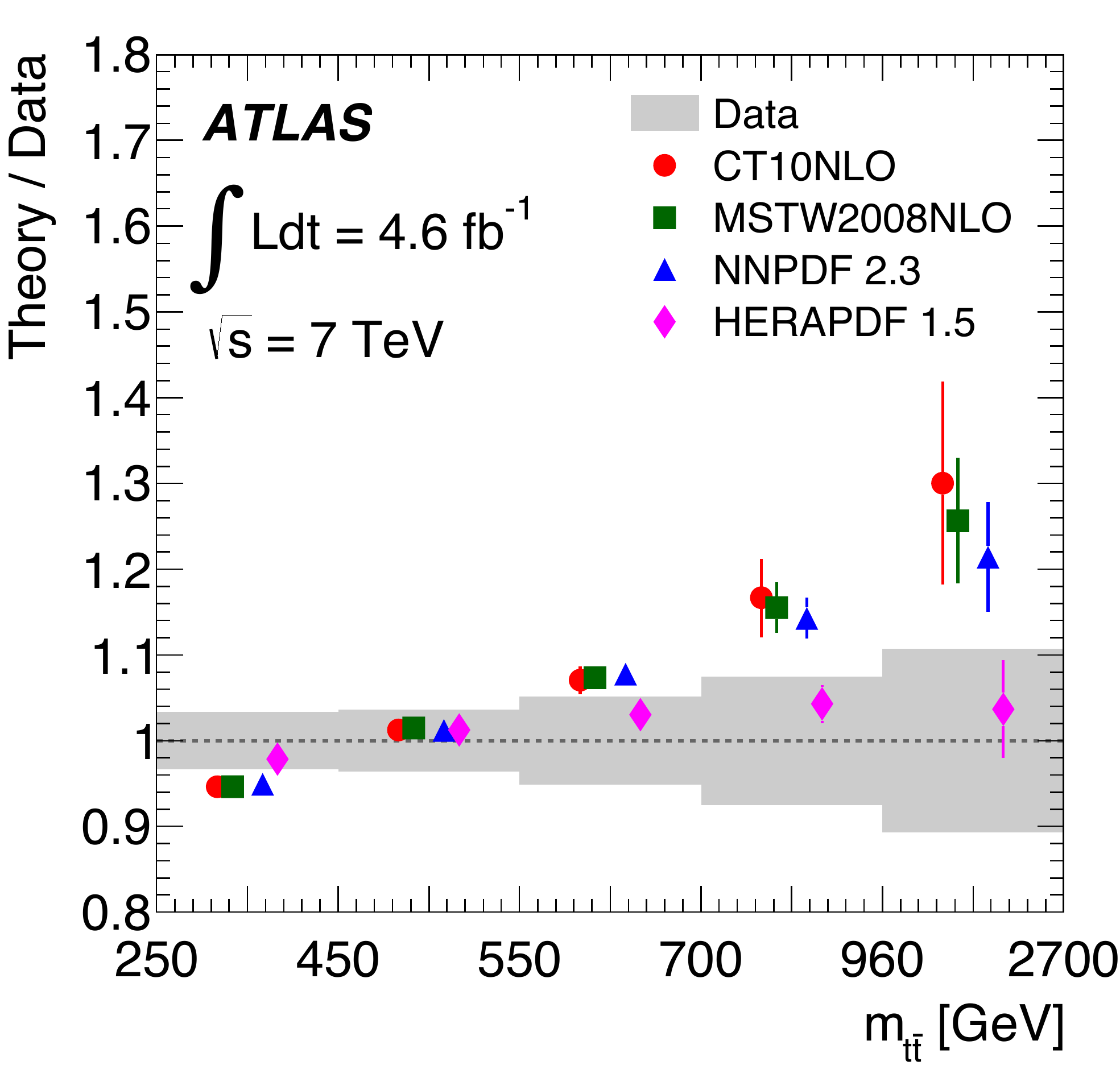}{0.50}{15}
\subfig{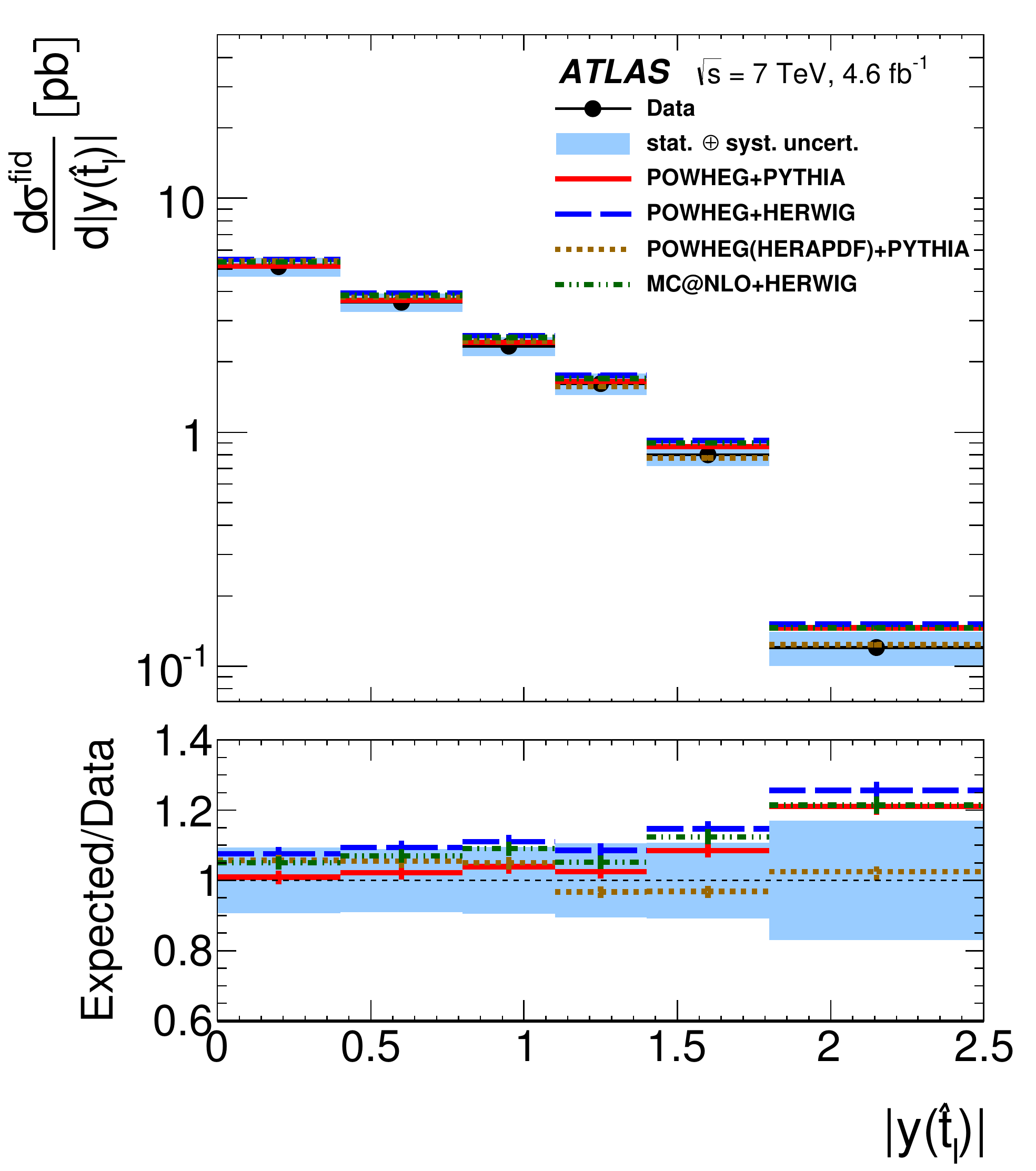}{0.50}{0}
\caption{Ratios of the NLO QCD predictions to the measured normalized
differential \xs\ for different PDF sets as a function of
$m_{\ttbar}$~\pcite{Aad:2014zka} (left).  Differential \ttbar \xs\ as a function
of the leptonic pseudo-top-quark rapidity~\cite{Aad:2015eia} (right).} 
\label{diff-7TeV}
\end{figure}

In a new approach, top-quark proxies ({\em pseudo-top-quarks}) were built from
the reconstructed objects or final state particles~\cite{Aad:2015eia}.  The
definition is chosen such that the experimental observable is strongly
correlated with the top-quark parton and suffers from a smaller model
dependence.  The differential distributions of the kinematic properties of the
pseudo-top-quarks, separately reconstructed in the leptonic and hadronic leg,
were compared with \MCatNLO, \Powheg~(\Fref{diff-7TeV}, right) and \Alpgen,
coupled to different parton-shower choices. The observables investigated with
the \SI{7}{\TeV} dataset showed highest sensitivity to the PDF set and
parton-shower model employed.

The top-quark transverse momentum was consistently found to be softer in data
than in the MC samples based on LO or NLO generators interfaced to parton
showers.  While this was also confirmed later in the \SI{8}{\TeV} dataset, the
agreement between the data and the \Powheg+\Herwig\ simulation for this variable
is substantially better than other simulations. In subsequent CMS analyses a
systematic uncertainty was added corresponding to the difference between the CMS
data and simulation in the top-quark \pT distribution. Since in all ATLAS
top-quark measurements a comparison between \Powheg+\Pythia\ and
\Powheg+\Herwig\ is performed, this discrepancy is considered to be covered by
the modelling systematics.

Most of the analyses were repeated at $\sqrt{s} =  \SI{8}{\TeV}$ with the larger
sample of events available at that energy, allowing almost an order of magnitude
more events to be selected.  The statistical precision together with
improvements in kinematic reconstruction algorithms and extended systematic
studies, lead to a significant reduction of the total uncertainties. CMS
analysed the full \SI{8}{\TeV} dataset in the single-lepton and dilepton
channels~\cite{Khachatryan:2015oqa}, as well as in the all-hadronic
channel~\cite{Khachatryan:2015fwh}.  The strategy largely followed earlier
analyses, with additional extracted distributions: the \pT of the top quark and
anti-quark in the \ttbar rest frame, of the leading and subleading top
(anti-)quark, $\Delta \phi(t, \bar{t})$, as well as the $b$-jet pair system
quantities $m_{\bbbar}$ and $p_{\text{T},\bbbar}$.  \Fref{diff-cms-atlas-8tev}
(top) illustrates that, with the precision of these measurements, differences
can be seen when comparing different models. For example the NLO+NNLL
calculation~\cite{Li:2013mia} does not describe the transverse momentum
distribution of the \ttbar pair. The top-quark transverse momentum compared to
the default MC simulation prediction \MadGraph+\Pythia\ 6 shows the same slope
as was found in \SI{7}{\TeV}, now confirmed in all three decay channels. 

\begin{figure}[htbp]
\subfig{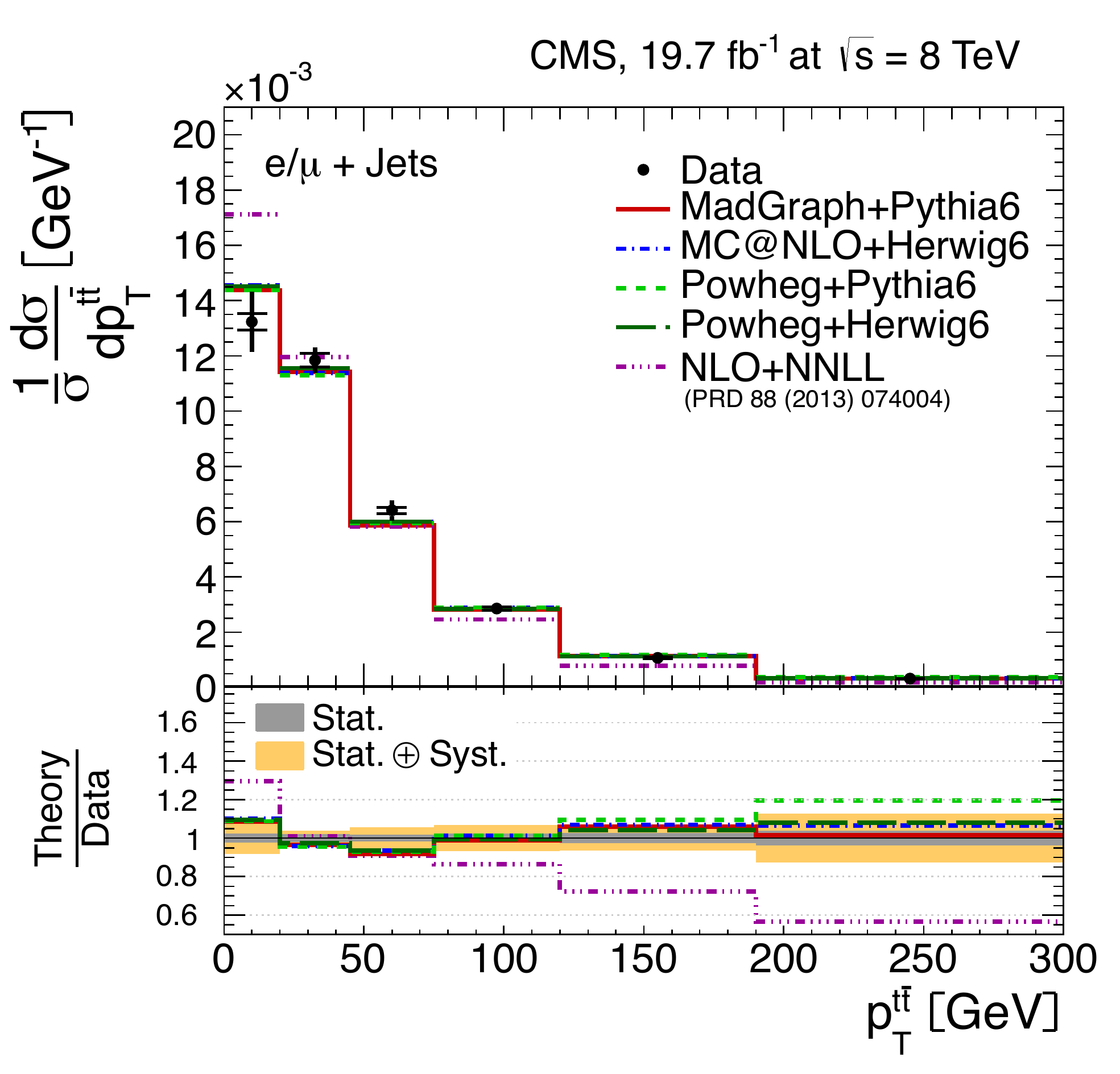}{0.5}{0}
\subfig{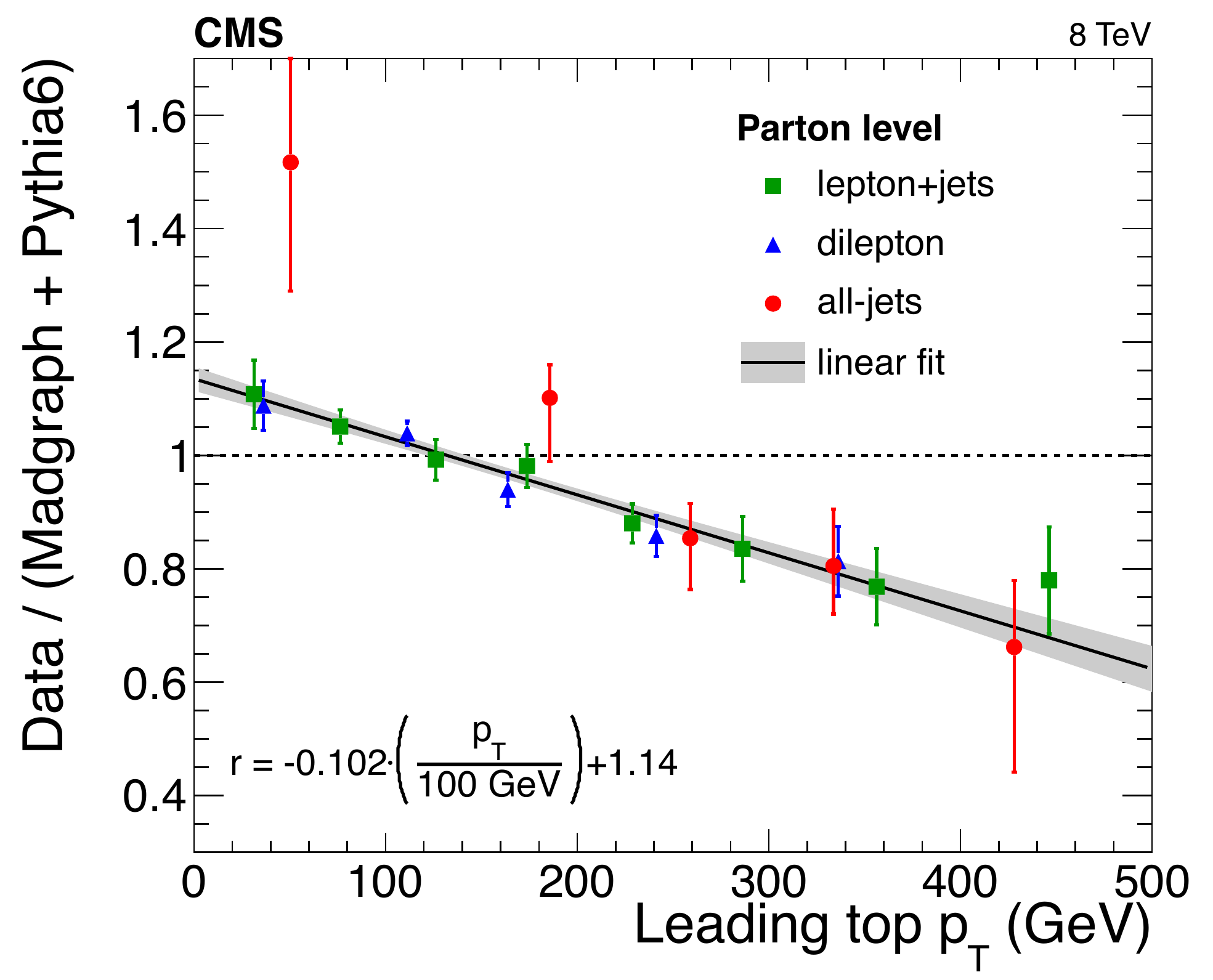}{0.5}{10}
\subfig{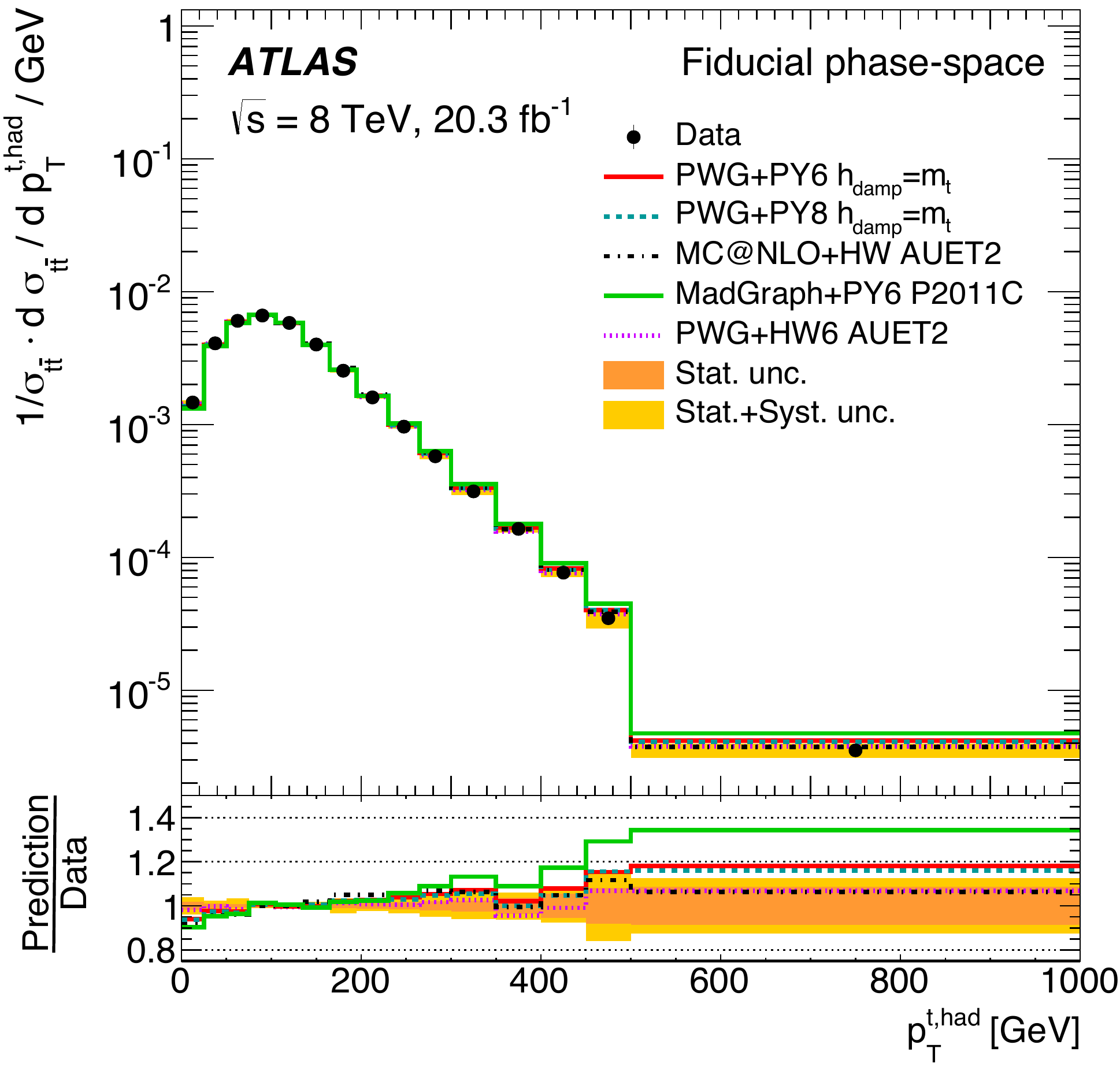}{0.44}{0}
\subfig{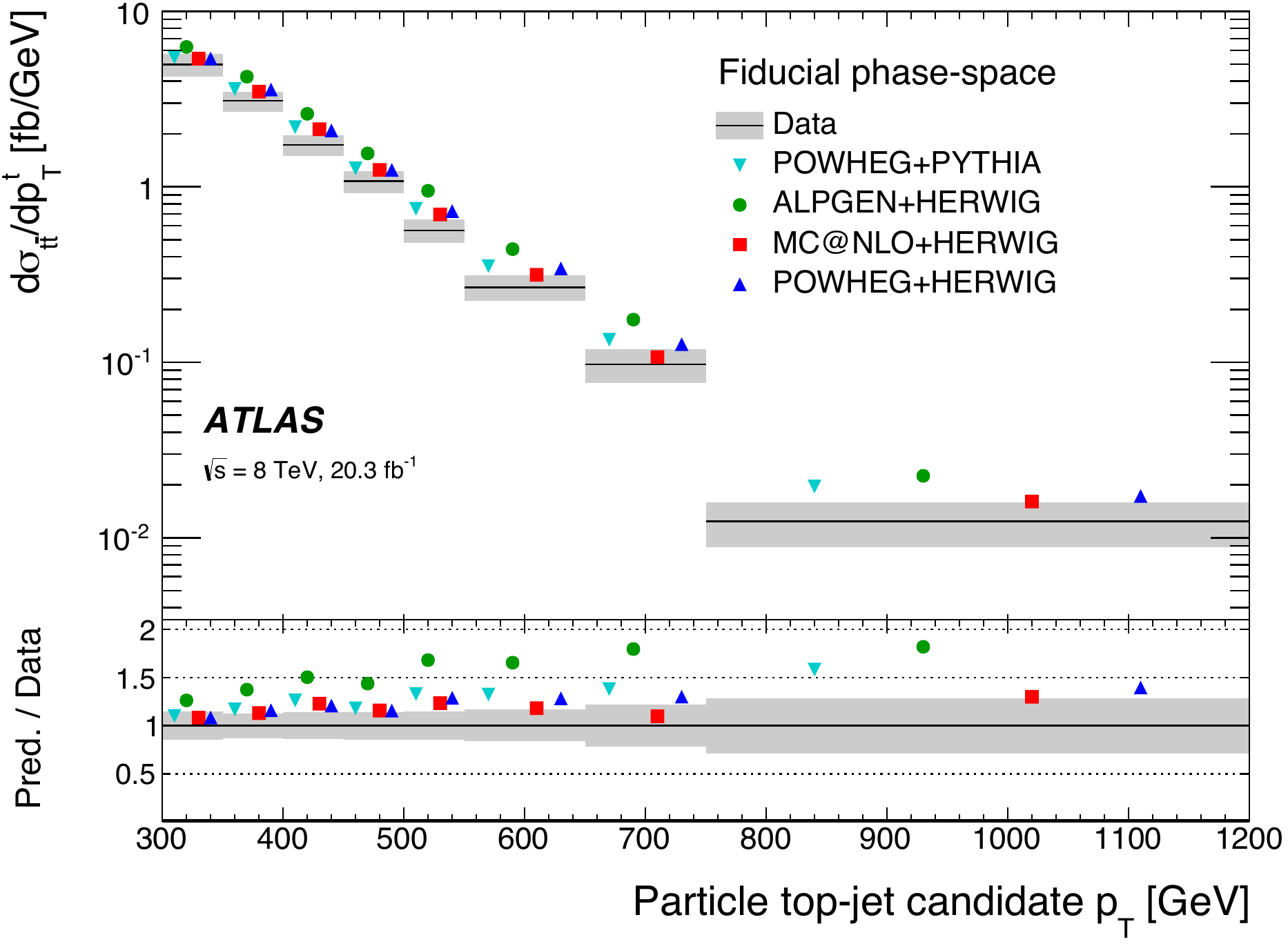}{0.56}{4}
\caption{Normalized differential \ttbar production \xs\ in the \ljets\ as a
function of the transverse momentum of the \ttbar
system~\cite{Khachatryan:2015oqa} (top left).  Data-to-theory ratio
(\MadGraph+\Pythia\ 6) at parton level as a function of the leading top-quark
\pT in different decay channels (additional figure of
\cite{Khachatryan:2015fwh}, top right).  Normalized differential \ttbar
production \xs\ as a function of the top \pT distribution in a fiducial
phase-space~\pcite{Aad:2015mbv} (bottom left) and the particle-level boosted top
candidate \pT~\pcite{Aad:2015hna} (bottom right).  The measurements are compared
to predictions of different generators and parton-shower programs.}
\label{diff-cms-atlas-8tev}
\end{figure}

Differential \xs s at \SI{8}{\TeV} were also measured by ATLAS in the \ljets, as
a function of the hadronic top-quark transverse momentum
(\Fref{diff-cms-atlas-8tev}, bottom) and rapidity, and as a function of the
mass, transverse momentum, and rapidity of the \ttbar system.  In addition, a
new set of observables was included, describing the hard-scattering interaction
($\chi^{\ttbar}$, $y_{\text{boost}}^{\ttbar}$) and sensitive to the emission of
radiation along with the \ttbar pair ($\Delta \phi^{\ttbar}$,
$|p_{\text{out}}^{\ttbar}|$, $H_{\text{T}}^{\ttbar}$,
$R_{Wt}$)~\cite{Aad:2015mbv}.  In general, the Monte Carlo predictions agree
with data in a wide kinematic region.  The rapidity distributions are not well
modelled by any generator under consideration, using the current settings and
parton distribution functions.  The level of agreement is improved when more
recent sets of parton distribution functions are employed, where \textsc{NNPDF
3.0 NLO}~\cite{Ball:2014uwa} is found to provide the best description.
\Fref{lhctop_ttdiff} shows comparisons of some differential distributions
obtained by ATLAS and CMS to the full NNLO calculation~\cite{Czakon:2015owf}.
ATLAS data appear to be well modelled by the full NNLO calculation, while the
CMS data show a small residual difference.

\begin{figure}[htbp]
\centering
\subfig{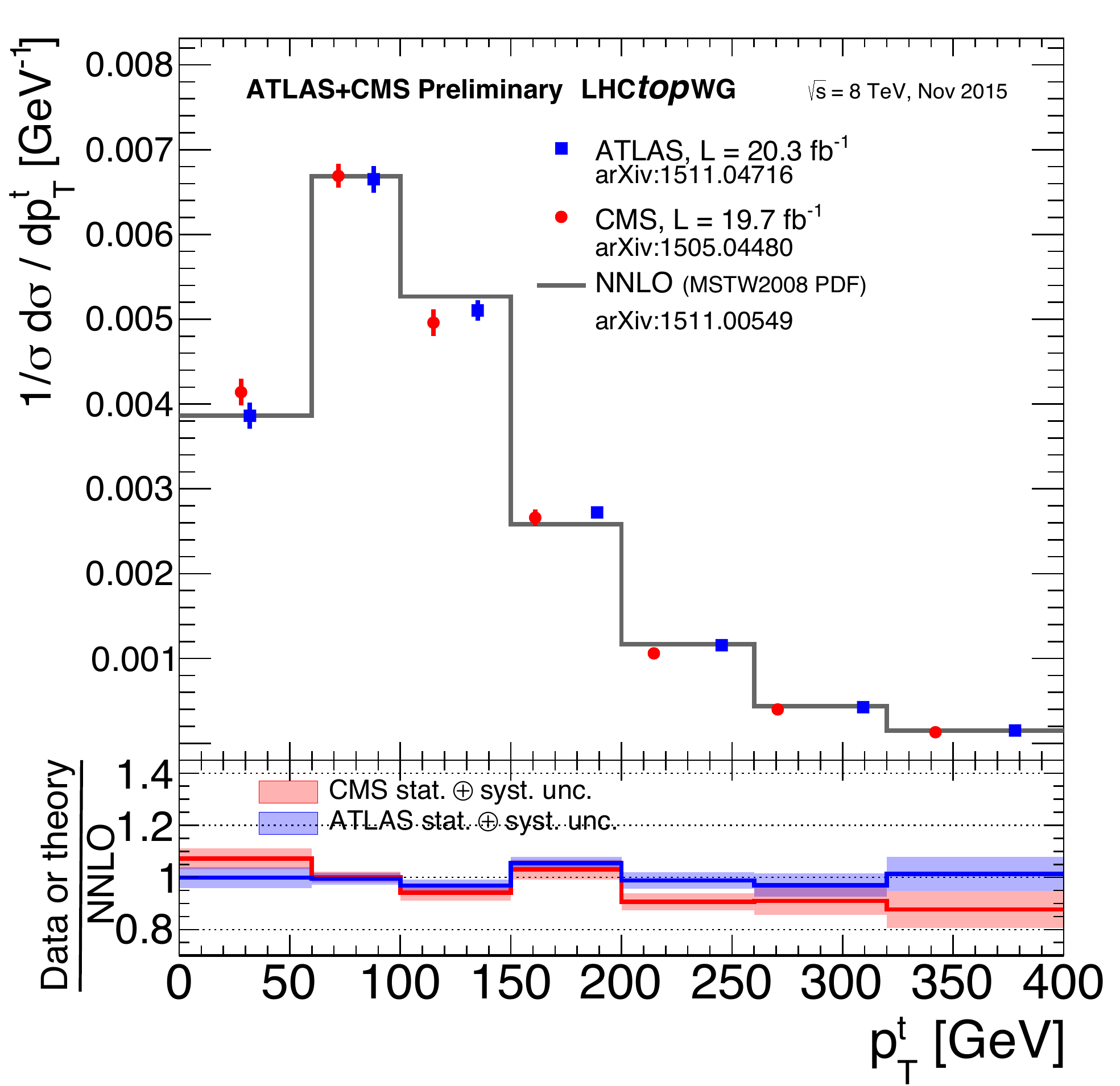}{0.47}{0}
\subfig{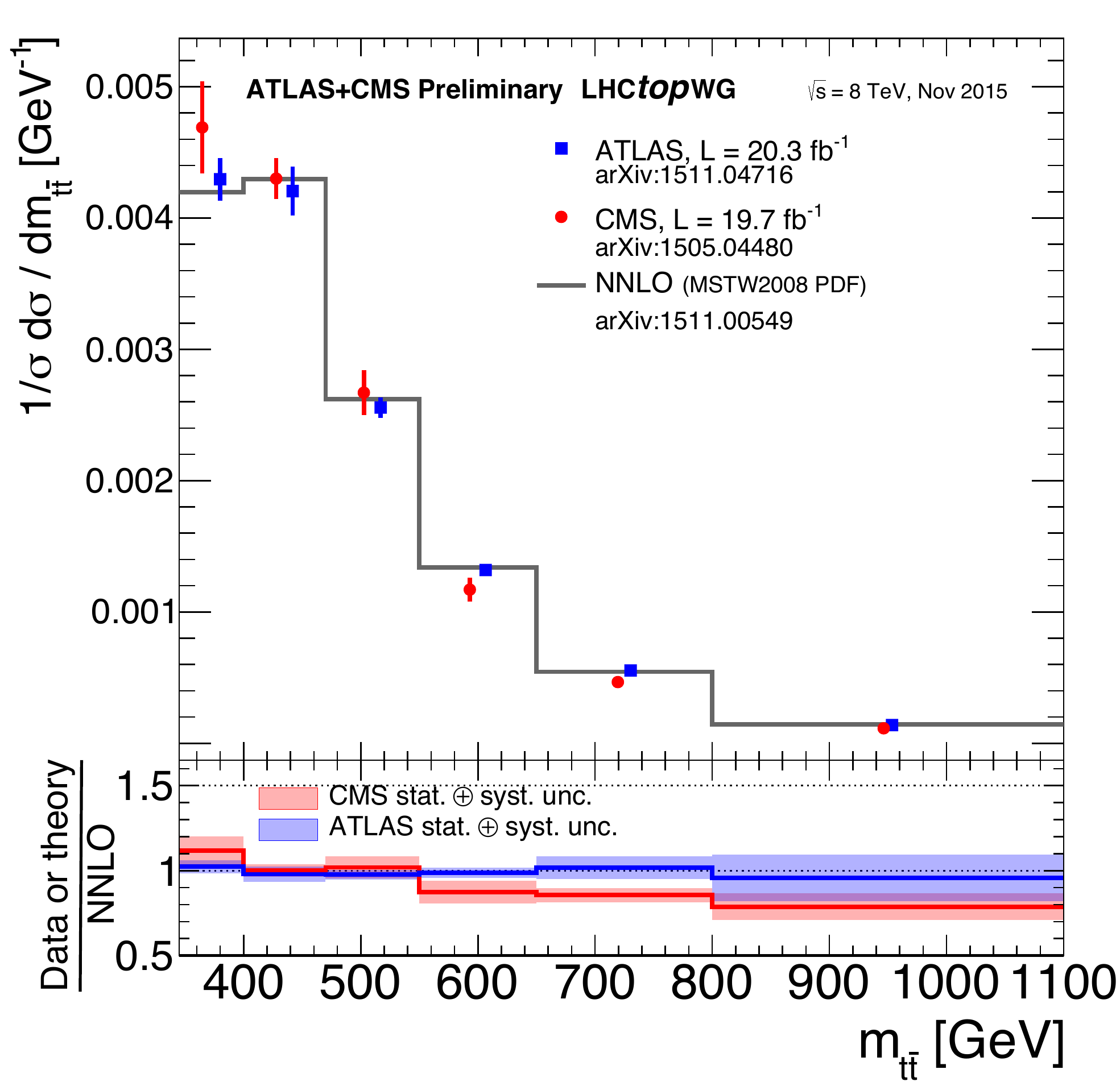}{0.47}{0}
\caption{
Full phase-space normalized differential \ttbar \xs\ as a function of the
transverse momentum of the top quark (left) and the invariant mass of the \ttbar
system (right).  The CMS and ATLAS results are compared to the NNLO calculation.
The shaded bands show the total uncertainty on the data measurements in each
bin. Measurements are from Refs.~\cite{Khachatryan:2015oqa, Aad:2015mbv}.}
\label{lhctop_ttdiff}
\end{figure}

The high top-quark \pT region has been the target of further detailed studies,
since a possible discrepancy could hint at physics effects beyond the Standard
Model and studies allow for an improved understanding of the proton PDF.  When
top quarks are produced with large Lorentz boost their decay products tend to be
more collimated and might escape standard reconstruction techniques, which
typically assume a minimum angular separation of the final state products in the
detector.  In high-\pT studies, the hadronically decaying top quark is
reconstructed as a single large radius-parameter ($R$) jet. Jet substructure
techniques are employed to identify such large-$R$ jets and tested for
compatibility with top-quark decays.  
A detailed discussion on high-\pT top-quark reconstruction can be found in 
Ref.~\cite{Schaetzel:2014kha}.
Measurements have been performed with the
full \SI{8}{\TeV} dataset by ATLAS~\cite{Aad:2015hna} and
CMS~\cite{Khachatryan:2016gxp}.  Large-$R$ jets, compatible with hadronic
top-quark decays were selected and differential \ttbar \xs s as a function of
the top-quark \pT and rapidity were measured, both at particle level in a
fiducial region and at parton level.  The measurements have a threshold of $300$
to \SI{400}{\GeV} and extend beyond \SI{1}{\TeV}. Experimental uncertainties at
particle level are of the order of $10-30\%$ and are dominated by the jet energy
scale uncertainty of large-$R$ jets.  Discrepancies between data and simulations
observed at low-$\pT$ appear to be confirmed in the boosted regime.  Different
PDF and parton-shower parameter settings are shown to improve the agreement
between data and simulation considerably (\Fref{diff-cms-atlas-8tev} bottom).

Recently top-quark events were also observed in the forward region ($2 < \eta < 4.5$)
by the LHCb experiment~\cite{Aaij:2015mwa} using \RunOne\ data.
Measurements in this phase space help constraining the gluon PDF at large Bjorken-$x$,
and to test differential NNLO calculations in an extended $\eta$ region.
A significant excess of events over the Standard
Model $Wb$ prediction with a high-\pT muon and a high-\pT $b$-jet was found.
Single top-quark and \ttbar production cannot be
distinguished and thus inclusive \xs s were extracted in a fiducial region at
$7$ and \SI{8}{\TeV} with uncertainties of $20 - 30\%$. Differential event
counts and a charge asymmetry as a function of the \pT of the $\mu - b$-jet system
were also reported, the latter being sensitive to physics beyond the SM.

\section{Single top-quark production}
\label{secsingle}

Single top quarks are mostly produced via charged-current electroweak
interactions. Three channels can be defined, based on the virtuality of the
\Wboson: \mbox{$s$-channel,} $t$-channel and associated $Wt$ channel.
Measurements of the \xs s are useful to directly extract the CKM matrix element
\Vtb, to verify QCD calculations, and to compare to PDFs.  Extensions of the
Standard Model might affect the three channels in different ways and thus it is
important to establish them separately.

\subsection{$t$-channel}

At the LHC, single top-quark production via the $t$-channel has the largest
production \xs. The final state is characterized by the presence of a light
quark recoiling against the top quark and an additional low-\pT $b$ quark, from
gluon splitting, which often is outside the acceptance of the detectors.
Because of the smaller number of jets and the smaller \xs\ with respect to
\ttbar production, backgrounds are generally more important than in the case of
\ttbar.  To suppress multijet events, only semileptonically decaying top quarks
are considered. At least one jet is required to be $b$-tagged, in order to
reduce the \Wjets\ background, at the cost of an increased sensitivity to the
correct modelling of heavy-flavour production associated with bosons. Finally it
is of particular importance to efficiently reject $c$ quarks from $W\!+c$ jet or
\ttbar decays and thus the $b$-tagging algorithms were often specifically tuned
to achieve this.

With only \ipb{35} of \SI{7}{\TeV} data CMS found evidence of $t$-channel
production at the \signif{3.5} level, complementing a two-dimensional maximum
likelihood fit with a multivariate analysis~\cite{Chatrchyan:2011vp}.  The two
discriminating variables are based on the angle between the lepton and the light
jet in the top-quark rest frame, effectively probing the polarization of the
top-quark, which is predicted to be almost 100\% for signal, and on the
pseudorapidity of the light jet, which is expected to be produced more forward
than the jets from the background processes.  For the second analysis a boosted
decision tree was trained with nearly 40 variables, including $\pT(\ell)$,
$m_{W\!jj}$, $\pT(jj)$, $\pT(b)$ and $m_{\ell \nu b}$. \Fref{t-chan-Vars} shows
the distributions of two of the discriminating variables. The signal was
modelled with \MadGraph+\Pythia, and background shapes were taken from
simulation, and, in the case of \Wjets, rescaled according to data-to-simulation
comparisons in control regions.  The results of the two analyses were
statistically combined to obtain an overall measurement of the \xs\ with an
uncertainty of $36\%$, compatible with predictions based on NLO+NNLL
perturbative QCD calculations~\cite{Kidonakis:2011wy}.

\begin{figure}[htbp]
\subfig{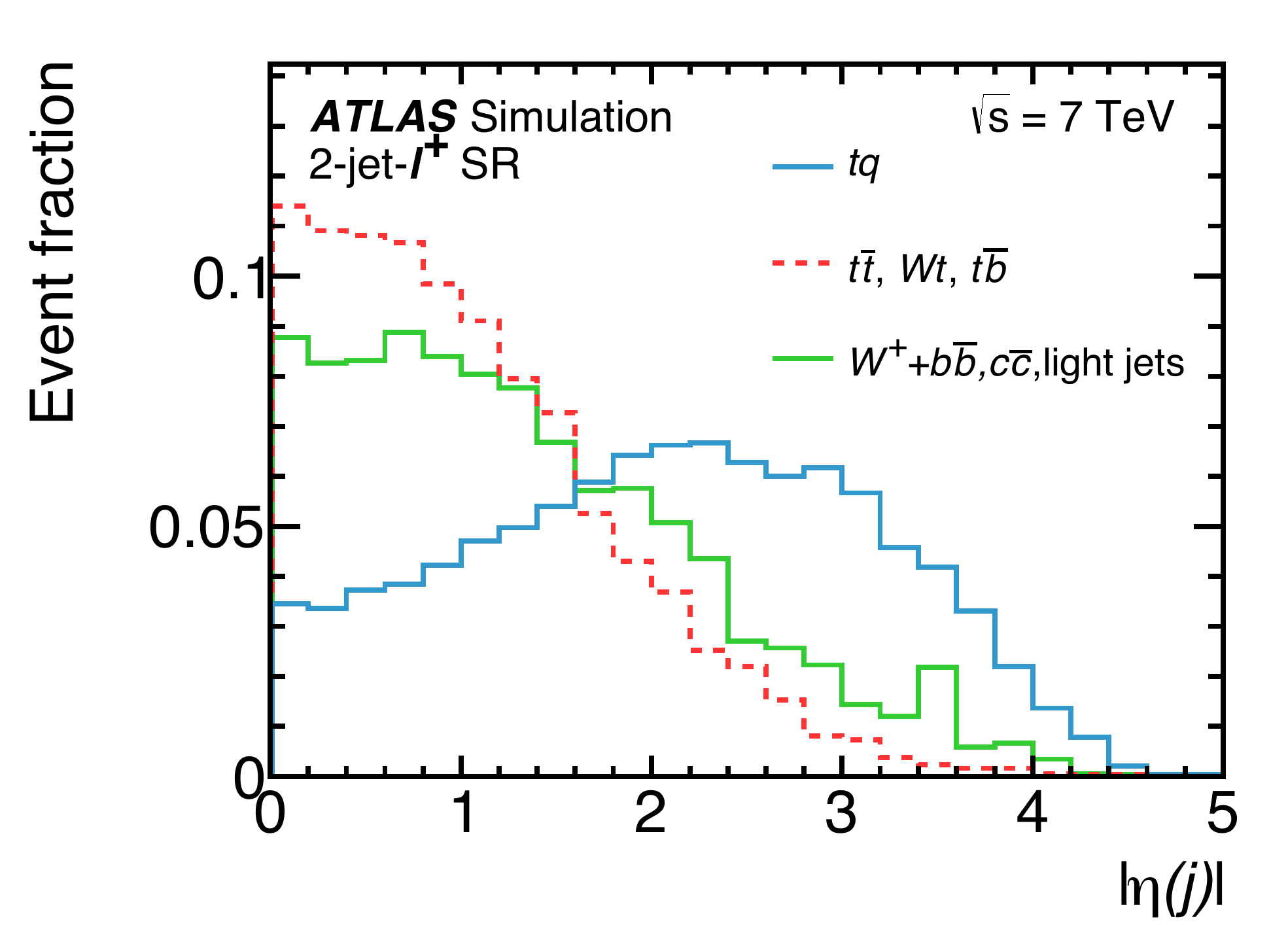}{0.5}{0}
\subfig{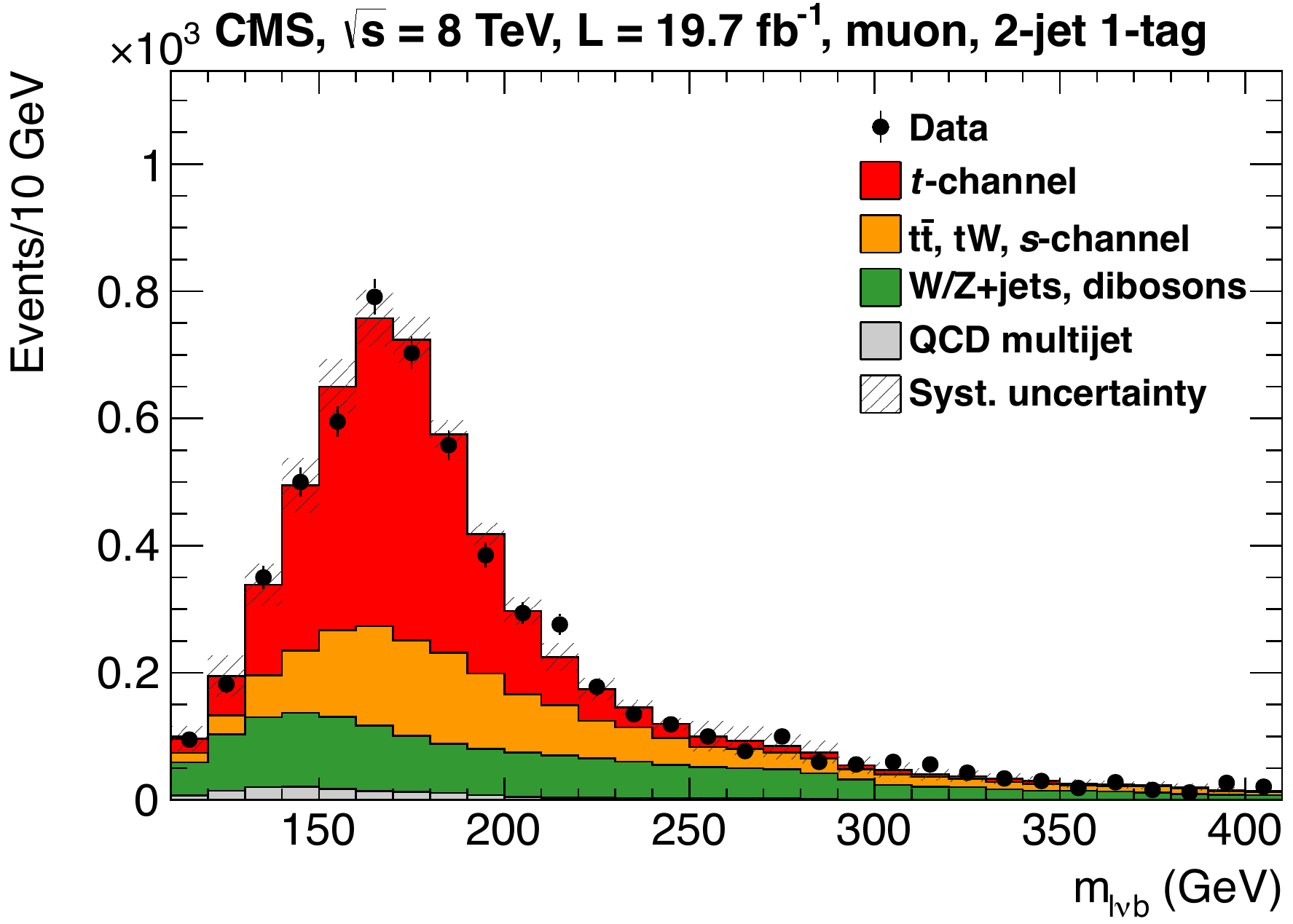}{0.5}{5}
\caption{Distributions of two discriminating variables used in single-top
\mbox{$t$-channel} analyses. Probability density of the absolute value of the
pseudorapidity of the untagged jet, $|\eta(j)|$, in the 2-jet-$\ell^+$ signal
region~\pcite{Aad:2014fwa} (left). Distribution of the reconstructed top-quark
mass, $m_{\ell \nu b}$, for muon decay channels, in the forward jet
region~\pcite{Khachatryan:2014iya} (right).} \label{t-chan-Vars}
\end{figure}

With $30$ to $40$ times more data, still at \SI{7}{\TeV}, both ATLAS and CMS
determined the inclusive \xs, reducing the uncertainties to
$24\%$~\cite{Aad:2012ux} and $9\%$~\cite{Chatrchyan:2012ep}, respectively.  In
the ATLAS analysis a neural network (NN) was trained and the result was
cross-checked with a cut-based analysis.  The signal was modelled using the
\AcerMC\ generator. The larger amount of data allowed the signal in the two
regions to be separated, depending on the number of reconstructed jets: at
leading order two jets are expected, while a third jet may arise from
higher-order processes.  The backgrounds with fake leptons were estimated using
a selection of jets that satisfy a subset of the electron identification
requirements.  The normalizations of the \Wjets\ and $W$+heavy-flavour processes
were directly determined in the fit of the NN discriminant. Two NNs were trained
in the $2$- and $3$-jet regions separately. The optimization resulted in using
12 and 18 variables, respectively, based on invariant masses, such as $m_{\ell
\nu b}$ or $m_{jj}$, the properties of the highest \pT untagged jet, $|\eta|$
and \ET, the spatial separation of reconstructed objects in the detector, and
the global energy deposition in the detector (\met, \HT). The cut-based analysis
used a subset of the variables and separated the sample further, based on the
charge of the reconstructed lepton: due to the difference in PDF of the incident
u and d quarks involved in the hard scattering, the ratio $R_t \equiv \sigma_t /
\sigma_{\bar{t}}$ is expected to be $R_t \sim 2$.  Systematic uncertainties
dominated the total uncertainty.  The limited knowledge of initial and final
state radiation and of the calibration of the $b$-tagging efficiency sum up to
$\sim\!\!20\%$ uncertainty on the final measurement.  The CMS analysis drew from
the previous experience. The two-dimensional likelihood was reduced to the usage
of $|\eta_j|$ to minimize reliance on the SM assumption for top-quark
polarization and after imposing a requirement on $m_{\ell \nu b}$ a neural
network was trained in addition. The final result was quoted as a combination of
the three results.  The single top-quark $t$-channel signal was modelled with
\Powheg.  Events were classified in six signal regions, based on the number of
jets and $b$-jets, including those with 4 jets, allowing a better constraint of
the systematic uncertainties that were implemented as nuisance parameters in the
fit. For the neural network about 40 variables were selected for the training,
with the following being ranked as the most discriminating: $|\eta_j|$, the
reconstructed transverse $W$ boson mass $m_{\rm T}$, $m_{jj}$ of the leading
jets, and the total transverse energy of the event.  The experimental
uncertainties and background estimates were both marginalised in the fit, while
the theoretical uncertainties were not.  The combination improved on the single
multivariate result, reducing the total uncertainty to $9\%$.  Some of the
discriminant output distributions for BDTs and NNs are shown in
\Fref{t-chan-MVA}.

\begin{figure}[htbp]
\subfig{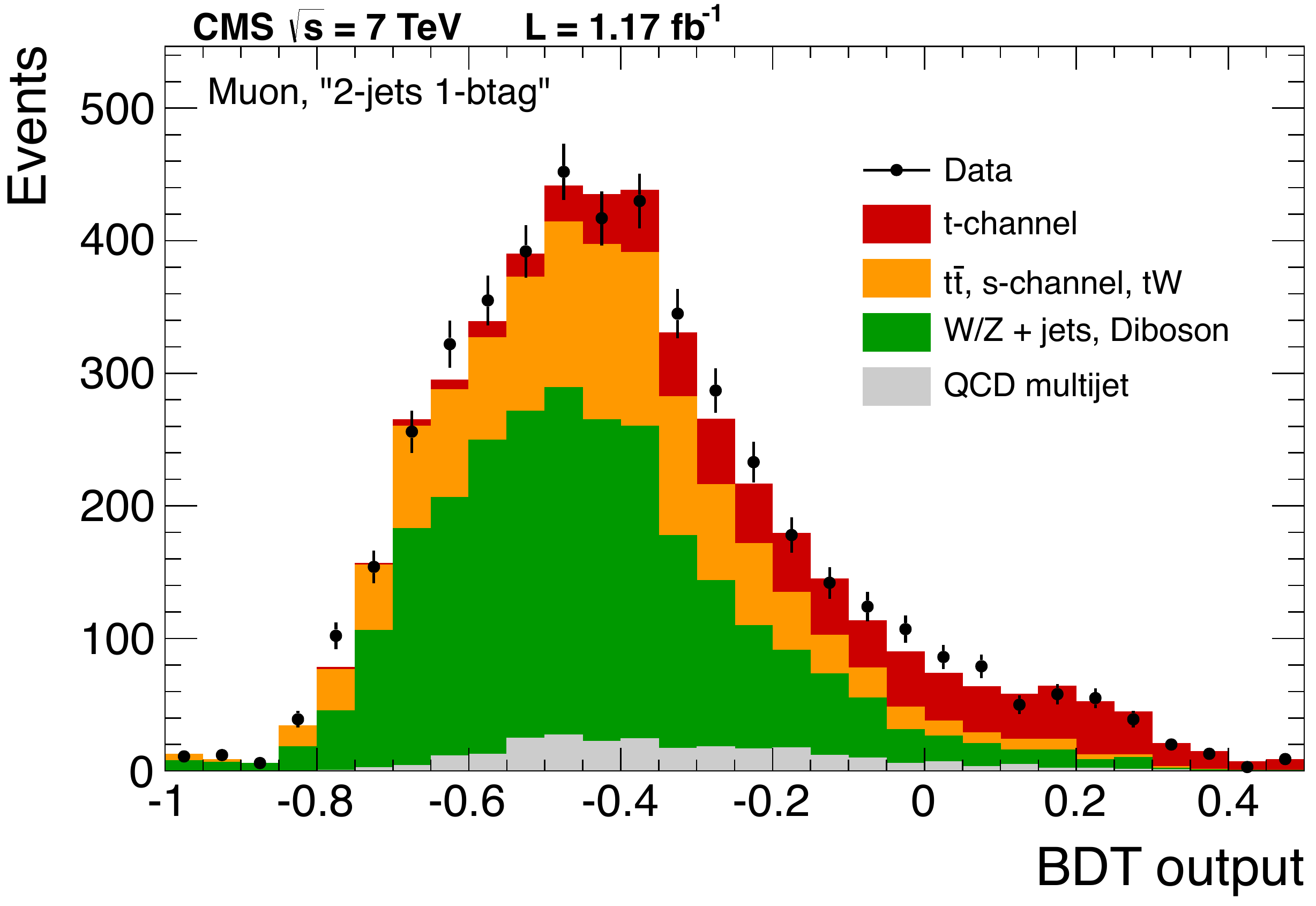}{0.5}{20}
\subfig{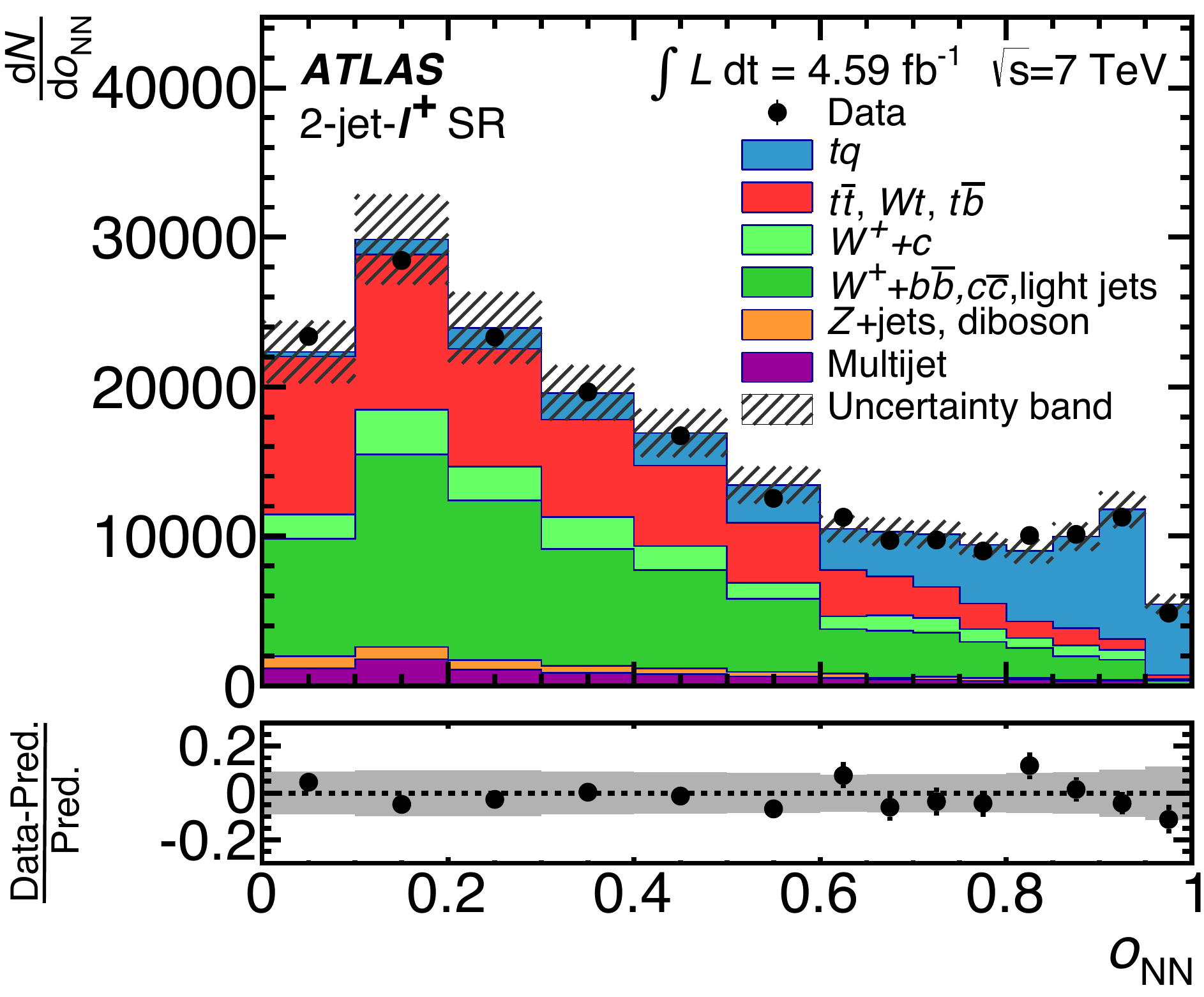}{0.5}{0}
\caption{Distributions of the BDT discriminator output in the muon channel for
the 2-jets 1-$b$-tag category. Simulated signal and background contributions are
scaled to the best fit results~\pcite{Chatrchyan:2012ep} (left).  Neural network
discriminant distribution normalized to the result of the binned
maximum-likelihood fit in the 2-jet-$\ell^+$ channel~\pcite{Aad:2014fwa}
(right).} \label{t-chan-MVA}
\end{figure}

The full \SI{8}{\TeV} dataset has been exploited by
CMS~\cite{Khachatryan:2014iya} to extract the single top-quark and anti-quark
\xs s, the ratios $\sigma(t, \SI{8}{\TeV})/\sigma(t, \SI{7}{\TeV})$ and \Rt, and
to compare these results with QCD
calculations~\cite{Kidonakis:2012db,Kidonakis:2013zqa} and with different PDF
predictions. The analysis only considered the $|\eta_j|$ distribution in the
2-jets 1-$b$-tag region to extract information on the single-top $t$-channel.
Here the \ttbar template from simulation was corrected with the help of data
measured at higher jet and $b$-tag multiplicities. The extracted values of \Rt
were compared with different PDFs, as shown in \Fref{t-chan-Rt} (left).

\begin{figure}[htbp]
\subfig{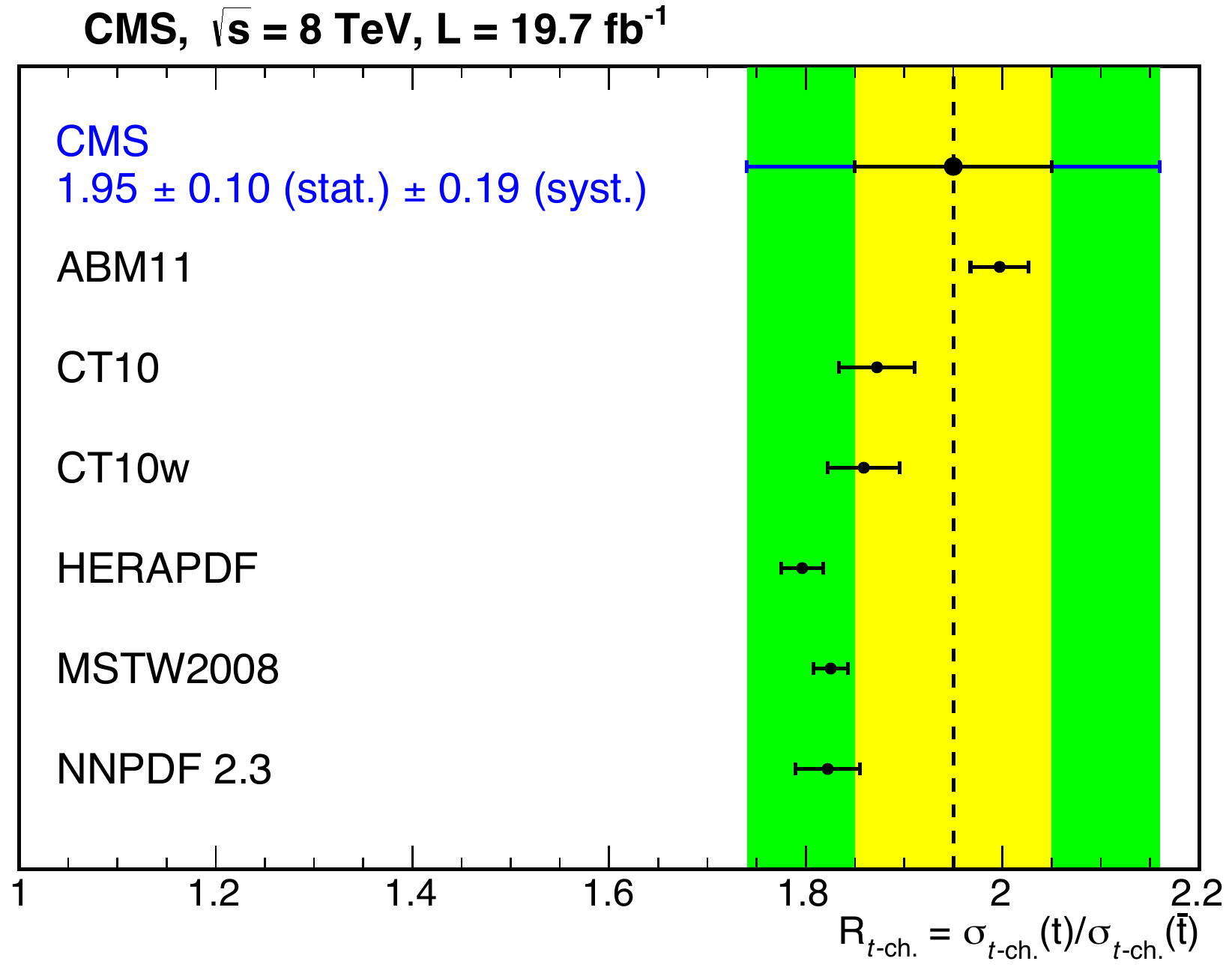}{0.48}{0}
\subfig{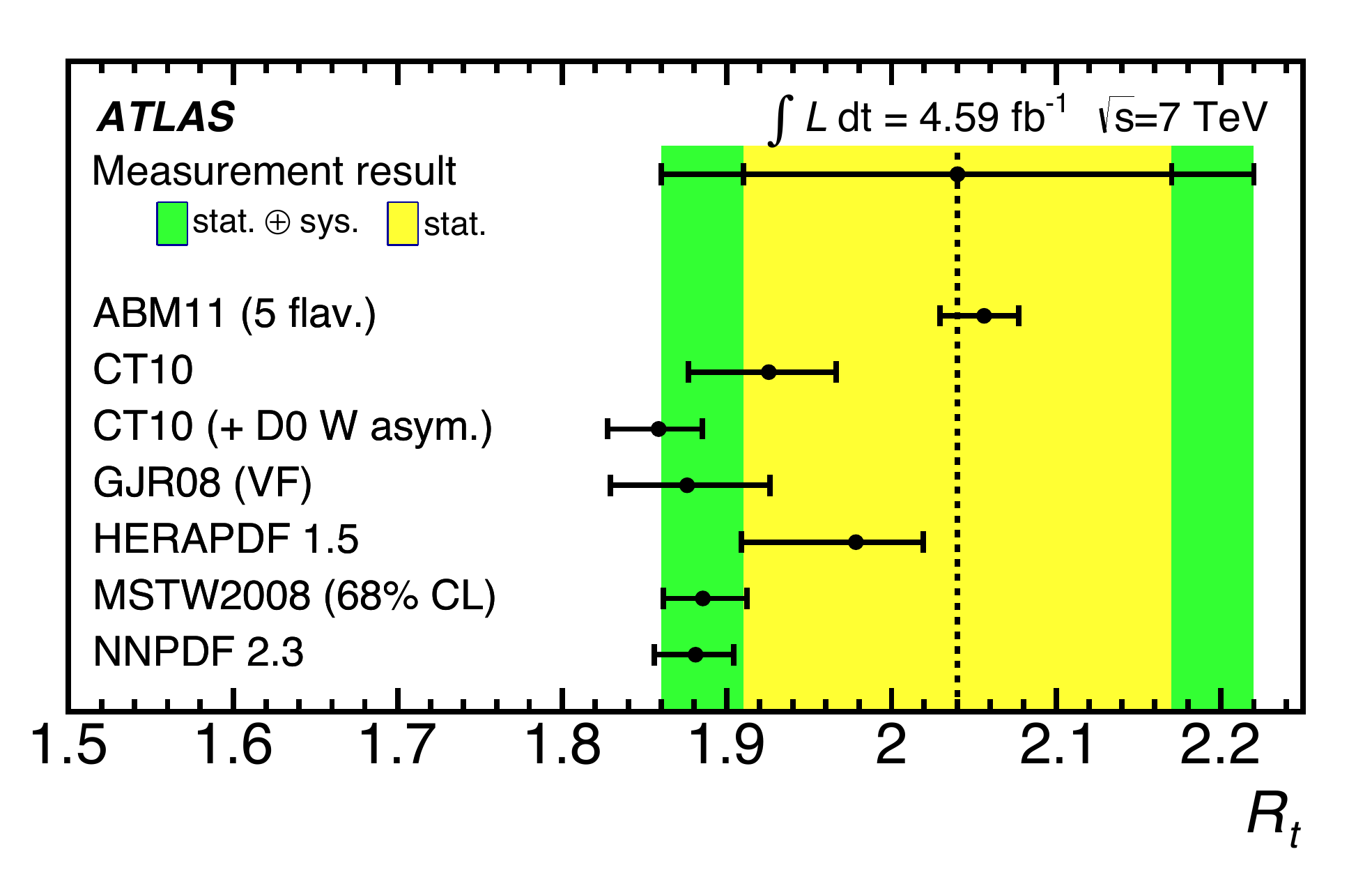}{0.52}{3}
\caption{Comparison between observed and predicted values of \Rt. The
predictions were calculated at NLO accuracy in the five-flavour scheme and given
for different NLO PDF sets. The uncertainty includes the uncertainty on the
renormalization and factorization scales, the combined internal PDF, and the
$\alpha_S$ uncertainty.  CMS, using \SI{8}{\TeV}
data~\pcite{Khachatryan:2014iya} (left) and ATLAS, using \SI{7}{\TeV}
data~\pcite{Aad:2014fwa} (CMS).} \label{t-chan-Rt}
\end{figure}

With the full \SI{7}{\TeV} dataset ATLAS~\cite{Aad:2014fwa} obtained
measurements of the \xs s and \xs\ ratios (\Fref{t-chan-Rt}, right).  A similar
strategy of training a NN was adopted, additionally taking advantage of the
difference in single top-quark and anti-quark production.  First single-top
$t$-channel differential \xs s were measured as a function of the top quark and
anti-quark \pT and rapidity after unfolding to the parton level and compared to
\MCFM\ predictions at NLO (\Fref{t-chan-diff}). 

\begin{figure}[htbp]
\subfig{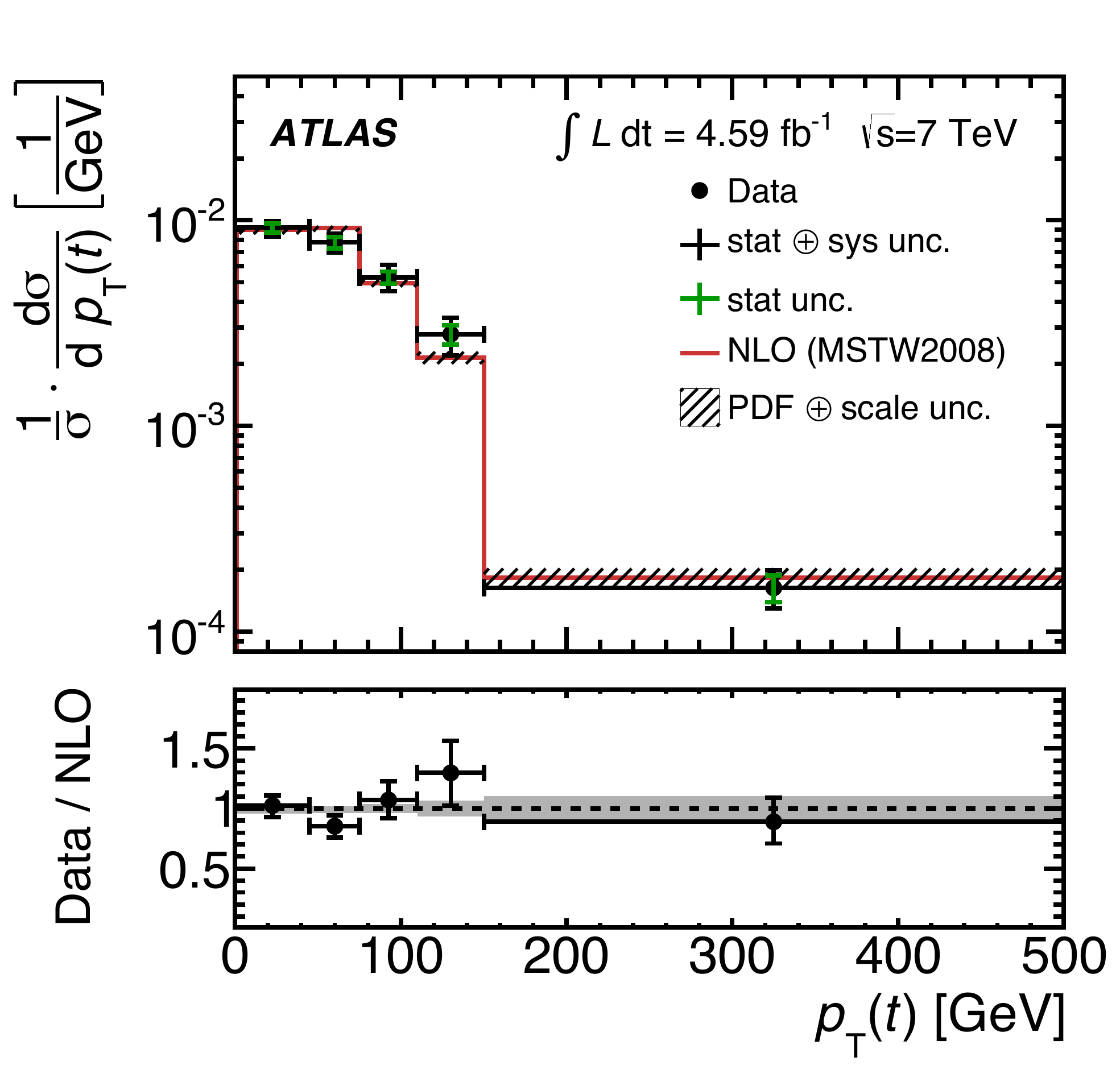}{0.5}{0}
\subfig{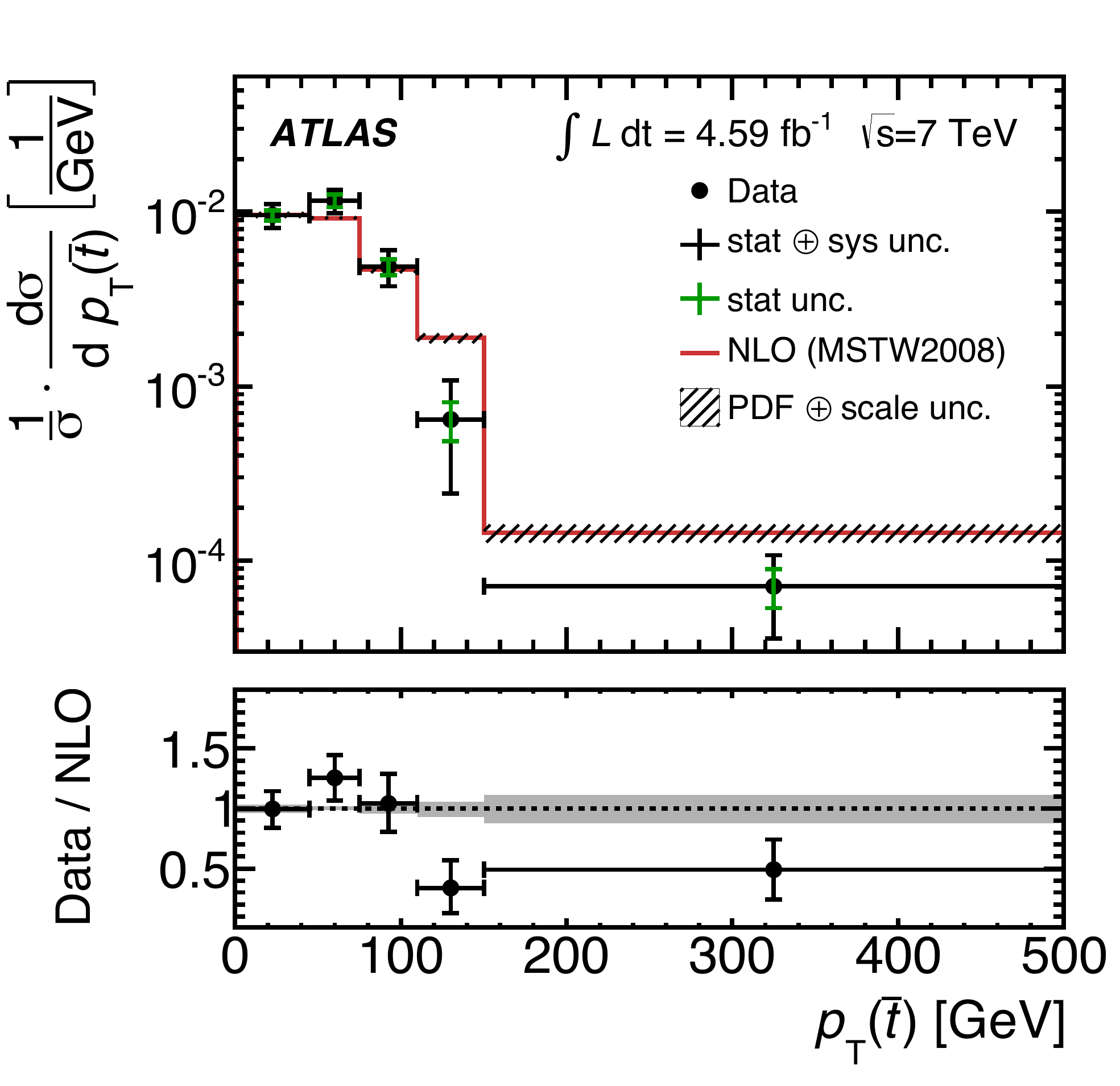}{0.5}{0}
\caption{Normalized differential \xs\ as a function of $\pT(t)$
compared to the QCD NLO calculation~\pcite{Aad:2014fwa} for top quarks (left)
and anti-quarks (right).}
\label{t-chan-diff}
\end{figure}

\subsection{$Wt$ channel}

Associated $Wt$ production is an interesting channel to study, as well, as its
sensitivity to new physics is complementary to the other channels and it
represents an important background in studies of the Higgs boson.  The
associated production of a single top-quark and a \Wboson\ involves the
interaction of a gluon and a $b$ quark, emitting an on-shell \Wboson. The final
state is thus the same as in \ttbar, but with one fewer $b$-jet. At NLO the $Wt$
and \ttbar diagrams interfere, but the effect is small and considered as a
modelling uncertainty by comparing two algorithms to avoid the double counting
of generated events~\cite{Frixione:2008yi, Tait:1999cf}. The predicted \xs\ is
calculated at approximate NNLO, with soft gluon resummation at
NNLL~\cite{Kidonakis:2010ux}.

The first evidence of $Wt$ production was reported by ATLAS with \ifb{2.05} of
\SI{7}{\TeV} data~\cite{Aad:2012xca} where only leptonic decays of the \Wboson s
were considered, yielding final states $Wt \to \ell \nu b \ell \nu$. The signal,
which was modelled with the \AcerMC\ generator, resides mostly in the 1-jet bin
after a dilepton event selection, as can be seen in \Fref{Wt-chan}.  The
contribution from \Zjets\ and events with fake leptons was estimated with
data-driven methods.  The large \ttbar dilepton background is difficult to
suppress and therefore a BDT using 22 input variables was constructed for each
of the channels $ee$, $\mu\mu$ and $e\mu$, separately. The most discriminating
variables were the magnitude of the vectorial sum of the transverse momenta of
the jet, the leptons and \met, and this variable normalized to the scalar sum of
the same transverse momenta.  These variables were also used to construct BDTs
in the 2-jet and 3 or more jet regions, in order to constrain the background
normalization.  A template fit was then performed to the three BDT output
distributions, allowing the $Wt$ yield to be extracted, and mitigating the
effect of systematic uncertainties affecting extra jet production, such as the
jet energy scale and resolution, as well as the modelling of initial and final
state radiation. The result of the measurement has a significance of
\signif{3.3} and the measured $Wt$ \xs\ is compatible with the prediction.

\begin{figure}[htbp]
\subfig{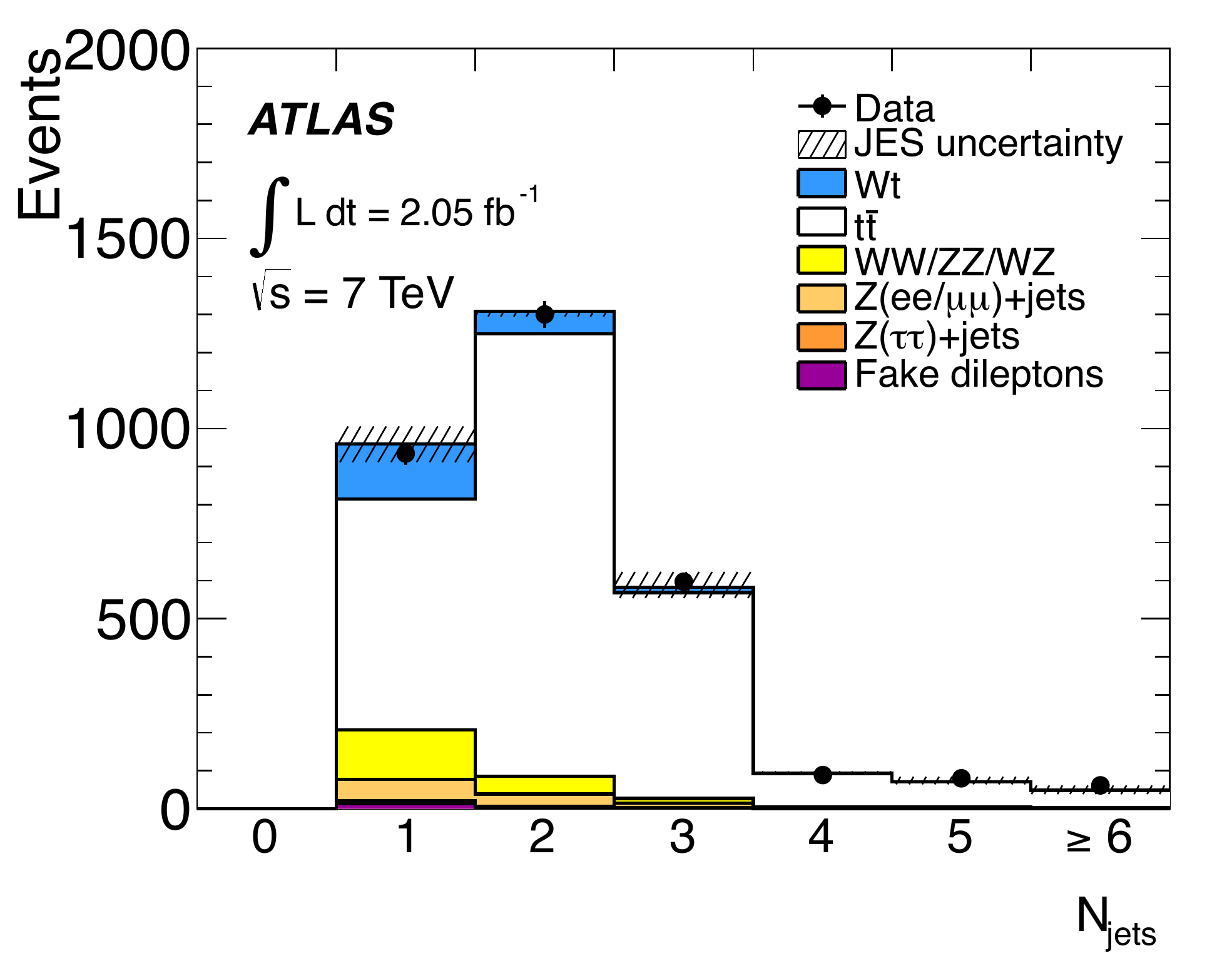}{0.5}{0}
\subfig{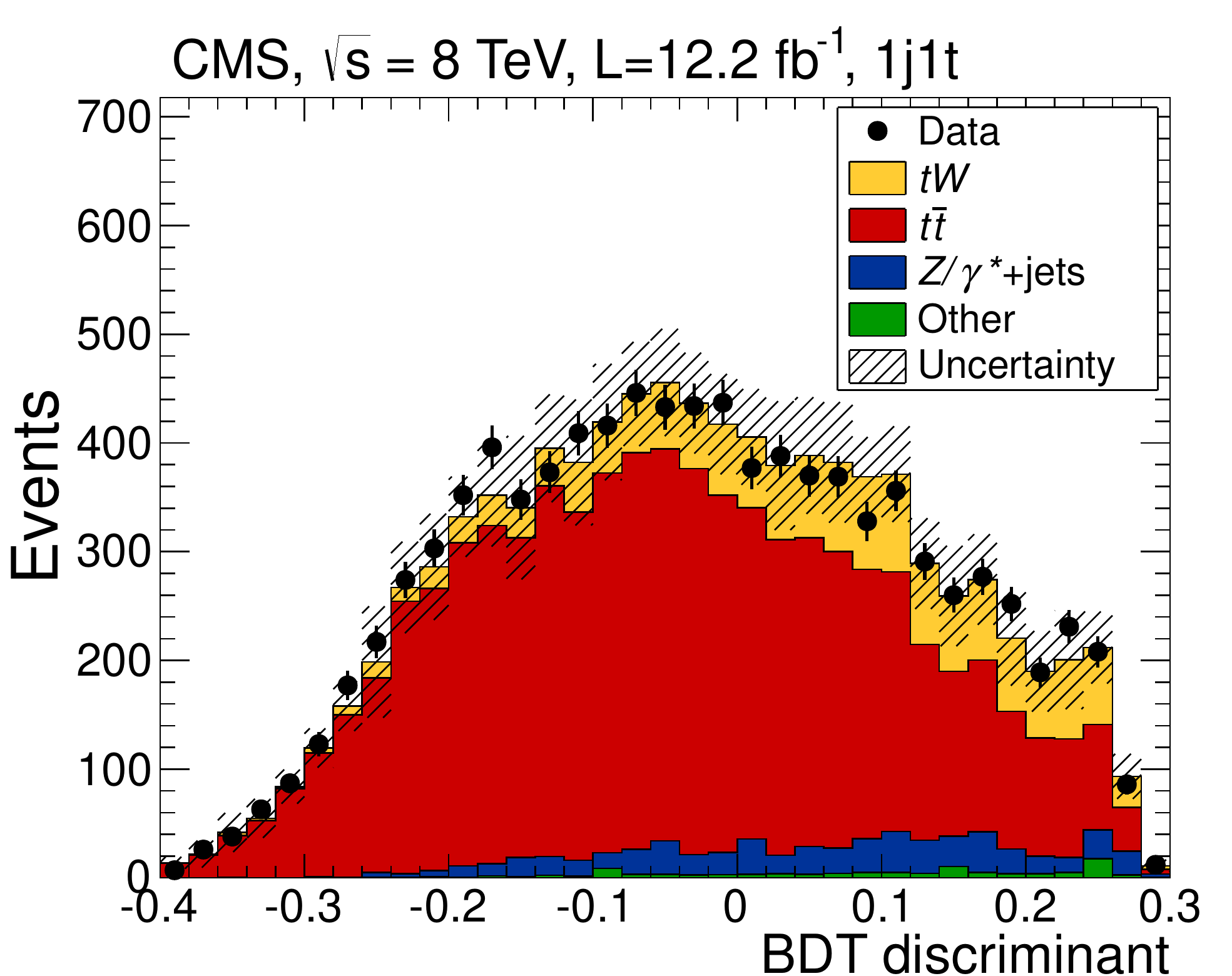}{0.5}{10}
\caption{Number of jets with $\pT > \SI{30}{\GeV}$ and $|\eta| < 2.5$ after the
selection; hatched bands show the jet energy scale (JES) uncertainty. The $Wt$
signal is normalized to the theory prediction~\pcite{Aad:2012xca} (left). The
BDT discriminant, in the signal region (1 jet, 1 top) for all final states
combined. The hatched band represents the combined effect of all sources of
systematic uncertainty~\pcite{Chatrchyan:2014tua} (right).} \label{Wt-chan}
\end{figure}

This result was confirmed by CMS with the full \SI{7}{\TeV} dataset, at the
\signif{4.0} level~\cite{Chatrchyan:2012zca}. The analysis strategy was similar,
with the addition of further refining the categorization based on the
$b$-tagging information from the selected jets, and correcting the \Zjets\
simulation to match the observed \met distribution with a data sideband region
to reduce mismodelling. The BDTs were trained with four input variables only,
and the analysis was cross-checked with a cut-based approach, using the $e\mu$
channel, only, obtaining comparable results. 

The definitive observation of this process was only possible with the
\SI{8}{\TeV} dataset. Using a partial dataset of \ifb{12.2} CMS reported a
significance of \signif{6.1}~\cite{Chatrchyan:2014tua} pursuing the same
analysis strategy as at \SI{7}{\TeV}. The BDT output distribution in the signal
region with one extra jet is shown in \Fref{Wt-chan}. The most recent and most
precise measurement was performed by the ATLAS collaboration, using the complete
\RunOne\ dataset at \SI{8}{\TeV}~\cite{Aad:2015eto}, with a significance of
\signif{7.7} and a measurement of the total \xs\ in agreement with the
prediction and with an uncertainty of $16\%$, dominated by systematic
uncertainties, for instance the amount of additional radiation.  The template
fit has been replaced by a profile likelihood fit to the BDT classifier output.
\Fref{Wt-chan2} (left) shows the most sensitive variable in the signal region
with one central $b$-tagged jet.  The measurement has been also performed in a
fiducial region, including \ttbar and $Wt$ contributions, with a relative
precision of $8\%$. This analysis treats $Wt$ and \ttbar as signal so that it
can be compared to full NLO $WbWb$ calculations in the future~\cite{Cascioli:2013wga}.

\begin{figure}[htbp]
\subfig{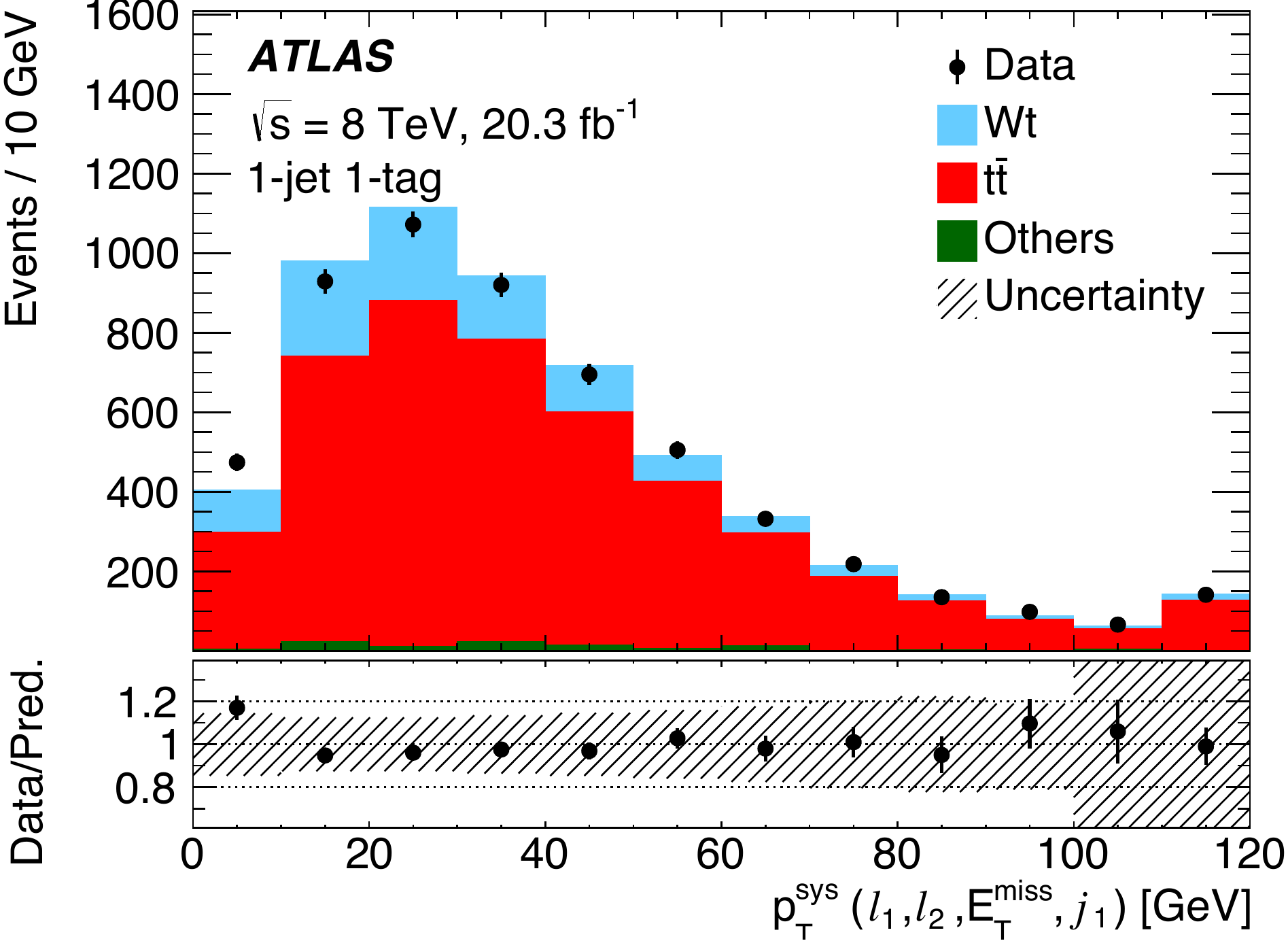}{0.52}{15}
\subfig{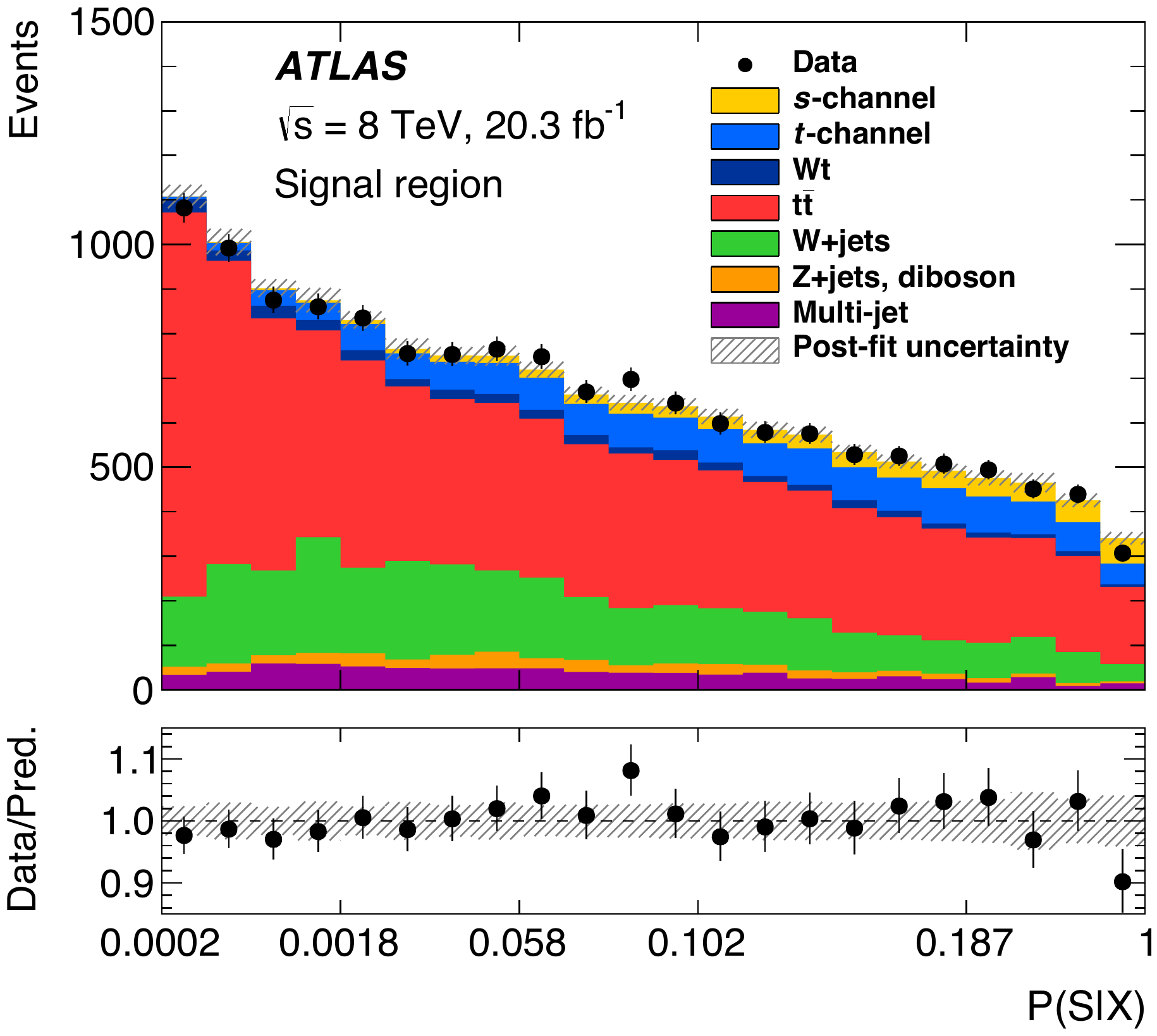}{0.48}{0}
\caption{Distribution of the most important BDT input variables for the
\mbox{1-jet} 1-tag region of the $Wt$ analysis: the \pT of the system of the
leptons, jet and \met~\pcite{Aad:2015eto} (left). The hatched area represents
the total uncertainties. Post-fit distribution of the ME discriminant in the
$s$-channel signal region~\pcite{Aad:2015upn} (right). The hatched bands
indicate the total uncertainty result including all correlations.}
\label{Wt-chan2}
\end{figure}

\subsection{$s$-channel}

Production of single top-quarks in the $s$-channel is strongly suppressed
relative to the backgrounds at the LHC, since it proceeds through
quark-antiquark annihilation, requiring a sea antiquark in the initial state.
The distinctive signature for this process is the presence of a top and a bottom
quark in the final state. The apparent \xs\ could be strongly modified by the
presence of a charged Higgs boson or an exotic $W^{\prime}$ boson that would
yield the same final state particles. The single-top \mbox{$s$-channel} process
has been calculated at NLO+NNLL~\cite{Smith:1996ij,Kidonakis:2010tc}.  

At \SI{8}{\TeV}, using a BDT analysis, ATLAS placed an upper limit of $2.6$
times the Standard Model prediction~\cite{Aad:2014aia} with a significance of
\signif{1.3} (\signif{1.4} expected, assuming the SM rate). The quoted \xs\
uncertainty was dominated by systematic effects, mainly the scale uncertainties
in \met and the jet energy.  Events with exactly two $b$-tagged jets were
selected and a BDT was trained to reduce the impact of the large remaining
\ttbar background. A similar BDT analysis was performed by CMS with the 7 and
\SI{8}{\TeV} datasets~\cite{Khachatryan:2016ewo}, obtaining a combined
significance of \signif{2.5} (\signif{1.1} expected).

The ATLAS dataset has been reanalysed with improved event selections,
simulation, calibration and employing a matrix element (ME) method to separate
the signal from the dominant \ttbar and $W$ boson production with heavy
flavour~\cite{Aad:2015upn}.  Theoretical calculations were used to compute a
per-event signal probability.  For each event, likelihoods were computed for
signal and each background process under consideration.  The likelihoods were
determined using the corresponding partonic \xs s and transfer functions, that
map the measured final state to the partons, taking into account permutations,
reconstruction efficiencies and resolution effects.  A discriminant was then
built as a ratio of these likelihood probabilities. The distribution of this
discriminant for the signal region is shown in \Fref{Wt-chan2} (right).  The
analysis obtained a significance of \signif{3.2} (\signif{3.9} expected) and a
measured \xs\ with a relative uncertainty of $35 \%$.  Approximately half of the
improvement was due to the change in method from BDT to ME. The ME approach is
statistically powerful and is less affected by the limited size of MC samples
available for training and calibrating the analysis.

The most precise single-top \xs\ measurements in each channel and experiment at
$7$ and \SI{8}{\TeV} are summarized in \Fref{SingleTopSummary} and compared to
predictions.

\begin{figure}[htbp]
\centering
\includegraphics[width=\textwidth]{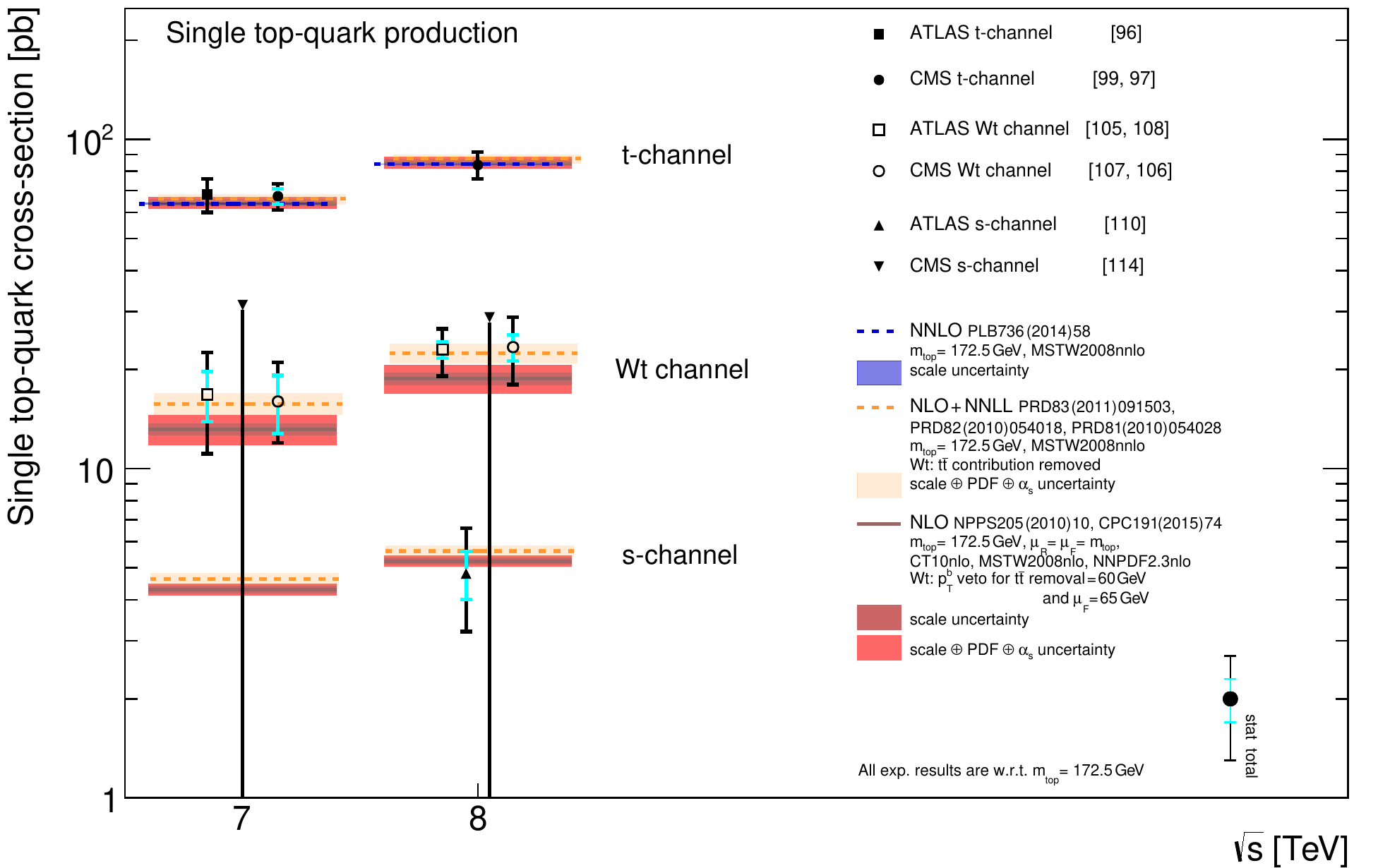}
\caption{Summary of inclusive \xs\ measurements at the LHC at $7$ and \SI{8}{\TeV} 
for the three single top-quark channels compared to theoretical predictions.}
\label{SingleTopSummary}
\end{figure}

\subsection{Determination of \Vtb from single top-quark measurements}

The three single top-quark production channels discussed in this section, are
directly related to the Cabibbo-Kobayashi-Maskawa (CKM) matrix element \Vtb, as
the predicted \xs\ is proportional to $\Vtb^2$. For the extraction the
assumption is that $\Vtb \gg \Vtd, \Vts$ and that the $tWb$ interaction is
purely left-handed as predicted by the Standard Model. The measurements are
independent on the assumption about the number of quark generations and about
the unitarity of the CKM matrix. The measured single-top \xs s in the different
channels were thus divided by the respective theoretical prediction to extract
$\Vtb^2$.  The most precise measurements are obtained in the $t$-channel with
uncertainties of $\sim\!\!5\%$, followed by the $Wt$ channel with uncertainties
of $\sim\!\!10\%$, and finally the $s$-channel with an uncertainty of
$\sim\!\!20\%$.  The \Vtb measurements from single-top inclusive \xs\
measurements at $7$ and \SI{8}{\TeV} are summarized in
\Fref{VtbSingleTopSummary}. 

\begin{figure}[htbp]
\centering
\includegraphics[width=\textwidth]{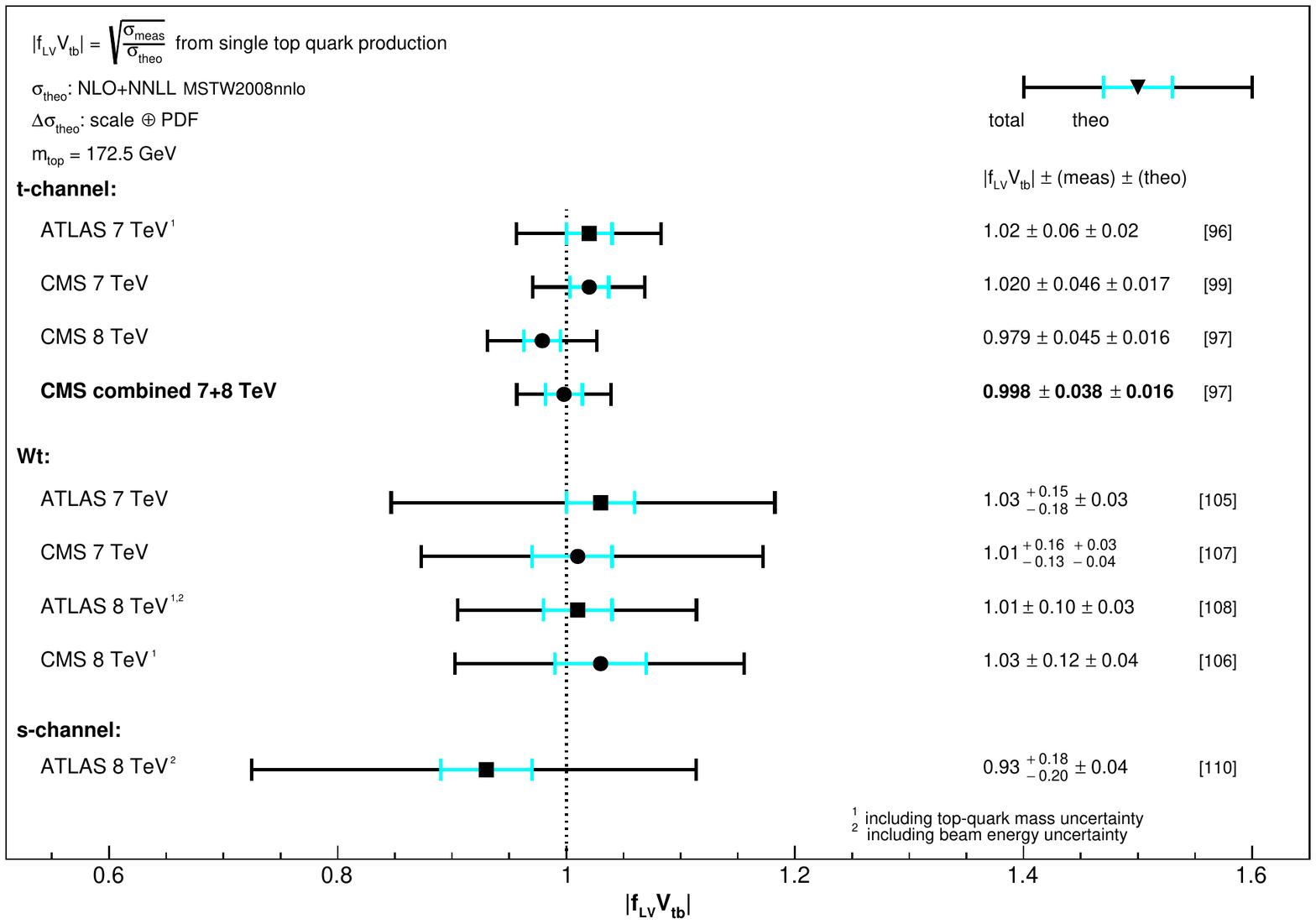}
\caption{Summary of \Vtb determinations from single-top cross-section
measurements at the LHC at $7$ and \SI{8}{\TeV}.} \label{VtbSingleTopSummary}
\end{figure}

\section{Top-quark mass}
\label{sec:mass}

Arguably the most distinctive property of the top quark is its mass, about 40
times larger than the mass of the second-heaviest known fermion, the $b$ quark.
As for all other fermion masses, it is a free parameter of the Standard Model.
Its value is therefore not predicted by the theory but can be inferred from
experimental measurements, either through direct measurements of the invariant
mass of its decay products, or indirectly through predicted relations with other
measured observables, such as the \ttbar production \xs\ at a given
centre-of-mass energy. 

\subsection{Early mass measurements at the LHC}

Traditionally the most precise measurements for the \topmass\ have relied on the
full kinematic reconstruction of selected top-quark pair events, and extraction
of the \topmass\ based on the invariant mass of its decay products. Employing
analysis techniques developed over many years at the Tevatron collider, the
first top mass measurements at the LHC appeared quickly after the start of data
taking. Designed to extract maximum possible statistical information from the
events, these techniques allowed a first hint of the top mass already in very
early LHC data, with just \ipb{3}~\cite{Khachatryan:2010ez}, and a first
measurement after \ipb{36} at \SI{7}{\TeV}~\cite{Chatrchyan:2011nb} using the
dilepton decay channel, as illustrated in \Fref{fig:FirstMass}.

\begin{figure}[htbp]
\subfig{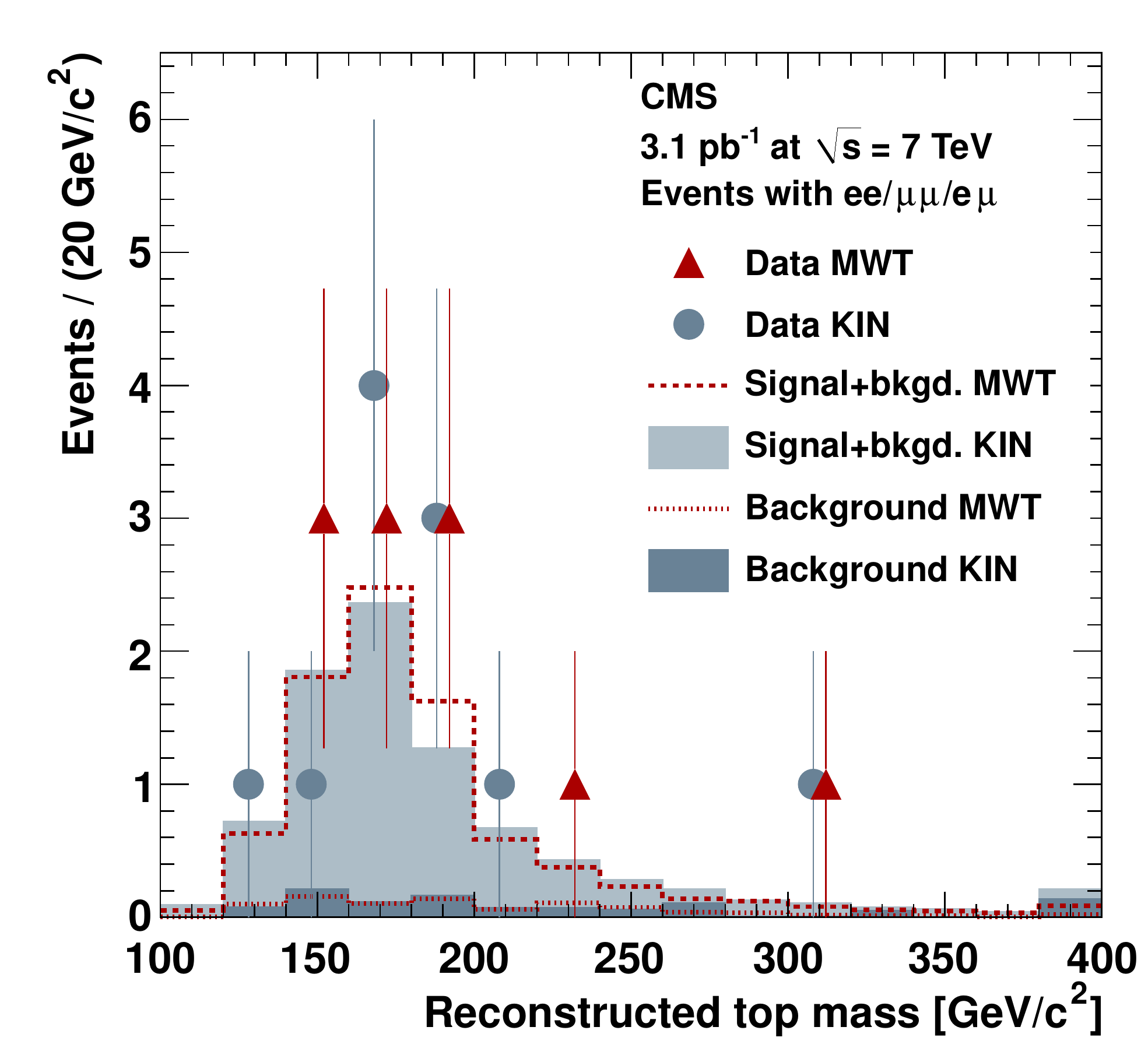}{0.51}{0}
\subfig{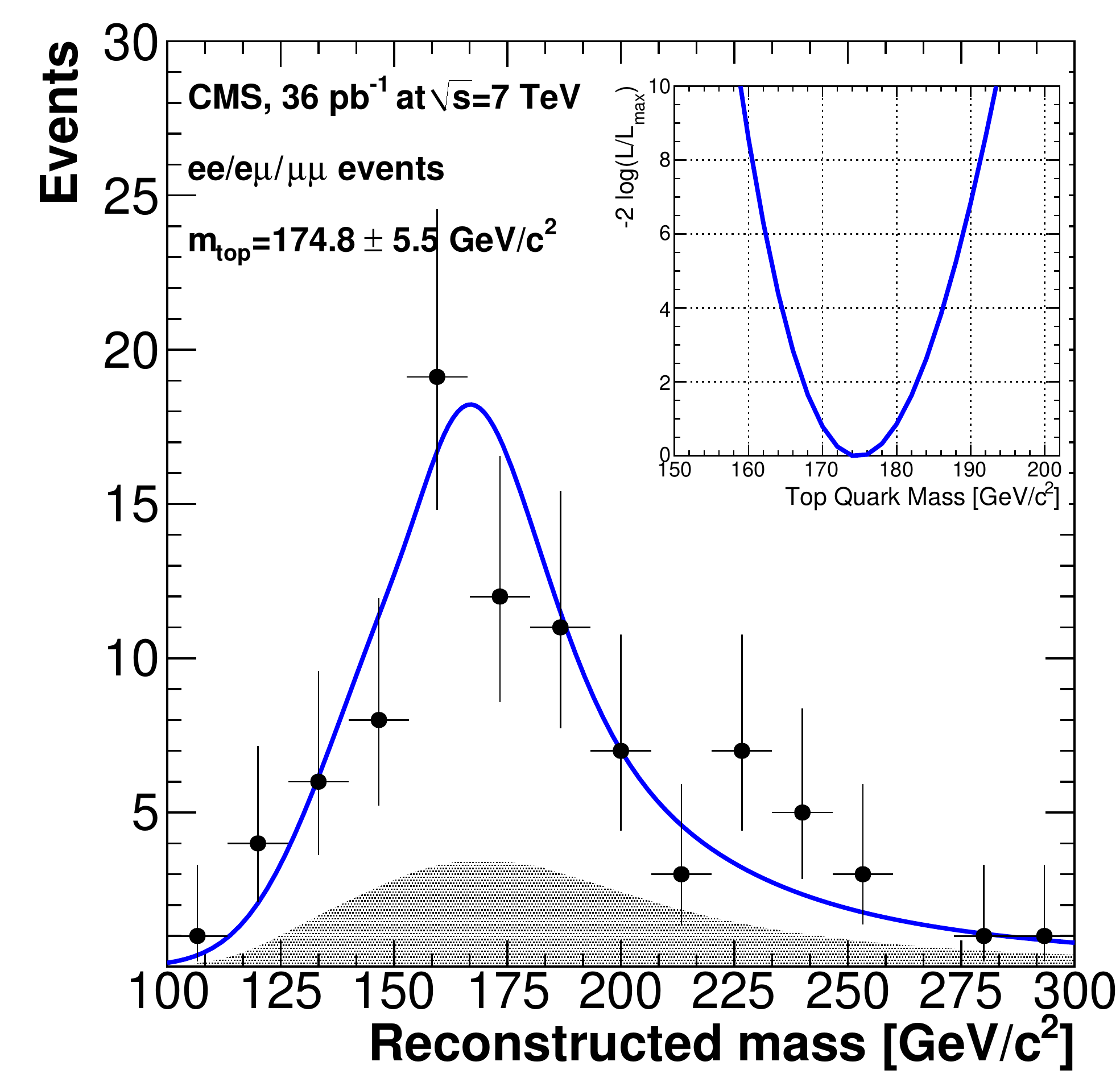}{0.49}{0}
\caption{Measurements of the \topmass\ in the dilepton channel, with
\ipb{3}~\pcite{Khachatryan:2010ez} (left) and \ipb{36}~\pcite{Chatrchyan:2011nb}
(right) of data at \SI{7}{\TeV}.} \label{fig:FirstMass}
\end{figure}

Other measurements soon followed, with more data and a better understanding of
the jet energy scale calibration. The \ljets\ has the advantage of combining
moderate backgrounds with a sizable branching fraction, and was employed by
ATLAS to perform the first measurement at the LHC with a relative precision
better than 2\%~\cite{ATLAS:2012aj} using \ifb{1} of data at \SI{7}{\TeV}.
Another advantage of the \ljets\ is that the \Wboson\ from one of the top quarks
decays hadronically, which gives the possibility to measure the jet energy scale
{\em in-situ}, in the selected \ttbar events. In \cite{ATLAS:2012aj} this was
done by introducing the $R_{32}$ variable, defined as the ratio of the
reconstructed mass of the three jets from the hadronic top decay $m_{bjj}$ and
the mass of the two jets out of the trijet system that are not $b$-tagged,
$m_{jj}$. Effectively this variable measures the ratio between the \topmass\ and
the \Wboson\ mass. Since the latter is known with excellent precision, this
variable reduces the sensitivity to the overall scale of the jet energies, and
it reduces event-by-event fluctuations due to the finite resolution of the
reconstructed $W$ mass $m_{jj}$. The \topmass\ was extracted from a template fit
of the $R_{32}$ distribution, shown in \Fref{fig:ATLAS_1_ifb}.

\begin{figure}[htbp]
\subfig{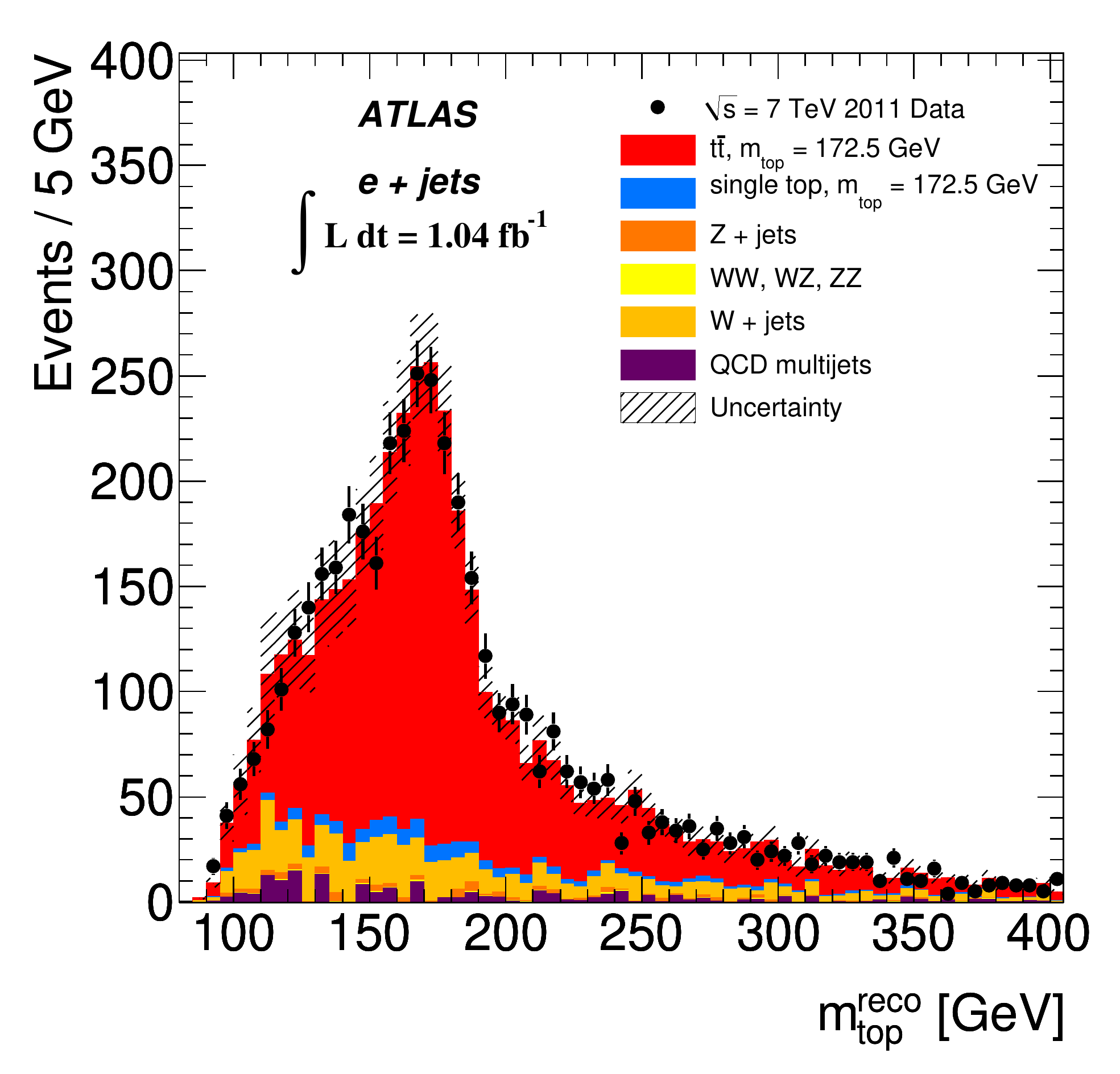}{0.47}{0}
\subfig{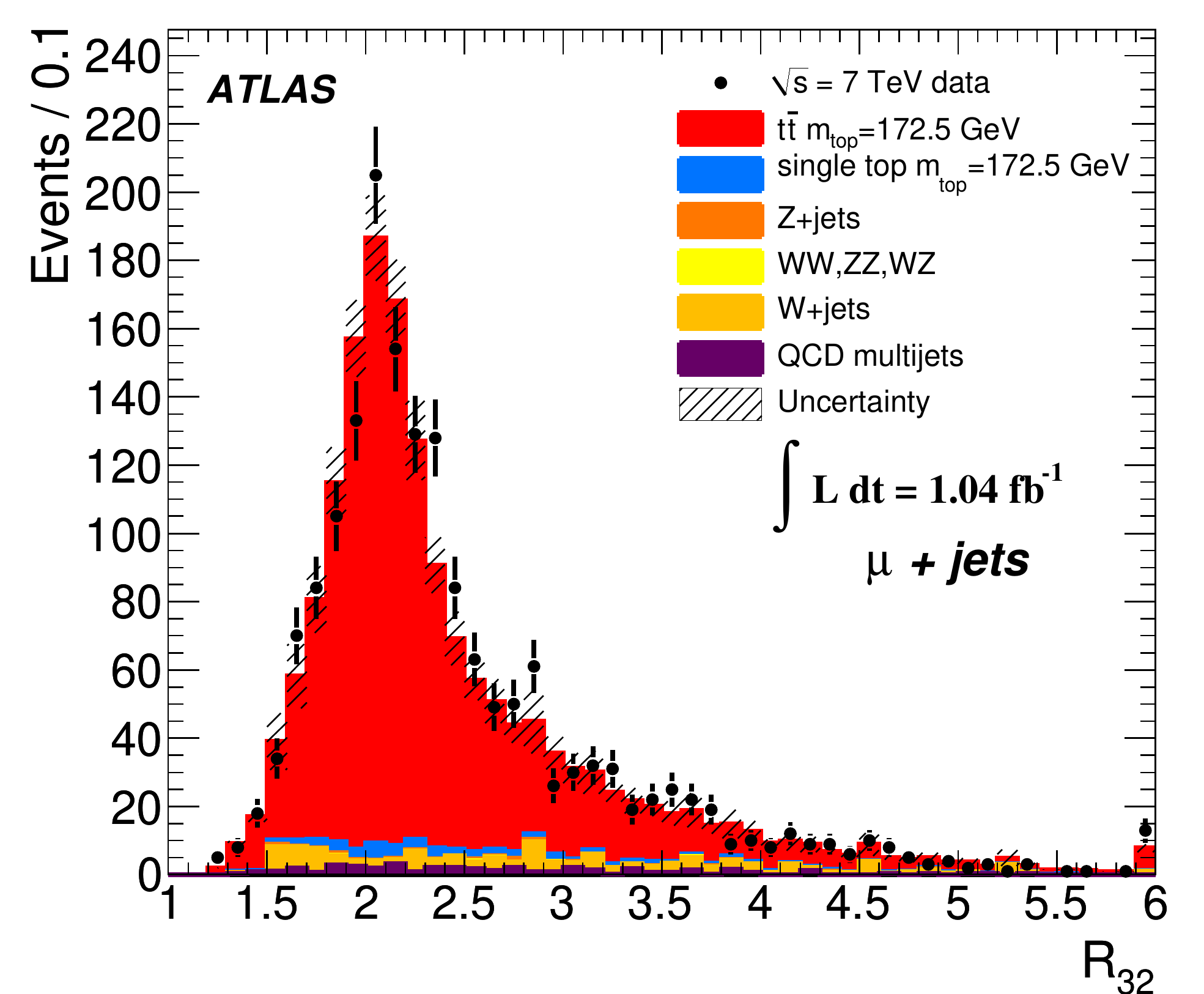}{0.53}{2}
\caption{Measurement of the \topmass\ in the \ljets\ by ATLAS
with partial \SI{7}{\TeV} data~\pcite{ATLAS:2012aj}.}
\label{fig:ATLAS_1_ifb}
\end{figure}

\subsection{Towards ultimate precision in the \ljets}

Using the full dataset available at \SI{7}{\TeV}, CMS performed a 2-dimensional
fit to determine an overall jet energy scale factor together with the
\topmass~\cite{Chatrchyan:2012cz}.  Templates of the reconstructed \Wboson\ mass
$m^{\rm reco}_W$ (\Fref{fig:SingleLeptonMass}, top left) were used together with
templates of the reconstructed \topmass\ $m^{\rm fit}_t$
(\Fref{fig:SingleLeptonMass}, top right) to construct an event-by-event {\em
ideogram} likelihood taking into account all jet assignments compatible with the
available $b$-tagging information. The event-by-event top mass reconstruction
was improved beyond the detector resolution by using a kinematic fit imposing
equality of the top quark and top anti-quark masses and constraining the
reconstructed \Wboson\ mass to the known value. This approach has the advantage
that one can explicitly check the compatibility of the fitted jet energy scale
factor with unity, the value corresponding to the original jet energy
calibration. While this allows the average overall jet energy scale to be
constrained, or more precisely the jet energy scale for light-flavour jets from
\Wboson\ decay, it does not constrain dependencies of the jet response due to
jets originating from quarks of different flavours, in particular $b$-jets. 

\begin{figure}[htbp]
\begin{center}
\includegraphics[width=0.44\textwidth]{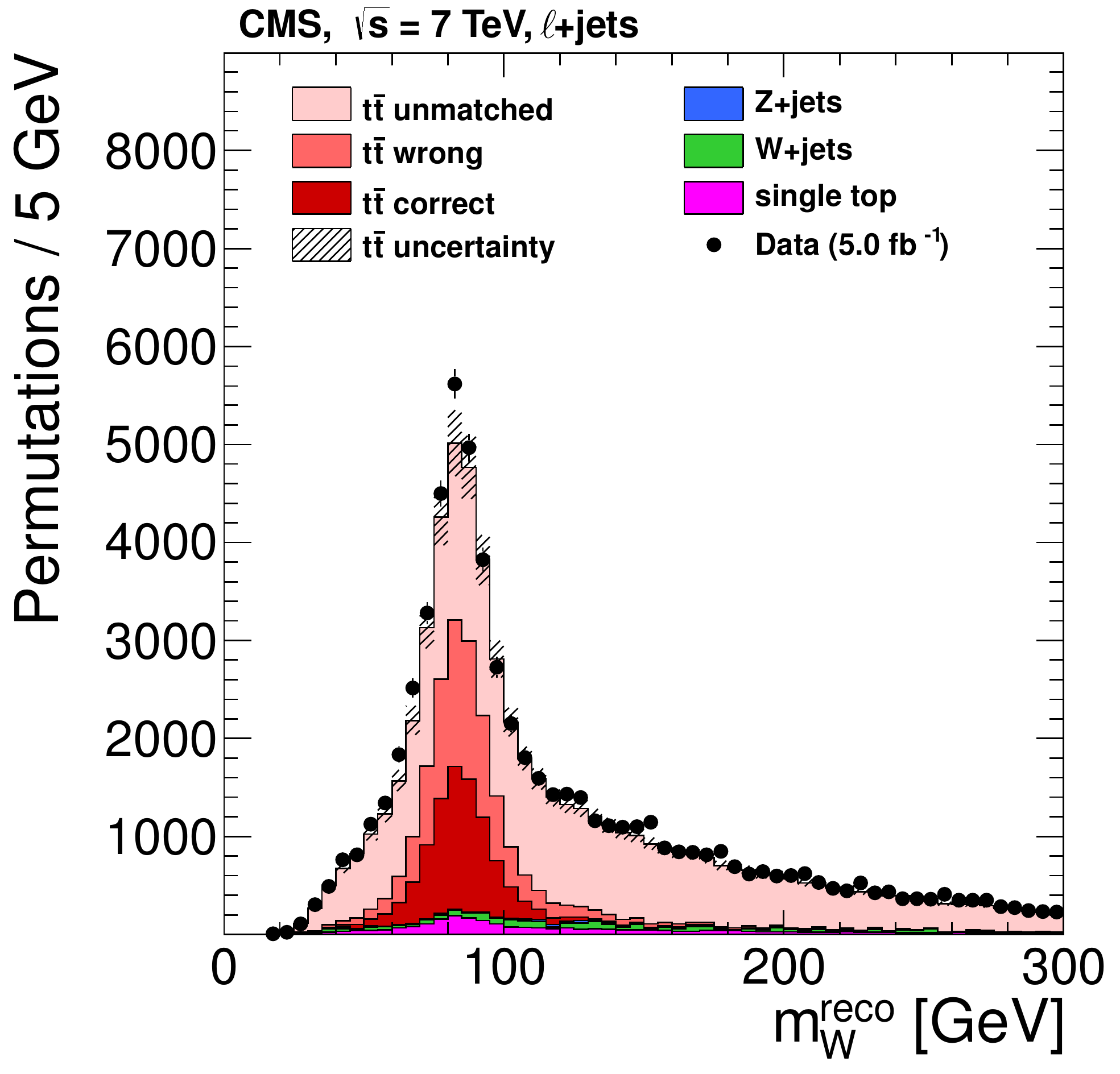}
\includegraphics[width=0.44\textwidth]{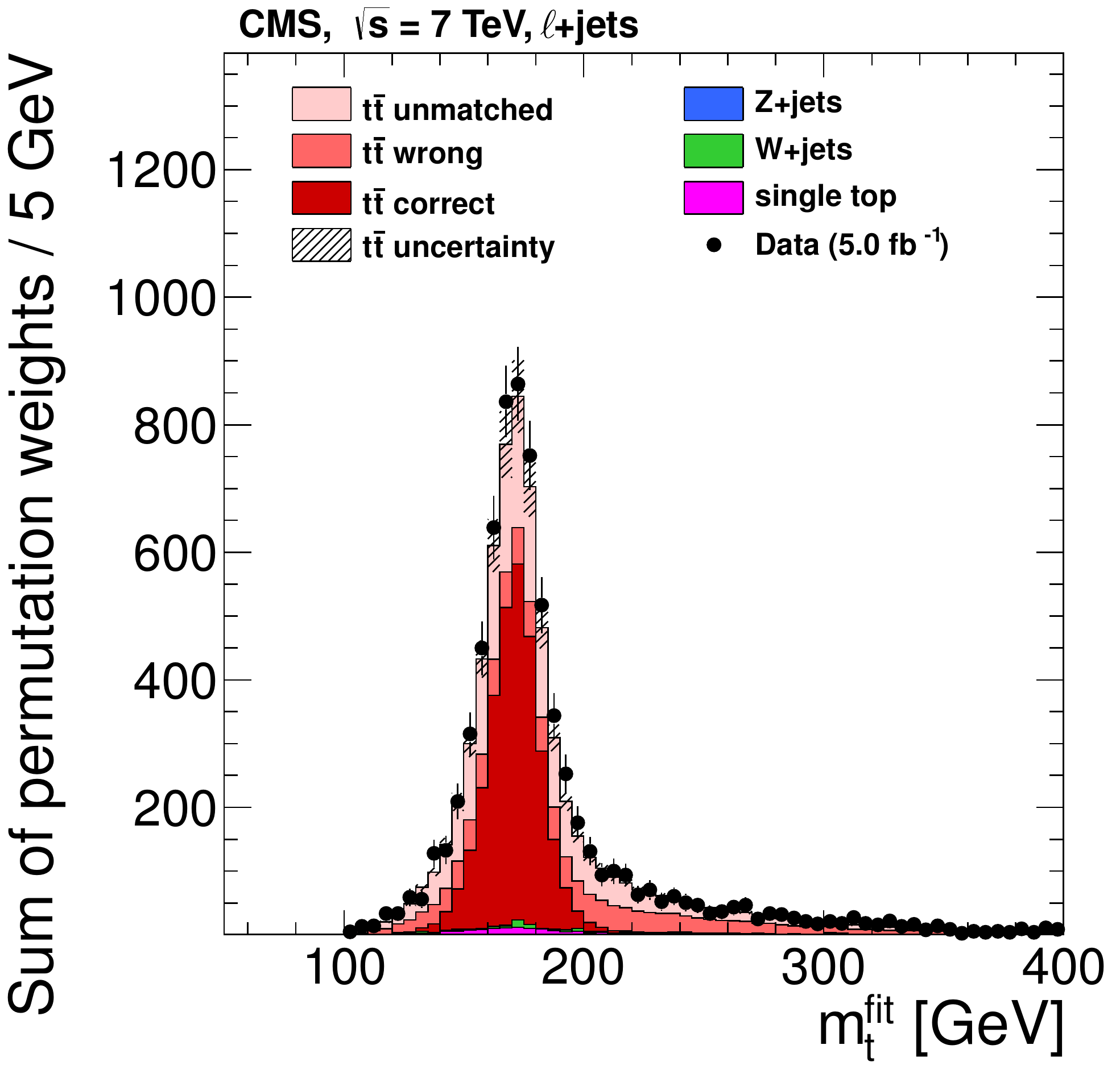}
\includegraphics[width=0.49\textwidth]{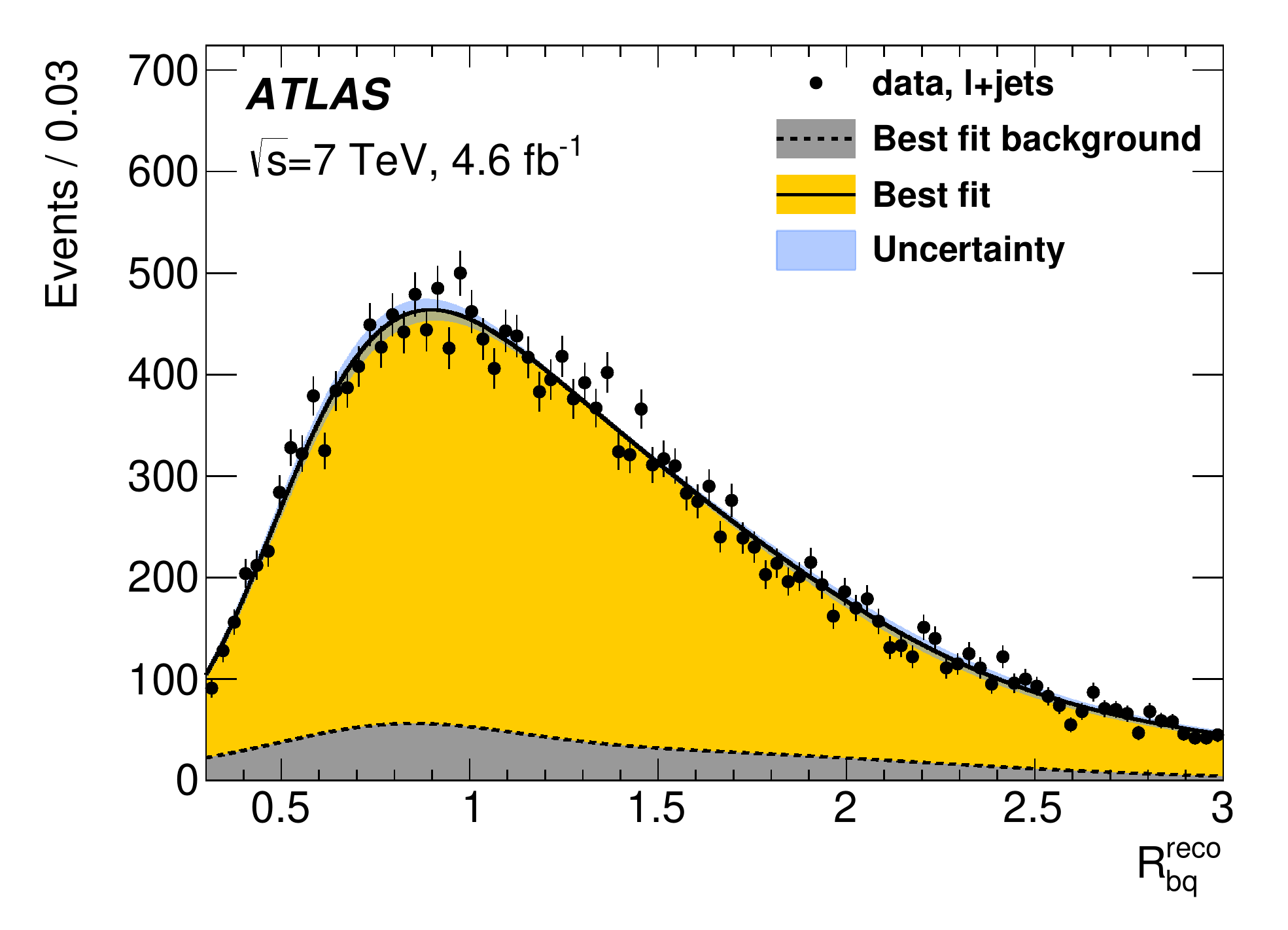}
\includegraphics[width=0.49\textwidth]{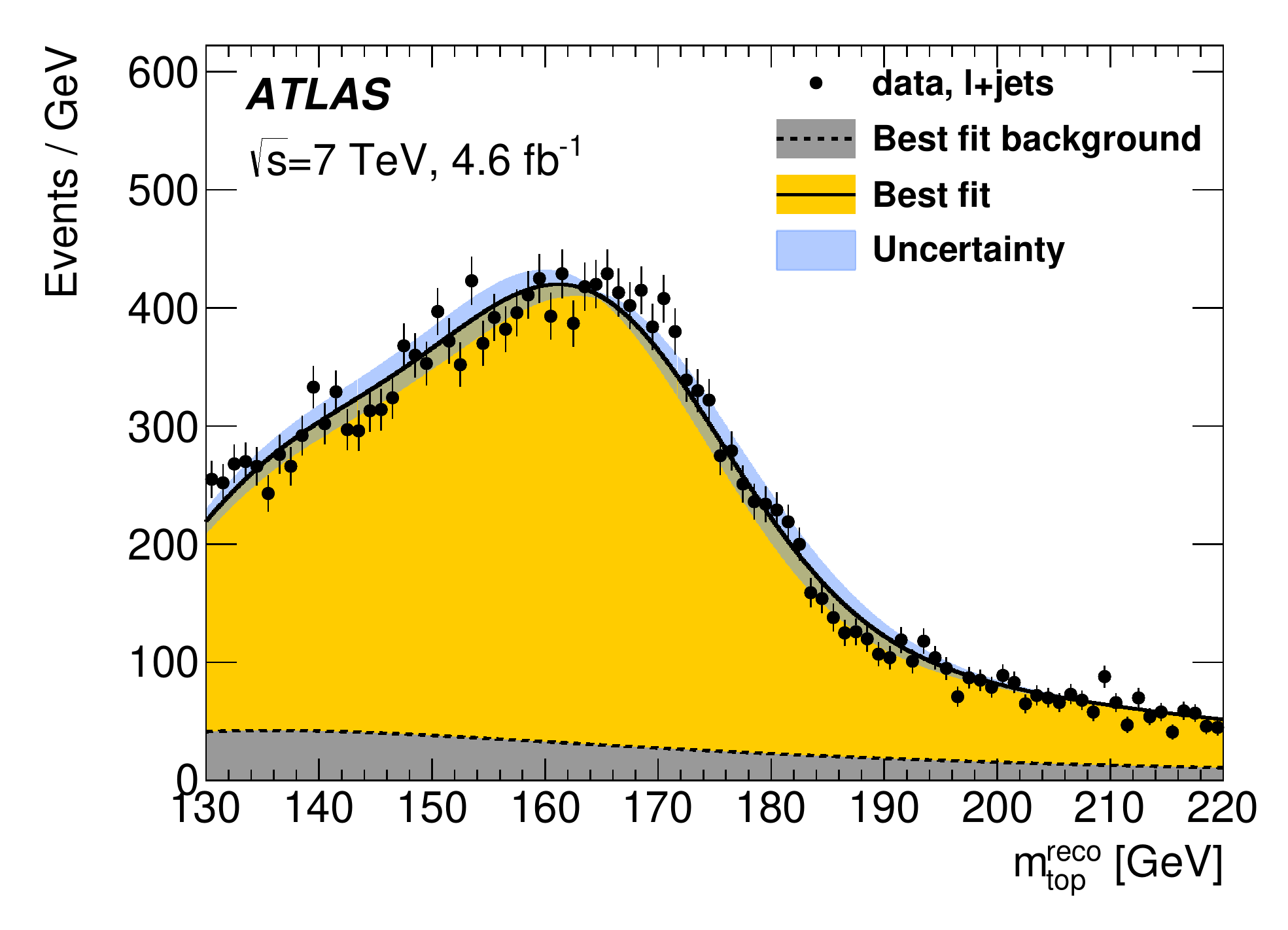}
\includegraphics[width=0.49\textwidth]{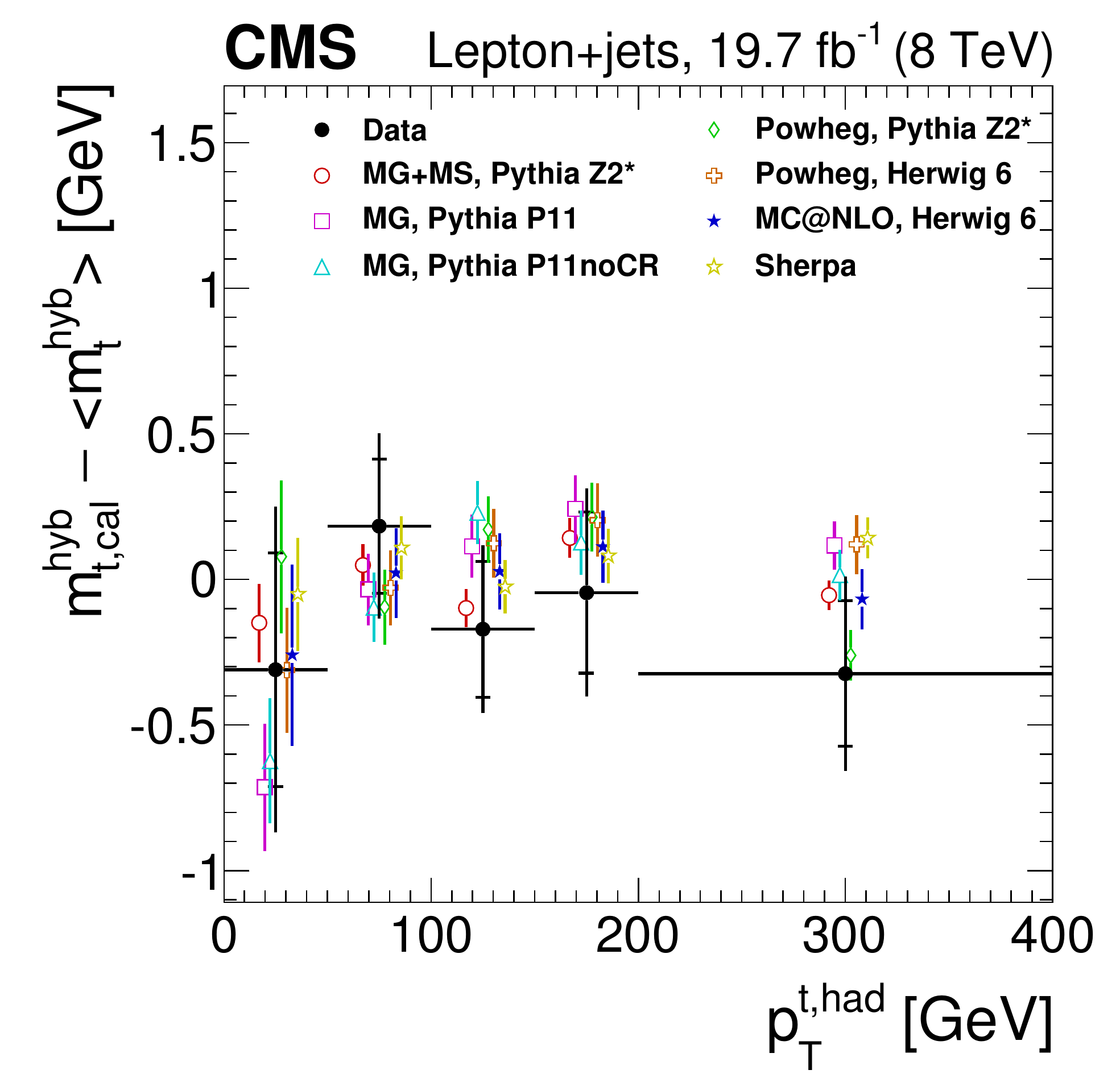}
\includegraphics[width=0.49\textwidth]{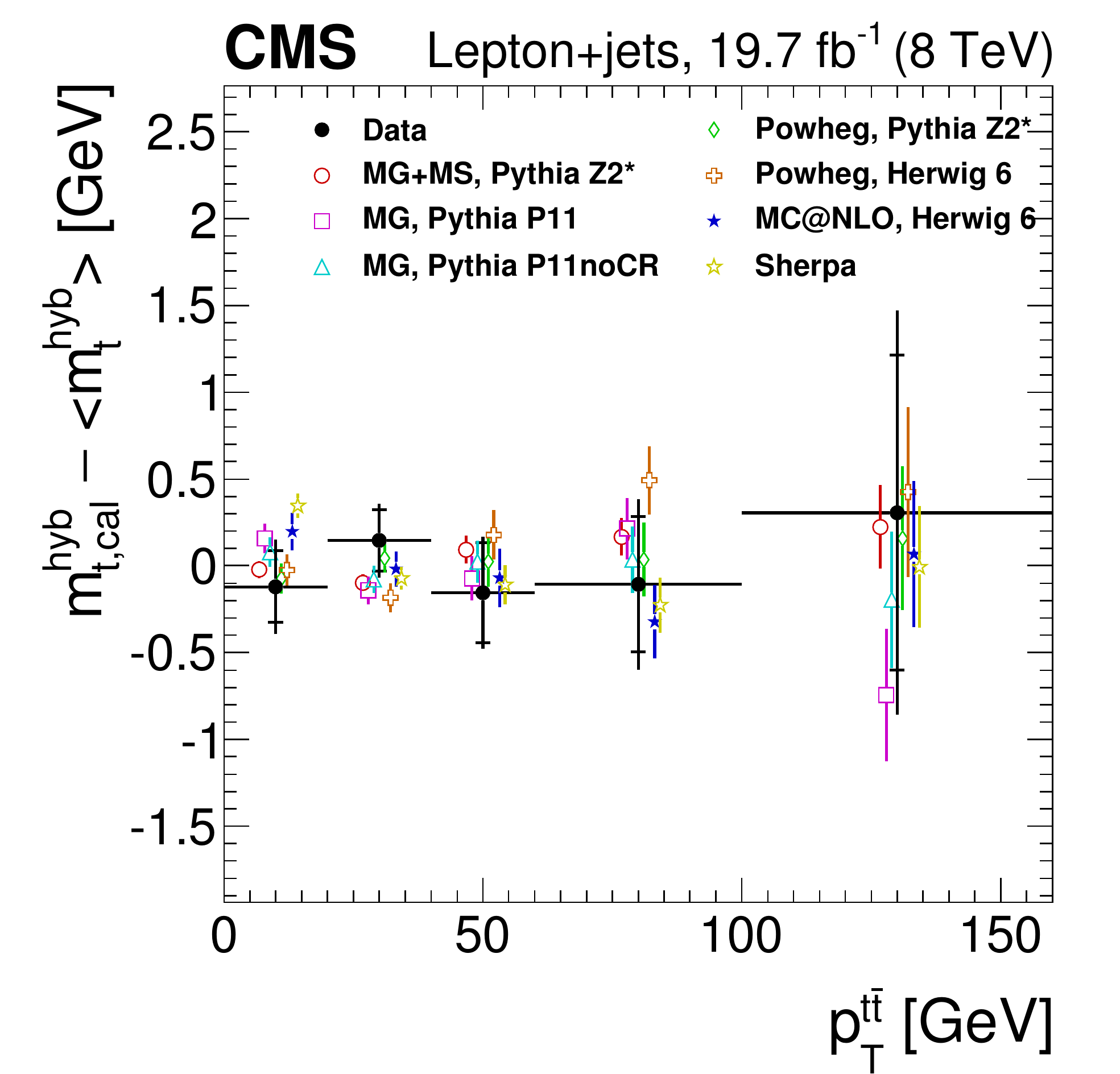}
\end{center}
\caption{Mass analysis in the \ljets\ using the full \SI{7}{\TeV} dataset in
CMS~\pcite{Chatrchyan:2012cz} (top) and ATLAS~\pcite{Aad:2015nba} (middle),
using the reconstructed mass of the \Wboson\ (top left) and transverse momentum
balance of $b$-tagged jets versus other jets (middle left) to improve the jet
energy scale calibration and reduce systematic uncertainty on the reconstructed
top mass, shown in plots on the right. The bottom plots show the stability of
the measured top mass as a function of kinematic event variables, using the
\SI{8}{\TeV} single-lepton analysis by CMS~\pcite{Khachatryan:2015hba}.}
\label{fig:SingleLeptonMass}
\end{figure}

The ATLAS analysis in the \ljets\ addressed this issue by performing a
3-dimensional fit of the \topmass, overall jet energy scale factor JSF, and an
additional scale factor bJSF describing any possible deviation between the jet
response in jets from light-flavour quarks and $b$ quarks~\cite{Aad:2015nba}. To
constrain bJSF, a new variable was introduced probing the transverse momentum
balance of $b$-tagged jets versus all other jets. A distribution of this
variable, $R^{\rm reco}_{bq}$ is shown in \Fref{fig:SingleLeptonMass} (middle
left), and the reconstructed top mass in \Fref{fig:SingleLeptonMass} (middle
right).  This analysis approach significantly reduces the otherwise dominant
systematic uncertainties due to $b$-jet energy calibration, at the cost of an
increased statistical uncertainty. With larger datasets at \SI{8}{\TeV} and
future LHC data beyond \RunOne, however, prospects are excellent for further
reduction of the overall measurement uncertainties.

At \SI{8}{\TeV} CMS employed the same ideogram likelihood technique as before,
but using a hybrid approach whereby the 2-dimensional fit and 1-dimensional fit
with fixed jet energy scale factor were combined with a weight, reflecting the
relative precision of the externally provided jet energy scale calibration and
the scale factor determined {\em in situ} in the selected \ttbar events.
Together with the advantage of the larger data sample this yielded the most
precise measurement of the top-quark mass to date, with a total relative
uncertainty of $3.0\,\permil$~\cite{Khachatryan:2015hba}.  The larger dataset
also allowed the stability of the measured \topmass\ to be tested as a function
of kinematic properties of the selected events. The difference in measured mass,
compared to the inclusive measurement, is shown in bins of the $p_{\rm T}$ of
the reconstructed top quark and of the \ttbar system in
\Fref{fig:SingleLeptonMass} (bottom). This study was done for eight variables,
and several generator setups, and no significant deviation in measured $m_t$ was
observed. The stability of the result in many corners of phase space is an
indication that the main physics effects that could possibly bias the top mass
measurement are accurately described by the MC simulation models used, and
properly corrected for in the analysis, at the current level of precision.

\subsection{State-of-the-art in the dilepton channel}

Thanks to the large \ttbar production \xs\ at the LHC, analyses in the dilepton
channel also yield sub-percent precision \topmass\ measurements, in spite of the
smaller branching fraction and presence of neutrinos.
CMS~\cite{Chatrchyan:2012ea} and ATLAS~\cite{Aad:2015nba} measured the \topmass\
in this channel at \SI{7}{\TeV}, already, and again at
\SI{8}{\TeV}~\cite{Khachatryan:2015hba}. While CMS still relied on full
kinematic reconstruction to obtain the best possible statistical precision, the
ATLAS measurement demonstrated that with the large number of \ttbar events
available at the LHC, simpler and potentially more robust variables can be
equally powerful. Using the invariant mass of the lepton from the \Wboson\ decay
and the corresponding $b$-jet, $m_{\ell b}$, a template fit was performed. The
advantage of using $m_{\ell b}$ is that the distribution of this variable can in
principle be predicted using fixed-order perturbative QCD calculations, opening
the door to measurements with a well-defined perturbative mass
scheme~\cite{Biswas:2010sa}.  The distributions for \SI{7}{\TeV} are shown in
\Fref{fig:MassDilepton}.

\begin{figure}[htbp]
\subfig{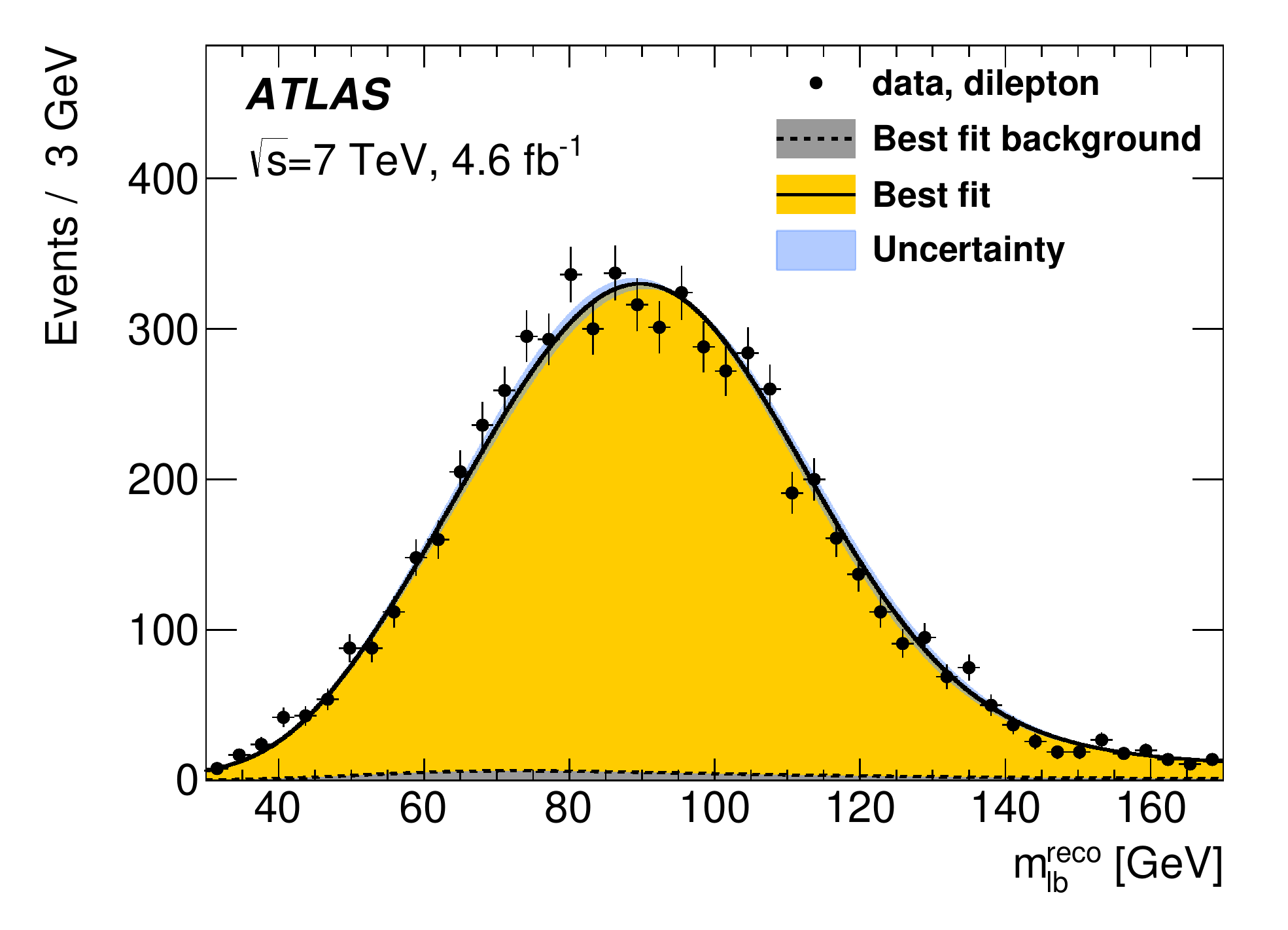}{0.515}{0}
\subfig{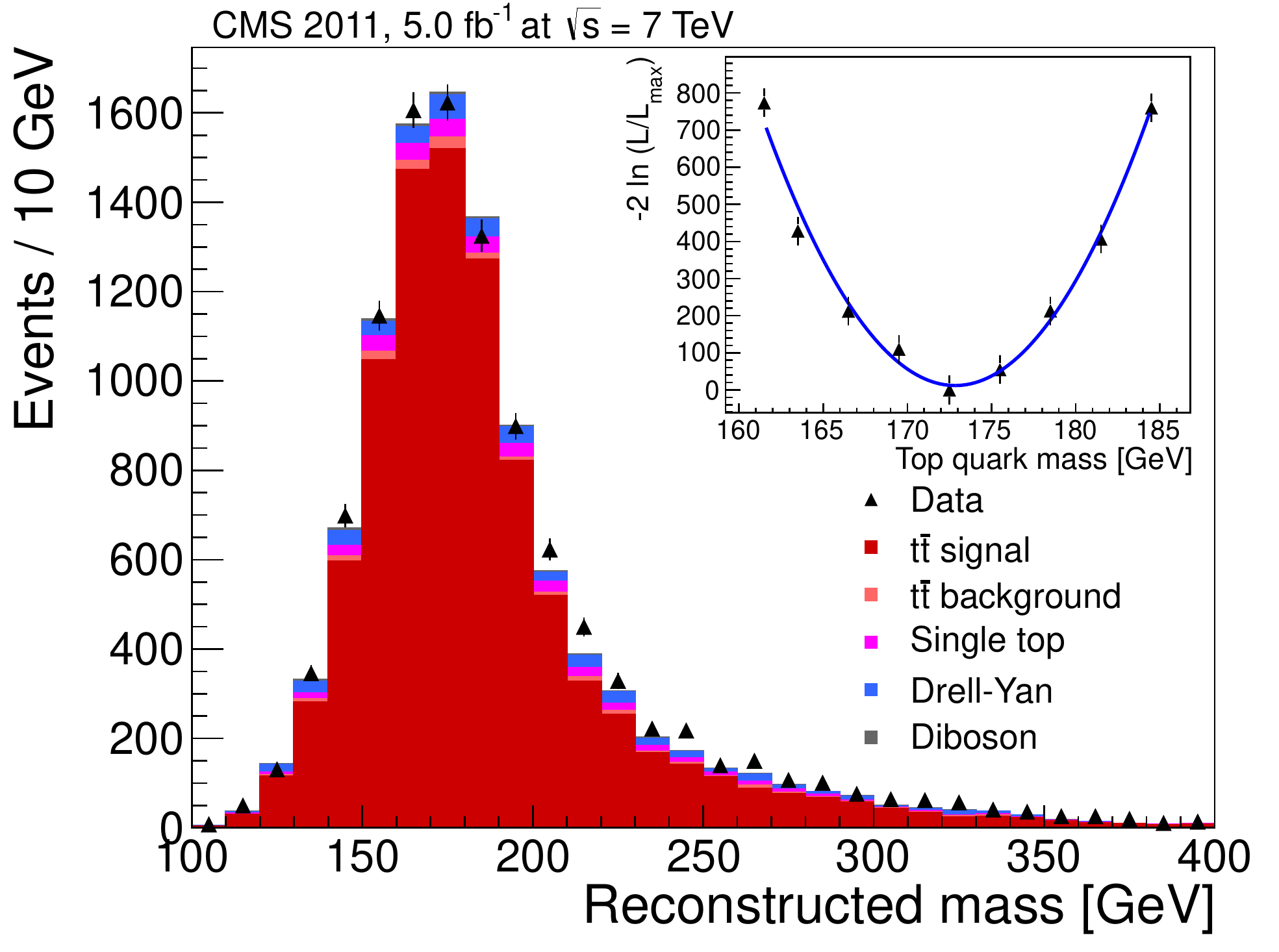}{0.485}{0}
\caption{Measurements of the \topmass\ in the dilepton channel by
ATLAS~\pcite{Aad:2015nba} (left) and CMS~\pcite{Chatrchyan:2012ea} (right).}
\label{fig:MassDilepton}
\end{figure}

\subsection{The all-hadronic channel}

At the LHC the \ttbar all-hadronic channel can also yield competitive mass
measurements, thanks to the more favourable signal-to-background
ratio~\cite{Aad:2014zea,Chatrchyan:2013xza}, compared to the Tevatron, albeit
with an increased number of additional jets from  ISR and FSR and corresponding
jet assignment ambiguities. The absence of neutrinos helps to achieve an
excellent event-by-event mass resolution.  To reduce the jet energy scale
uncertainty, ATLAS used a template fit to the $R_{32}$ variable introduced
before in the \ljets, as shown in \Fref{fig:Allhad} (left). In the case of CMS
the same ideogram likelihood fit as in the \ljets\ was performed, including the
\emph{in situ} constraint of the jet energy scale factor, but in the end a more
precise expected result was obtained at \SI{7}{\TeV} using a traditional
1-dimensional fit, keeping the jet energies fixed to the CMS standard jet
calibration obtained from other sources, yielding the fitted mass distribution
shown in \Fref{fig:Allhad} (right).  At \SI{8}{\TeV} CMS used the same hybrid
fit as in the \ljets, combining the 1- and 2-dimensional fits to obtain the best
precision.  The significant background from QCD multijet events poses two major
challenges for analyses in the all-hadronic channel. The jet \pT thresholds in
the trigger are high, forcing offline reconstruction selection thresholds to be
even higher, or alternatively requiring a very good understanding of the trigger
efficiency close to the trigger thresholds. Secondly, for the modelling of the
multijet background it is typically not feasible to generate Monte Carlo samples
with sufficiently large number of events, and the accuracy of the modelling of
multijet events is also hard to establish.  Therefore, dedicated methods were
developed to derive the background shapes from data events. The CMS analysis
employed a side-band technique, where the shape of the background was estimated
from events with six or more jets that did not pass the $b$-tagging
requirements~\cite{Chatrchyan:2013xza}, and the probability for individual jets
to be $b$-tagged is applied on a jet-by-jet basis.  The ATLAS analysis relied on
six regions, defined by two variables with minimal correlation, to determine the
shape, normalization and uncertainty on the normalization~\cite{Aad:2014zea}.

\begin{figure}[htbp]
\subfig{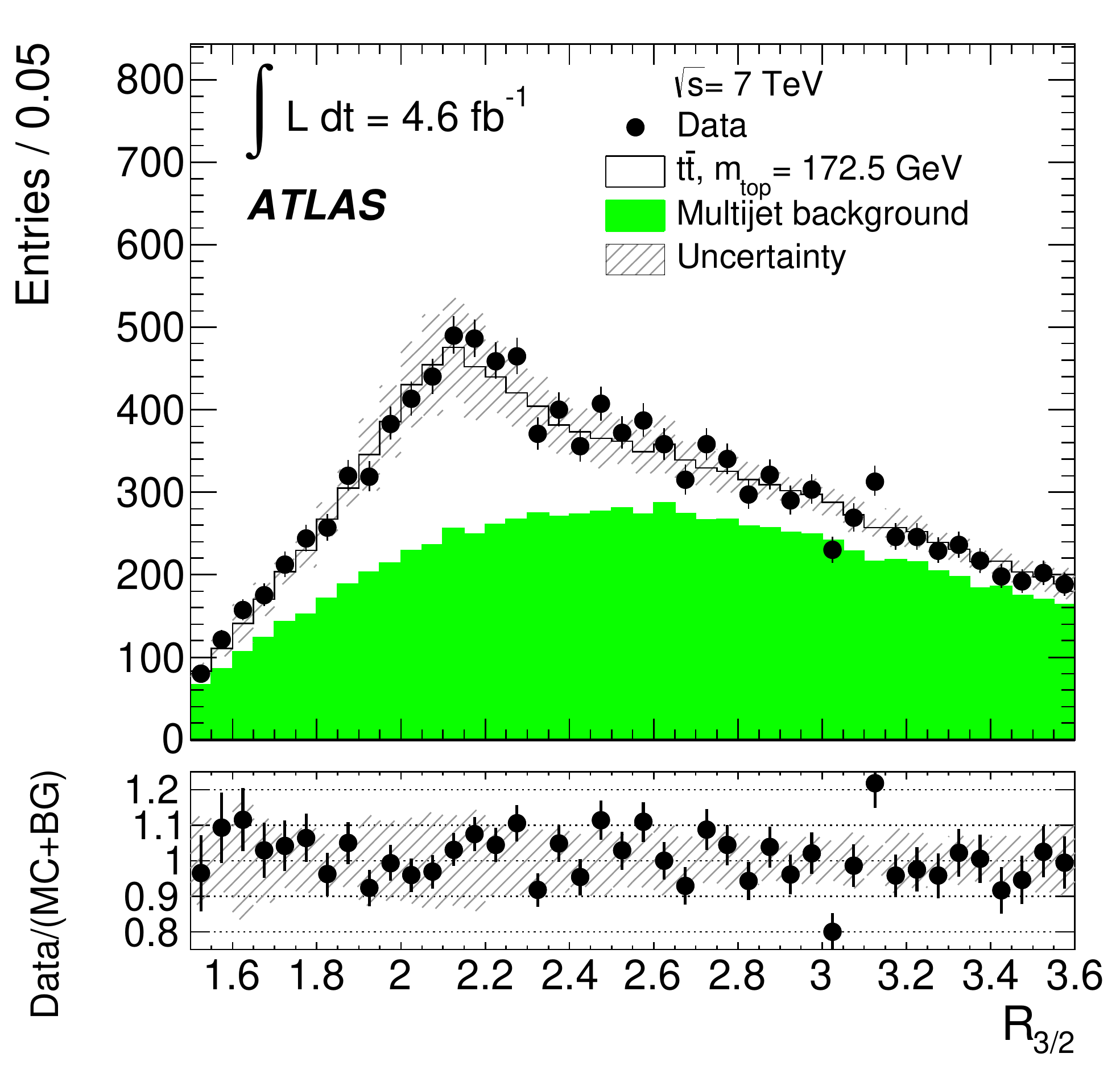}{0.52}{0}
\subfig{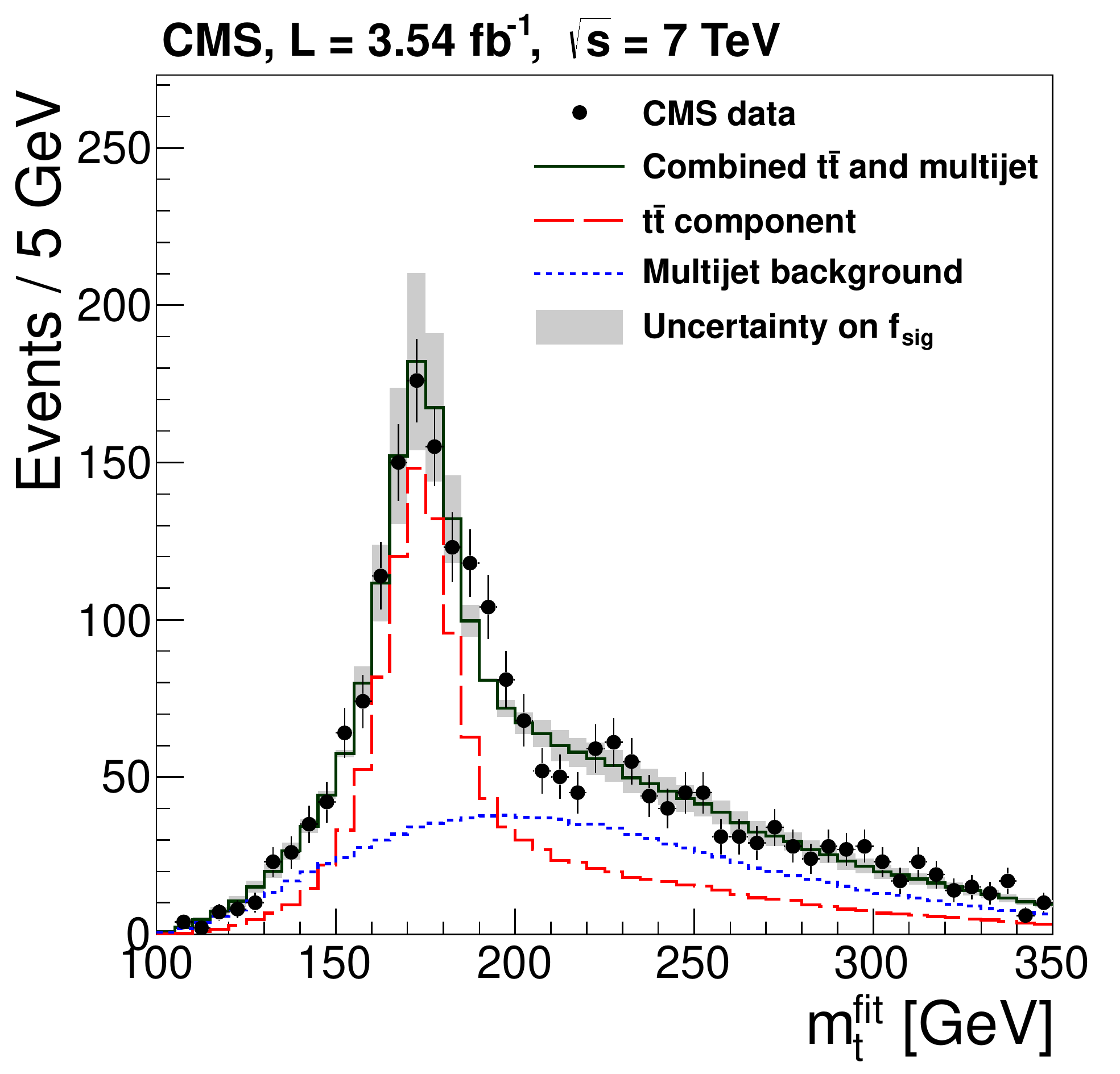}{0.52}{5}
\caption{Measurements of the \topmass\ in the all-hadronic channel by
ATLAS~\pcite{Aad:2014zea} (left) and CMS~\pcite{Chatrchyan:2013xza} (right).}
\label{fig:Allhad}
\end{figure}

\subsection{Difference between the top quark and top anti-quark mass}

Both ATLAS and CMS have measured the difference between the top quark and top
anti-quark mass, using data collected at \SI{7}{\TeV} in the \ljets. A genuine
difference in mass between particle and anti-particle would indicate a violation
of $CPT$ symmetry, invalidating a fundamental assumption of any quantum field
theory.  Also, for $pp$ collisions at the LHC, it is not excluded that QCD
effects might have a slightly different influence on the top quark and top
anti-quark decay products in the events. Thus, it is important to test that such
differences are negligible, or at least properly modelled by the event
simulation. 

The ATLAS analysis~\cite{Aad:2013eva} used a variable that measures the
difference between the top quark and anti-quark mass on an event-by-event basis,
shown in \Fref{fig:MassDifference} (left). The approach adopted by
CMS~\cite{Chatrchyan:2012uba} was to use the charge of the semileptonically 
decaying top (anti-)quark to decide whether the decay was from a top quark or
anti-quark, and to reconstruct the mass from the hadronic decay of 
the other (anti-)quark in the same event, as shown for the top anti-quark 
sample in \Fref{fig:MassDifference} (right). Thus the
masses were extracted separately for the top quark and anti-quark samples using
the ideogram technique.  Both experiments measured a difference compatible with
equality of the masses, with a precision of $0.5 -$\SI{0.7}{\GeV}.  Many of the
systematic uncertainties cancel by taking the difference, or are at least
strongly reduced. The measurements are still limited by the statistical
precision, and have good potential to become more precise from the analysis of
the larger data samples recorded at a centre-of-mass energy of \SI{8}{\TeV} or
to be recorded in future LHC runs.

\begin{figure}[htbp]
\subfig{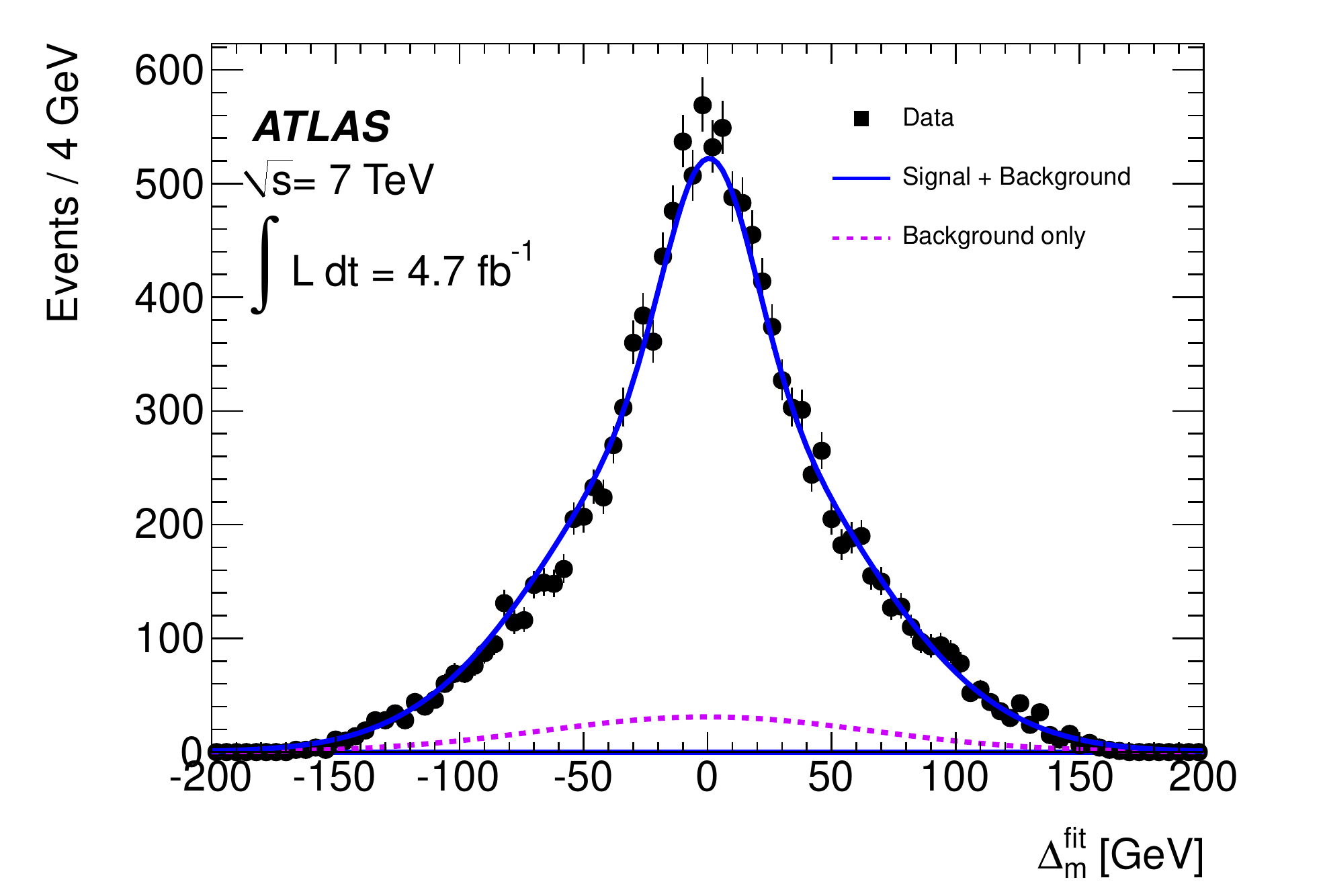}{0.535}{0}
\subfig{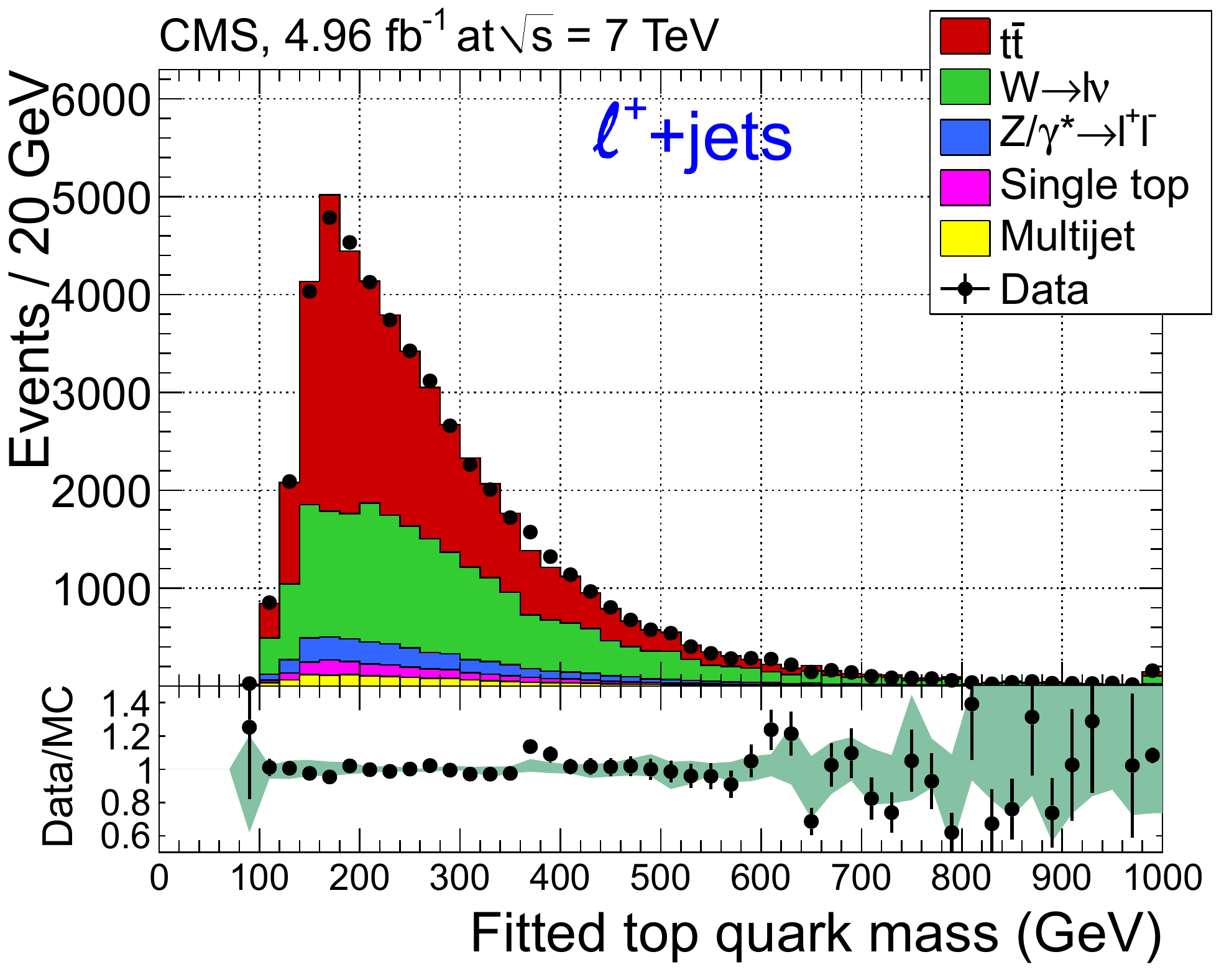}{0.465}{0}
\caption{Difference of top quark and anti-quark mass per event, estimated by
ATLAS~\pcite{Aad:2013eva} (left), and reconstructed mass from the hadronically
decaying top anti-quarks in the sample with positively charged leptons for
CMS~\pcite{Chatrchyan:2012uba} (right).} \label{fig:MassDifference}
\end{figure}

\subsection{World average combination}

The four experiments ATLAS, CDF, CMS and D0 prepared a combined result of the 
most precise \topmass\ measurements available as of March 2014, yielding 
the following world-average value of the \topmass~\cite{ATLAS:2014wva}:
\begin{equation*}
m_t = 173.34 \pm \SI{0.76}{\GeV}. 
\label{eq:waMass}
\end{equation*}

As can be seen in \Fref{fig:WorldAverageMass} the world average value is in good agreement 
with the results obtained at the LHC, including the measurements that appeared more recently.

\begin{figure}[htbp]
\begin{center}
\includegraphics[width=\textwidth]{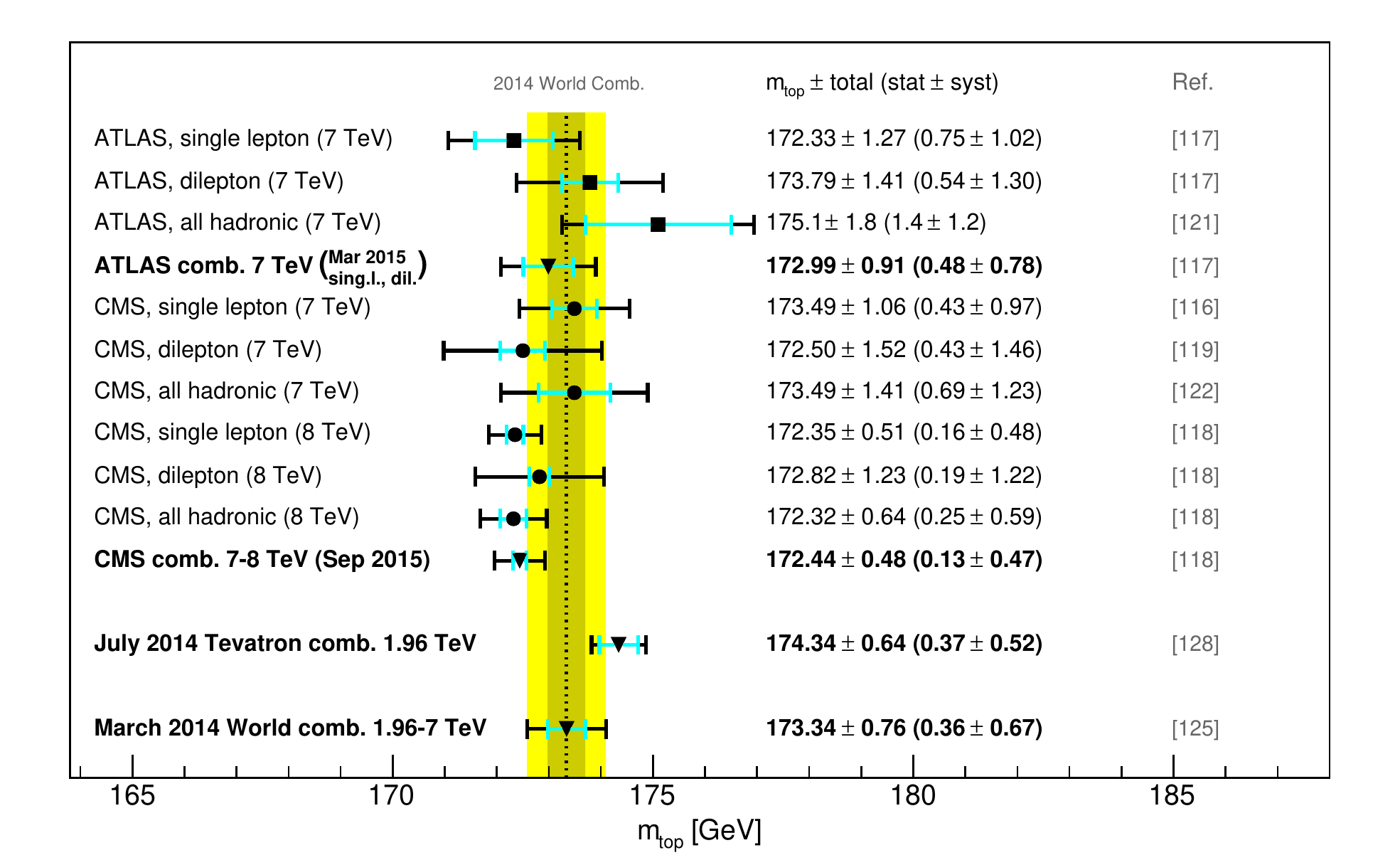}
\end{center}
\caption{Summary of the most precise LHC \topmass\ measurements by ATLAS and CMS 
in each channel, obtained at 7 and \SI{8}{TeV} and their combinations, compared to the 
2014 world average~\cite{ATLAS:2014wva}.}
\label{fig:WorldAverageMass}
\end{figure}

The latest ATLAS measurements~\cite{Aad:2015nba,Aad:2014zea} using the full 
\SI{7}{\TeV} dataset were not included in the world average, but are in good agreement. 
The combination of the results in the dilepton and \ljets\ 
yields~\cite{Aad:2015nba} $m_t = 172.99 \pm \SI{0.91}{\GeV}$.

The latest CMS measurements in all decay channels at
\SI{8}{\TeV}~\cite{Khachatryan:2015hba} used analysis methods that are very
similar to the \SI{7}{\TeV} analyses. Thanks to larger datasets, an improved jet
energy calibration, and small improvements in the methods, the final uncertainty
was significantly reduced in spite of a more complete treatment of systematic
uncertainties. The overall combined CMS result becomes $m_t = 172.44 \pm
\SI{0.48}{\GeV}$; a bit lower in value but in agreement with the world average.

Also at the Tevatron new top-quark mass measurements were published after March
2014.  In particular the measurement by D0 using the \mbox{Run II} data set of
\ifb{9.7} in the \ljets\ yielded a very precise result of $m_t = 174.98 \pm
\SI{0.76}{\GeV}$~\cite{Abazov:2014dpa,Abazov:2015spa}. Including this new
measurement the updated Tevatron mass combination, $m_t = 174.34 \pm
\SI{0.64}{\GeV}$~\cite{Tevatron:2014cka}, appears to have converged to a
somewhat higher value than the latest LHC measurements as shown in
\Fref{fig:WorldAverageMass}. A careful evaluation of correlations between
sources of systematic uncertainty for the latest measurements at the Tevatron
and the LHC is underway, in preparation for the next update of the world average
combination. 

\subsection{Interpretation of the top quark mass measurements}

With a precision below 0.5\% in the best results it is important to define
exactly what quantity is measured. All measurements discussed so far, and
included in the combinations, are calibrated using Monte Carlo simulations and
thus measure the Monte Carlo top mass parameter as a matter of definition. A
proper relation between this Monte Carlo mass parameter and the SM $m_t$
parameter used in field theoretical calculations is lacking. For the \topmass\
the choice of a renormalization scheme for the theoretical mass definition is
known to have large effects on the numerical value of the obtained mass. The
difference between two popular schemes, \MSbar and the pole mass, is as large as
\SI{10}{\GeV}. Once the mass is known in a short-distance scheme like the \MSbar
scheme, however, it can be transferred to a different short-distance mass scheme
with good precision, typically well below \SI{50}{\MeV}~\cite{Marquard:2015qpa}.
However, the pole mass scheme is not a short-distance scheme and it is affected
by the so-called renormalon ambiguity which limits the precision of its
translation to other mass schemes to about \SI{70}{\MeV} at
best~\cite{Marquard:2015qpa,Beneke:2016cbu}.  It has been argued that the Monte
Carlo mass parameter is expected to be close in numerical value to the pole mass
definition, with an unknown offset that is thought to be up to \SI{1}{\GeV}, and
that it would be preferable to avoid the use of the pole mass scheme
all-together and directly relate the Monte Carlo mass to a short distance mass
definition like the MSR mass~\cite{Hoang:2014oea,Moch:2014tta,Hoang:2008yj}. For
a proper choice of scale parameter this scheme could have a numerical value
close to the pole mass, without suffering from renormalon ambiguities.

To establish quantitatively the size of a possible offset between the Monte
Carlo mass and a suitable theoretical mass definition it is useful to compare
the Monte Carlo prediction for a physical observable to a prediction from a
first-principles QCD calculation with corrections up to the level of stable
particles after radiation and hadronisation. In principle the offset can be
different for different observables, so the calibration would have to be
performed for each observable of interest. 

One could also argue that in practice the differences between the MC mass and a
pole mass definition are related to perturbative and non-perturbative QCD
corrections that are to first approximation described by the MC simulations and
to some extent already covered by systematic uncertainties related to
theoretical modelling already assigned. If this is true, different methods using
different observables should yield top mass results that are compatible within
the assigned uncertainties. 

Either way, a possible bias in the mass definition could depend in principle on
the observable, and it is therefore useful to measure the top mass with as many
different observables as possible, and to explore alternative methods that allow
the use of a well-defined QCD calculation without the use of a Monte Carlo
program as intermediate step. 

\subsection{Non-standard methods}

In pursuit of a more complete picture of the universality of top mass
measurements, several alternative methods have been pursued at \RunOne. 

The CMS collaboration has documented the kinematic endpoint method in the
dilepton channel\cite{Chatrchyan:2013boa}, which uses purely kinematic
predictions for the endpoint of lepton-related distributions. 

Both experiments have also extracted the top-quark pole mass by comparing the
\ttbar \xs\ calculated at NNLO+NNLL accuracy with the measured \xs s at 7 and
\SI{8}{\TeV}~\cite{Chatrchyan:2013haa,Aad:2014kva,Khachatryan:2016mqs}.  The
analyses were already described in \Sref{sec:ttbar} and the results are
illustrated in \Fref{fig:PoleMass}.  While this is a theoretically very clean
method to extract the \topmass, the precision is less than what can be achieved
with the direct measurements of the invariant mass of the top decay products.
Methods that promise to yield a mass measurement with improved precision, in a
well-defined pole-mass scheme are the proposed $\ttbar+$jet
method~\cite{Alioli:2013mxa} or the use of $m_{\ell b}$~\cite{Biswas:2010sa}
with a fixed-order QCD prediction. Applying the $\ttbar+$jet method to the
\SI{7}{\TeV} dataset, the ATLAS collaboration obtained a measurement of the
top-quark mass in the pole mass scheme~\cite{Aad:2015waa} with a precision of
\SI{2.1}{\GeV}, with a potential to obtain further improvements using the larger
\SI{8}{\TeV} dataset. 

\begin{figure}[htbp]
\subfig{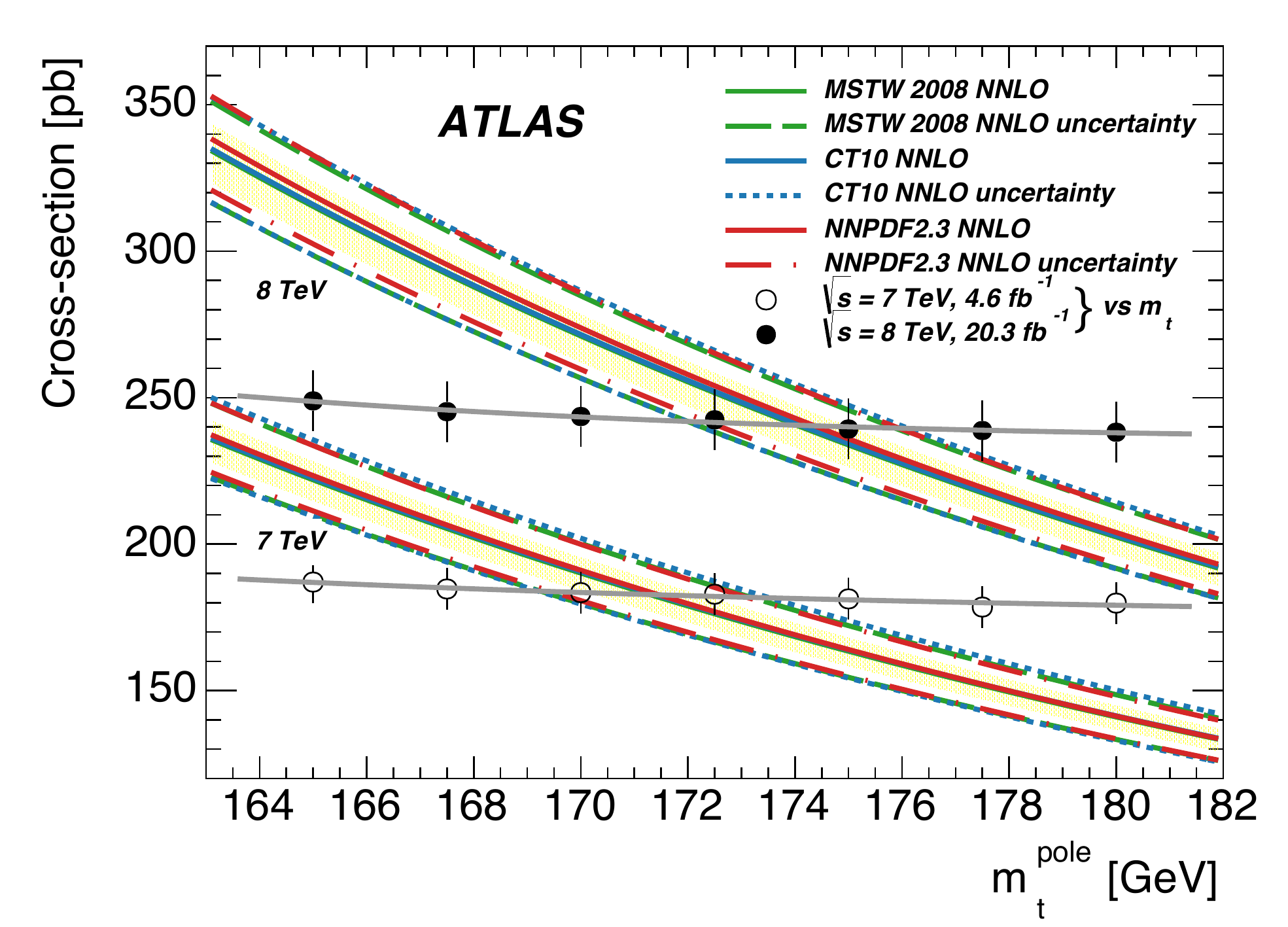}{0.5}{0}
\subfig{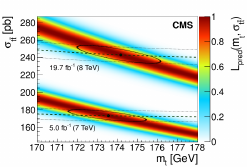}{0.5}{7}
\caption{Extraction of the top-quark pole mass from the measured inclusive
\ttbar \xs\ by ATLAS~\pcite{Aad:2014kva} (left) and
CMS~\pcite{Khachatryan:2016mqs} (right).} \label{fig:PoleMass}
\end{figure}

Several other methods have also been proposed with the underlying goal of
measuring the \topmass\ using different observables, thus obtaining
complementary information. Ultimately this will help to obtain the most accurate
measurement of the top mass and also provide limits on scenarios in which
physics beyond the Standard Model might affect some observables more or
differently than others.  For example, the use of only leptonic
variables~\cite{Frixione:2014ala} has been proposed to minimize potentially
poorly understood and modelled effects of non-perturbative QCD. Other proposals
advocate the use of the $b$-jet energy spectrum~\cite{Agashe:2012bn}, the
invariant mass of a $J/\psi$ meson and an isolated
lepton~\cite{Kharchilava:1999yj} or the boost of the $b$-jets extracted from the
transverse $b$-hadron decay length measured with charged particles 
only~\cite{Hill:2005zy}. The
latter approach was implemented initially by CDF and extended by CMS to use the
invariant mass of the $b$-jet secondary vertex and the lepton from \Wboson\
decay~\cite{Khachatryan:2016wqo}, still using only charged particles in the
measurement, thus avoiding systematic uncertainties related to full
reconstruction of jets.  The results are promising and indicate a potential to
reach sub-\si{\GeV} precision with this method in the near future. 

An overview of existing LHC measurements with alternative observables, compared
to the world average top mass result is shown in \Fref{fig:MassAlternative}. So
far all results agree with each other and with the more precise LHC measurements
from full reconstruction. With uncertainties of \SI{1.5}{\GeV} or larger it is
too early to resolve hypothesized sub-\si{GeV} differences related to the choice of
the observable and of the mass scheme definition.

\begin{figure}[htbp]
\begin{center}
\includegraphics[width=\textwidth]{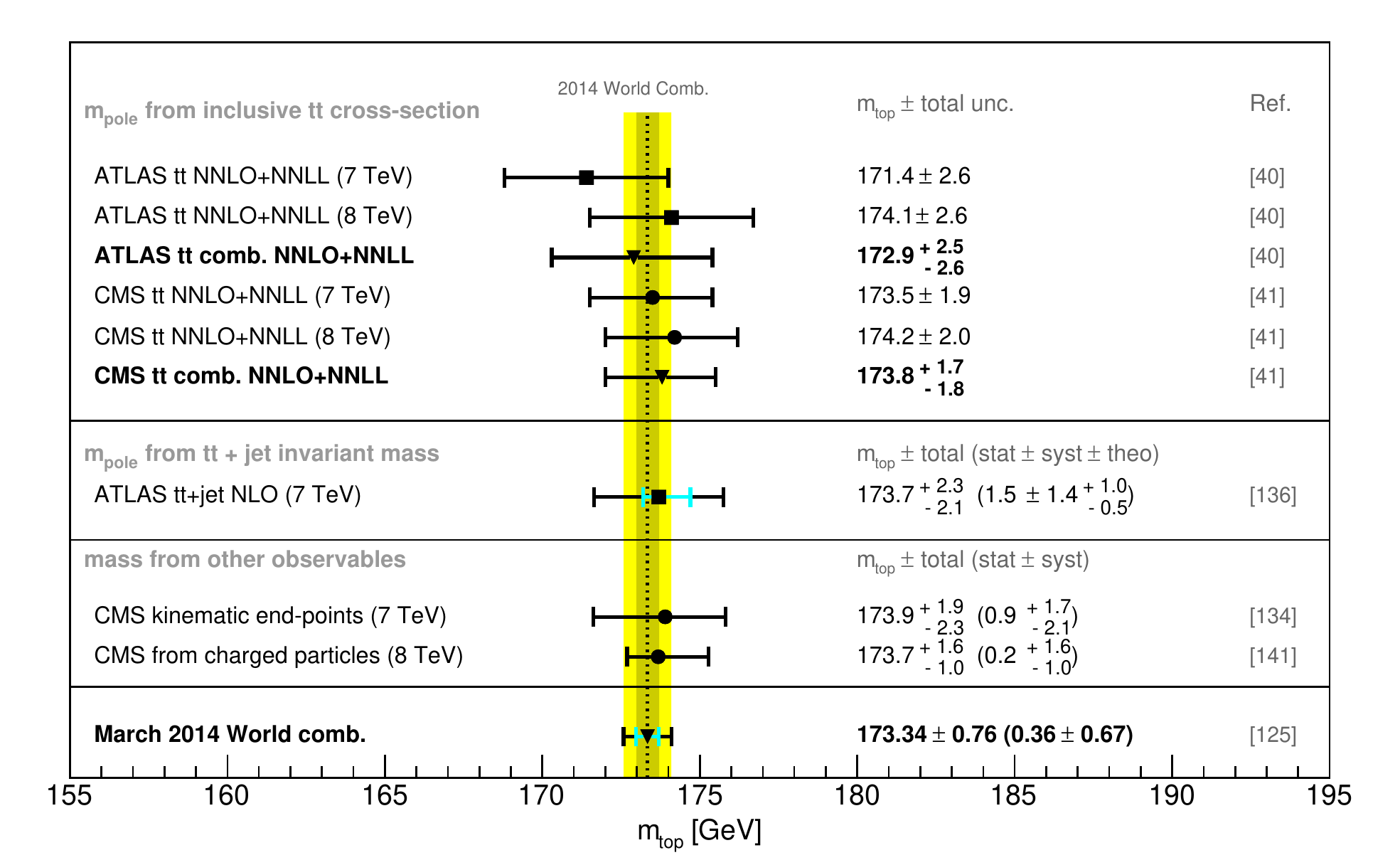}
\end{center}
\caption{Summary of \topmass\ measurements using alternative methods, compared to the 2014
top mass world average.} \label{fig:MassAlternative}
\end{figure}

The measurement of the \topmass\ and its theoretical interpretation is clearly a
very active field and there is still potential for improvements, from further
analysis of \RunOne\ data, theoretical developments, as well as from the future
LHC data.

\section{Top-quark spin}

The average lifetime of top quarks is a factor of $\sim\!\!10$ smaller than the
hadronization timescale ($1/\Lambda_{\mathrm{QCD}} \sim \SI{e-24}{s}$), which in
turn is smaller than the spin decorrelation timescale
$m_\mathrm{t}/\Lambda_{\mathrm{QCD}}^2 \sim \SI{3e-21}{s}$. This means that the
top-quark states maintain their spin from the time of production to the time of
decay, and information about the spin state is preserved in the distributions of
the top-quark decay products. 

\subsection{Top-quark polarization}

In the case of \ttbar production, which happens predominantly through the strong
interaction, parity conservation in QCD implies that the top quarks are produced
with zero longitudinal polarization. Small corrections from the weak interaction
have a negligible effect.  A deviation from this SM prediction would be a clear
indication of physics beyond the Standard Model. At \SI{7}{\TeV}, the ATLAS
collaboration performed a measurement combining the dilepton and \ljets
s~\cite{Aad:2013ksa} and CMS in the dilepton channel~\cite{Chatrchyan:2013wua,
Khachatryan:2016xws}, shown in \Fref{fpolarisation}, all confirming negligible
polarization of top quarks in \ttbar production, in agreement with the SM
prediction.

Production of single-top quarks, on the other hand, proceeds through the weak
interaction, leading to a strong polarization of the top (anti-)quarks. Using
single top-quark events produced through the $t$-channel mode, clear indications
of this polarization were found by CMS at
\SI{8}{\TeV}~\cite{Khachatryan:2015dzz}, albeit with a somewhat smaller strength
than predicted by the SM at NLO (\Fref{fpolarisation}, bottom right). 

\begin{figure}[htbp]
\subfig{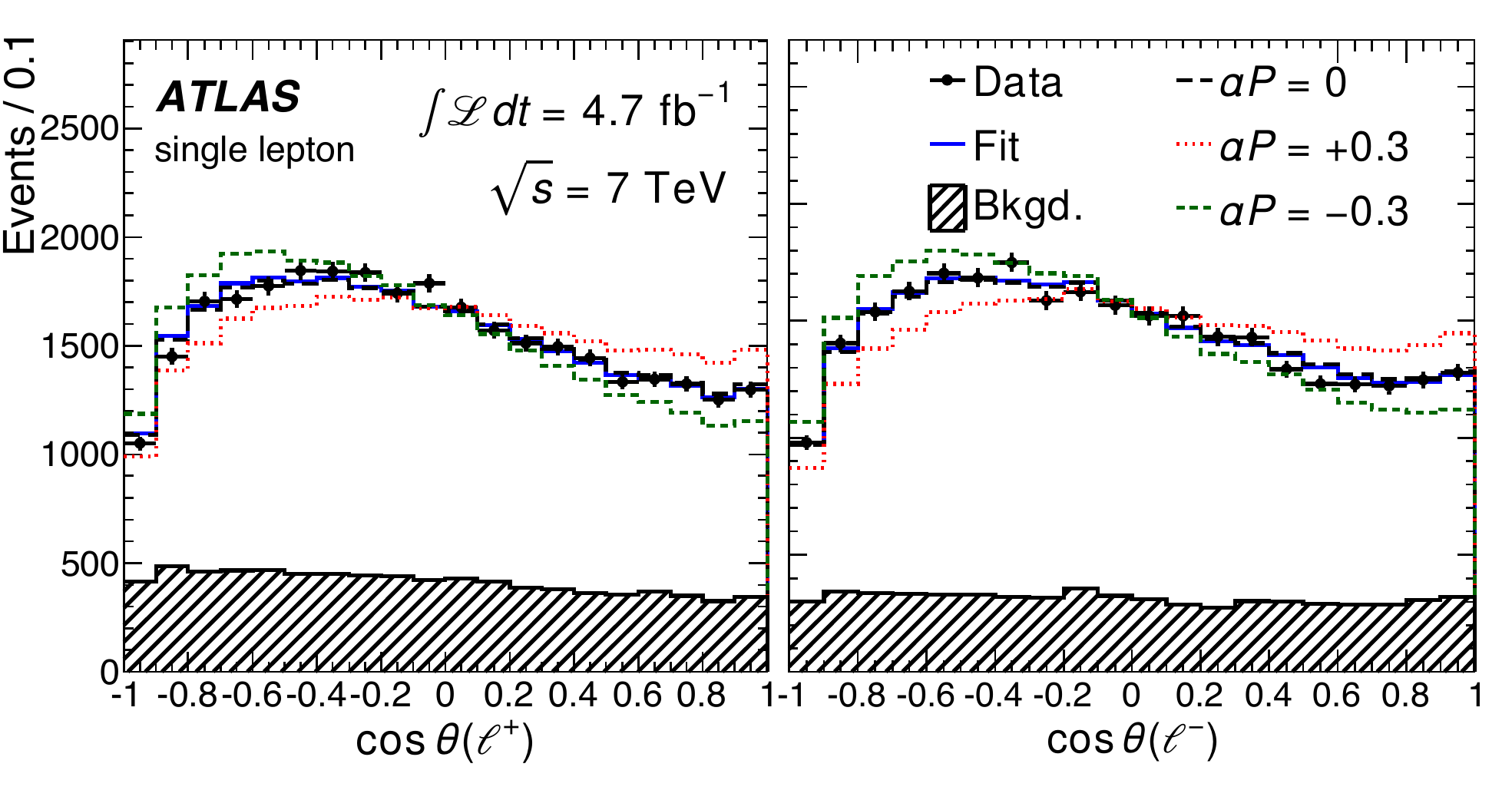}{1.0}{0}
\subfig{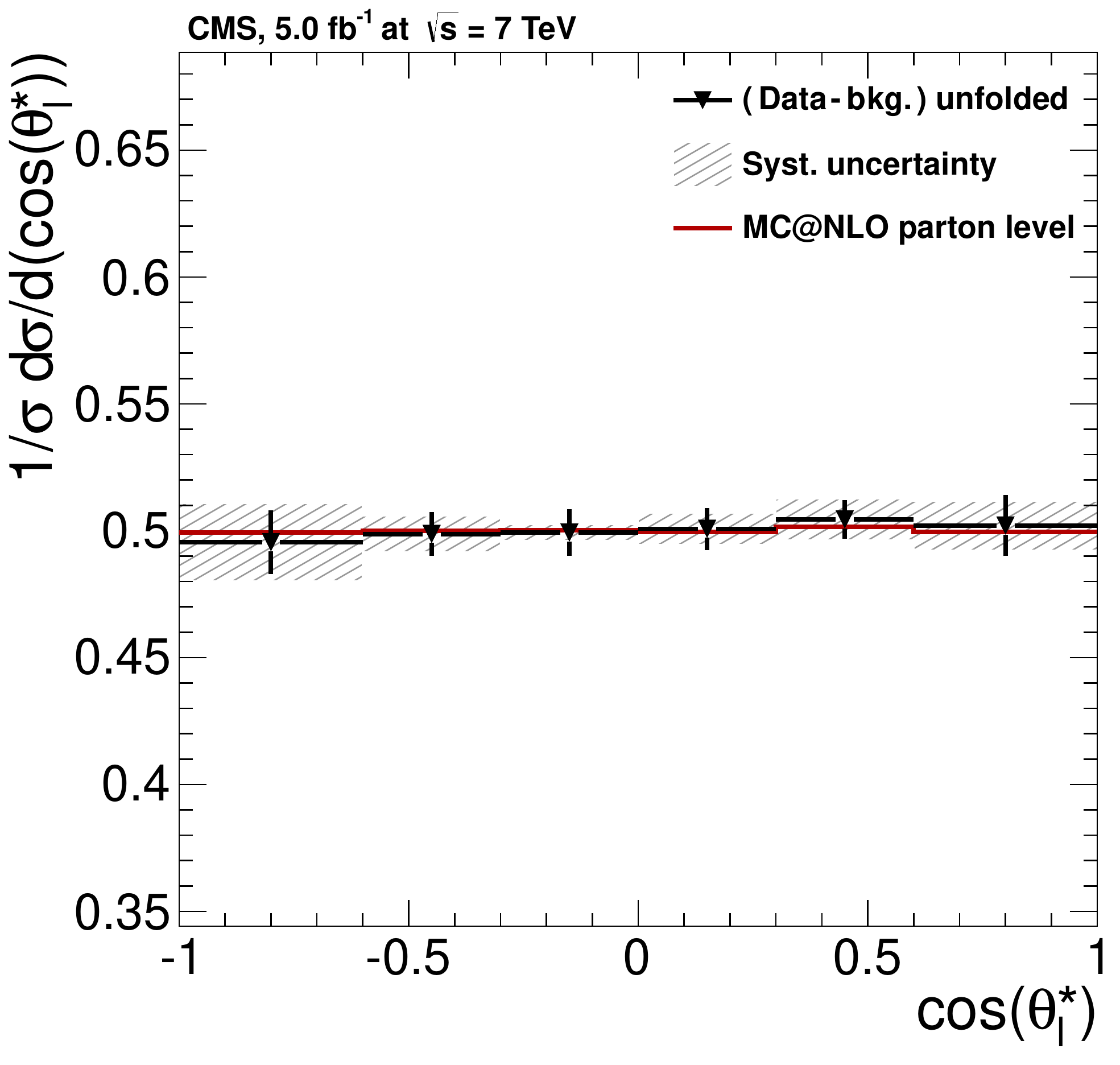}{0.5}{0}
\subfig{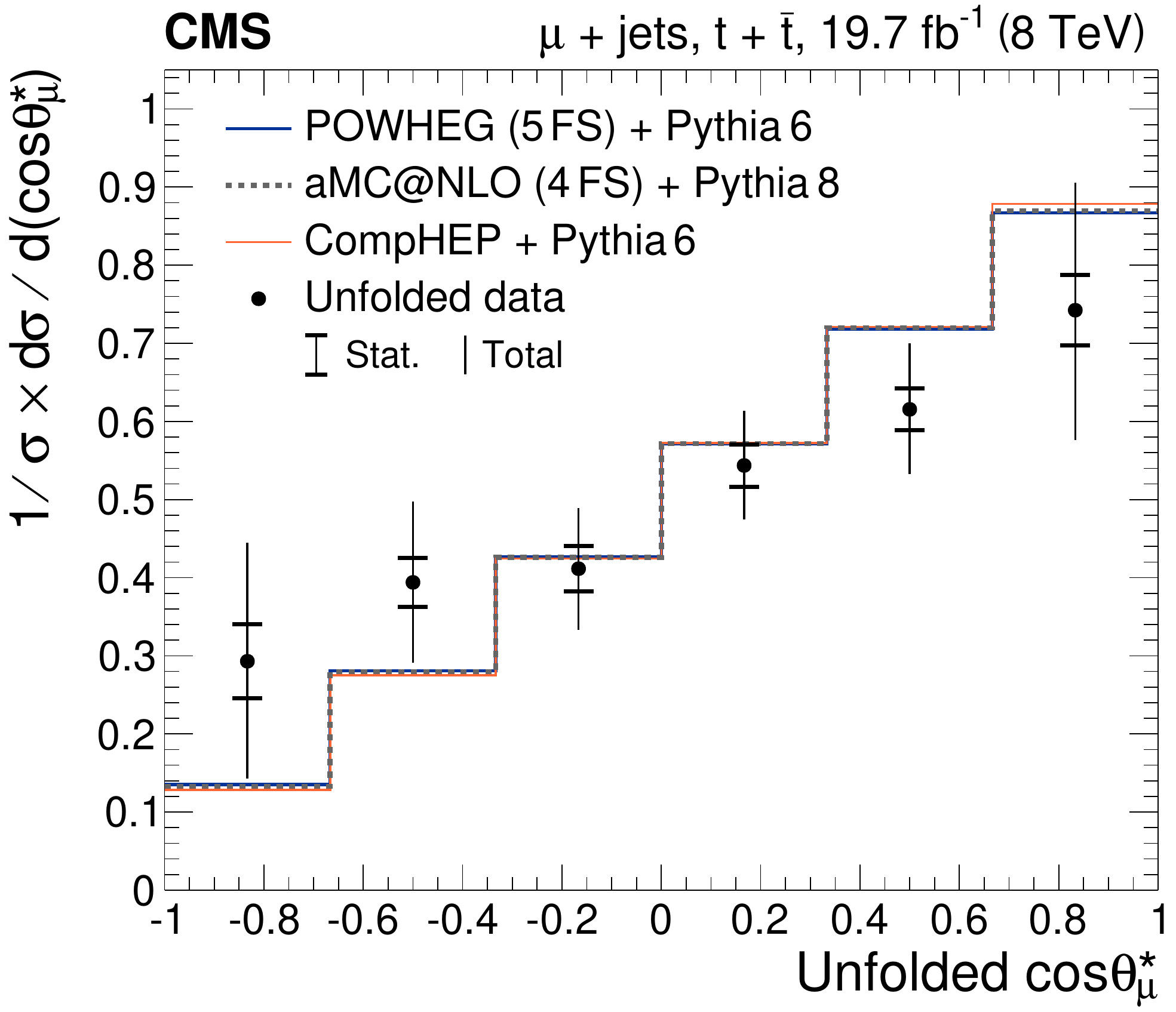}{0.5}{10}
\caption{Top: Result of the template fit to $\cos
\theta_{\ell}$~\pcite{Aad:2013ksa}, the polar angle of the lepton with respect
to the quantization axis in the \ttbar \ljets, compared to different
polarization hypotheses.  Positively charged leptons are shown on the left, and
negatively charged leptons on the right.  Bottom: Measurement of top-quark
polarization in the dilepton channel confirming the absence of top polarization
in \ttbar events on the left~\pcite{Chatrchyan:2013wua}, and measurement of
top-quark polarization in single-top $t$-channel production, showing some
evidence of polarization, though with a strength slightly below the SM
prediction\protect\cite{Khachatryan:2015dzz}.  \label{fpolarisation}}
\end{figure}

\subsection{Top-quark pair spin correlations}

While top quarks in \ttbar production are unpolarized, their spins are expected
to be strongly correlated. One can measure the \ttbar spin correlation strength
$A$ by studying the angular correlations between the decay products, where 
\begin{equation*}
A=\frac{(N_{\uparrow\uparrow}+N_{\downarrow\downarrow}) -
(N_{\uparrow\downarrow} +
N_{\downarrow\uparrow})}{(N_{\uparrow\uparrow}+N_{\downarrow\downarrow})+
(N_{\uparrow\downarrow} + N_{\downarrow\uparrow})} 
\end{equation*} 
is the asymmetry between the number of \ttbar pairs with aligned and
anti-aligned spins. The value of $A$ depends on the spin quantization axis
chosen and on the production mode. 

At the LHC, the dominant production mode is gluon fusion, with a small
contribution from \qqbar annihilation which is reduced further at higher
collision energies, leading to slightly different values of $A$ at different
values of $\sqrt{s}$. 

Deviations from the correlation strength predicted by the SM can be quantified
by considering a mix of SM correlated events and uncorrelated events, and
defining the fraction $f$ of \ttbar events with the SM prediction of spin
correlation as 
\begin{equation*} 
f = \frac{N^{\ttbar}_\mathrm{SM}}{N^{\ttbar}_\mathrm{SM}
+ N^{\ttbar}_\mathrm{uncor}}, \label{eq.Deff} 
\end{equation*}
where $N^{\ttbar}_\mathrm{SM}$ is the number of SM \ttbar events, and
$N^{\ttbar}_\mathrm{uncor}$ represents the number of events with uncorrelated
\ttbar spins. The top quark and anti-quark in the uncorrelated \ttbar events are
expected to decay spherically.  In the assumption that there are only SM and
uncorrelated \ttbar events, the physical range of the parameter $f$ is
restricted to the interval from 0 to 1, with $f = 1$ corresponding to a sample
of \ttbar events produced by the SM and $f = 0$ indicating a sample with
uncorrelated top pairs. An unconstrained template fit is often performed,
allowing values of this parameter outside this interval.

At the LHC, the ATLAS collaboration performed the first measurement that
excluded the uncorrelated hypothesis by more than \signif{5}~\cite{ATLAS:2012ao}
using a template fit to the difference in azimuthal angle between the two
oppositely charged leptons ($\Delta \phi$) in the dilepton final state at
\SI{7}{\TeV}. Subsequent measurements analysed a variety of angular variables,
either performing template fits~\cite{Aad:2014pwa} or presenting unfolded
distributions after correcting for detector efficiency and resolution
effects~\cite{Chatrchyan:2013wua,Aad:2015bfa}.  Measurements in the dilepton
channel were also performed at \SI{8}{\TeV} by both
collaborations~\cite{Aad:2014mfk,Khachatryan:2016xws}.
 
Measurements in the single-lepton final state at the LHC have been performed
employing a more complete reconstruction of the \ttbar events.  The ATLAS
collaboration performed a measurement based on the opening angle distributions
between the decay products of the top quark and anti-quark~\cite{Aad:2014pwa} at
$\sqrt{s} = \SI{7}{\TeV}$, while CMS used the \ttbar single-muon channel
employing full reconstruction of the events and a matrix element method at
$\sqrt{s} = \SI{8}{\TeV}$~\cite{Khachatryan:2015tzo}.  In this approach, the
likelihood of an observed event to be produced by a given theoretical model is
calculated, taking into account the full kinematic configuration of the
top-quark decay products in every event. The likelihood ratio of the sample
allowed the correlated and uncorrelated hypothesis to be distinguished, see
\Fref{fspincorr} (left), and a template fit was performed to extract a value of
$f$ best fitting the data.

\begin{figure}[htbp]
\subfig{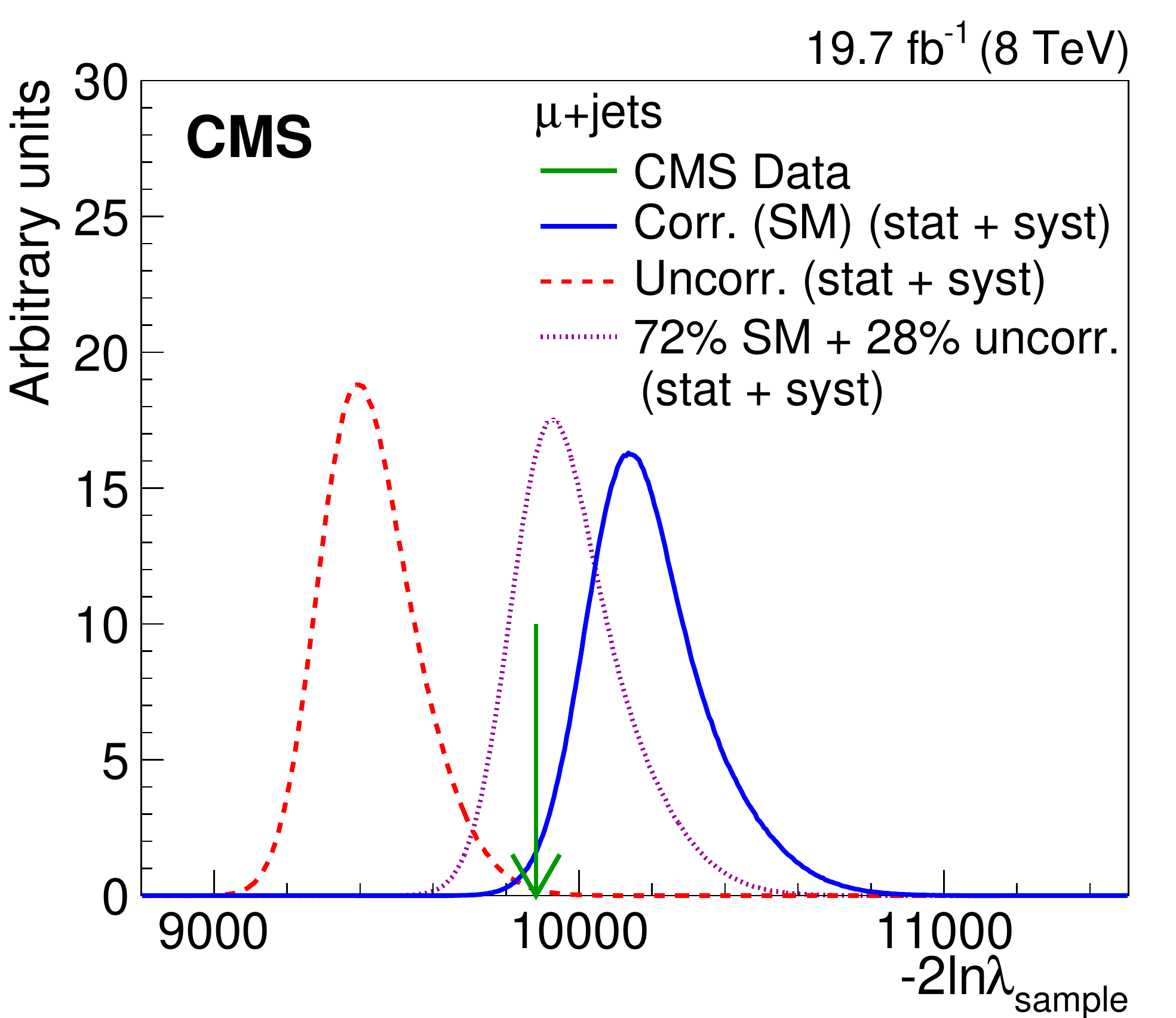}{0.5}{8}
\subfig{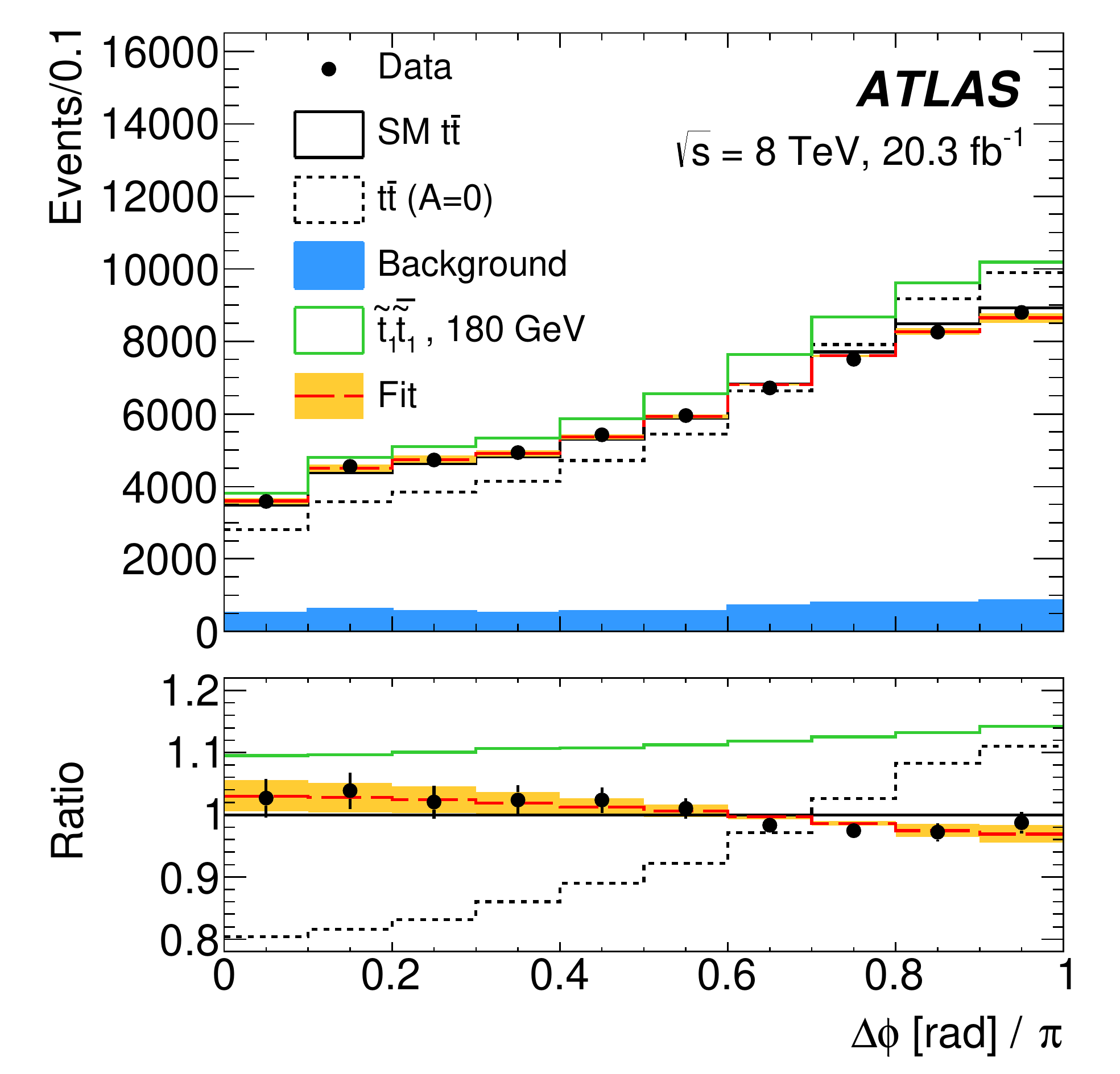}{0.5}{0}
\caption{Correlation likelihood observed by
CMS~\protect\cite{Khachatryan:2015tzo} compared to the expectation from MC
samples using SM spin correlation or zero correlation (left).  Reconstructed
$\Delta \phi$ distribution for the sum of the three dilepton channels. The
predictions for SM \ttbar production is compared to the no spin correlation
hypothesis.  The prediction for $\tilde{t}_1\bar{\tilde{t}}_1$ production
($m_{\tilde{t}_1} = \SI{180}{\GeV}$ and $m_{\tilde{\chi}^0_1} = \SI{1}{\GeV}$)
normalized to the NLO \xs\ including next-to-leading-logarithm corrections is
also shown~\protect\cite{Aad:2014mfk} (right).  \label{fspincorr}}
\end{figure}

\begin{figure}[htbp]
\begin{center}
\includegraphics[width=\textwidth]{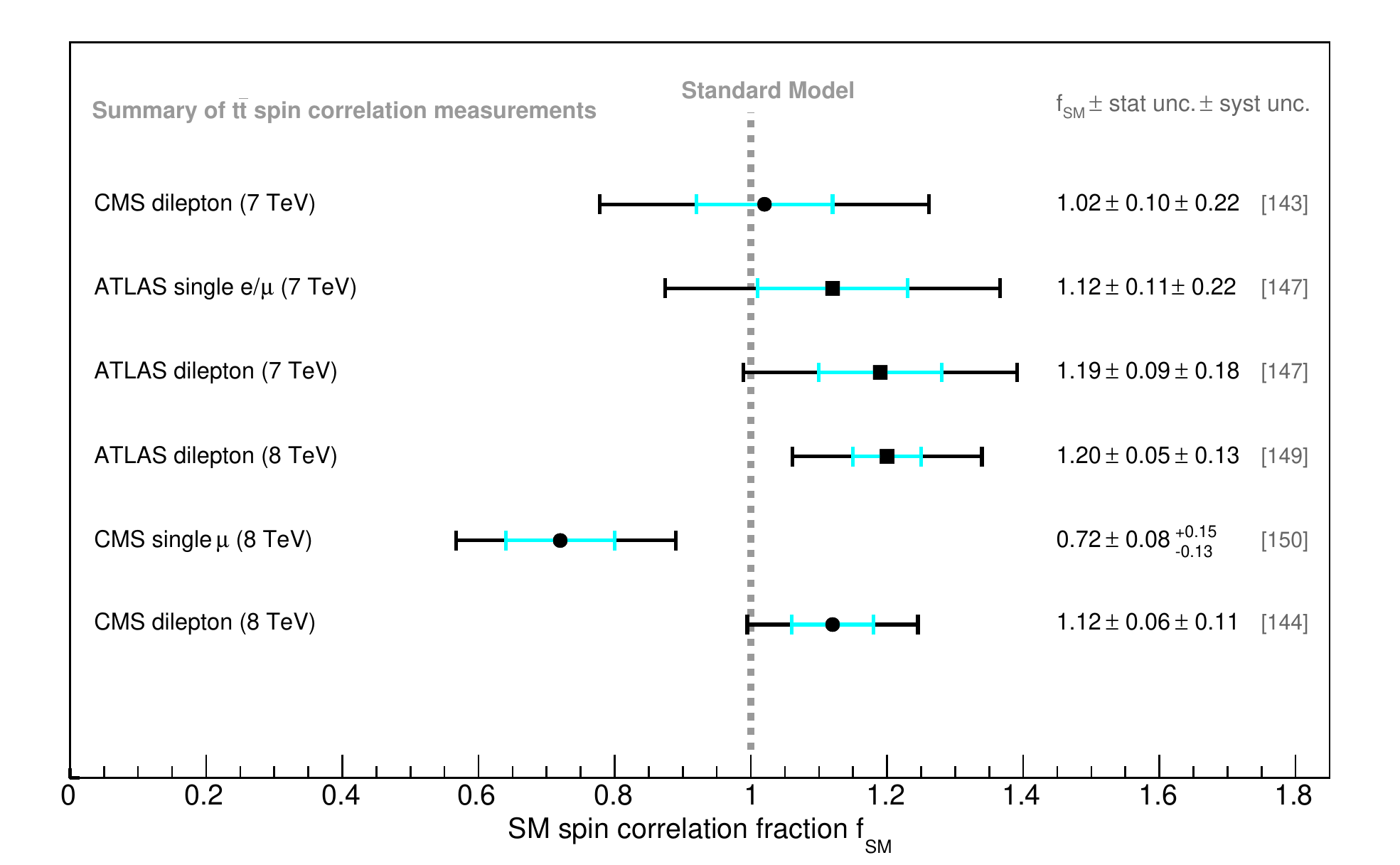}
\end{center}
\caption{Summary of spin correlations measurements.}
\label{summary_spincorr}
\end{figure}

A summary of spin polarization and correlation results in \RunOne\ is given in
\Fref{summary_spincorr}. The most precise results were obtained in the dilepton
channel at \SI{8}{\TeV}. Both ATLAS and CMS measured a spin correlation slightly
stronger than expected, although compatible with the SM prediction, in most
cases.  As the presence of physics beyond the SM in certain scenarios would
reduce the observed spin correlation strength, these measurements tend to
disfavour this class of new physics models. The ATLAS spin correlation
measurement was interpreted to exclude the existence of top squarks
($\tilde{t}_1$) with a mass between the \topmass\ and \SI{191}{\GeV} at 95\%
confidence level~\cite{Aad:2014mfk}, assuming a 100\% branching ratio for
$\tilde{t}_1 \to t \tilde{\chi}^0_1$ and $m_{\tilde{\chi}^0_1} = \SI{1}{\GeV}$,
as illustrated in~\Fref{fspincorr} (right).  In a more general approach, the CMS
collaboration set limits on anomalous top-quark chromo moments, in particular on
the real part of the chromo-magnetic dipole moment and the imaginary part of the
chromo-electric dipole moment~\cite{Khachatryan:2016xws}.

\section{Charge asymmetry}

The production of top-quark pairs via gluon fusion and quark annihilation is
predicted to be symmetric under charge conjugation at leading-order in QCD.  Due
to the interference between initial and final state gluon radiation and the
interference between Born and box diagrams a net asymmetry arises at
next-to-leading order for \qqbar and $qg$ induced \ttbar production. At the
Tevatron the initial asymmetric charge configuration of \ppbar collisions causes
top quarks to be preferentially emitted in the direction of the incoming quark
and top anti-quarks in the direction of the incoming anti-quarks. The charge
asymmetry is measured as a forward-backward asymmetry \AFB and is computed at
next-to-next-to-leading order accuracy to be $(9.5 \pm
0.7)\%$~\cite{Czakon:2014xsa}. At the LHC the $pp$ collisions provide a
symmetric initial state and a small charge asymmetry is induced by the momentum
difference of the valence and sea quarks: valence quarks carry on average a
larger fraction of the proton momentum than sea anti-quarks, hence top quarks
are produced slightly more forward and top anti-quarks are produced slightly
more centrally.  The charge asymmetry \Atop is then conveniently defined as 

\[ \Atop =
\frac{N(\Delta |y| > 0) - N(\Delta |y| < 0)}{N(\Delta |y| > 0) + N(\Delta |y| <
0)}, 
\] 
where $\Delta |y| \equiv |y_t| - |y_{\bar{t}}|$. The SM prediction for the
charge asymmetry has been calculated at NLO QCD and including electroweak
corrections to be $\sim\!\!1.1\%$~\cite{Kuhn:2011ri,Bernreuther:2012sx} at
\SI{8}{\TeV}.  The interest in precisely measuring charge asymmetries in
top-quark pair production at the LHC has grown considerably after the CDF and D0
collaborations reported measurements of $A_{FB}$ that were significantly larger
than the SM predictions.  However, most recent measurements at the Tevatron and
calculations at NNLO QCD~\cite{Czakon:2016ckf} were found to be in much better
agreement.  Determinations of the charge asymmetry at the LHC are shown in
Table~\ref{ca-table} and discussed in this section.

\fulltable{\label{ca-table}Measurements of \Atop and \Alep by the two LHC
experiments at 7 and \SI{8}{\TeV} in the single-lepton and dilepton channels,
compared to theoretical predictions at NLO QCD, including NLO electroweak
corrections~\cite{Kuhn:2011ri, Bernreuther:2012sx}.} 
\begin{tabular}{@{}ccccr@{$\;\pm\;$}lcc}
\br
Exper. & $\sqrt{s}$ & Luminosity & Channel & \multicolumn{2}{c}{$\Atop\,(\%)$} &
Ref. & Theory (\%)\\
\mr
CMS   & \SI{7}{\TeV} & \ifb{1.09}   & \sglept & $-1.3$  & $ 2.8 ^{+2.9}_{-3.1}$ & \cite{Chatrchyan:2011hk} & \multirow{6}{*}{$1.23 \pm 0.05$} \\
ATLAS & \SI{7}{\TeV} & \ifb{1.04}   & \sglept & $-1.9$  & $ 2.8 \pm 2.4$        & \cite{ATLAS:2012an} \\
CMS   & \SI{7}{\TeV} & \ifb{5.0}    & \sglept & $0.4$   & $ 1.0 \pm 1.1$         & \cite{Chatrchyan:2012cxa} \\
ATLAS & \SI{7}{\TeV} & \ifb{4.7}    & \sglept & $0.6$   & \hspace{3ex}$1.0$                 & \cite{Aad:2013cea} \\
CMS   & \SI{7}{\TeV} & \ifb{5.0}    & dilepton& $-1.0$  & $ 1.7 \pm 0.8$        & \cite{Chatrchyan:2014yta} \\
ATLAS & \SI{7}{\TeV} & \ifb{4.6}    & dilepton& $2.1$   & $ 2.5 \pm 1.7$         & \cite{Aad:2015jfa} \\ \\
CMS   & \SI{8}{\TeV} & \ifb{19.6}   & \sglept & $0.10$  & $ 0.68 \pm 0.37$      & \cite{Khachatryan:2015oga} \\
CMS   & \SI{8}{\TeV} & \ifb{19.6}   & \sglept & $0.33$  & $ 0.26 \pm 0.33$      & \cite{Khachatryan:2015mna} \\
ATLAS & \SI{8}{\TeV} & \ifb{20.3}   & \sglept & $0.9$   & \hspace{3ex}$0.5$        & \cite{Aad:2015noh} & $1.02 - 1.11$ \\ 
CMS   & \SI{8}{\TeV} & \ifb{19.5}   & dilepton& $1.1$   & $ 1.1 \pm 0.7$        & \cite{Khachatryan:2016ysn} \\ 
ATLAS & \SI{8}{\TeV} & \ifb{20.3}   & dilepton& $2.1$   & \hspace{3ex}$1.6$        & \cite{Aad:2016ove} \\ \\
ATLAS & \SI{8}{\TeV} & \ifb{20.3}   & boosted & $4.2$   & \hspace{3ex}$3.2$            & \cite{Aad:2015lgx} & $1.60 \pm 0.04$ \\ \\
& & & & \multicolumn{2}{c}{$\Alep\,(\%)$} &\\
\mr
CMS   & \SI{7}{\TeV} & \ifb{5.0}    & dilepton& $0.9$ & $ 1.0 \pm 0.6$        & \cite{Chatrchyan:2014yta} & \multirow{2}{*}{$0.70 \pm 0.03$} \\
ATLAS & \SI{7}{\TeV} & \ifb{4.6}    & dilepton& $2.4$ & $ 1.5 \pm 0.9$         & \cite{Aad:2015jfa} \\ \\
CMS   & \SI{8}{\TeV} & \ifb{19.5}   & dilepton& $0.3$ & $ 0.6 \pm 0.3$         & \cite{Khachatryan:2016ysn} & \multirow{2}{*}{$0.64 \pm 0.03$} \\
ATLAS & \SI{8}{\TeV} & \ifb{20.3}   & dilepton& $0.8$ & \hspace{3ex}$0.6$        & \cite{Aad:2016ove} \\
\br
\end{tabular}
\endfulltable

With the first \ifb{} of $pp$ collision data both experiments measured the
charge asymmetry in the \ljets~\cite{Chatrchyan:2011hk, ATLAS:2012an} with
similar statistical and systematic uncertainties and total uncertainties of
$\sim\!\!4\%$. The \ttbar system was reconstructed either with a kinematic fit
or by choosing the best match of a \Wboson\ candidate with a jet and, in order
to be able to compare the resulting asymmetry with predictions, the measured
$\Delta |y|$ distributions were extracted via iterative or regularized
unfolding, accounting for acceptance and detector effects.  At large values of
$m_{\ttbar}$ possible new physics effects are expected to be enhanced, but
within uncertainties no trend could be observed when measuring the asymmetry as
a function of this variable.  With the full \SI{7}{\TeV}
dataset~\cite{Chatrchyan:2012cxa, Aad:2013cea} the uncertainty of the
measurements was reduced and additional unfolded differential distributions were
extracted as a function of $|y|$, \pT and the invariant mass of the \ttbar
system, or requiring a minimum velocity of the \ttbar system along the beam
axis.  Results have been compared with SM expectations and BSM models such as
colour-octet axigluons or a model featuring an effective axial-vector coupling
of the gluon. \Fref{ca7} illustrates the measurements of the charge asymmetry at
\SI{7}{\TeV} in the \ljets.

\begin{figure}[htbp]
\subfig{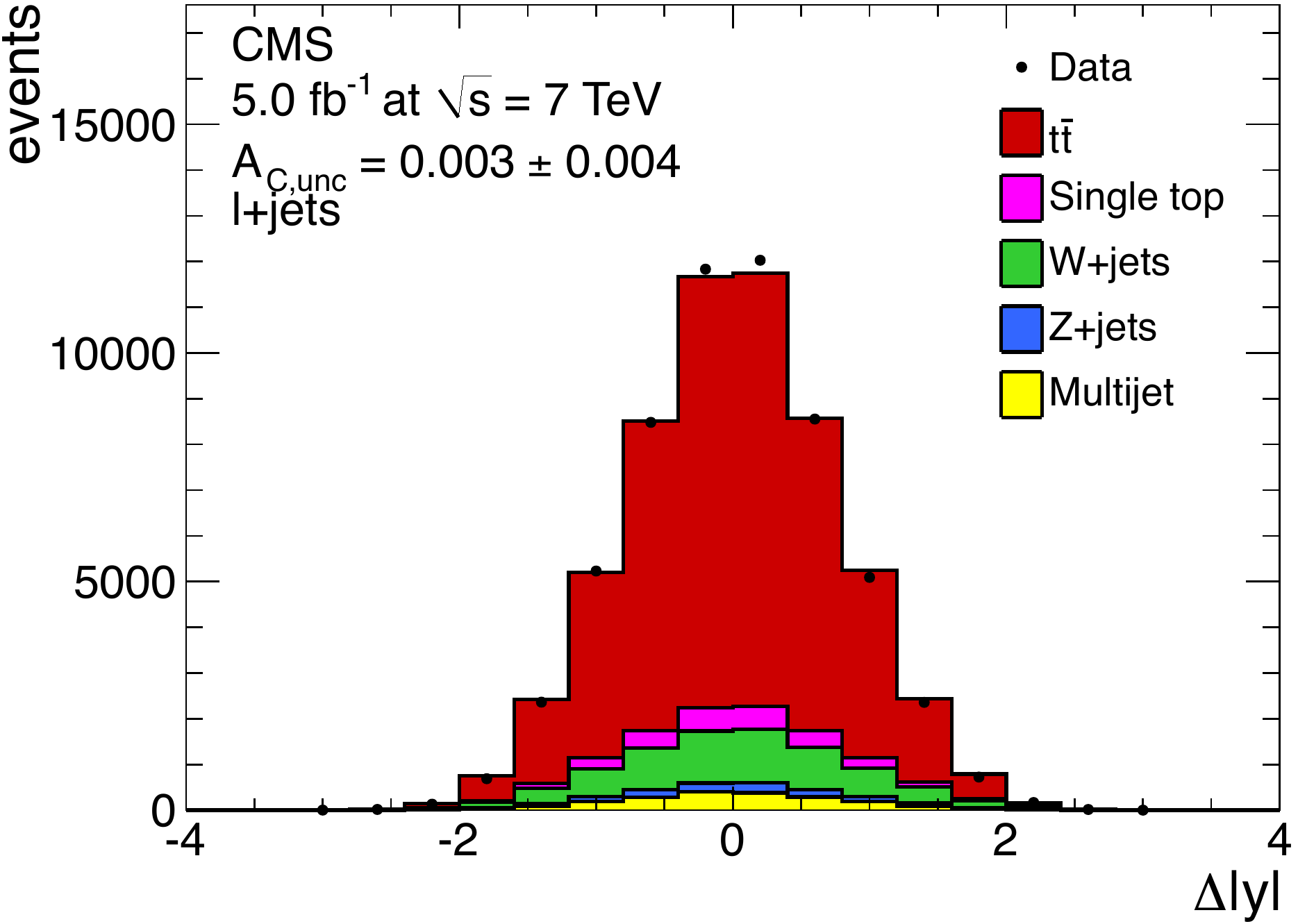}{0.5}{0}
\subfig{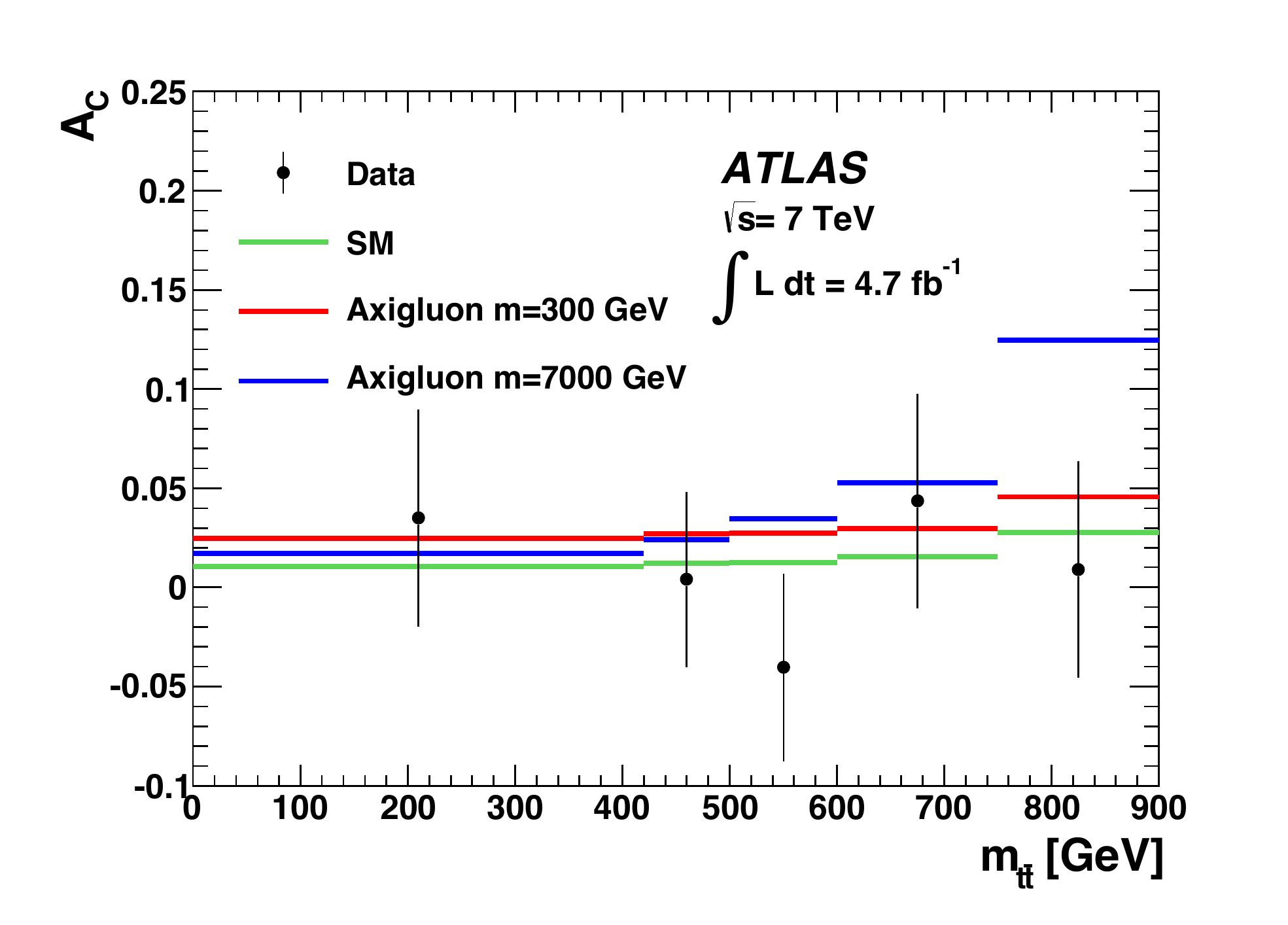}{0.5}{0}
\caption{Measurements of the charge asymmetry with \SI{7}{\TeV} data. Data
compared to simulation normalized in control regions for the distribution as a
function of $\Delta |y|$ \pcite{Chatrchyan:2012cxa} (left) and distribution of
$A_C$ as a function of $m_{\ttbar}$ after unfolding, compared with the SM
predictions (green lines) and the predictions for a colour--octet axigluon with
a mass of \SI{300}{\GeV} (red lines) and \SI{7000}{\GeV} (blue
lines)~\pcite{Aad:2013cea} (right).} \label{ca7}
\end{figure}

Measurements of the inclusive and differential charge asymmetry were also
performed at \SI{8}{\TeV} with very similar
techniques~\cite{Khachatryan:2015oga, Aad:2015noh}.  The inclusive \ttbar
production charge asymmetry has been measured with uncertainties of $0.5 -
0.8\%$, dominated by statistical uncertainties. CMS also provided an inclusive
measurement in a fiducial region.  The same dataset was re-analysed by CMS, with
the alternative approach of using a template fit instead of
unfolding~\cite{Khachatryan:2015mna}.  More events with a lower purity were
selected, while to reconstruct the \ttbar kinematics and to determine the sample
composition alternative methods were used.  For the template construction a
transformed variable was used, $\Upsilon \equiv \tanh \Delta |y|$, that has the
same asymmetry \Atop, but is bounded. The background composition after event
selection was determined with a likelihood fit.  The reconstructed distribution
of $\Upsilon$ was then fit to templates with symmetric and asymmetric components
to extract the net asymmetry (\Fref{ca8}, left).  Both measurement techniques,
the Bayesian unfolding~\cite{Aad:2015noh} and the template
method~\cite{Khachatryan:2015mna} are dominated by statistical uncertainties and
thus further improvements are to be expected with the upcoming larger datasets
at the LHC at 13 and \SI{14}{\TeV}. It has to be noted, however, that the charge
asymmetry at larger collision energies is expected to be reduced by
approximately a factor of two between 7 and \SI{14}{\TeV}, as the fraction of
\qqbar-induced \ttbar production in $pp$ collisions decreases with
centre-of-mass energy.  The measurements of \Atop at the LHC and of \AFB at the
Tevatron are compared to the expectation of the Standard Model and of several
extensions in \Fref{ca8} (right). 

\begin{figure}[htbp]
\subfig{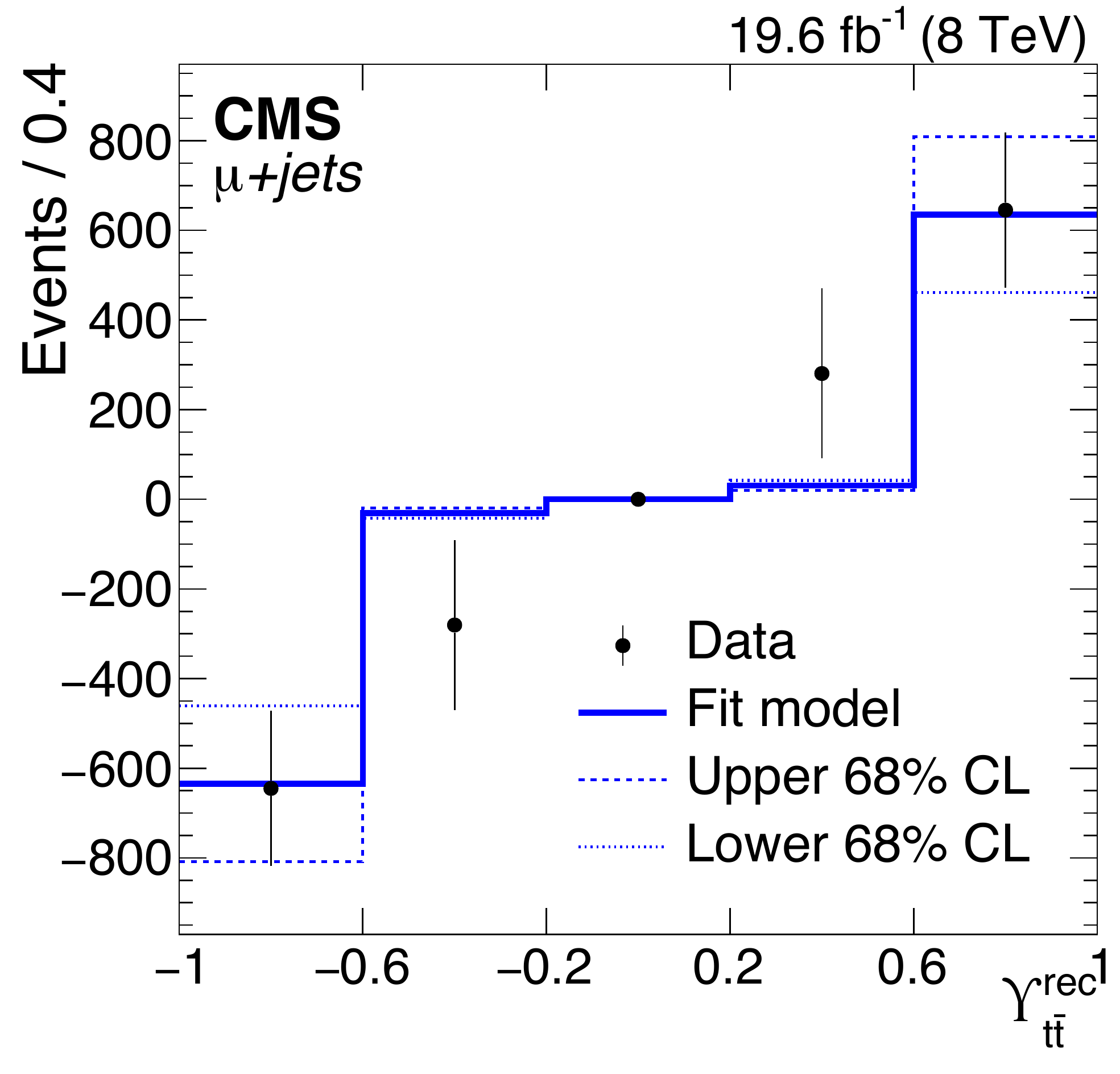}{0.5}{0}
\subfig{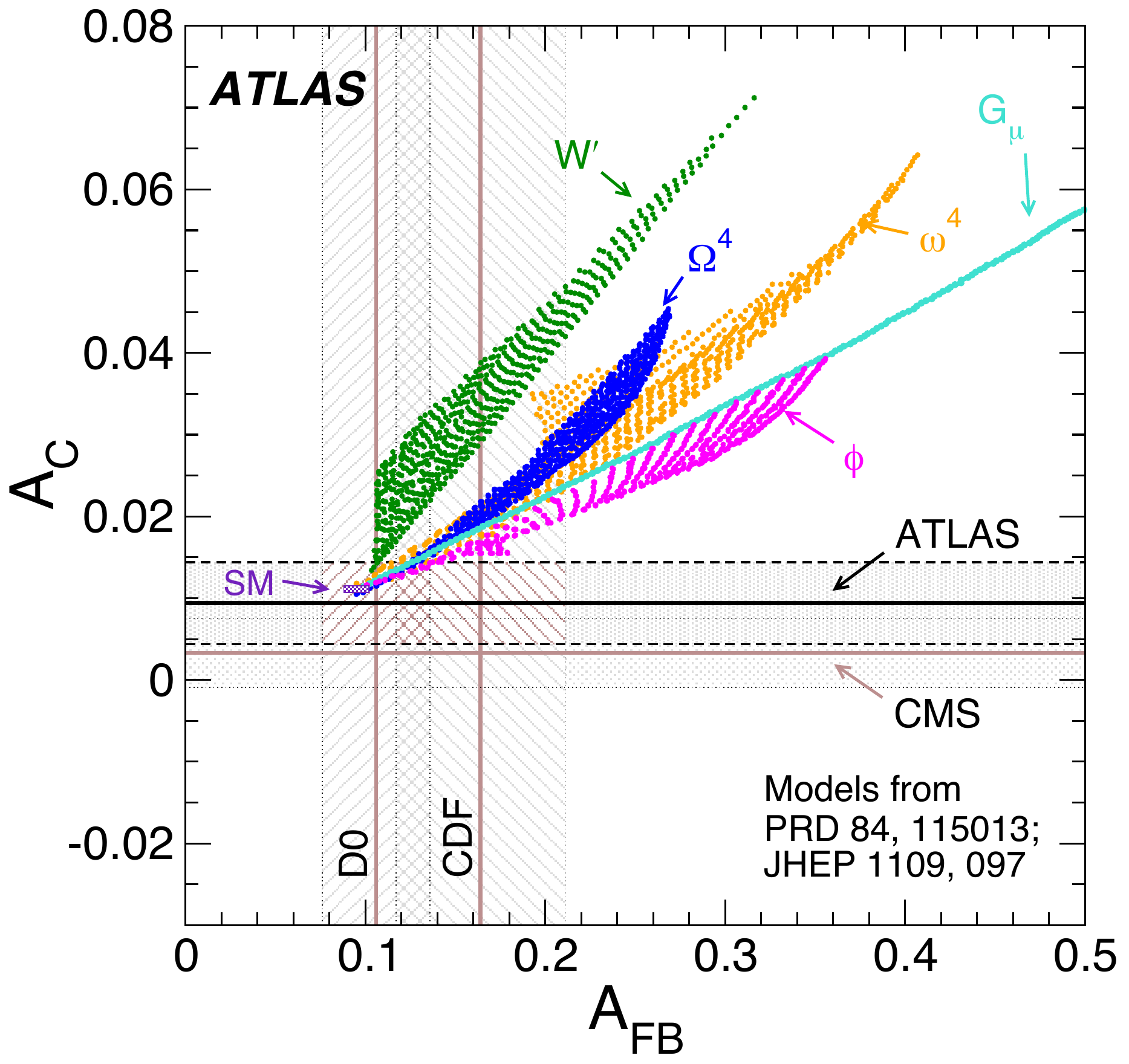}{0.5}{0}
\caption{Measurements of the charge asymmetry with \SI{8}{\TeV} data.  Template
fit as a function of the reconstructed $\Upsilon$ variable in the \ttbar
single-muon channel~\pcite{Khachatryan:2015mna} (left).  A representative choice
of models is shown in the summary figure including a $W^{\prime}$ boson
exchanged in the $t$-channel, a heavy axigluon $G_{\mu}$ exchanged in the
$s$-channel, a scalar isodoublet $\phi$, a colour-triplet scalar $\omega^4$, and
a colour-sextet scalar $\Omega^4$~\pcite{Aad:2015noh} (right).  The current
measurements are in good agreement with the SM expectation.  \label{ca8}}
\end{figure} 

Extensions of the SM with heavy particles can predict a significantly enhanced
charge asymmetry at high \ttbar invariant masses at the LHC.  Using
reconstruction techniques specifically designed for the decay topology of highly
boosted top quarks, ATLAS also determined the charge asymmetry in a fiducial
region with $m_{\ttbar} > \SI{750}{\GeV}$ and $-2 < \Delta|y_t| < 2$ in bins of
$m_{\ttbar}$~\cite{Aad:2015lgx} finding no significant discrepancy compared to
the SM expectation.

For the \ttbar dilepton channel an additional asymmetry can be defined, \Alep,
based on the difference in absolute rapidity of the two leptons, $\Delta |\eta|
\equiv |\eta_{\ell^+}| - |\eta_{\ell^-}|$, as

\[
\Alep = \frac{N(\Delta |\eta| > 0) - N(\Delta |\eta| < 0)}{N(\Delta |\eta| > 0)
+ N(\Delta |\eta| < 0)}.  
\]

Since the directions of the leptons do not fully follow the direction of the
parent top quarks and anti-quarks, the predicted value of \Alep is smaller than
the prediction for \Atop. On the other hand, \Alep is based on the precise
measurement of reconstructed lepton directions and does not rely on jet and
\ETmiss quantities, so there is no need for a full event reconstruction.  Both
\Atop and \Alep are sensitive to possible new physics effects arising in \ttbar
production, with contributions from processes such as axigluon or $W^\prime$ and
$Z^\prime$ exchange typically predicting larger values of \Atop or \Alep, or
both.  With the \SI{7}{\TeV}~\cite{Chatrchyan:2014yta, Aad:2015jfa} and
\SI{8}{\TeV}~\cite{Khachatryan:2016ysn, Aad:2016ove} datasets CMS and ATLAS
measured \Atop and \Alep, including unfolded distributions as a function of
\ttbar system quantities and simultaneous extraction of both inclusive
parameters. A systematic uncertainty below $0.3\%$ is attained in this channel
for the measurement of \Alep in a fiducial volume.  All measurements are
consistent with expectations within uncertainties (\Fref{ca-dil}).

\begin{figure}[htbp]
\subfig{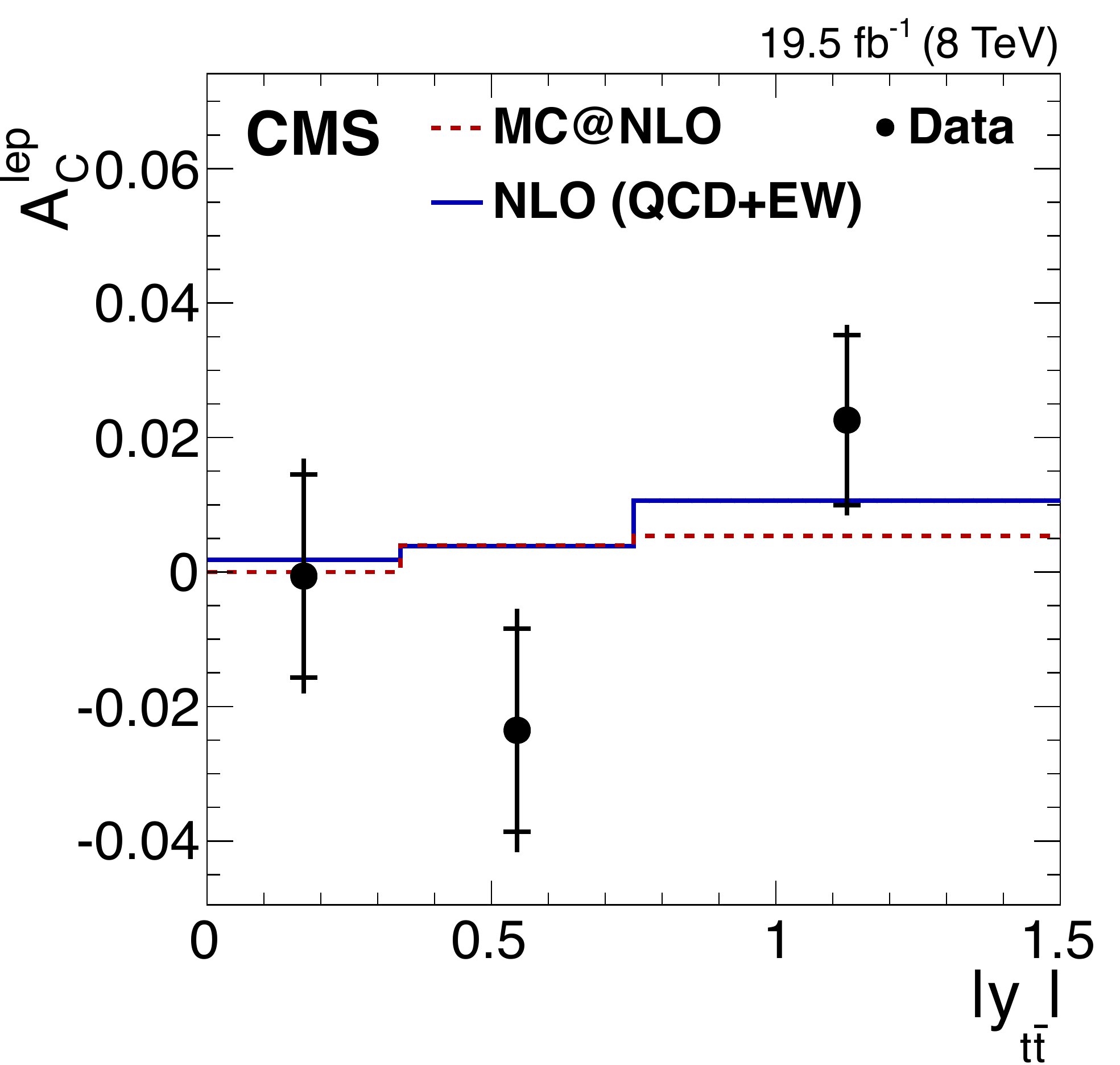}{0.5}{0}
\subfig{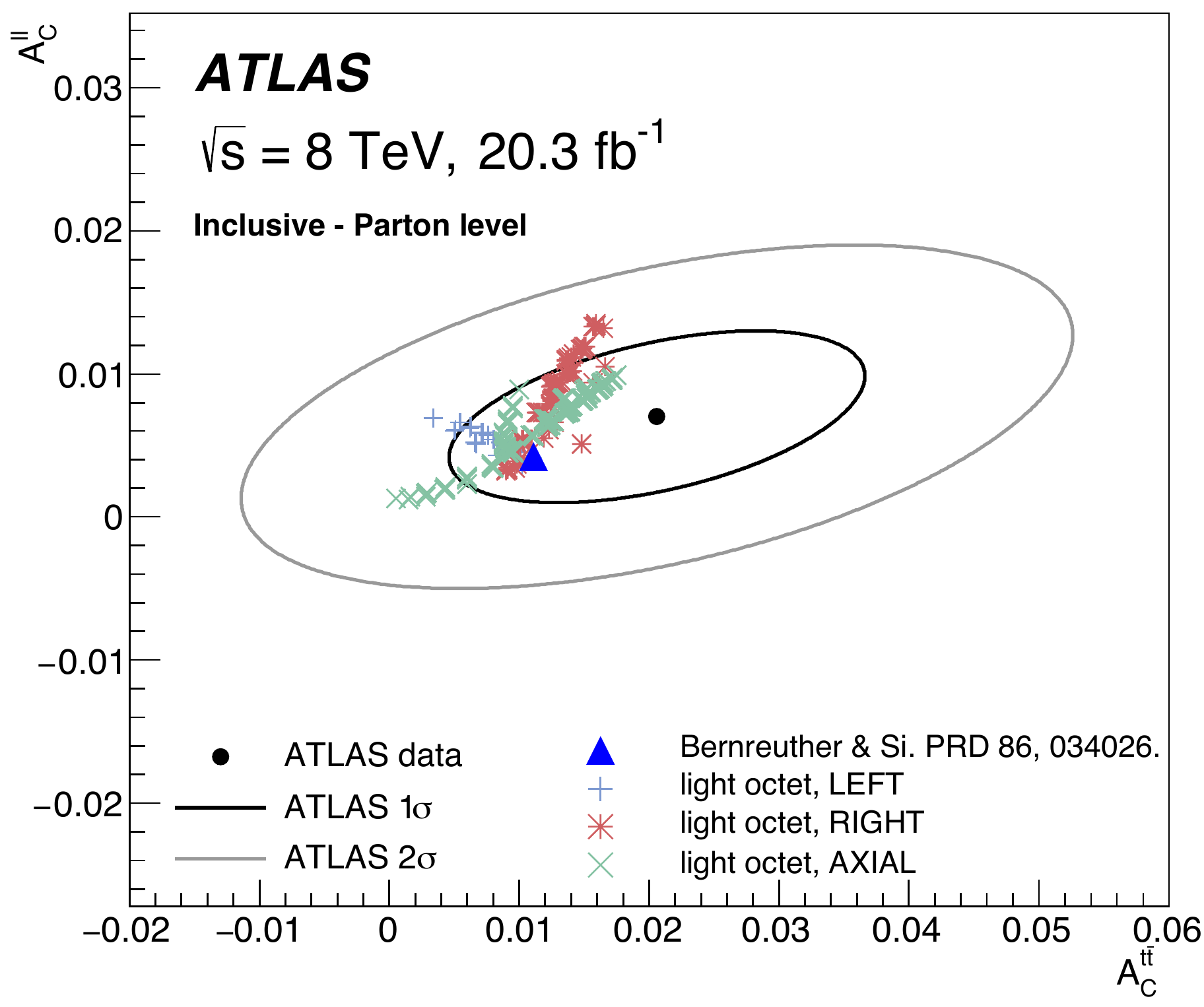}{0.5}{10}
\caption{Measurements of the leptonic-based charge asymmetry.  Dependence of the
unfolded \Alep values on $|y_{\ttbar}|$ compared to parton-level
predictions~\pcite{Khachatryan:2016ysn} (left).  Comparison of the measured
\Alep and \Atop values, along with \signif{1} and \signif{2} uncertainty
contours, to various predictions~\pcite{Aad:2016ove} (right).} \label{ca-dil} 
\end{figure}

\section{Top-quark couplings}
\label{sec:couplings}

\subsection{The $tWb$ vertex}
The $tWb$ vertex is characterized by the electroweak $V-A$ structure. The
polarization of the $W$ bosons produced in top-quark decays can be longitudinal,
left- or right-handed. The angular distribution of the top-quark decay products,
boosted into the $W$ boson rest frame, where $\thetastar \equiv
\angle(\vec{p}_{\ell^\pm},-\vec{p}_b)$, is given by
\[
\frac{1}{\sigma} \frac{\rmd\sigma}{\rmd\cos \thetastar} = 
                 \frac{3}{4}\sin^2 \thetastar\;F_0 
               + \frac{3}{8}(1 - \cos \thetastar)^2\;F_L 
               + \frac{3}{8}(1 + \cos \thetastar)^2\;F_R.
\]
The helicity fractions $F_0$, $F_L$ and $F_R$ are the fractions of events with a
particular polarization. At leading order, their values are predicted to be
approximately 70\%, 30\% and 0\%, respectively, and have been calculated at NNLO
QCD accuracy with uncertainties at the sub-percent
level~\cite{Czarnecki:2010gb}. 

In the \ttbar \ljets\ a constrained kinematic fit is employed in order to
improve the reconstruction of the $W$ boson rest frame. The helicity fractions
are either directly extracted from template fits to the $\cos \thetastar$
distribution or indirectly from the calculation of asymmetries based on unfolded
spectra of the same variable. Fits are performed under the condition
$F_0+F_L+F_R=1$, or, in order to further decrease uncertainties, the additional
assumption $F_R = 0$.  Measurements have been performed with the \SI{7}{\TeV}
dataset in \ttbar topologies, including the dilepton channel~\cite{Aad:2012ky}
or considering also the hadronic top-quark decay~\cite{Chatrchyan:2013jna}.  CMS
also performed measurements with the \SI{8}{\TeV} dataset in the \ttbar
\ljets~\cite{Khachatryan:2016fky} and in a selection enhancing the single-top
$t$-channel contribution~\cite{Khachatryan:2014vma}.  The modelling of the
top-quark kinematics and the reconstruction of the rest frame have an important
impact on the overall uncertainty on these measurements.  \Fref{fig_wheli}
illustrates one of the measurements and shows the summary of the measurements
performed at the LHC \RunOne.

\begin{figure}[htbp]
\subfig{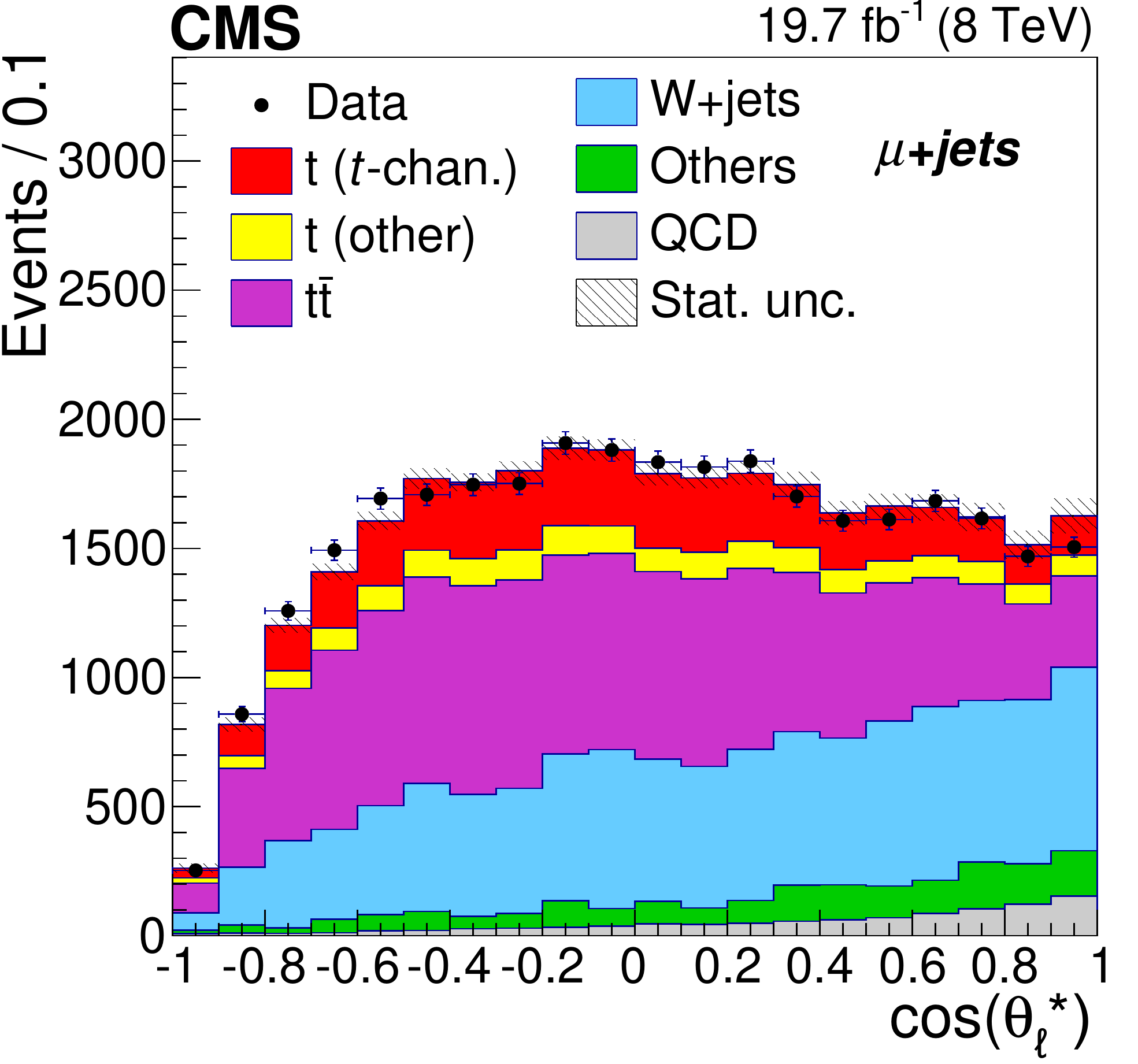}{0.45}{10}
\subfig{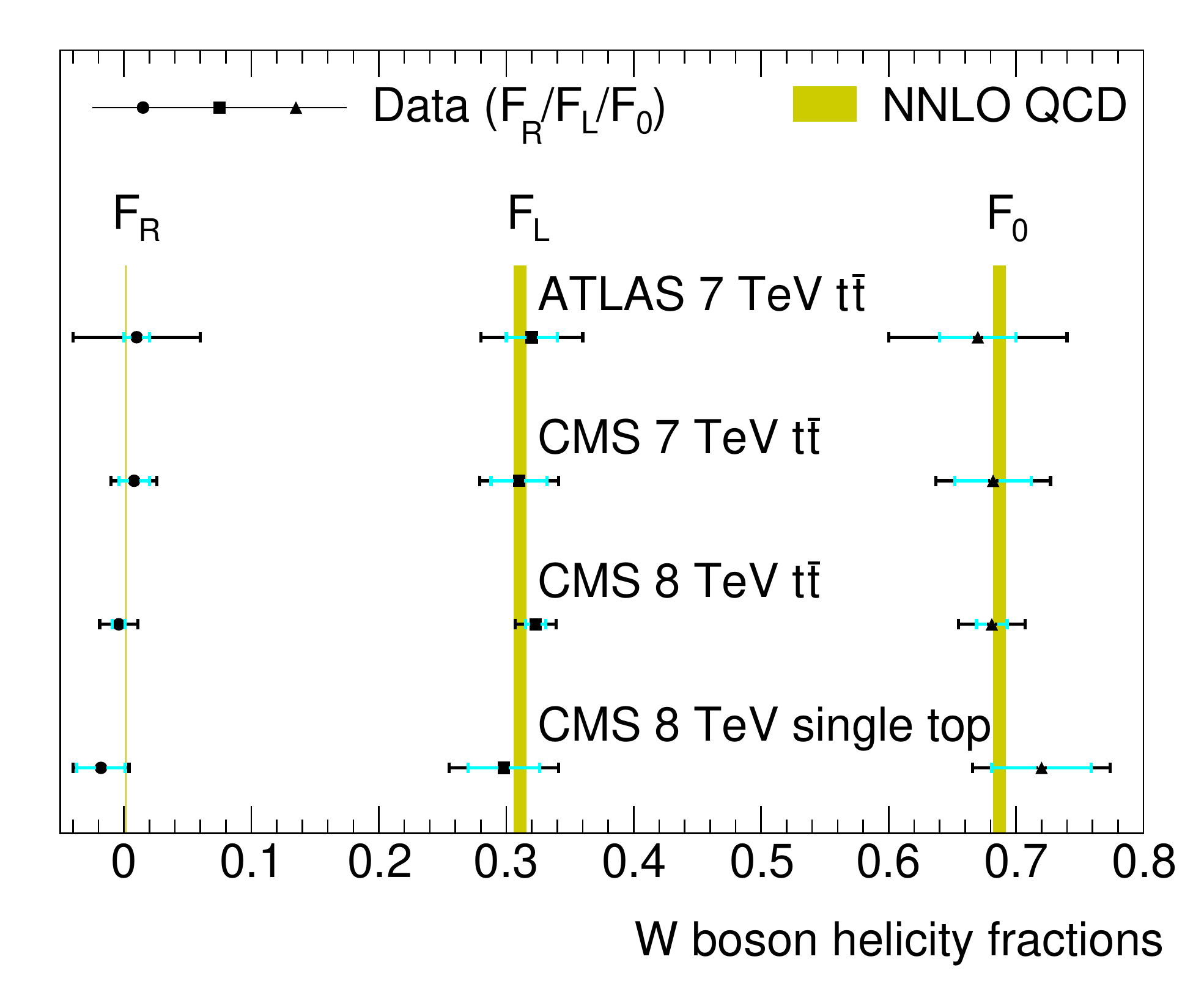}{0.55}{0}
\caption{Distribution of $\cos \thetastar$ for a single-top enhanced event
selection used to fit for the helicity fractions~\pcite{Khachatryan:2014vma}
(left).  Summary of the LHC \RunOne\ measurements of $F_R$, $F_L$ and $F_0$
(right).} \label{fig_wheli}
\end{figure}

Results are compatible with the Standard Model expectations and are interpreted
in terms of constraints on anomalous $tWb$
couplings~\cite{AguilarSaavedra:2008zc, AguilarSaavedra:2009mx} using the most
general dimension-six Lagrangian
\[
\fl \qquad
{\cal L}_{tWb} = 
- \frac{g}{\sqrt{2}} \bar{b} \gamma^\mu (V_L P_L + V_R P_R) t W_{\mu}^- 
- \frac{g}{\sqrt{2}} \bar{b} \frac{\rmi \sigma^{\mu \nu} q_{\nu}}{m_W} (g_L P_L
+ g_R P_R) t W_{\mu}^- + h.c.,
\]
where $P_{L,R}$ are the projection operators. In the Standard Model and at tree
level, the complex constants are $V_L = V_{tb}$ and $V_R = g_L = g_R = 0$.
Assuming $V_L = 1$ and $V_R = 0$ the real parts of $g_R$ and $g_L$ can be
constrained as illustrated in \Fref{fig_Wano} (left).  By measuring double
differential angular distributions in single-top events,
ATLAS~\cite{Aad:2015yem} additionally set a limit on the imaginary part of
$g_R$, consistent with SM expectations (\Fref{fig_Wano}, right).

\begin{figure}[htbp]
\subfig{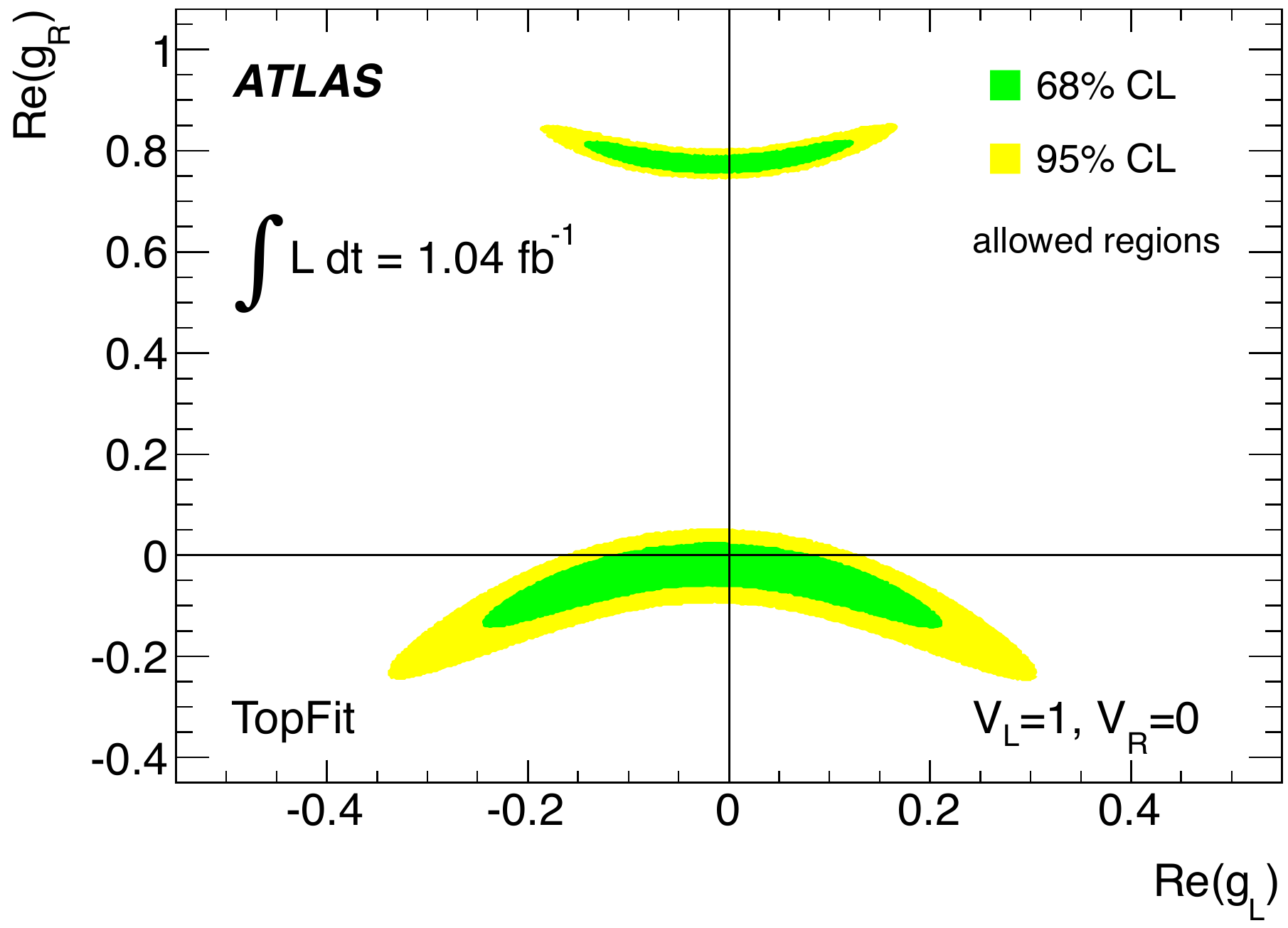}{0.5}{0}
\subfig{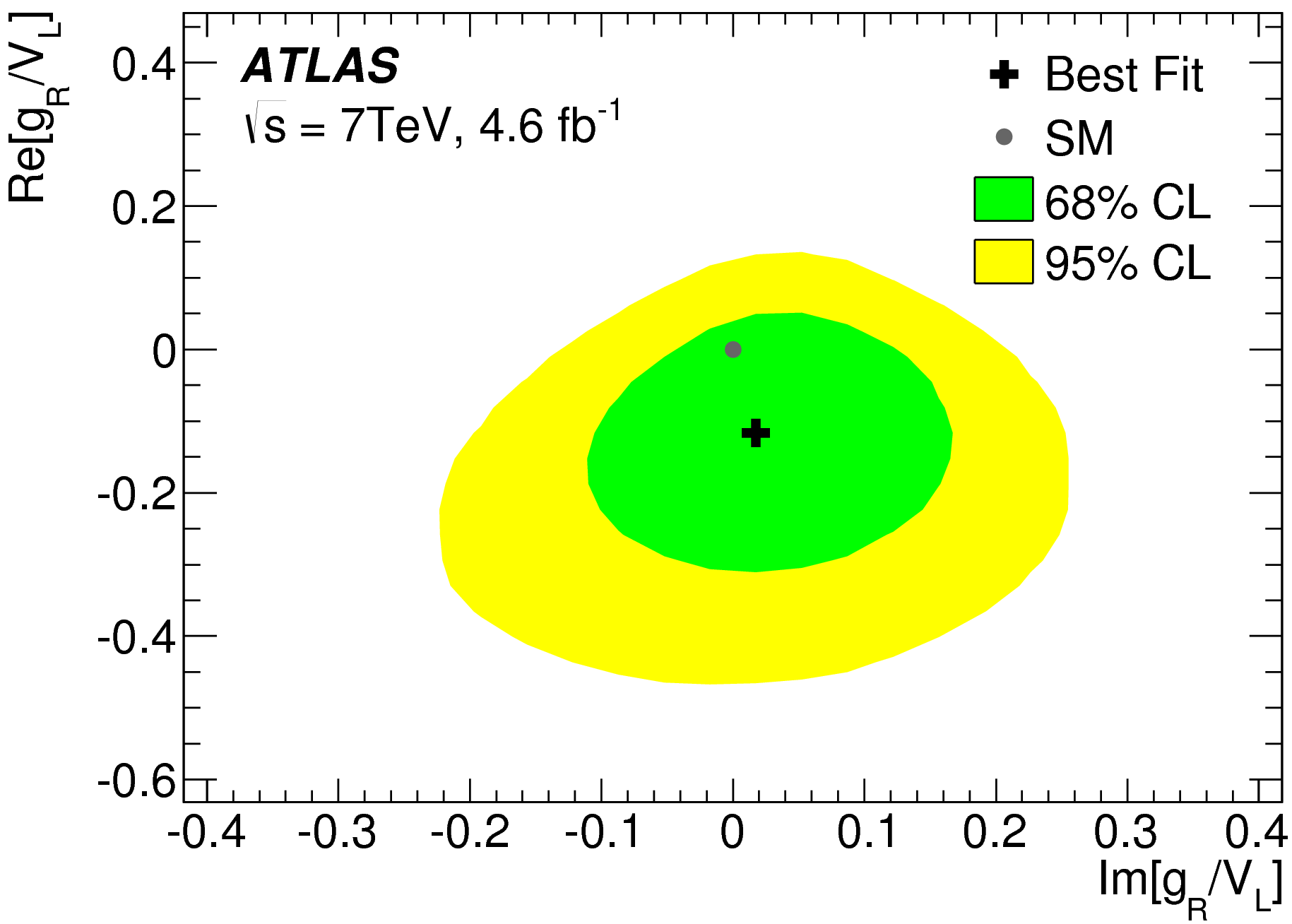}{0.5}{0}
\caption{Allowed regions for the $tWb$ anomalous couplings $g_L$ and
$g_R$~\pcite{Aad:2012ky} (left) and for the real and imaginary part of the ratio
$g_R / V_L$~\pcite{Aad:2015yem} (right).} \label{fig_Wano}
\end{figure}

The extraction of \Vtb from measurements in the single-top channels has been
discussed in \Sref{secsingle}.  Indirect access to the value of \Vtb can also be
obtained by studying the branching fraction ratio $R_b \equiv {\cal B} (t\to Wb)
/ {\cal B} (t\to Wq)$ in \ttbar events, where $q = d,s,b$. Under the assumption
of a three-generation CKM matrix $R_b = \Vtb^2$. With the full \SI{8}{\TeV}
\ttbar dilepton sample CMS extracted the $b$-quark content from the
distributions of the number of jets and $b$-tagged
jets~\cite{Khachatryan:2014nda}. Multijet control samples are used to determine
the $b$-tagging efficiency. Using the likelihood fit technique a measurement of
$R_b$ was performed, with dominant systematic uncertainties mostly due to the
$b$-tagging efficiency. From this $\Vtb = 1.007 \pm 0.016$ was extracted,
corresponding to $\Vtb > 0.975$ at 95\% confidence level if $\Vtb<1$ is imposed.
This result has been combined with the measurement of $t$-channel single-top
production to extract the total decay width $\Gamma(t\to Wb) / {\cal B} (t \to
Wb) = 1.36^{+0.14}_{-0.11}\;\si{\GeV}$, compatible with SM expectations.

\subsection{Couplings to the photon}

In the Standard Model the couplings of the top quark to the neutral electroweak
vector bosons ($X^0$), the photon and the $Z$ boson, are described by the
effective vertex~\cite{Hollik:1998vz}
\[
\Gamma_{\mu}^{\ttbar X^{0}} = - \rmi e \left[ \gamma_{\mu} ( F_1^V
+ \gamma_5 F_1^A ) + \frac{\sigma_{\mu \nu}}{2 m_t} (q + \bar{q})^{\nu} ( \rmi
F_2^V + \gamma_5 F_2^A ) \right],
\]
where $q$ and $\bar{q}$ are the outgoing momenta of the top quark and
anti-quark, $F_1^V (F_1^A)$ is the vector (axial vector) form factor and $F_2^V
(F_2^A)$ is related to the magnetic (electric) dipole form factor.  For the
photon, at tree level, $F_2^{V} = F_2^{A} = 0$ and $F_1^{V,A}$ are related to
the electric charge and the weak angle $\theta_W$. The first step to study the
structure of the $t\gamma t$ vertex is to observe associated \ttga production.
In the modelling of this process the interference between radiative top-quark
production and decay processes needs to be taken into account, as photons can be
emitted by any initial or final state charged particle. To avoid divergences,
requirements are imposed on the transverse momenta of the produced photons, as
well as on the invariant mass or $\Delta R$ distance of pairs of objects.  The
\ttga \xs\ has been calculated at NLO accuracy in QCD~\cite{Melnikov:2011ta}.

Top-quark pair production in association with a photon has been observed by
ATLAS with the \SI{7}{\TeV} dataset~\cite{Aad:2015uwa}, requiring photons in the
central detector region with a minimum energy in the transverse plane of
\SI{20}{\GeV} and not close to a reconstructed lepton or jet. The \xs\
measurement was performed in a fiducial phase space.  The extraction of the \xs\
proceeded via a binned likelihood template fit to a photon track-isolation
variable (\Fref{fttgamma}, left).  Signal and background template histograms
were derived from simulation and data. With about $360$ selected events in the
\ljets\ the no-signal hypothesis was excluded with a significance of
\signif{5.3} and the fiducial \xs\ was measured with a total uncertainty of
$\sim\!\!30\%$, dominated by the jet energy scale uncertainty, and is consistent
with calculations using LO simulations rescaled to NLO.

\begin{figure}[htbp]
\subfig{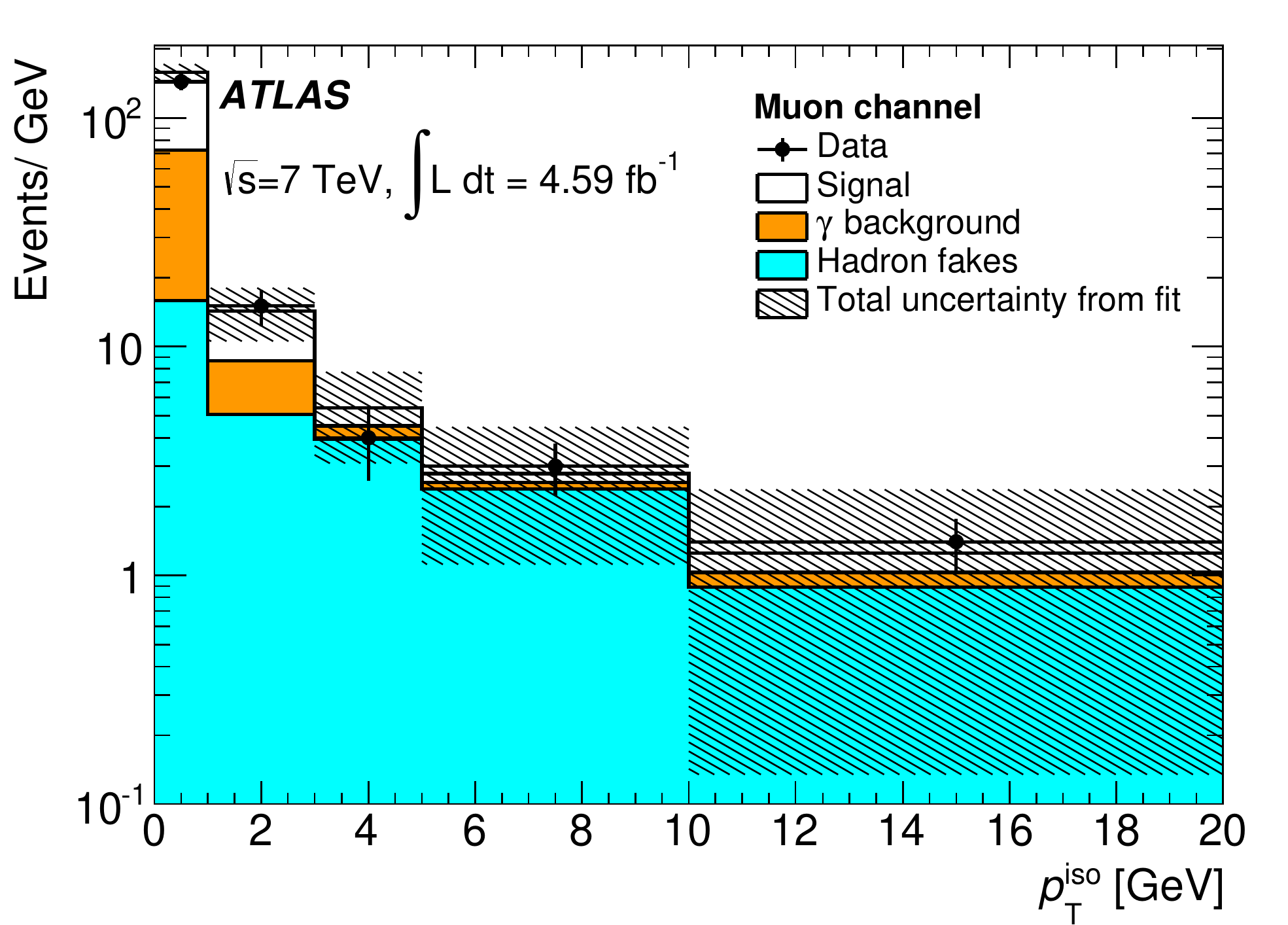}{0.55}{0}
\subfig{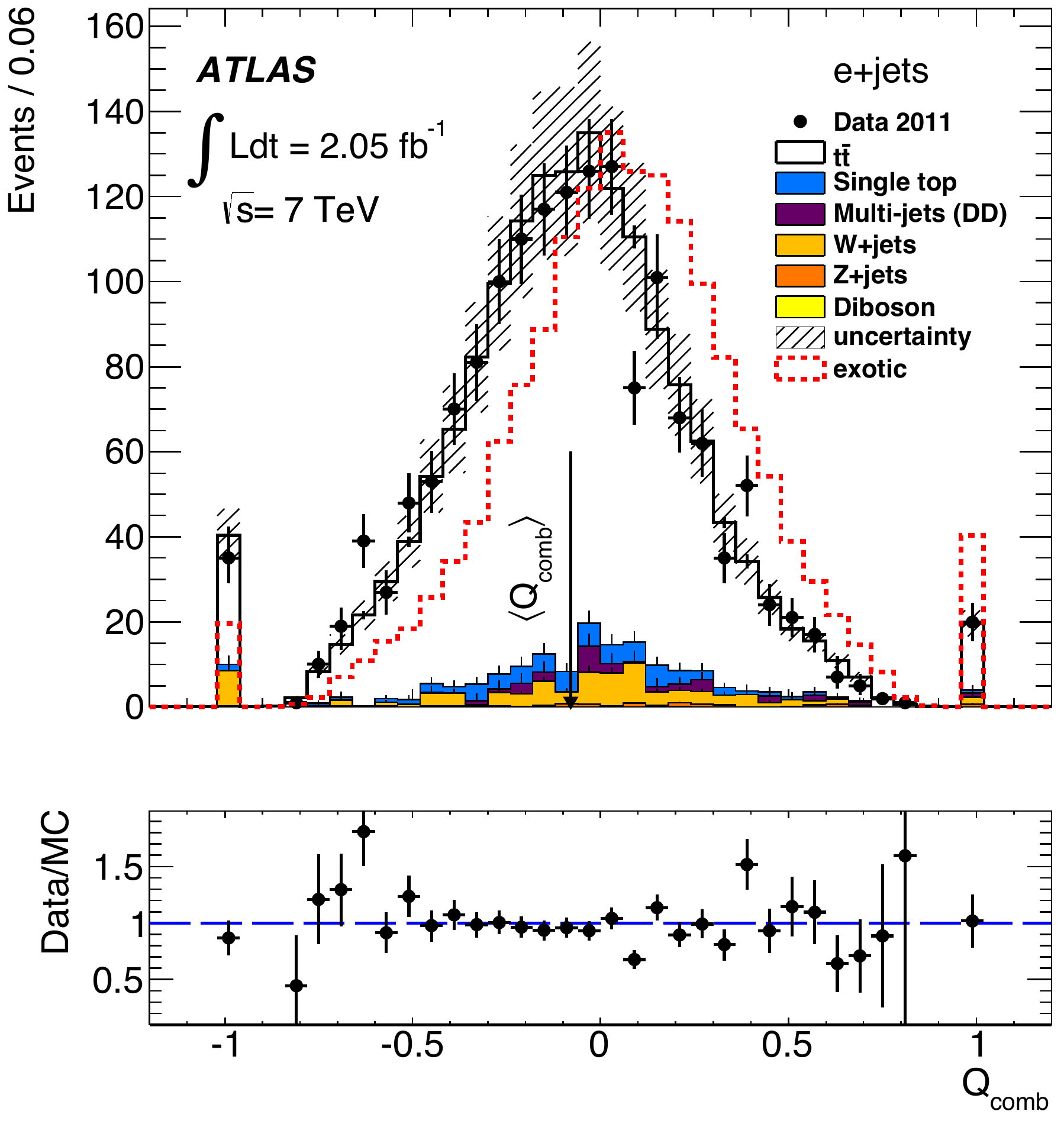}{0.45}{-15}
\caption{Result of the combined likelihood fit to extract the $\ttbar \gamma$
cross-section, using the track-isolation distributions as the discriminating
variable, for the single-muon channel~\pcite{Aad:2015uwa} (left).  Distribution
of the combined charge $Q_{\textrm{comb}} = q_b \cdot q_{\ell}$, in
single-electron final states.  Data strongly prefer the SM top charge rather
than an exotic model with charge $4/3e$~\pcite{Aad:2013uza} (right).}
\label{fttgamma}
\end{figure}

The measurement of the top-quark electric charge is related to the $t\gamma t$
vertex. In the Standard Model the top quark is expected to have an electric
charge of $+2/3$ in units of the electron charge magnitude. Besides the standard
\ttbar decay with $t \to W^+ b$ and $\bar{t} \to W^- \bar{b}$ the alternative
hypothesis of a $-4/3$ charge is conceivable, resulting in $t \to W^- b$
and $\bar{t} \to W^+ \bar{b}$. While the charge of the \Wboson\ can be determined on
an event-by-event basis from its leptonic decay, the $b$-quark charge is not
directly accessible from the reconstructed jet. In Ref.~\cite{Aad:2013uza} a
method to statistically assign a charge to the quark initiating a $b$-jet was
developed, based on a weighted average of the charge of the up to ten tracks
with the highest transverse momentum associated to the jet. An algorithm was
devised to optimize the efficiency and purity related to the pairing of the two
identified $b$-jets with the reconstructed lepton in the \ljets, based on the
invariant mass $m_{\ell b}$. The ATLAS collaboration measured the top-quark
charge to be $0.64 \pm 0.08$, excluding the alternative model by more than
\signif{8} (\Fref{fttgamma}, right).

\subsection{Couplings to the $Z$ boson}

Associated production of \ttbar with a $Z$ boson probes the $t$-$Z$ coupling.
The production \xs\ $\sigma_{\ttZ}$ is simultaneously extracted with
$\sigma_{\ttW}$ in channels with two or more leptons, since the two processes
are experimentally intertwined.  A variety of new physics models can alter the
prediction of these \xs s.  With the \SI{7}{\TeV} data~\cite{Chatrchyan:2013qca}
and then in a first analysis of the \SI{8}{\TeV}
dataset~\cite{Khachatryan:2014ewa} CMS reported evidence of \ttZ production,
with significances of $3.3$ and \signif{3.1} respectively.  But the luminosities
and energies available at the LHC ought to allow this process to be clearly
observed and more channels have been included and the analysis strategies
refined.

Using four signatures (opposite-sign dilepton, same-sign dilepton, trilepton,
and tetralepton) and performing a simultaneous maximum likelihood fit to
\SI{8}{\TeV} data, ATLAS extracted both \xs s, \ttW and \ttZ, with significances
of $5.0$ and \signif{4.2}, respectively, over the background-only
hypothesis~\cite{Aad:2015eua}.  \Fref{attV} shows the fit to the various signal
and control regions.  The sensitivity to the \ttZ\ process comes from the tri-
and tetralepton channels. Important irreducible Standard Model backgrounds that
produce three or four leptons are diboson ($WZ$, $ZZ$) events, determined
through fits in control regions, and associated single top-quark production
($tZ$, $WtZ$). The measured \xs\ is $\sigma_{\ttZ} = 176^{+58}_{-52}$ fb, with
uncertainties dominated by the statistical component.

\begin{figure}[htbp]
\centering
\includegraphics[width=1.015\textwidth]{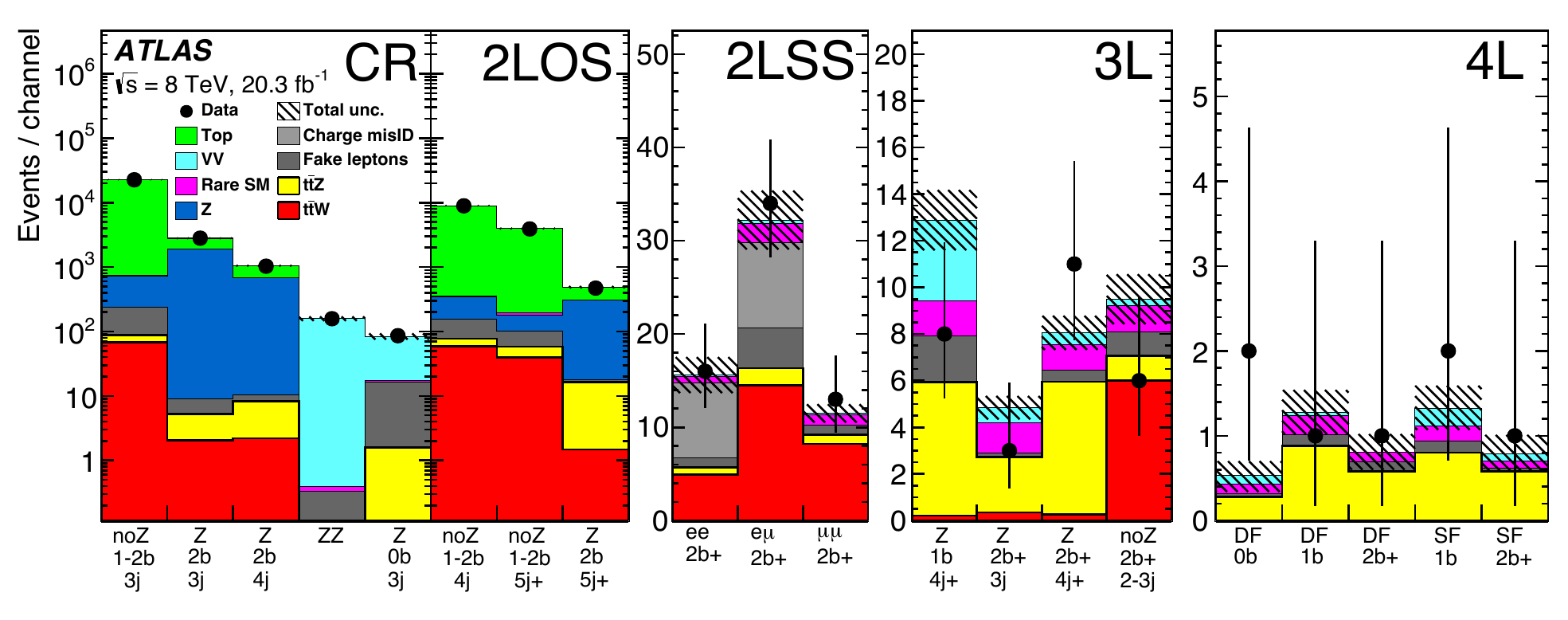}
\caption{Maximum likelihood fit in 15 signal and 5 control regions for the four
channels of the ATLAS measurement~\protect\cite{Aad:2015eua} to extract
$\sigma_{\ttZ}$ and $\sigma_{\ttW}$.} \label{attV}
\end{figure}

CMS also determined the \xs s of the \ttZ and \ttW processes at \SI{8}{\TeV},
with significances of $4.8$ and \signif{6.4}, respectively, using the same final
states~\cite{Khachatryan:2015sha}. The analysis strategy aimed at reducing the
statistical uncertainty by lowering the requirements on the quality of the
reconstructed objects, at the expense of larger systematic uncertainties. A full
reconstruction of pre-selected events was attempted, by matching the
reconstructed objects in the detector to the decaying $W$ and $Z$ bosons, and to
the top quarks. A linear discriminant helped to determine the best permutation
in matching jets and leptons. Signal was separated from background by means of
several BDTs, one in each channel and jet multiplicity, exploiting the values of
the linear discriminant and other kinematic quantities (\Fref{cttV}).  The
measured \xs\ is $\sigma_{\ttZ} = 242^{+65}_{-55}$ fb.  The result was used to
place constraints on the axial and vector components of the $t$-$Z$ coupling and
on dimension-six operators in an effective field theory framework.

\begin{figure}[htbp]
\subfig{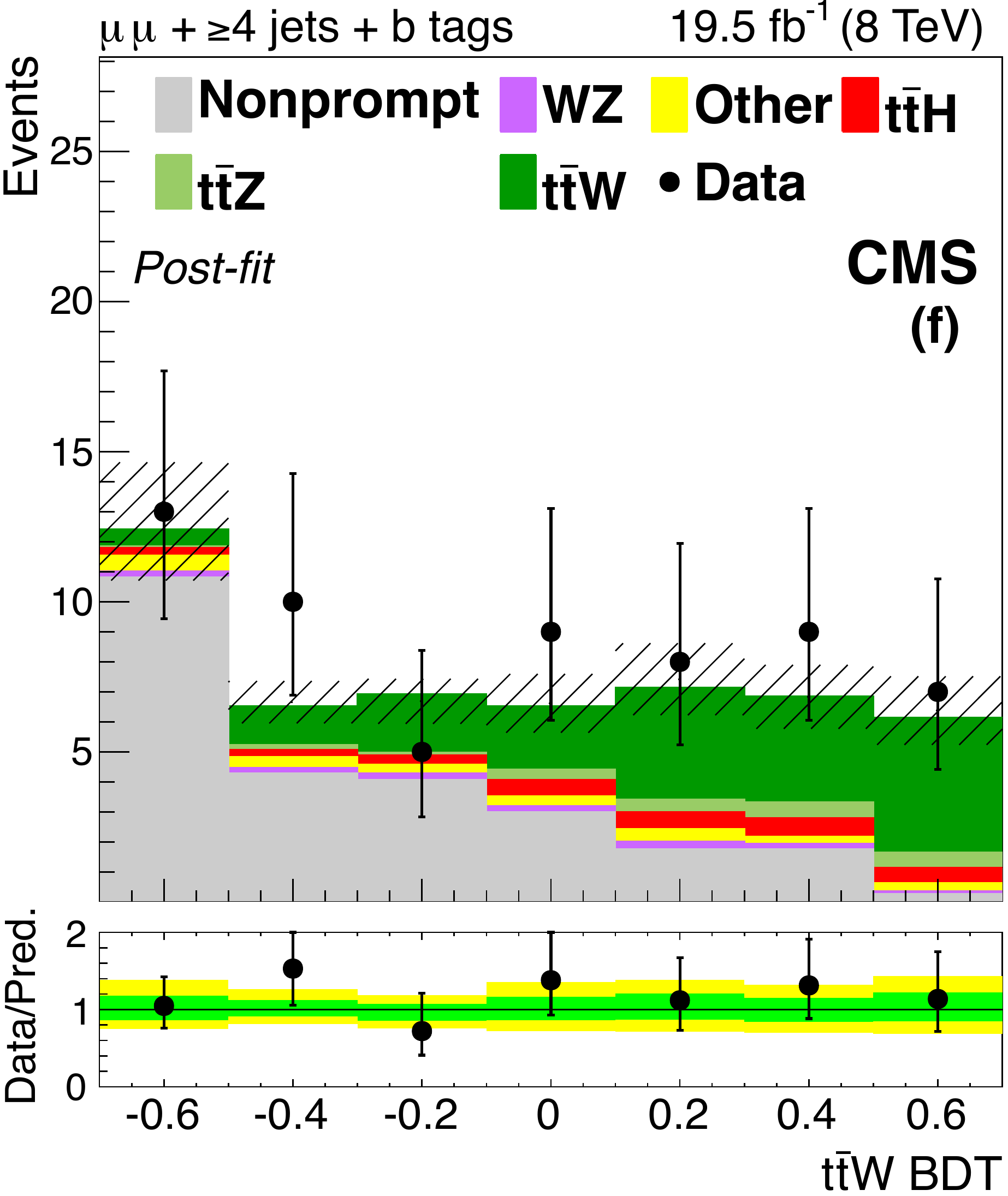}{0.30}{0}
\subfig{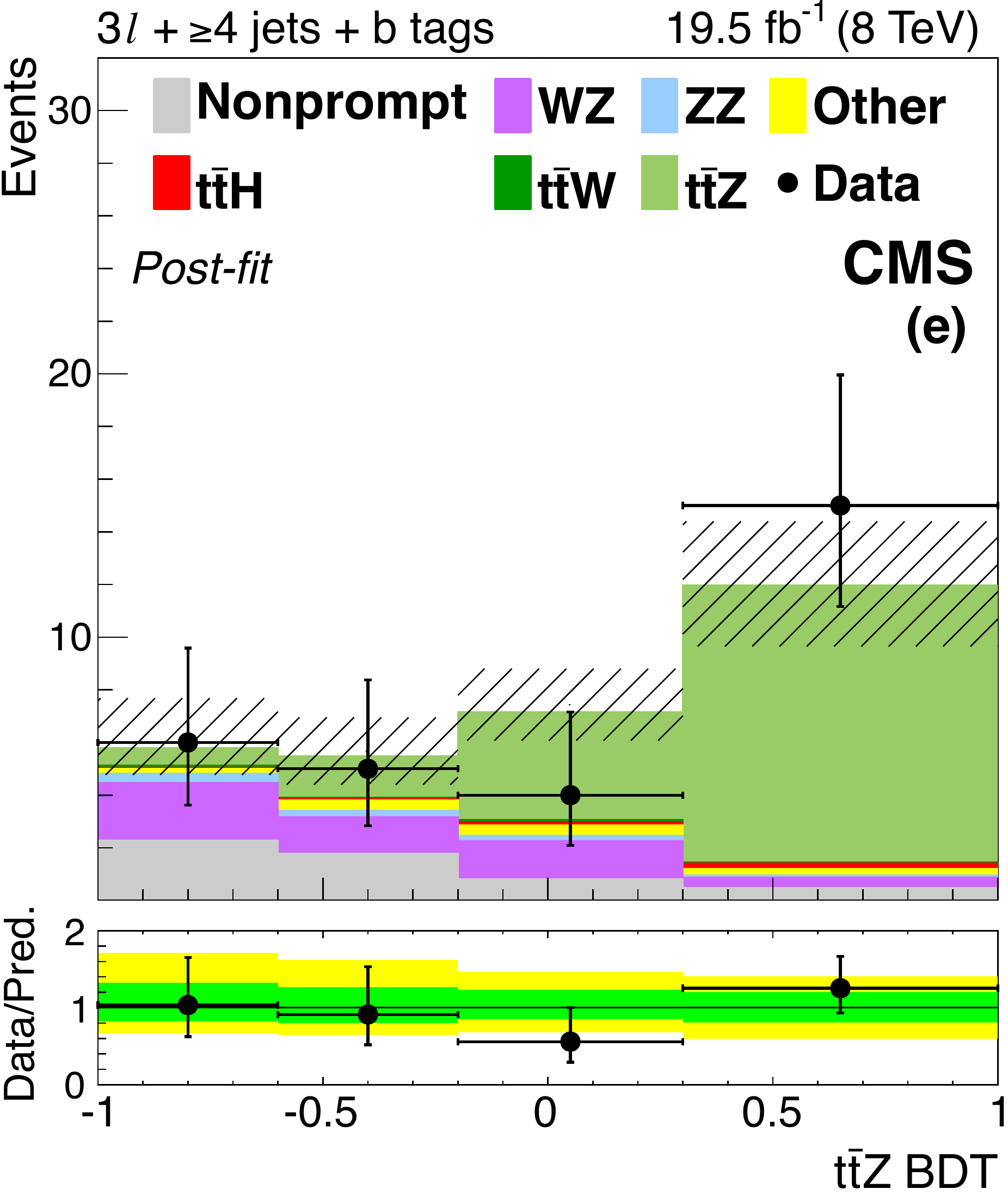}{0.30}{0}
\subfig{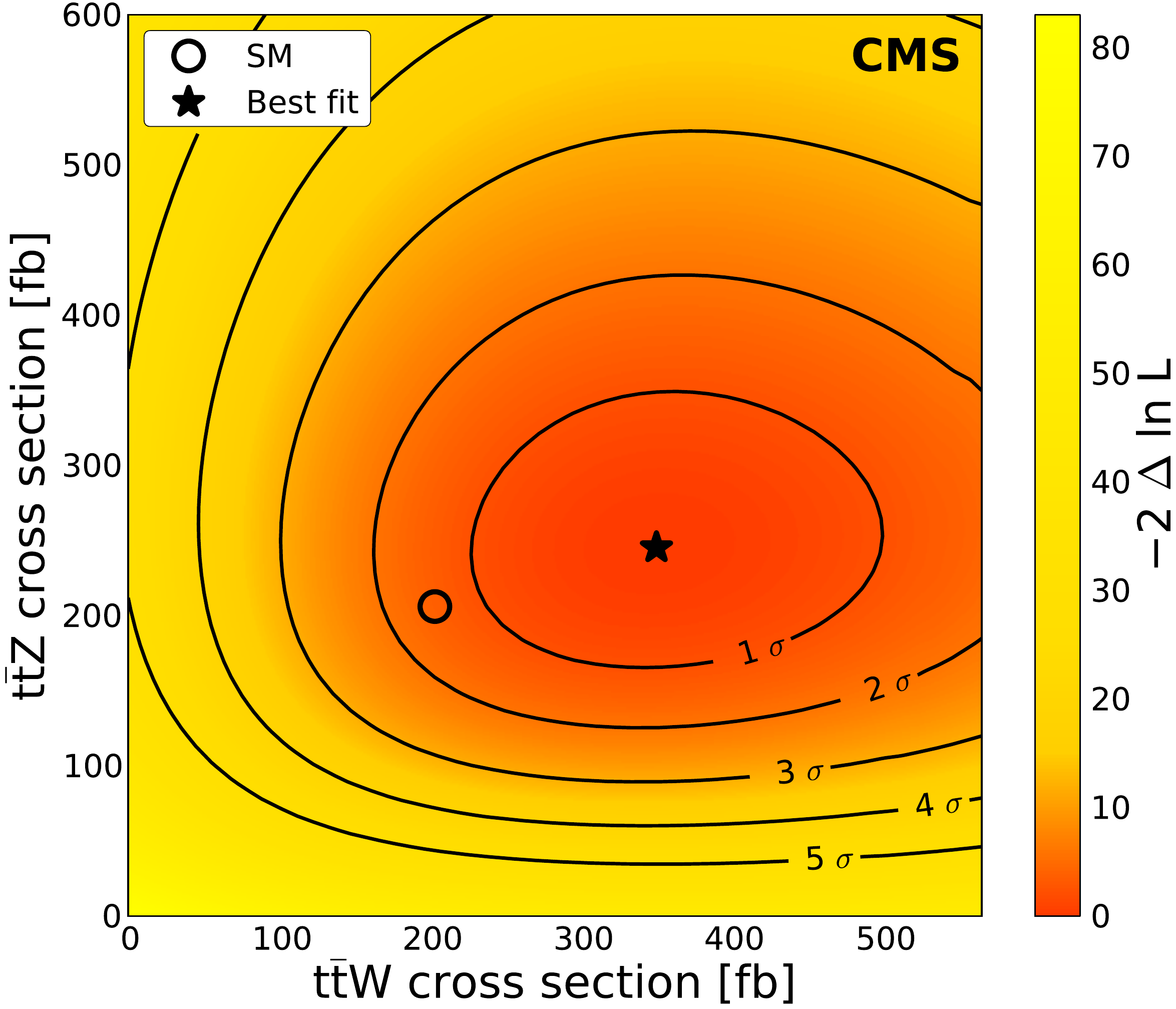}{0.40}{0}
\caption{BDT distributions in two sample channels targeting the extraction of
\ttW (left) and \ttZ (centre), and result compared to Standard Model predictions
(right) for the CMS measurement~\protect\cite{Khachatryan:2015sha}.}
\label{cttV}
\end{figure}

\subsection{Couplings to the Higgs boson}

Due to the large mass of the top quark the Yukawa coupling to the Higgs boson is
expected to be very large, $y_t = 0.9956 \pm 0.0043$, using current values for
$G_F$ and $m_t$~\cite{pdg2014}. While the effects of this coupling can be
inferred indirectly from Higgs boson production through gluon fusion and Higgs
decay to a pair of photons through loops, the associated production process \ttH
will allow this coupling to be probed directly and unambiguously. However, the
expected total \xs\ of $\sigma_{\ttH} = 129^{+12}_{-16}$ fb at
\SI{8}{\TeV}~\cite{Dittmaier:2011ti} is significantly smaller than that of
\ttbar production, which is an important background.

Due to the small production \xs\ and the large backgrounds, no single decay mode
of \ttH is expected to be sensitive in analyses using \RunOne\ data.  Instead,
several possible final states have been explored by ATLAS and CMS. They are
broadly organized in three categories depending on the targeted Higgs boson
decays: $H\to\bbbar, H\to\gamma\gamma$ and $H \to$ leptons, the latter mainly
targeting $H\to WW$, $H\to ZZ$ and $H\to\tau\tau$ decays.

The \ttH, $H \to \bbbar$ channel is characterized by very large combinatorial
back\-grounds, reconstruction inefficiencies and small signal-to-background
ratios. An important non-reducible background is the associated $\ttbar\bbbar$
production. CMS and ATLAS reported results in the single and dilepton
channels~\cite{Khachatryan:2015ila, Aad:2015gra} with the \RunOne\ dataset, also
exploiting a matrix-element technique. ATLAS also tackled the very challenging
all-hadronic channel, which required dedicated data-driven techniques to
estimate the multijet background~\cite{Aad:2016zqi}.  On the other hand, the
\ttH, $H\to\gamma\gamma$ channel is very clean but has a very small branching
ratio. The measurements exploit a data-driven continuum background modelling,
similar to that used in $H\to\gamma\gamma$ analyses. ATLAS and CMS reported
results in channels with or without extra leptons produced by the top-quark
pair~\cite{Aad:2014lma, Khachatryan:2014qaa}.  The remaining Higgs boson decays
are collected in final states with multiple leptons, categorized according to
the number and charge of electrons, muons and hadronically decaying tau leptons
in order to optimize the signal-to-background ratio. CMS and ATLAS analysed
\RunOne\ data also in this channel~\cite{Khachatryan:2014qaa, Aad:2015iha}.
\Fref{ttH} shows data compared to expectation in the different channels. 

\begin{figure}[htbp]
\subfig{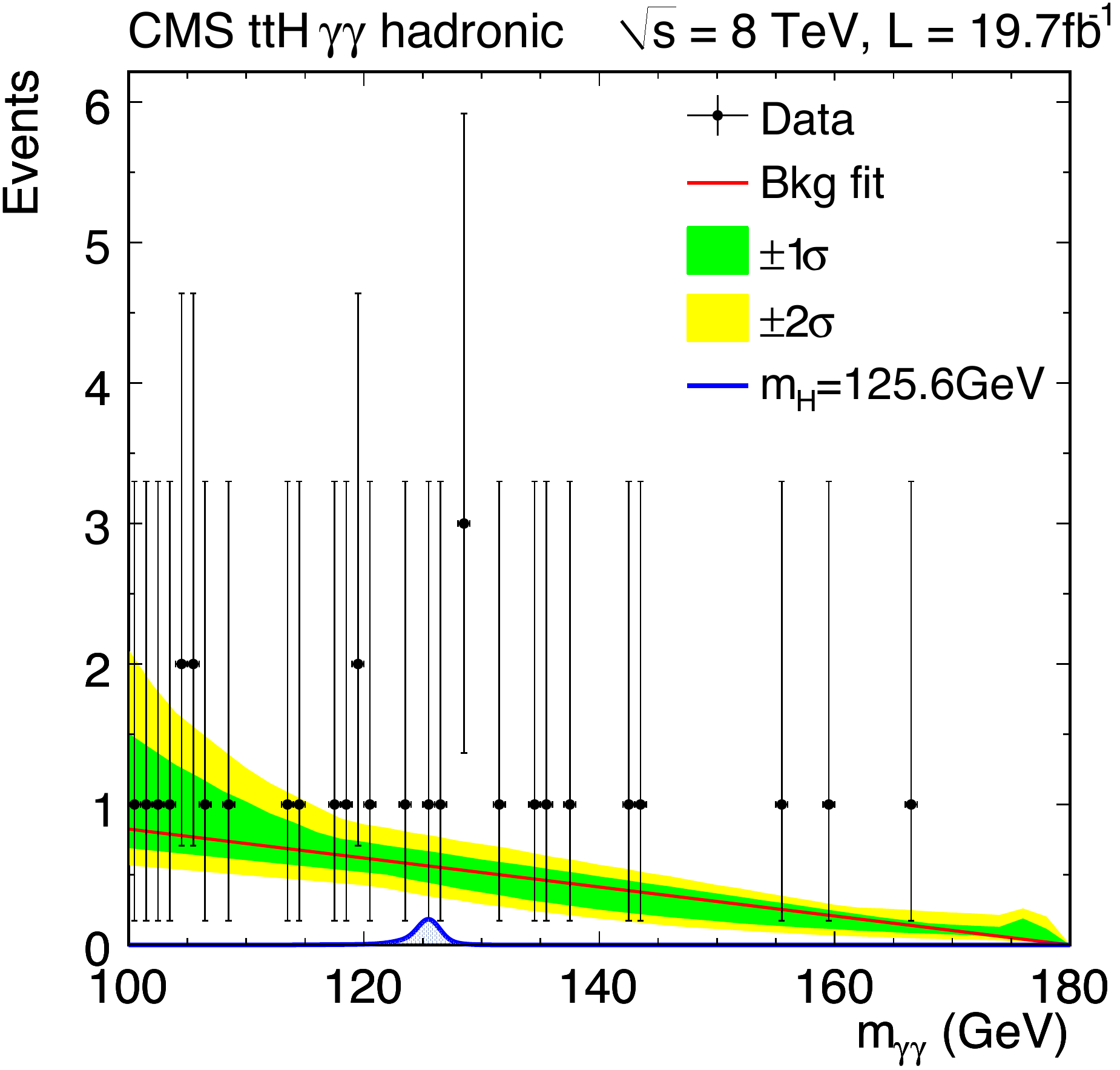}{0.335}{0}
\subfig{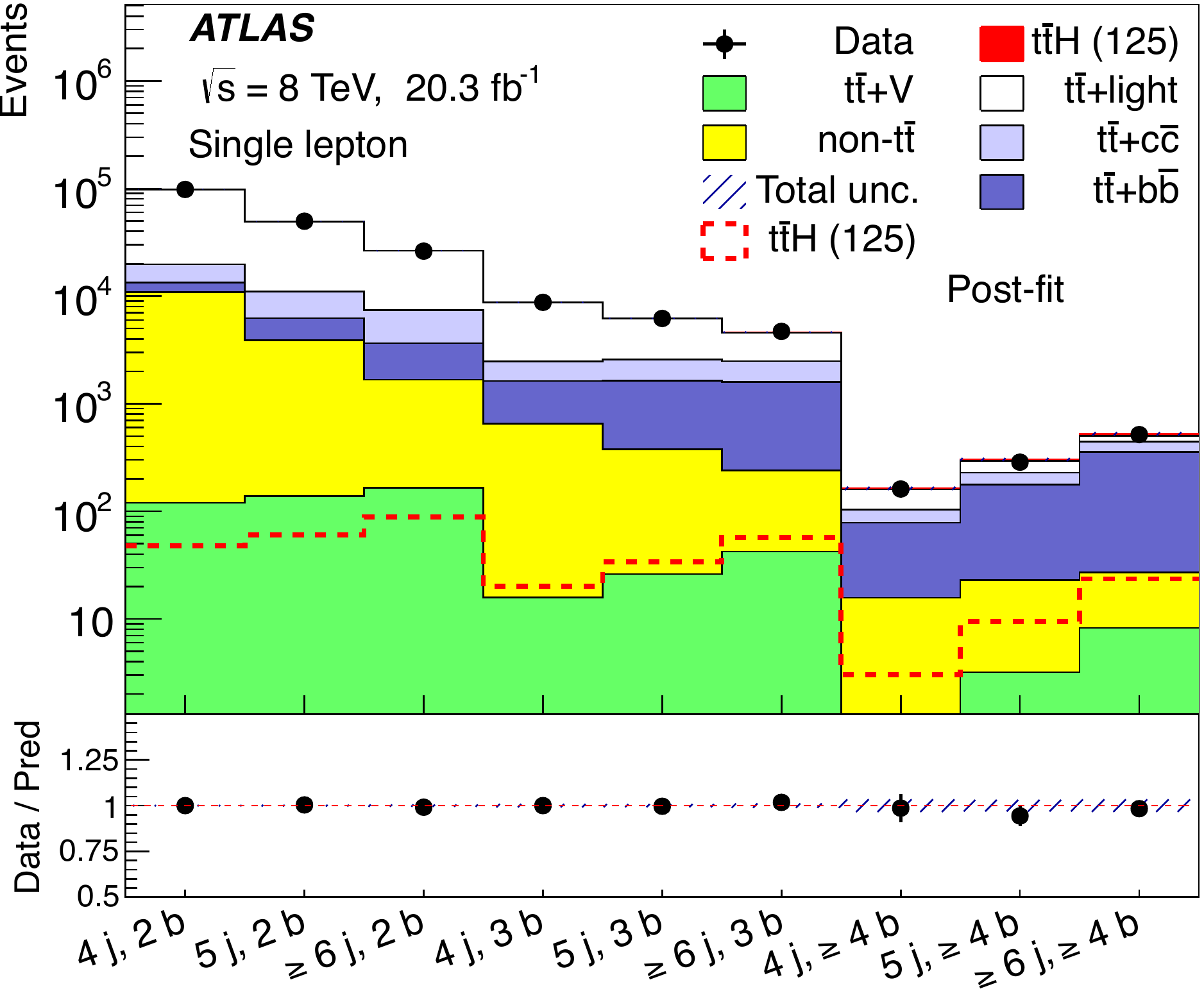}{0.375}{0}
\subfig{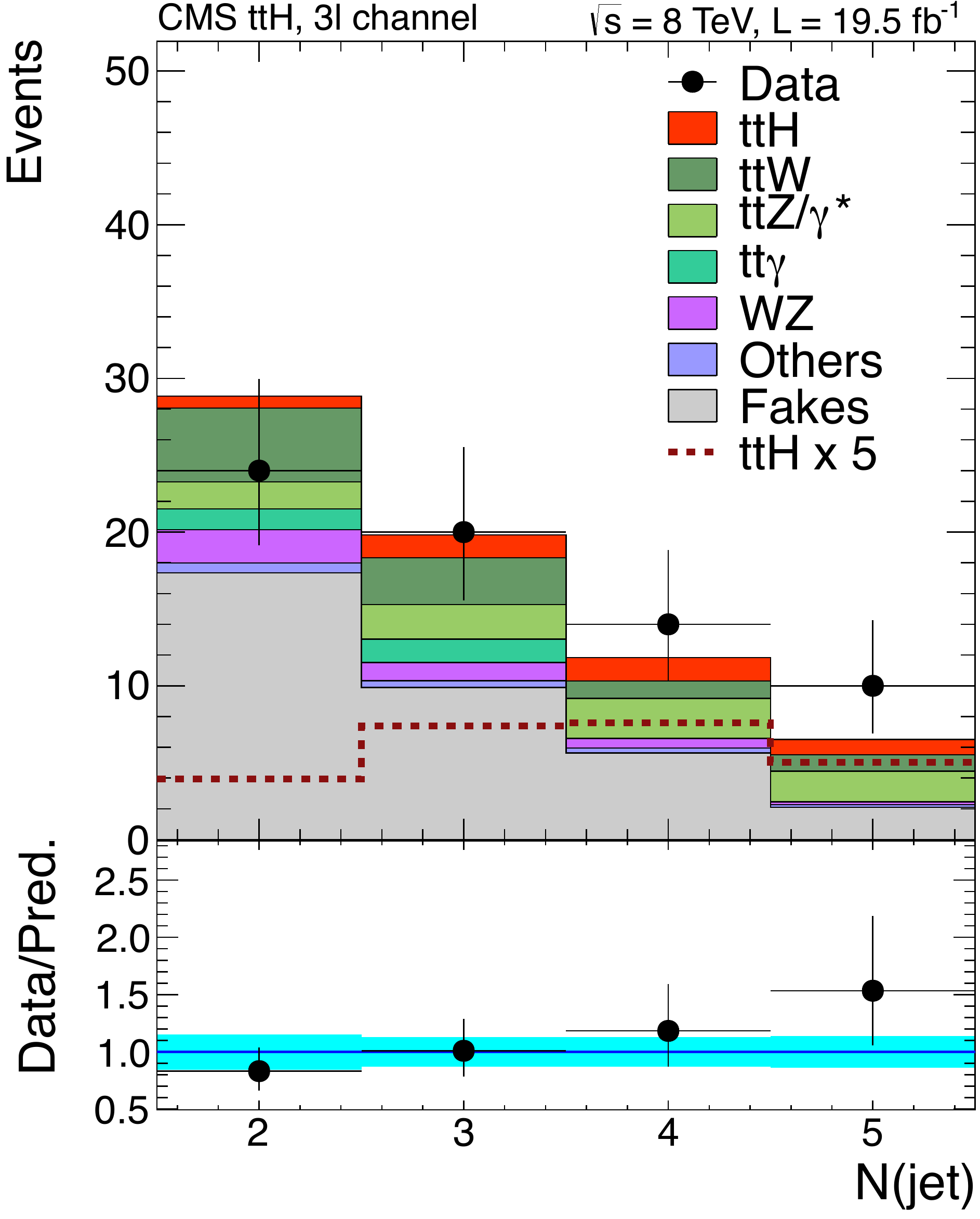}{0.27}{0}
\caption{Invariant mass of the $H \to \gamma\gamma$ candidate
(left)~\pcite{Khachatryan:2014qaa}, post-fit result showing signal and
background composition in the different \ttH, $H\to \bbbar$ categories in the
\ljets\ (centre)~\pcite{Aad:2015gra} and number of jets in the trilepton channel
(right)~\pcite{Khachatryan:2014qaa}.} \label{ttH}
\end{figure}

Both experiments have combined their \ttH measurements in the framework of a
global Higgs couplings extraction~\cite{Khachatryan:2014jba, Aad:2015gba,
Aad:2016zqi}.  Table~\ref{tab:ttH} shows the results obtained in the different
channels and in the combination.

\fulltable{\label{tab:ttH}Measurements of the \ttH signal strength $\mu$,
defined as $\mu \equiv \sigma_{\ttH}(\textrm{measured}) /
\sigma_{\ttH}(\textrm{SM})$, in the different decay channels of the Higgs boson
(rows) and of the top-quark pair (columns). The overall combinations of the
signal strengths are shown in the last row~\pcite{Aad:2016zqi,
Khachatryan:2014jba}.}
\begin{tabular}{@{}c@{\hspace{5ex}}ccc@{\hspace{7ex}}ccc}
\br
& \multicolumn{3}{c}{ATLAS} & \multicolumn{3}{c}{CMS}\\
\mr

$H \to \gamma\gamma$  & \multicolumn{3}{c}{\aerr{1.3}{2.6}{1.7}} 
                      & \multicolumn{3}{c}{\aerr{2.7}{2.6}{1.8}}\\[2ex]
& $+0\ell$ & $+1\ell$ & $+2\ell$ 
& $+0\ell$ & $+1\ell$ & $+2\ell$\\
\mr

$H \to \bbbar$ & $1.6 \pm 2.6$ & $1.2 \pm 1.3$ & $2.8 \pm 2.0$ 
& & \aerr{1.7}{2.0}{1.8} & \aerr{1.0}{3.3}{3.0}\\[0.9ex]
$\ell^{\pm}\ell^{\pm}$ & \aerr{2.8}{2.1}{1.9} & \aerr{2.8}{2.2}{1.8} & \aerr{1.8}{6.9}{2.0} 
& \aerr{5.3}{2.1}{1.8} & \aerr{3.1}{2.4}{2.0} & \aerr{-4.7}{5.0}{1.3}\\[0.9ex]
$\tau_{\mathrm{had}} (\tau_{\mathrm{had}})$ & & \aerr{-9.6}{9.6}{9.7} & \aerr{-0.9}{3.1}{2.0}
& & \multicolumn{2}{c}{\aerr{-1.3}{6.3}{5.5}}\\
\mr
Combined & \multicolumn{3}{c}{$1.7 \pm 0.8$} &
\multicolumn{3}{c}{\aerr{2.9}{1.0}{0.9}}\\
\br
\end{tabular}
\endfulltable

The coupling of top quarks to Higgs bosons can also be probed in the single-top
channel with an additional Higgs boson ($tH$). The process is expected to be
suppressed by a factor of ten with respect to $\ttH$ production, due to
destructive interference between diagrams.  Since $\ttH$ effectively probes
$y_t^2$, it is interesting to also measure the $tH$ process, which could be
dramatically enhanced, if for instance $\kappa_t \equiv y_t/y_t^{\textrm{SM}} =
-1$.  ATLAS did not observe such a large enhancement in the $H\to\gamma\gamma$
channel~\cite{Aad:2014lma}, and determined $\kappa_t \in (-1.3,+8.0)$ at 95\%
confidence level.  CMS performed a dedicated analysis to search for $t$-channel
$tH$ production in all Higgs decay modes. Depending on the assumed
$H\to\gamma\gamma$ branching ratio, upper limits of $600$ to \SI{1000}{fb} were
obtained for exotic $tH$ production~\cite{Khachatryan:2015ota}, a factor of 3 to
5 larger than the expected \xs\ in the case of $\kappa_t = -1$. 

The precise measurement of the top Yukawa coupling, as well as the couplings to
other bosons, will continue to be an important goal for the LHC programme and is
currently vigorously pursued by both collaborations.

\section{Exotic top-quark decays}

Top quarks predominantly decay via the charged weak current transition $t \to
Wb$. The decays $t \to Ws$ and $t \to Wd$ are suppressed by the corresponding
CKM matrix elements and expected to occur at the $0.2\%$ and \SI{8e-5}{} level,
respectively.  With the large number of top quarks produced at the LHC it is
interesting to study decays that are either highly suppressed in the Standard
Model, like flavour-changing neutral current (FCNC) transitions, $t \to Zq$, $t
\to \gamma q$, $t \to gq$ or $t \to Hq$, or require the presence of new
particles such as a charged Higgs boson that appears in SUSY extensions, $t \to
H^+ q$, or that would be a first indication for baryon number violation, $t \to
\bar{b} \bar{c} \ell^+$, or lepton flavour violation, $t \to e \mu
q$~\cite{Davidson:2015zza}.

\subsection{Search for top-quark decays via flavour-changing neutral currents}

While top-quark decays to light down-type quarks are CKM-suppressed, the
transition to an up-type quark is even more suppressed due to the GIM
mechanism~\cite{Glashow:1970gm} with predicted branching fractions of ${\cal
O}(10^{-14})$, not directly accessible to the LHC experiments. Such decay modes
could be significantly enhanced in extensions of the Standard Model, for
instance due to the presence of additional virtual particles in the penguin
loops. The effect of additional heavy particles on top-quark FCNCs can be
described by dimension-six gauge-invariant operators.  Searches for FCNC
interactions involving top quarks have been performed at the LHC probing the
$tgq$, $tZq$ and, more recently, the $t\gamma q$ and $tHq$ vertices.

At hadron colliders top-quark decays $t \to gq$ cannot be easily distinguished
from other multijet signatures. Thus the search strategy for such anomalous
couplings is to consider the production and decay $gq \to t \to b \ell \nu$,
yielding a final state of exactly one semileptonically decaying top quark, with
no extra jets. As substantial SM background is present after the event
selection, neural-network based multivariate analyses have been performed by
ATLAS using a partial \SI{7}{\TeV} dataset~\cite{Aad:2012gd} and the complete
\SI{8}{\TeV} dataset~\cite{Aad:2015gea}.  The most discriminating variables rely
on the fact that top quarks produced through FCNC have low \pT, the $W$ bosons
produced in top-quark decays are boosted, and the produced top
quark-to-antiquark ratio differs between FCNC and SM processes. The
neural-network output as shown in \Fref{fcnc} (left) was then used in a binned
likelihood fit to set limits on anomalous couplings, finding $\kappa_{tug} <
\SI{5.8e-3}{} \Lambda$ and $\kappa_{tcg} < \SI{1.3e-2}{} \Lambda$, where
$\kappa_{tqg}$ are the multiplicative factors of the strong coupling constant in
the tensor term of the effective Lagrangian and $\Lambda$ is related to the mass
scale above which the effective theory breaks down, expressed in \si{\TeV}.

\begin{figure}[htbp]
\subfig{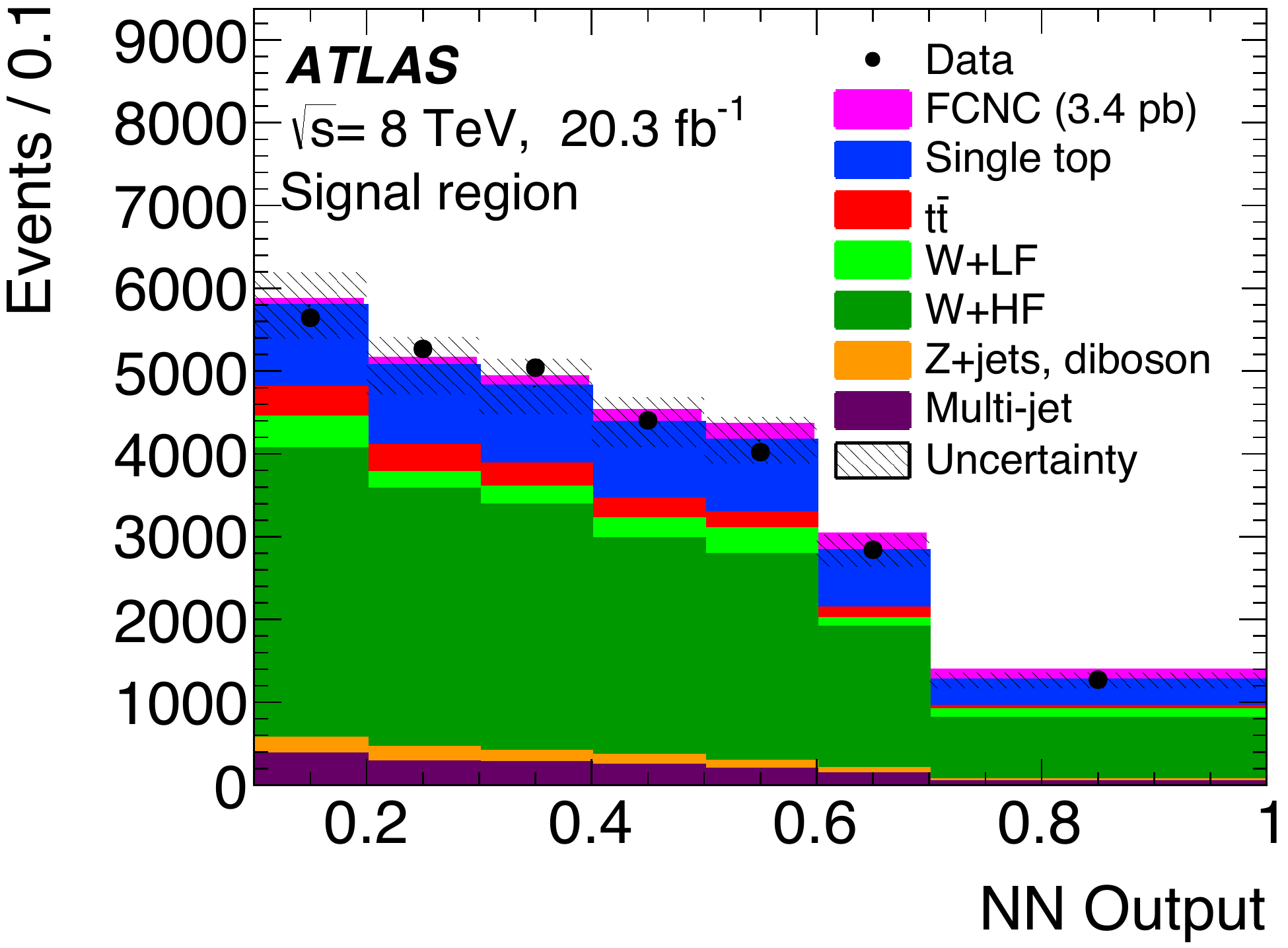}{0.55}{3}
\subfig{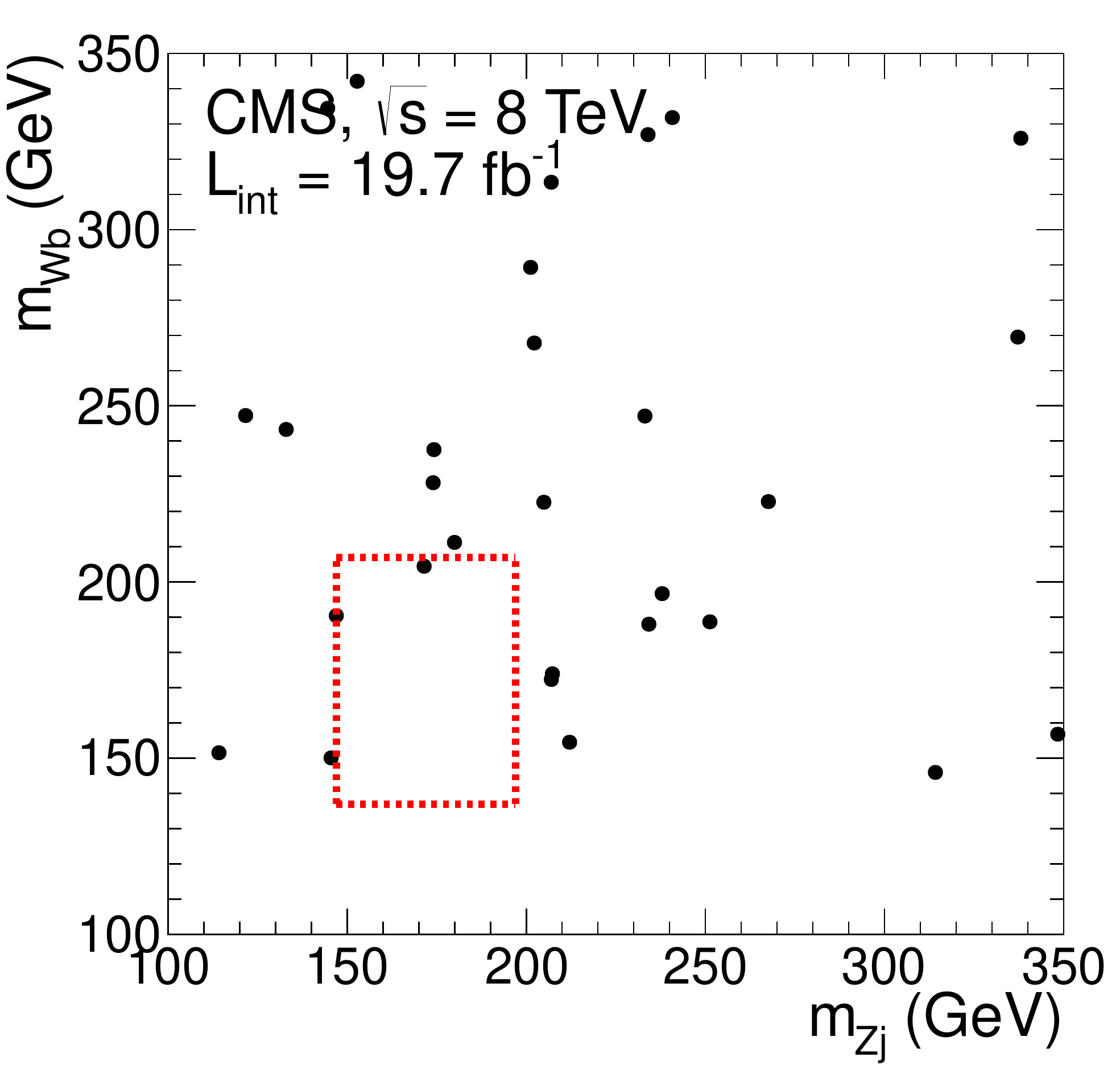}{0.45}{0}
\caption{Neural-network output comparing selected data with expectation,
including single-top production via FCNC with an assumed \xs\ of
\SI{3.4}{pb}~\pcite{Aad:2015gea} (left).  Two-dimensional scatter distribution
after event selection of $m_{W\!b}$ vs.~$m_{Zj}$ in the search for $t\to Zq$
decays~\pcite{Chatrchyan:2013nwa} (right).} \label{fcnc}
\end{figure}

The top-quark decay to a $Z$ boson and an up-type quark has also been searched
for. Upper limits on the decay rate were obtained by ATLAS with a partial
dataset of \SI{7}{\TeV} data~\cite{Aad:2012ij} and by CMS with the full
\SI{7}{\TeV} dataset~\cite{Chatrchyan:2012hqa}, as well as by CMS with the
complete \RunOne\ dataset~\cite{Chatrchyan:2013nwa} and by ATLAS with the full
\SI{8}{\TeV} data~\cite{Aad:2015uza}.  Given the small expected branching
fraction, searches are performed in the channel, where only one top quark decays
through FCNC, $\ttbar \to ZqWb \to \ell\ell q \ell \nu b$. The final state
considered features three leptons, which is particularly advantageous to reduce
the Standard Model backgrounds, at the expense of a small signal efficiency.
Because of the requirement of three leptons, low lepton \pT thresholds could be
used.  For the reconstruction of the \Zboson\ the pair of opposite-sign
same-flavour leptons with $m_{\ell\ell}$ closest to $m_Z$ was chosen. At least
two jets were required and the \mbox{$b$-tagging} information was explicitly
used when the size of the dataset allowed for it.  Selected events were required
to be kinematically compatible with the FCNC decay, for instance using the
top-quark, \Wboson\ and \Zboson\ mass constraints to define a $\chi^2$.
Alternatively, all boson-jet combinations for the \ttbar decay were examined
after requiring a large sum of \MET and \pT of leptons and jets, and the
top-quark pair with the largest azimuthal separation was chosen.  Signal events
were modelled with the \TopReX~\cite{toprex}, \MadGraph\ or
\Protos~\cite{AguilarSaavedra:2010rx} event generators.  The contribution of
events with fake leptons was estimated from data, considering the different
possible combinations of lepton flavour and electric charge of three leptons,
and extrapolating the measured values from side-band regions using the
prediction from simulation.  The $b$-tagging information was used to reduce the
backgrounds from Drell-Yan and diboson production.  The FCNC signal and the
\ttbar and \ttV processes have different number of $b$-jets and this was
exploited to estimate the various contributions by measuring the number of
events in different $b$-tag multiplicities.  \Fref{fcnc} (right) shows the
two-dimensional signal region for the \SI{8}{\TeV} search by CMS.

The $t\gamma q$ vertex can be conveniently probed by searching for single
top-quark production in association with a photon. With the \SI{8}{\TeV} dataset
CMS searched for this signature in final states with an isolated high-\pT muon,
at least one jet, but at most one $b$-tagged jet, and missing transverse
momentum~\cite{Khachatryan:2015att}.  The signal was simulated with \Protos\
with a minimum \pT requirement of \SI{30}{\GeV} for the associated photon.
Events were classified as signal candidates if the reconstructed invariant mass
$m_{\mu\nu b}$ was compatible with the \topmass. Several discriminating
variables were found to be useful to reject the remaining background, dominated
by the $W\gamma+$jets process, and were combined in BDTs, one for each of
$tu\gamma$ and $tc\gamma$, whose distribution were then fitted to extract the
signal component (\Fref{fcnc2}, left).  The \pT of the photon candidate is the
single most powerful variable. As no evidence for signal was found, limits on
the anomalous couplings $\kappa_{tq\gamma}$ of $3-10\%$ were placed and
translated into limits on the branching fractions $t\to q\gamma$.  The
measurement was also performed in a restricted phase space, corresponding to the
acceptance of the detector, including a requirement on the transverse momentum
and angular distances of the photon of $\pT > \SI{50}{\GeV}$, $\Delta
R(\gamma,\mu) > 0.7$, and $\Delta R(\gamma,\textrm{jet}) > 0.7$.

\begin{figure}[htbp]
\subfig{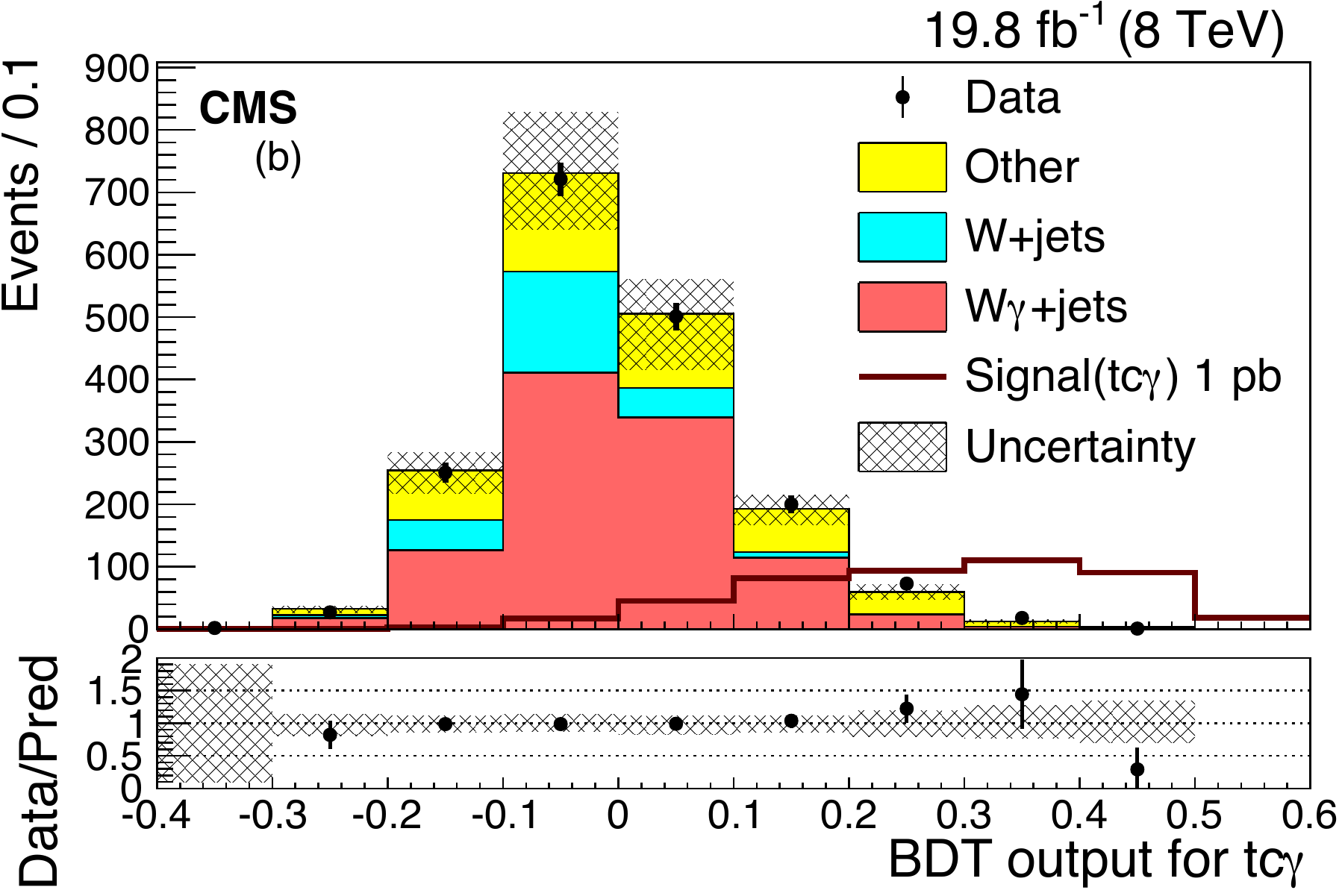}{0.5}{0}
\subfig{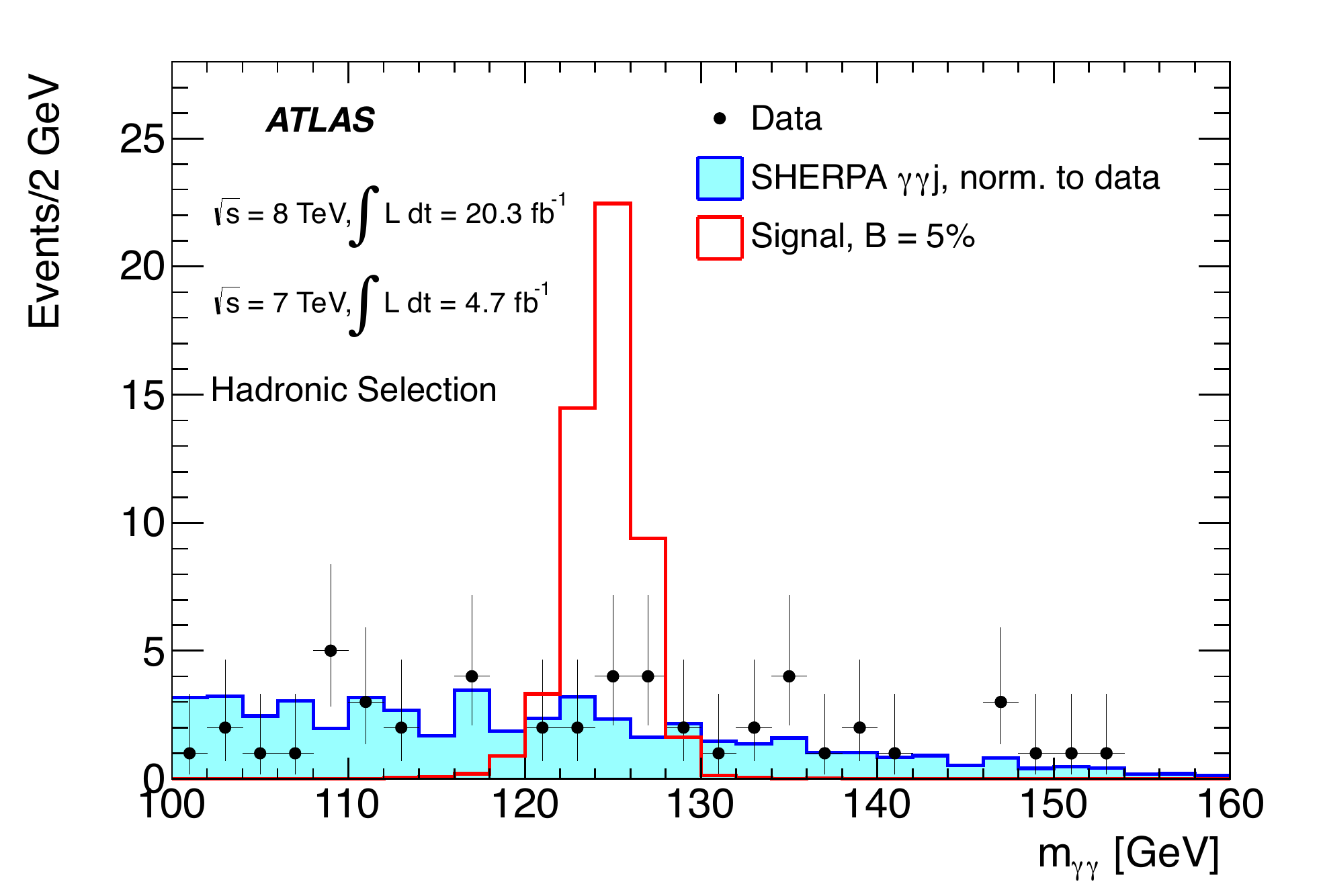}{0.5}{0}
\caption{BDT output distribution for the data, backgrounds and the expected
$tc\gamma$ signal for an assumed \xs\ of \SI{1}{pb}~\pcite{Khachatryan:2015att}
(left).  Distribution of the invariant mass of the two photons,
$m_{\gamma\gamma}$, for events passing the full hadronic selection in a search
for the decay $t \to q H$.  The non-resonant production is simulated using {\sc
Sherpa} and normalized to data~\pcite{Aad:2014dya} (right).} \label{fcnc2}
\end{figure}

While the decay $t \to qH$ is also strongly suppressed in the Standard Model at
the level of $10^{-17} - 10^{-15}$, extensions can predict large enhancements,
up to $\sim\!\!10^{-3}$ for the $t \to cH$ branching fraction.  With the large
\ttbar samples collected during \RunOne, several searches have been performed
for the process $\ttbar \to HqWb$, classified according to the Higgs boson
decay. For the $H \to \gamma\gamma$ channel the disadvantage of the small
branching fraction is compensated by the very small backgrounds and good mass
resolution of the photon pair (\Fref{fcnc2}, right).  The searches were not
sensitive to the difference between the $u$ and $c$ quark in the final state.
ATLAS~\cite{Aad:2014dya} and CMS~\cite{Chatrchyan:2014aea} obtained upper limits
on the branching fraction $t \to q H$ of $0.79\%$ and $0.69\%$, respectively.
CMS~\cite{Chatrchyan:2014aea} and ATLAS~\cite{Aad:2015pja} also considered $t
\to qH$ decays in final states with at least three reconstructed leptons,
including hadronically decaying tau leptons, targeting the Higgs boson decays to
$WW$, $\tau \tau$ and $ZZ$.  Events were divided into mutually exclusive
categories based on the number of leptons, number of \tauhad, the number and
invariant mass of opposite-sign same-flavour lepton pairs, and, in the case of
CMS, the number of $b$-tagged jets, \met and \HT. These searches take advantage
of a larger branching ratio of the Higgs boson decay, but do not feature a
clearly-reconstructed Higgs-boson mass peak.  CMS obtained an upper limit on the
branching ratio $t \to cH$ of $1.3\%$, and, combining this measurement with the
$H \to \gamma\gamma$ analysis, $0.56\%$~\cite{Khachatryan:2014jya}, while ATLAS
obtained upper limits of $0.79\%$ and $0.78\%$, for $t \to cH$ and $t\to uH$,
respectively.  Finally, the $H \to \bbbar$ decay has also been considered,
together with a combination of all channels by ATLAS~\cite{Aad:2015pja}. The
high multiplicity of $b$-jets and kinematic differences between the signal and
the abundant background, dominated by $\ttbar$ production, were exploited in the
construction of a likelihood discriminant. Here also, no significant excess of
events above the expectation from background was observed and upper limits on
the branching fractions of $0.56\%$ and $0.61\%$ for $t \to cH$ and $t \to uH$
were extracted, respectively. The combination of all channels in ATLAS yielded
upper limits of $0.46\%$ and $0.45\%$ for the two decay modes.

\subsection{Search for top-quark decays to charged Higgs bosons}

Additional Higgs particles are predicted in several extensions of the Standard
Model, including a pair of charged Higgs bosons $H^{\pm}$, as for instance in
the two Higgs doublet models (2HDM)~\cite{Lee:1973iz}. In such models a
top-quark pair could decay through $\ttbar \to H^+ b W^{-}\bar{b}$, if $m_{H^+}
< m_t - m_b$.  In the MSSM scenario and for large $\tan \beta$ values the
charged Higgs boson is expected to preferentially decay via $H^+ \to \tau^+
\nu_{\tau}$, effectively enhancing the tau lepton production in \ttbar decays
with respect to other leptons, while for $\tan \beta < 1$, the decay mode $H^+
\to c\bar{s}$ is expected to dominate.  ATLAS and CMS searched for both decay
modes assuming charged Higgs boson masses of $80$ to \SI{160}{\GeV} for tauonic
decays and $90$ to \SI{150}{\GeV} for hadronic decays. The limits are quoted
assuming ${\cal B} (H^+ \to \tau^+ \nu_{\tau}) = 1$ or ${\cal B} (H^+ \to c
\bar{s}) = 1$, respectively.

Tauonic charged Higgs decays from top quarks have been searched for in \RunOne\
at 7 and \SI{8}{\TeV}.  For the \SI{7}{\TeV} analyses three final states were
considered~\cite{Aad:2012tj}, where discriminating variables could be identified
exploiting the different topologies of signal and background.  In the
$\taulep+$jets channel, leptons produced in tau lepton decays have to be
discriminated against those stemming directly from a $W$ boson decay.  Two
useful variables are the invariant mass $m_{b\ell}$, where the choice of the
$b$-tagged jet is based on a kinematic constraint, and the transverse mass
$m_{\rm T} (H)$~\cite{Gross:2009wia}, which, on an event-by-event basis,
represents a lower bound on the mass of the leptonically decaying charged boson
($H^+$ or $W^+$), produced in the top-quark decay.  In the $\tauhad+$$\ell$
channel the missing transverse momentum is expected to be larger for $H^+$
decays. Finally in the $\tauhad+$ jets channel the transverse mass $m_{\rm
T}(\tau, \mathrm{miss})$ is related to the mass of the leptonically decaying
charged boson, depending on the parent particle of the tau lepton (see
\Fref{chargedHiggs}, left).  A similar strategy was also pursued
in~\cite{Chatrchyan:2012vca} using a partial \SI{7}{\TeV} dataset.  The ratio of
$\tau \ell / \ell \ell$ events in \ttbar decays, allowed for cancellations in
the systematic uncertainties and a slight increase in the
sensitivity~\cite{Aad:2012rjx}.  The best sensitivities were achieved when
considering the full \SI{8}{\TeV} dataset~\cite{Aad:2014kga,
Khachatryan:2015qxa}.  No signal was observed in any of these searches, allowing
upper limits to be set on the branching fraction ${\cal B}(t \to H^+ b)$ of
$1.2-0.15\%$ for a mass range of $\SI{80}{\GeV} < m_{H^+} < \SI{160}{\GeV}$. The
results were also interpreted in the context of different MSSM benchmark
scenarios and used to set exclusion limits in the $\tan \beta$ vs $m_{H^+}$
plane. 

\begin{figure}[htbp]
\subfig{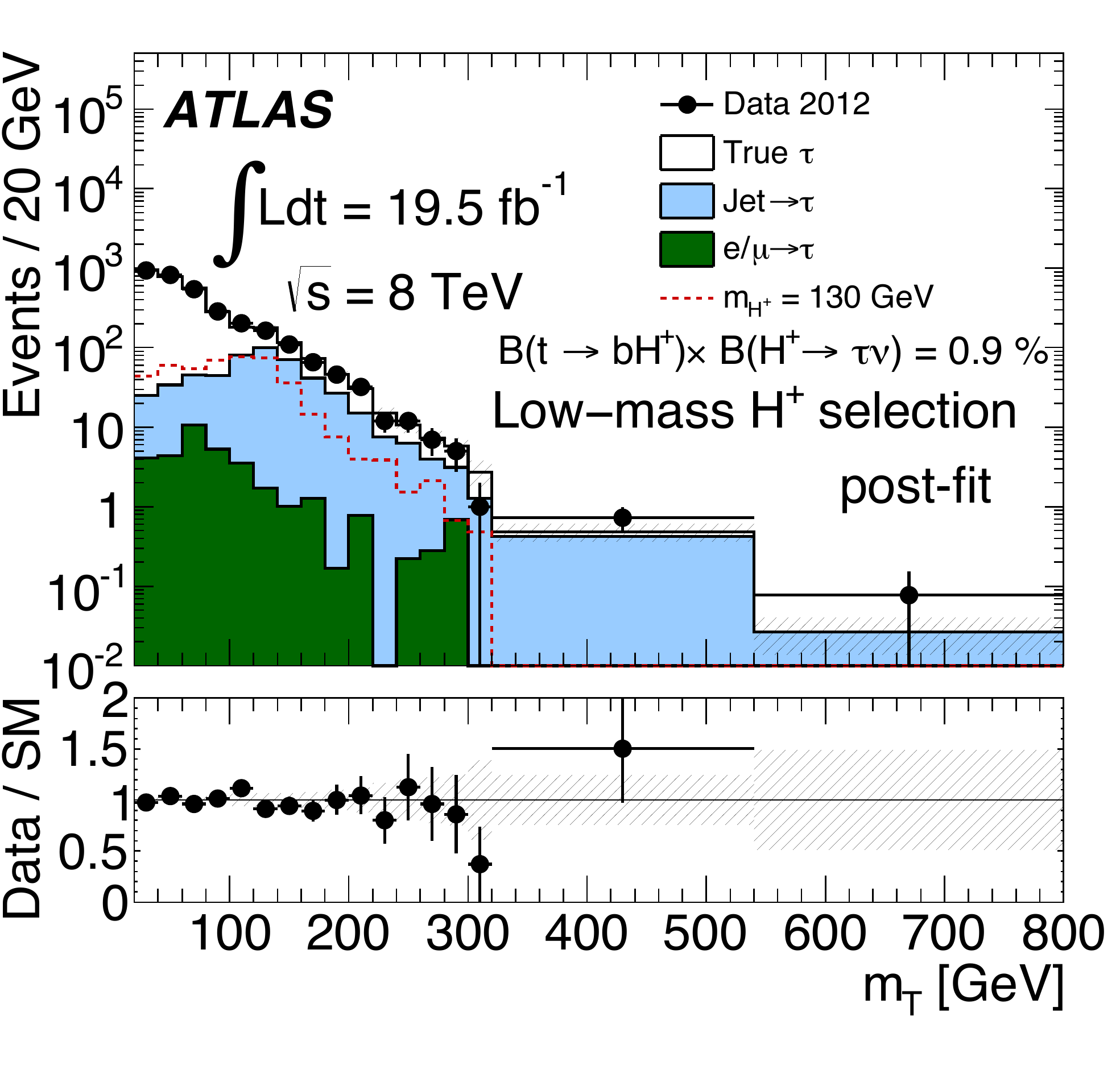}{0.50}{0}
\subfig{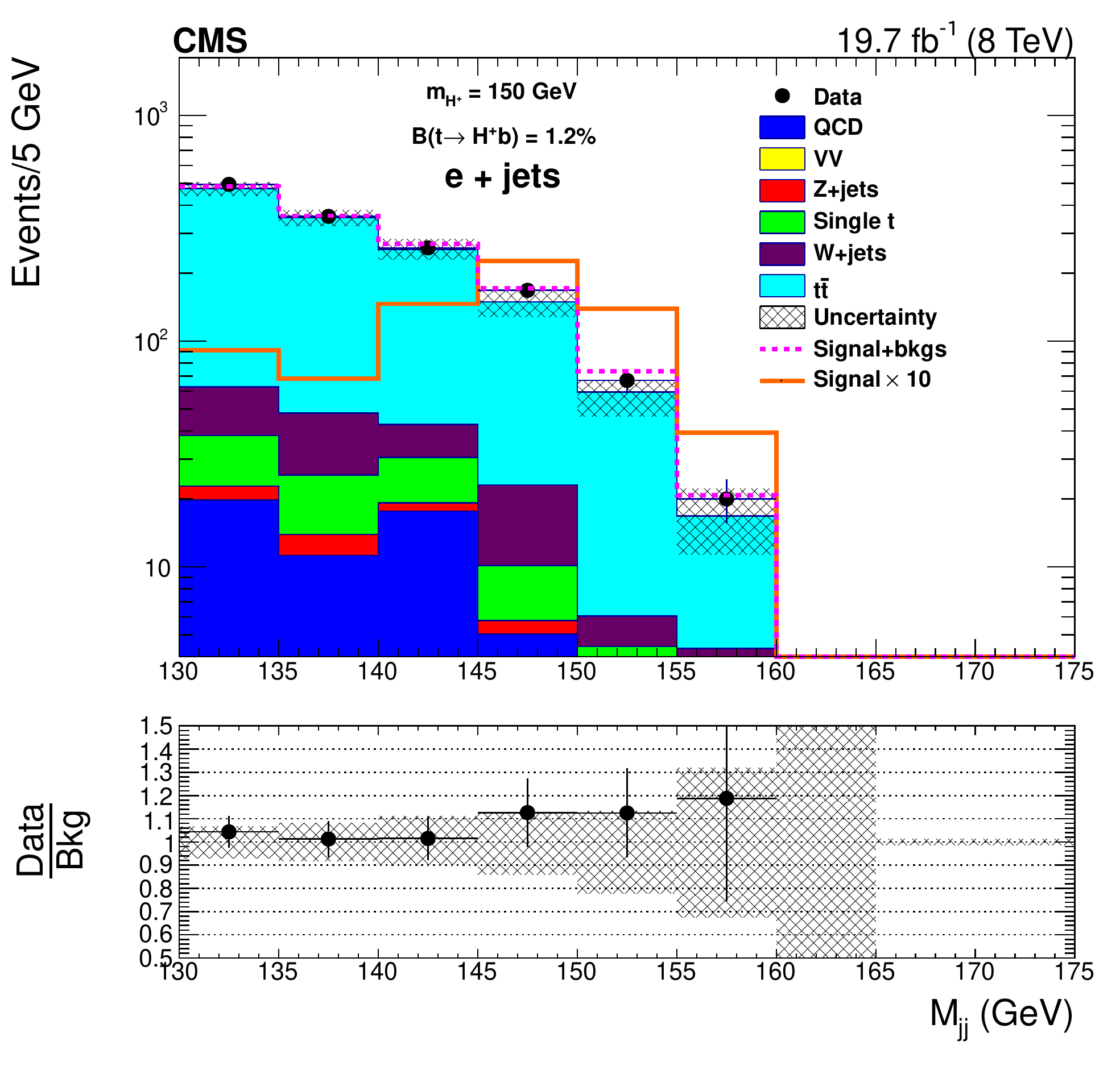}{0.50}{0}
\caption{Distribution of $m_{\rm T}$ after all selection criteria in the
$\tauhad+$jets channel. Data are compared to the SM scenario and with the
addition of an hypothetical charged Higgs boson with a mass of
\SI{130}{\GeV}~\pcite{Aad:2014kga} (left).  Dijet mass distribution of the
hadronically decaying boson for the electron+jets channel using background
templates including the expected yield in the presence of
signal~\pcite{Khachatryan:2015uua} (right).} \label{chargedHiggs}
\end{figure}

For small values of $\tan \beta$ the decay $H^+ \to c\bar{s}$ is expected to
dominate and dedicated searches have been conducted in this channel as well.
The final state is identical to a single-lepton \ttbar decay, with the
difference that the two light jets come from a $H^+$ rather than a $W$ boson
decay. ATLAS performed a search for this process with the \SI{7}{\TeV}
dataset~\cite{Aad:2013hla}. The \ttbar system was fully reconstructed via a
kinematic fit and the dijet invariant mass of the two jets assigned to the
hadronic $W$ boson decay was used to search for deviations from the SM
expectation. Upper limits on the branching fraction ${\cal B}(t \to H^+ b)$  of
$5-1\%$ were obtained for $\SI{90}{\GeV} < m_{H^+} < \SI{150}{\GeV}$.  With the
larger \SI{8}{\TeV} dataset CMS repeated this analysis obtaining a better
sensitivity, but data did not allow the upper limits to be
improved~\cite{Khachatryan:2015uua} (\Fref{chargedHiggs}, right).

\subsection{Search for baryon number violating top-quark decays}

The existence of baryon number violating (BNV) interactions is a necessary
condition for the baryon asymmetry of the Universe. While within the Standard
Model the baryon number is conserved, in many extensions BNV interactions can
occur. Thus far there is no experimental evidence for baryon number violation in
the many processes where it has been searched for. From the proton lifetime, a
stringent indirect limit on the branching fraction $t \to \bar{b} \bar{c}
\ell^+$ of $10^{-27}$ can be placed~\cite{Hou:2005iu}. However it has been
pointed out that cancellations could allow a large
enhancement~\cite{Dong:2011rh}, motivating a direct search for such decays.

If one of the top quarks in \ttbar production undergoes a BNV transition, the
final state of interest consists of one lepton, five quarks and no neutrino.
Using the full \SI{8}{\TeV} dataset CMS did not find evidence of such decays and
placed upper limits on $t \to \ell$+2 jets of 0.15\% at the 95\% confidence
level~\cite{Chatrchyan:2013bba}. The signal was modelled using \MadGraph\ and
the main background was SM \ttbar production.

A summary of experimental limits on exotic top-quark decays is shown
in~\Fref{exot_summary}.

\begin{figure}[htbp]
\centering
\includegraphics[width=\textwidth]{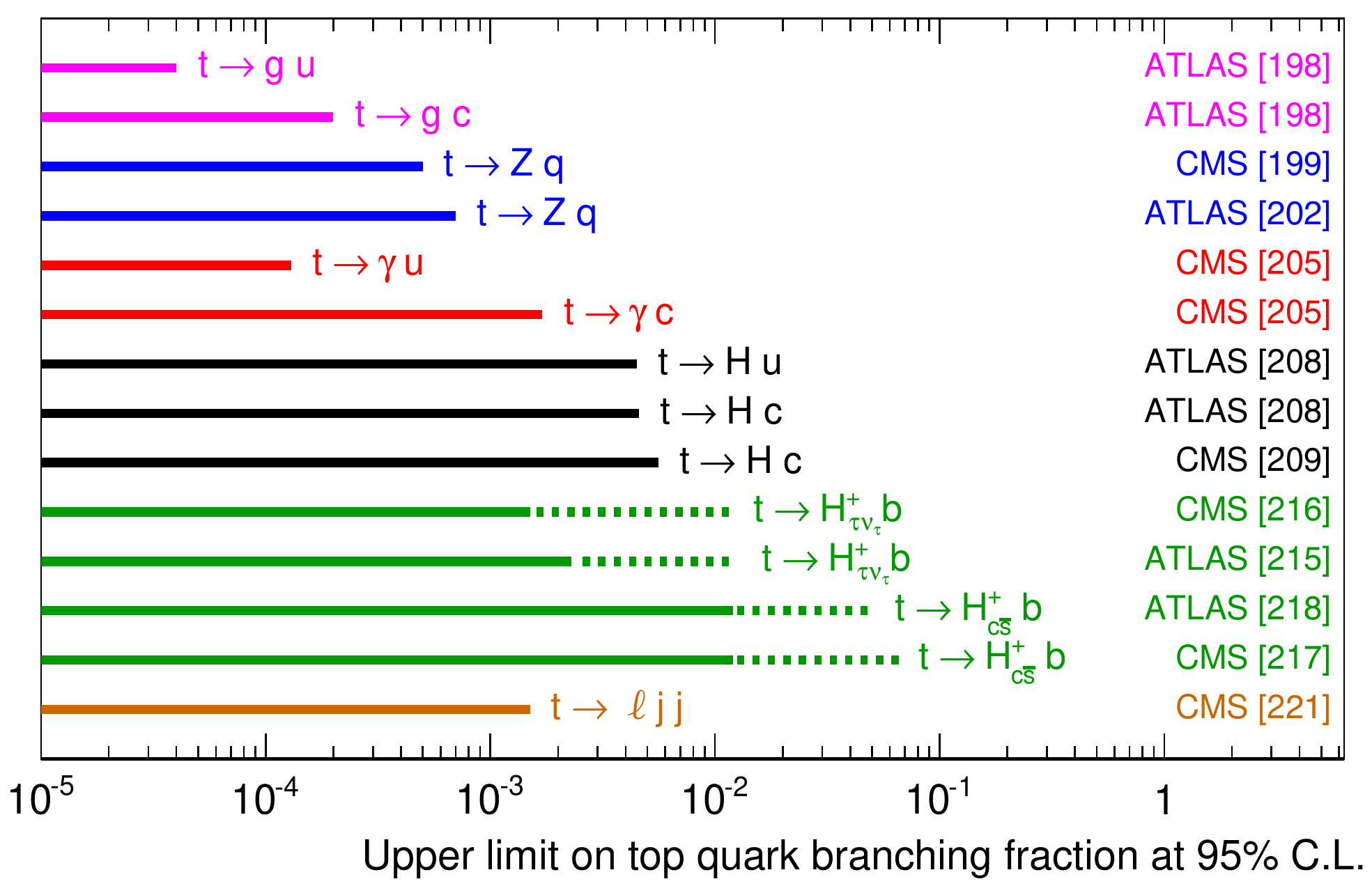}
\caption{\label{exot_summary} Summary of searches for exotic top-quark decays
through flavour changing neutral currents ($g$, $Z$, $\gamma$, $H$), to a
charged Higgs boson decaying to $\tau \nu_{\tau}$ or $c\bar{s}$ for \mbox{$80 <
m_{\tau \nu_{\tau}} < \SI{160}{\GeV}$} or \mbox{$90 < m_{c \bar{s}} <
\SI{150}{\GeV}$}, and violating baryon number conservation.  For each channel
and each LHC experiment the best limit on the corresponding branching fraction
is shown, together with the reference.}
\end{figure}

\section{Conclusions and outlook}

A comprehensive programme of top-quark physics has been established at the LHC
with 7 and \SI{8}{\TeV} proton-proton collision data collected from 2010 to
2012, with most of the analyses finalized in time for this review.

Measurements of \ttbar production provide stringent tests of perturbative QCD
and have reached a good level of maturity with the large data samples available.
The inclusive top-quark pair production is measured in the $e\mu$ channel with
an uncertainty of less than $4\%$, a better precision than the prediction at
NNLO, while measurements in several other channels complete the consistent
picture.  Differential measurements related to the top-quark or \ttbar system,
as well as to the number of additional jets, allow the validity of the
calculations and the simulation programs employed at hadron colliders to be
tested in more depth.  The high-\pT region is also explored, where top-quark
decay products start to merge and boosted techniques improve the reconstruction
efficiency.  In general, data and predictions match well, with some
discrepancies that arise from the choices of the parameters in the Monte-Carlo
programs and the limited accuracy of the calculations at NLO.  The large dataset
of \ttbar events has also been used to calibrate the $b$-tagging
algorithms~\cite{Chatrchyan:2012jua, Aad:2015ydr}, to determine the shapes of
quark-induced jets from top-quark decays~\cite{Aad:2013fba}, and to measure the
pull angle distribution~\cite{Aad:2015lxa}, describing the orientation of the
radiation from jets originating from a $W$ boson in \ttbar events and interpreted
as a measurement of colour flow structure. 

Single-top quark production has been observed at the LHC in the $t$- and $Wt$
channels, and there is first evidence also of $s$-channel production. The CKM
matrix element \Vtb has been determined with an uncertainty of $4\%$ and is
compatible with unity. The large $t$-channel dataset starts to be useful to also
extract other quantities, related to the simulation of the production process or
properties of the top quark. 

Many measurements of the properties of the top quark have considerably reduced
uncertainties compared to the pre-LHC era. The \topmass\ has been measured in
several channels and also with innovative techniques.  Overall, a total
uncertainty of less than \SI{500}{\MeV} has been achieved for conventional
methods, and an uncertainty of \SI{2}{\GeV} for direct measurements of the pole
mass.  The polarization of top quarks in \ttbar and single-top events has been
measured, and the \ttbar spin-correlations established with $\sim\!\!10\%$
uncertainty.  Motivated also by initial discrepancies found at the Tevatron, an
extensive set of measurements has been carried out to determine the \ttbar
charge asymmetry.  The results are compatible with both the Standard Model
prediction and zero asymmetry.

Associated production of top-quark pairs with additional bosons or quarks have
smaller production \xs s and start to become accessible with the \RunOne\
dataset.  Production of \ttga, \ttW and \ttZ have all been observed, while upper
limits have been placed on \ttH and $\ttbar\ttbar$.  Understanding additional
heavy-flavour production is important, for instance as a background to \ttH, and
is therefore a matter of detailed study. Many of these measurements represent
the first step towards the determination of the top-quark couplings, however a
larger dataset will be needed to start being sensitive to modifications that
might enter due to BSM physics.  

Searches for exotic top-quark decays are limited by the statistical power of the
datasets collected. Flavour-changing neutral currents, involving transitions
with a $Z$, $g$, $\gamma$ or $H$ boson have been searched for, as well as decays
to charged Higgs, or decays forbidden by baryon number conservation. No evidence
of any of these processes has been found and thus improved limits could be
established. 

LHC Run-2 has started in 2015 with a $pp$ collision energy of \SI{13}{\TeV},
targeting an integrated luminosity of about \ifb{100} by the end of 2018,
corresponding to one order of magnitude more top-quark events than those
collected so far.  First results with this new dataset are
emerging~\cite{Khachatryan:2015uqb} and eventually the larger dataset will give
access to yet another new realm of precision. With the new data, and further
progress in experimental and theoretical methods, measurements will reach
unprecedented precision, and top quarks will become even more powerful probes
for new physics searches.

Ultimately, after the high-luminosity upgrade of the LHC, the experiments are
expected to collect \SI{3}{\per\atto\barn} of $pp$ collisions at a
centre-of-mass energy of \SI{14}{\TeV}. The sensitivity of several top-quark
measurements has been estimated under these conditions~\cite{Agashe:2013hma,
CMS:2013zfa}.  The detectors will need to be adapted accordingly in order to
cope with the increased occupancy, levels of radiation and increased amount of
collisions per bunch crossing.  Millions of top quarks will be produced with
very large \pT and therefore techniques to distinguish such top-jets from
lighter quark-induced jets will become more and more important.  Techniques such
as those described in Ref.~\cite{Aad:2016pux}, like the HEPTopTagger, shower
deconstruction, or taggers using substructure variables, will be essential in
order to recover efficiencies at high top-quark transverse momentum and enhance
the sensitivity for new physics. 

More than twenty years after its discovery, top-quark physics has only recently
entered the precision era. With millions of top-quark events the production \xs
s, the intrinsic properties and decay mechanisms have been studied in detail. If
there is a connection between physics beyond the Standard Model and the top
quark, this has so far remained hidden.  With higher collision energies,
improved detectors, refined reconstruction techniques and much larger datasets
many areas of top physics will reach unprecedented levels of precision, allowing
to perhaps uncover new physics at play.

\section*{Acknowledgements} 
The authors gratefully acknowledge Maria Aldaya, Roberto Chierici, Maria Jos\'e
Costa, Pedro Ferreira da Silva, Andrea Giammanco, Richard Hawkings, Alison
Lister, Mark Owen, and Efe Yazgan for their comments and suggestions to the
manuscript.

The work of M.C. is funded by the European Research Council under the European
Union's Seventh Framework Programme ERC Grant Agreement n.  617185.  

\section*{References}


\providecommand{\newblock}{}

\end{document}